\documentclass[12pt]{article}
\usepackage[english]{babel}

\usepackage{amsmath}
\usepackage{amssymb}
\usepackage{amsfonts}
\usepackage{amsthm}
\usepackage{graphics,graphicx}
\usepackage{color}
\usepackage{epstopdf}
\usepackage{cite}
\usepackage{hyperref}
\usepackage{tikz}

\date{\today}

\title{Averaging-based approach to toughness homogenisation for radial hydraulic fracture}
\author{G. Da Fies$^{(1)}$, M. Dutko$^{(1,2)}$, D. Peck$^{(2*)}$ 
\\
{\it $^{(1)}$Rockfield Ltd, Swansea, UK}
\\
{\it $^{(2)}$Department of Mathematics, Aberystwyth University, }\\
{\it Aberystwyth, Wales, United Kingdom}
\\
{\it (*) Corresponding author: dtp@aber.ac.uk}
}

\begin{document}

\maketitle

\begin{abstract}

The homogenisation of the fracture toughness is considered in the context of a propagating hydraulic fracture. The radial (penny-shape) model is utilized, in order to incorporate the impact of the viscosity-toughness regime transition over time. A homogenisation strategy based upon temporal-averaging is investigated. This approach incorporates the instantaneous fracture velocity, meaning that it should remain effective in the case of step-wise crack advancement. The effectiveness of the approach is demonstrated for periodic toughness distributions, including those which are unbalanced, utilizing a highly accurate solver. 

\end{abstract}

\section{Introduction}

Hydraulic fracturing (HF) driven by pressurised fluid is a phenomenon observed in many natural and industrial processes. For example, natural processes of fluid fracturing due to the magma and volcanic gas flow are well known (see e.g. \cite{Unwin2021,Aiuppa2021}). The hydraulic fracturing phenomenon also occurs in medicine during fluid injection in human tissue or fluid-driven rupture of articular cartilage associated with osteoarthritis or hypertension. The latter receives significant attention from both medical practitioners and modellers, due to the significant health and social issues associated with aortic dissection \cite{Brunet2021,Criado2011,Holzapfel2019}.

Prediction of fluid driven fracture evolution is also an important step in the design and safety assessment of many technological applications and industrial processes.  One highly relevant technological application related to environmental issues is CO2 sequestration (capturing and storing of atmospheric CO2 in underground reservoirs). Here, assessment of possible fracturing of the storage reservoir due to the CO2 injection must be performed during the design stages to ensure it maintain its integrity.

A similar technological process where fluid driven hydraulic fracturing is often used is in the stimulated extraction of hydrocarbons from low porosity reservoir rock which would otherwise be economically unviable (for overviews, see e.g. \cite{Economides2000,King2010,Bai2011}). There are many design variables defining successful hydraulic fracturing (HF), such as achieved production, fracture geometry and the effective stimulated reservoir volume. The tight/low permeability unconventional reservoirs encountered during hydraulic fracturing are characterised by highly complex structures, due to the depositional history and subsequent deformation related to the tectonic movement, glaciation, and other factors. 

This heterogeneity can be observed in all of the processes described above. For example, the reservoir properties can be measured by core-logs and production data at almost every scale exhibition strong heterogeneity; the aortic wall segment comprises several layers of intima/media and adventitia with various material properties. When predicting process of HF evolution using computational methods it is clear, however, that it is neither time or cost-effective for computational models of hydraulic fracturing to incorporate this level of detail, and some upscaling techniques are almost always used. 

This problem of homogenising material parameters is far from unique to hydraulic fracture. Effective homogenisation strategies for the various properties of composites (for instance elastic, thermal, electrical etc.) are well developed and often have rigorous physical justifications (see e.g. \cite{Charalambakis2010} and references therein). A number of formulas and procedures to evaluate the average composite properties have been proposed using a variety of approaches. Importantly, while most provide slightly different results, these all lie within the Hashin-Strickman bounds \cite{HASHIN1963127}. 
In practice, many approaches are developed to ensure the results of both the original system and homogenised approximation are identical for a specific type of loading. Away from this type of loading, it is hoped (or demonstrated) that they will provide an acceptable approximation.

Unfortunately it is well known, as documented by Kachanov and his colleagues (see e.g. \cite{Kachanov1994,Kachanov2010}), that homogenisation of the fracture toughness does not make physical sense, at least in the case of microfractures. On the other hand, this issue is increasingly recognised as crucial for successful HF practice, as heterogeneous (mostly laminated) underground structures with varied toughness are always present. Any homogenisation strategies for dealing with this also have the constraint that solvers incorporating them must be able to provide data in real-time to be effectively utilized by practitioners. It is still not clear how to simultaneously homogenise the elastic properties while also homogenising the fracture toughness (see e.g. \cite{HOSSAIN201415}).

This raises the question: is it possible to build a consistent homogenisation theory for the fracture toughness that is able to predict the resulting crack growth in real-time and with an acceptable level of accuracy, at least for a very restricted class of practical problems and processes? If one knows that such a homogenisation is not possible then, for a specific type of composites under predefined loading conditions, can some approximate averaging be developed that can provide  practitioners with acceptable results (for some specific settings)? Moreover, if this is achieved for a small number of optimal (in some sense) solutions, is it possible to verify those few solutions in more detail by computing the problems and process solutions considering the materials microstructure?

While the answers to these questions are extremely important to the practice of hydraulic fracture, as has been documented by many researchers (see e.g.\linebreak \cite{DONTSOV2021108144}, in general up to this point a naïve averaging has been implemented to approximate the toughness (moving averages, averages on one computational cell, etc) without any guarantee on the accuracy of the results obtained. Recently, concentrated attempts have been made address a notion the averaged toughness in the case of the fluid driven fracture \cite{Gaspare2022,DONTSOV2021108144,DONTSOV2022110841}) providing some numerical comparisons. A phenomenological and well-balanced approach used in \cite{DONTSOV2021108144} allowed a simple formula for the averaged toughness to be proposed for the case of a step-wise toughness distribution. This has since been extended to the development of an algorithm for computing the fracture evolution within a domain including inhomogeneity of the toughness, in-situ stress and fluid leak-off for a plane strain crack \cite{DONTSOV2022110841} . Meanwhile, in \cite{Gaspare2022} this approach was verified showing when the results become acceptable from the practical point of view, not only for periodic stepwise but also for sinusoidal toughness distribution, by using a very accurate plane strain (KGD) solver \cite{Gaspare2020}. When providing such analysis/comparison, the most difficult aspect is simulating the process in the required level of details concerning the material microstructure. In the case of HF, the total fracture length can be hundreds of meters, while some of the process data available is the scale of several centimetres.  

\begin{figure}[b!]
 \begin{minipage}{0.35\textwidth}
  \centering
  \includegraphics[width=1\textwidth]{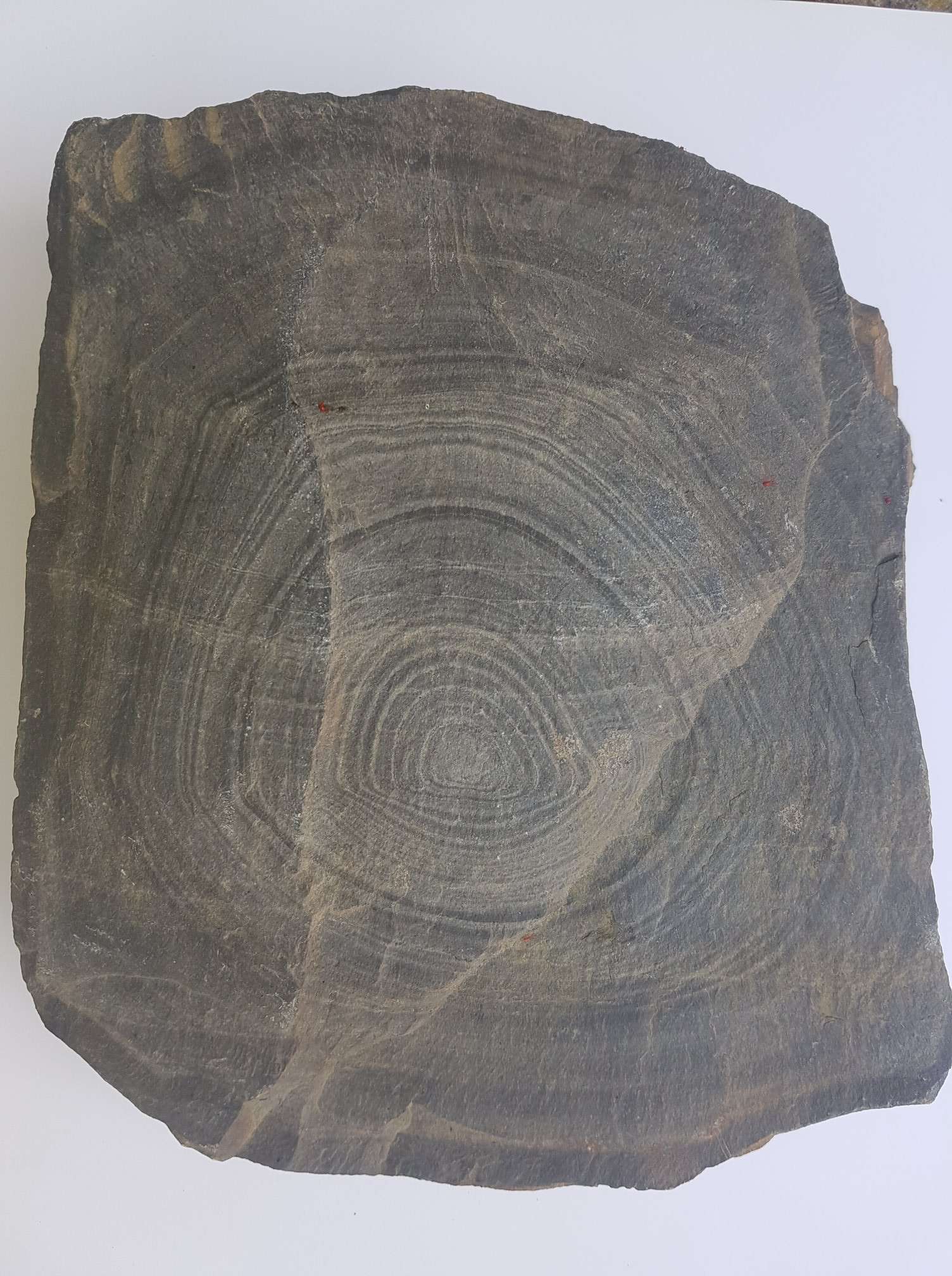}
  \put(-190,230) {{\bf (a)}}
  \end{minipage}
  \hspace{12mm}
  \begin{minipage}{0.6\textwidth}
   \centering
   \includegraphics[width=1\textwidth]{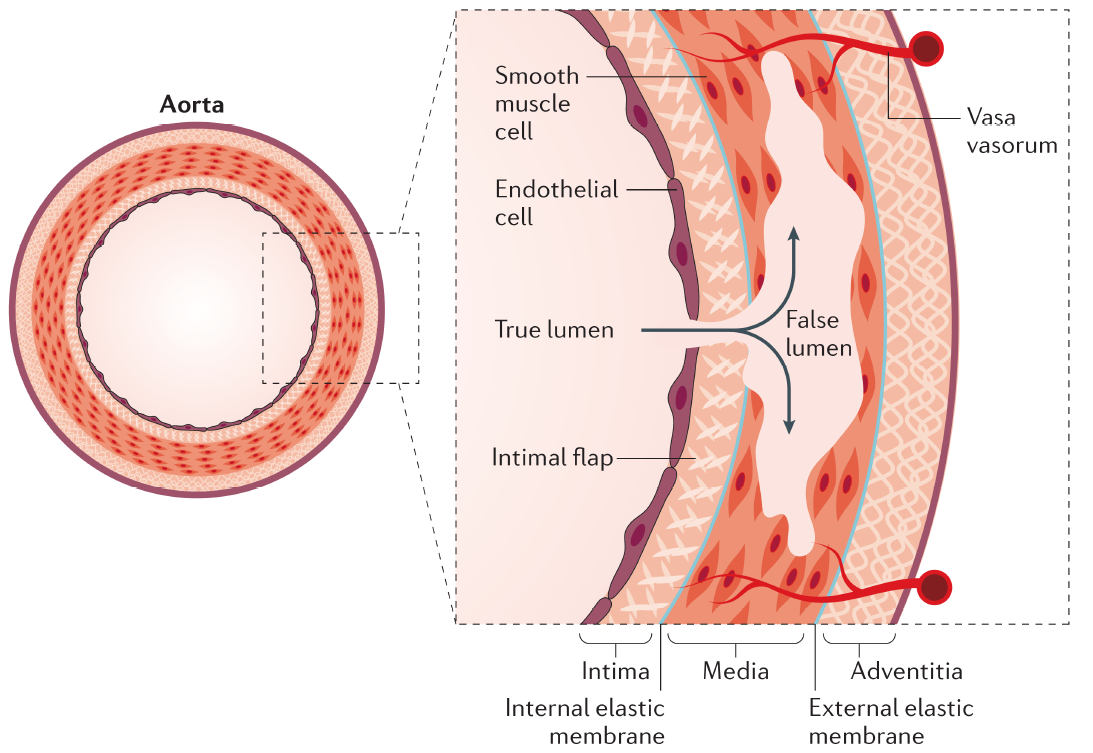}
    \put(-300,230) {{\bf (b)}}
  \end{minipage}
  \caption{Examples of radial heterogeneity in nature. Radial heterogeneity in (a) rock, (b) aortic dissection within a (heterogeneous) aortic wall (reproduced from \cite{Nienaber2016}). Such radial heterogeneity can also be observed in wood (tree rings).}  
  \label{Intro_radial}
 \end{figure}

The aim of this paper is to extend the approach proven be efficient for the plane strain case with periodic toughness to the radial inhomogeneity. Even though such toughness heterogeneity can be considered as an exceptionally rare case (although not impossible, such as fractures in wood (tree rings), stone, or arterial walls, see Fig.~\ref{Intro_radial}), the aim in this instance is to prove the concept rather than to claim a specific problem is solved. Moreover, as the approach proposed in this paper and in the plane strain case is based upon the concept of (temporal) averaging, rather than the maximum toughness assumption, it can be used for a wider class of composites and heterogeneity distributions (not only periodic, but also the random toughness distributions). Crucially, the measure used to define the homogenised toughness also incorporates the instantaneous crack velocity $v(L)$, meaning that it should remain effective even during the step-wise crack extension observed in HF (see e.g. \cite{CAO201724,Schrefler2019a,Schrefler2019b} and references therein), or when there is also inhomogeneity of the in-situ stress, fluid leak-off or other parameters.

This work is organised as follows. First, the problem of radial hydraulic fracture with inhomogeneous toughness is formulated, with the model, description of the toughness distribution, method of parameterising the fracture regime, and numerical solver all outlined. In the next section, the differing homogenisation strategies for the toughness are proposed, and a thorough investigation conducted for the cases where the toughness distribution is balanced, unbalanced, or has only a minimally thick layer of high-toughness. A summary of results is then provided at the end.


\section{Problem formulation}\label{Sect:2}

\subsection{The radial model with inhomogeneous toughness}

\begin{figure}[t]
  \centering
  \begin{tikzpicture}[scale=1.4]
    \draw[black] (-4,2.5) .. controls (1,1.35) and (2,0.9) .. (2.85,0); 
  \draw [black,thick,dotted,->] (-4,0) -- (-4,3); 
      \node at (-4,3.25) {$z$}; 
      \node at (-4.35,1.3) {$q_0$}; 
  \draw [black,->] (-4,0) -- (3.75,0); 
      \node at (4,0) {$r$}; 
  \draw[black,->] (-1,1.1) -- (-1,1.6); 
    \draw[black,->] (-0.8,1.05) -- (-0.8,1.55); 
      \draw[black,->] (-0.6,1) -- (-0.6,1.5); 
       \node at (-0.8,0.8) {$p$};  
 \draw[black,<->] (-2.2,0.05) -- (-2.2,2); 
  \node at (-2,1) {$w$}; 
  \draw[black,<->] (-3.9,-0.1) -- (2.85,-0.1); 
  \node at (-0.5,-0.4) {$L$}; 
  \draw[blue,thick,->] (-3.99,2.1) -- (-3.45,2.1); 
    \draw[blue,thick,->] (-3.99,1.6) -- (-3.45,1.6); 
      \draw[blue,thick,->] (-3.99,1.1) -- (-3.45,1.1); 
        \draw[blue,thick,->] (-3.99,0.6) -- (-3.45,0.6); 
         \draw[blue,thick,->] (-3.99,0.1) -- (-3.45,0.1); 
         \node at (-3.5,1.3) {{\tiny Fluid flow}}; 
 \draw[black,thick,dotted] (-1.5,3) -- (-1.5,1.9); 
  \node at (-2.75,2.75) {{\tiny Material 1}};
 \draw[black,thick,dotted] (1,3) -- (1,1.15); 
  \node at (-0.25,2.75) {{\tiny Material 2}};
 \draw[black,thick,dotted] (3.5,3) -- (3.5,0); 
 \node at (2.25,2.75) {{\tiny Material 1}};
  \end{tikzpicture}
  \caption{Schematic illustrating one quadrant of a cross-section of the radial (penny-shaped) model for arbitrary angle $\theta$, passing through layered materials with periodic (step-wise) toughness.}
   \label{PennyRad-Diag02}
\end{figure}
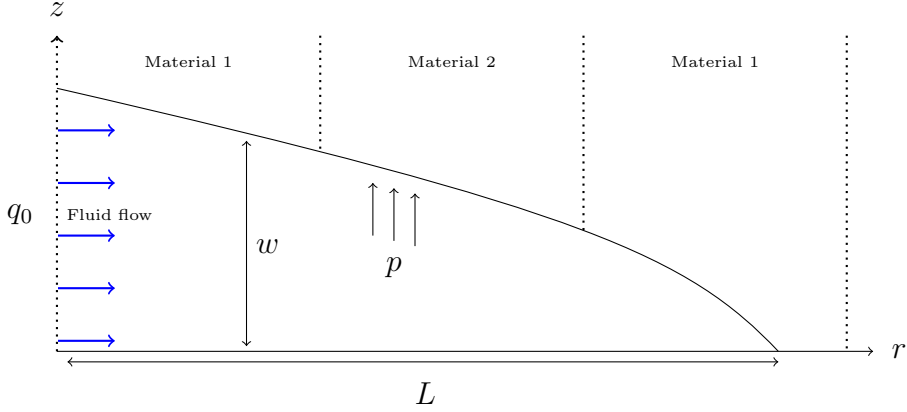

We consider the radial model of hydraulic fracture, whereby an axisymmetric (penny-shaped) crack filled with Newtonian fluid is expanding within an (elastic) solid domain. The fracture geometry is described by it's length, $L(t)$, and height, $w(r,t)$. The Newtonian fluid is pumped in at the centre, $r=0$, at a known rate $q_0 (t)$. As we are primarily concerned with the initial, storage dominated, regime, we can assume that the fluid leak-off is negligible without a loss of generality (i.e.\ the rock is assumed to be impermeable, for an overview of different fracture propagation regimes, see e.g.\ \cite{Peirce2008}). The effect of the shear stress and turbulent fluid flow on the fracture geometry can also neglected (see e.g. \cite{Lecampion2019,Peck2022a,Peck2022ab,Wrobel2017}). A diagram showing one quadrant of the domain cross-section is given in Fig.~\ref{PennyRad-Diag02}. The resulting formulation is almost identical to that of the classical radial formulation (see e.g. \cite{Peck2018a}), with some modification to allow for the variable, axisymmetric, material toughness $K_{Ic}(r)$ (the form of $K_{Ic}$ is outlined in Sect.~\ref{Sect:MaterTough}).

The continuity (fluid mass) equation is given by:
\begin{equation}\label{continuity1}
\frac{\partial w}{\partial t} + \frac{1}{r}\frac{\partial (r q)}{\partial r} = 0, \quad t>0 , \quad 0<r<l(t),
\end{equation}
where $q(r,t)$ is the fluid flow rate. Integrating over the space-time domain yields the global fluid balance equation:
\begin{equation}
2\pi \int_0^{L(t)} r \left( w(t,r) - w(0,r) \right) \, dr =  \int_0^t q_0 (\tau ) \, d\tau .
\end{equation}
The (Newtonain) fluid flow is assumed to be laminar, and as such $q$ is given by the Poiseuille equation:
\begin{equation} \label{fluidflow1}
q = -\frac{1}{M} w^3 \frac{\partial p}{\partial r} , \quad t>0 , \quad 0<r<l(t),
\end{equation}
where $M=12\mu$ and $\mu$ is the fluid viscosity. \\

In practical application, it is preferable not to use \eqref{fluidflow1} as it degenerates at the crack tip (where $w\to 0$, and $dp/dr \to -\infty$). Instead, we utilize the approach based on the speed equation, previously introduced and used in e.g.\ \cite{Linkov_3,Wrobel2015,Peck2018a,Peck2018b}. We begin by defining the fluid velocity:
\begin{equation}\label{eq:3}
v = \frac{q}{w} = -\frac{1}{M}  w^2 \frac{\partial p}{\partial r} , \quad t>0 , \quad 0<r<l(t).
\end{equation}
In the absence of fluid lag, and assuming the fluid leak-off is bounded, the fluid velocity matches that of the crack tip, yielding the speed equation:
\begin{equation}\label{eq:4}
\frac{dL}{dt} = v\left(t,L(t) \right) , \quad t>0.
\end{equation}
Note that, while \eqref{eq:3} still degenerates at the crack tip, we can utilize \eqref{eq:4} to evaluate the fluid velocity at this point. As such, with proper application of the tip asymptotics, all issues related to the degeneration at the crack tip are eliminated.

While the above equations did not require any modification to incorporate the variable fracture toughness, the remaining equations require some small modification. 

The crack extension is considered in terms of Linear Elastic Fracture Mechanics, taking the form of the Irwin criterion:
\begin{equation}
\label{toughness_cond}
K_I (t) = K_{Ic} \left( L(t) \right),
\end{equation}
where $K_I$ is the mode-I stress intensity factor, given by:
\begin{equation}
K_I (t) = 2\sqrt{\frac{L(t)}{\pi}} \int_0^{L(t)} \frac{p(t,s)}{\sqrt{L^2 (t) - s^2}} \frac{s}{L(t)} \, ds.
\end{equation}
It should be noted that criterion \eqref{toughness_cond} means that the fracture does not stop expanding at any point, although its rate of growth can decrease close to zero. This can largely be assumed, as we do not decrease the pumping rate during the process. It may however have a small effect on the result during the first instance of the crack tip encountering a significantly tougher layer.

Finally, the solid and fluid phases are related by the elasticity equation, which is taken in the form presented in \cite{Peck2018a}:
\begin{equation}
\label{elasticity}
w(t,r)=\underbrace{\frac{8L(t)}{\pi E'}\int_0^{L(t)} K\left(\frac{s}{L(t)},\frac{r}{L(t)}\right)\frac{\partial p}{\partial s}(t,s)\,ds}_{w_1 (t,r)}+\underbrace{\frac{4}{E'}K_{IC}(L(t))\sqrt{\frac{L^2(t)-r^2}{\pi L(t)}}}_{w_2 (t,r)},
\end{equation}
where:
\begin{equation} \label{Kernel1}
K(\eta , \xi)=
\begin{cases}
\eta \left[E\left(\arcsin(\eta)|\frac{\xi^2}{\eta^2}\right)-E\left(\arcsin(\frac{\eta}{\xi})|\frac{\xi^2}{\eta^2}\right)\right], \quad \eta<\xi, \\
\eta \left[E\left(\arcsin(\eta)|\frac{\xi^2}{\eta^2}\right)-E\left(\frac{\xi^2}{\eta^2}\right)\right], \quad \quad \quad \quad \quad \eta>\xi,
\end{cases}
\end{equation}
and $E$ is the incomplete elliptic integral of the second kind. 

Note that in \eqref{elasticity}, the term $w_1 (t,r)$ describes the effect of the (viscous) fluid pressure on the fracture walls, while  $w_2 (t,r)$ describes the impact of the material toughness. As such, the ratio of the size of these two terms can be used as a rough measure for whether the fracture is in the toughness, transient or viscosity dominated regime at a specific point in time. This approach to parameterising the fracture regime will be used in the investigation to follow, similar to that utilized in \cite{Gaspare2022}, with the details provided in Sect.~\ref{Sect:delta}.

\subsection{Form of the material toughness}\label{Sect:MaterTough}

Equations \eqref{continuity1}-\eqref{Kernel1} now provide a complete description of the relations between the various aspects of the radial hydraulic fracture in the case with (pre-defined) heterogeneous material toughness. The aim now is to investigate the effect of inhomogeneous toughness on the fracture behaviour, and the relative effectiveness of various proposed homogenisation strategies.

For the sake of simplicity the toughness is assumed to be axisymmetric in nature, such that the value of $K_{Ic}(r)$ is independent of the angle $\theta$. The domain is assumed to consist of layered materials, such that the toughness is periodic in space, with period $X$. The distribution of the toughness over this period is primarily defined in terms of its' maximum and minimum values $0<K_{min}<K_{max}<\infty$ where:
\begin{equation} \label{KIC_const}
K_{min} = \min_{0\leq r < X} K_{Ic} (r) , \quad K_{max} = \max_{0\leq r < X} K_{Ic} (r).
\end{equation}

To investigate the effect of differing distributions, two different distributions are considered: step-wise and sinusoidal. The former is the most physically realistic, representing a transition between two distinct rock layers. The latter, where the toughness distribution is a sinusoidal oscillation between $K_{min}$ and $K_{max}$, provides a useful counterpoint that both allows the effect of a smooth transition between the maximum and minimum regions to be examined, while also representing a limiting case of a highly layered material of varying toughness. Comparison of these two opposing cases will therefore allow for a full investigation of the result of different periodic toughness distributions. 

In addition to taking step-wise and sinusoidal toughness distributions, the concept of unbalanced layering is also considered. In this case the distribution remains periodic, however it is skewed such that the average toughness (over the spacial period) is decreased. This aspect of the distribution is defined by a parameter, $0<h<1$, given over the period $X$ as:
\begin{equation}\label{defh}
h = \frac{1}{K_{max}-K_{min}} \left[ \int_0^X K_{Ic} (r) \, dr - K_{min} \right] .
\end{equation}
The case with $h=0.5$ represents the case of balanced layering, which $h<0.5$ represents the case where the maximum toughness layer makes up a smaller proportion of the material than the minimum toughness layer. A similar parameter was used by the authors in preliminary investigations of the KGD model \cite{ARMA2021}. Note that in the step-wise case, this parameter represents the relative width of the tougher layer compared to the period. For the `sinusoidal' distribution, the toughness over the period $X$ is taken in the form:
$$
K_{Ic} (x) = A + B \exp\left( C\sin\left(\frac{\pi x}{X} \right) \right),
$$ 
where constants $A$, $B$, $C$ are taken to ensure that the the distribution has the correct values of $K_{max}$, $K_{min}$ and $h$, via conditions \eqref{KIC_const}, \eqref{defh}. This distribution was chosen in order to ensure that the toughness only takes the values $K_{max}$ and $K_{min}$ at a single point within each period.

\begin{figure}[t!]
\centering
\includegraphics[width=0.45\textwidth]{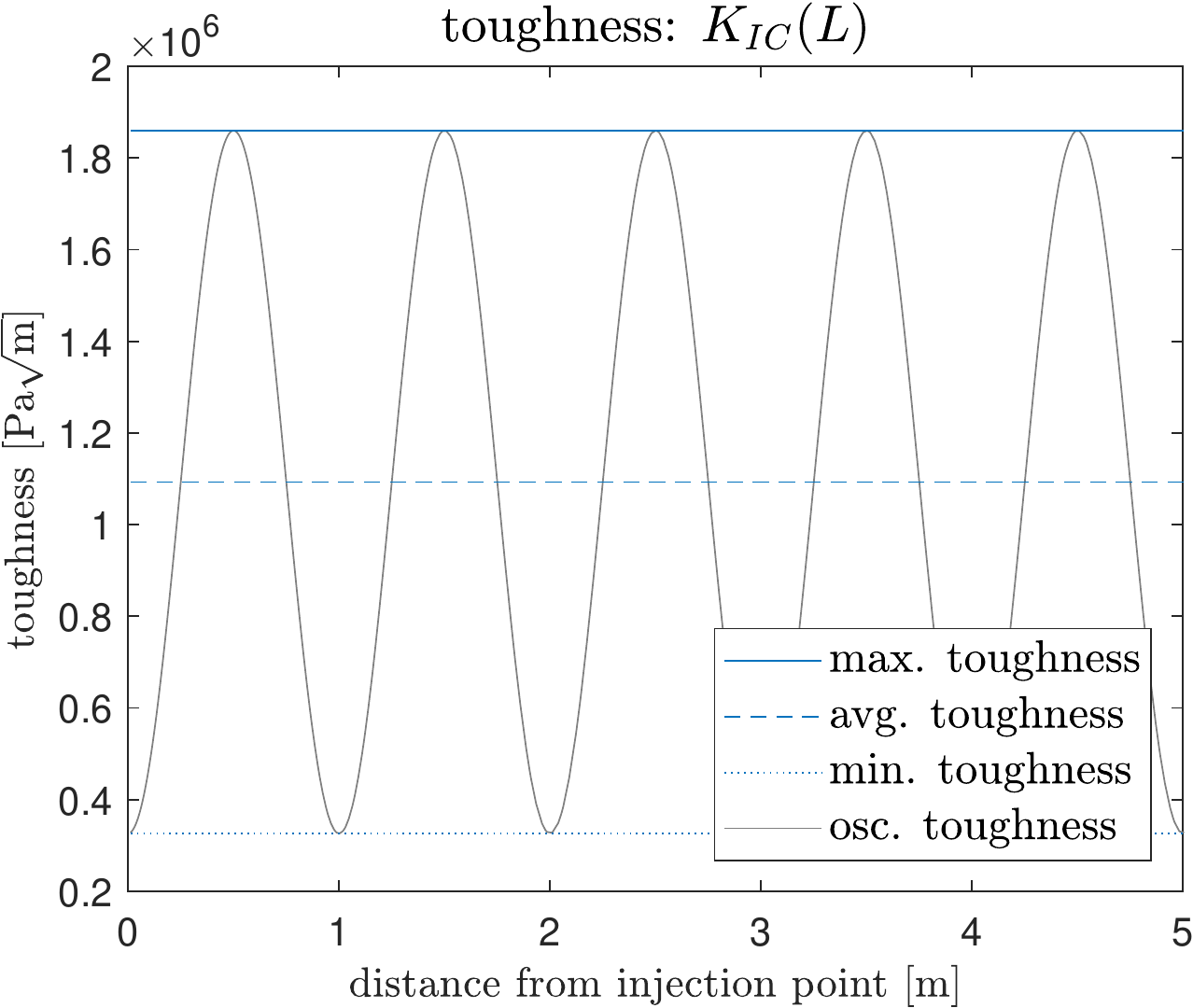}
\put(-225,155) {{\bf (a)}}
\hspace{12mm}
\includegraphics[width=0.45\textwidth]{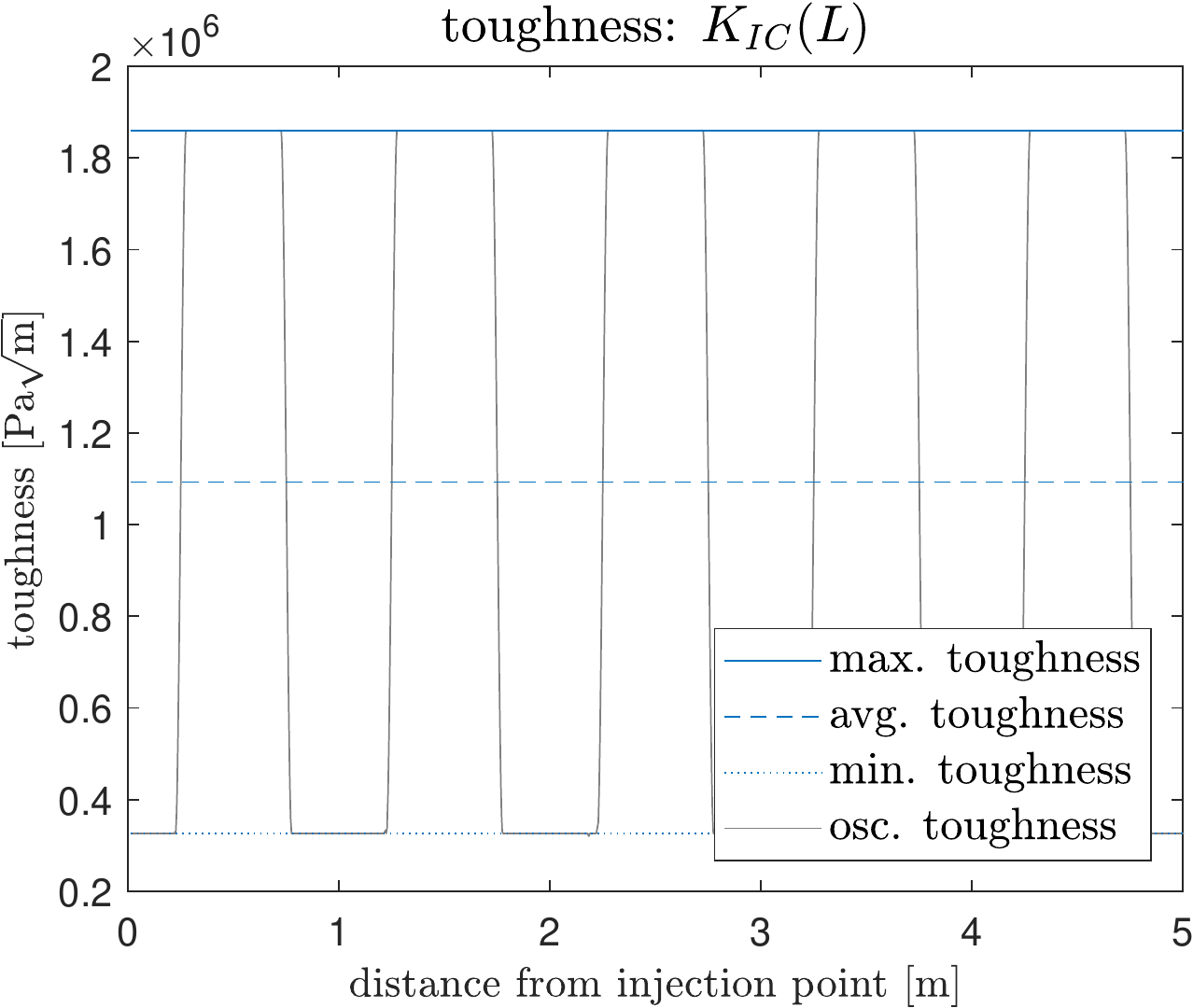}
\put(-225,155) {{\bf (b)}}
\\
\includegraphics[width=0.45\textwidth]{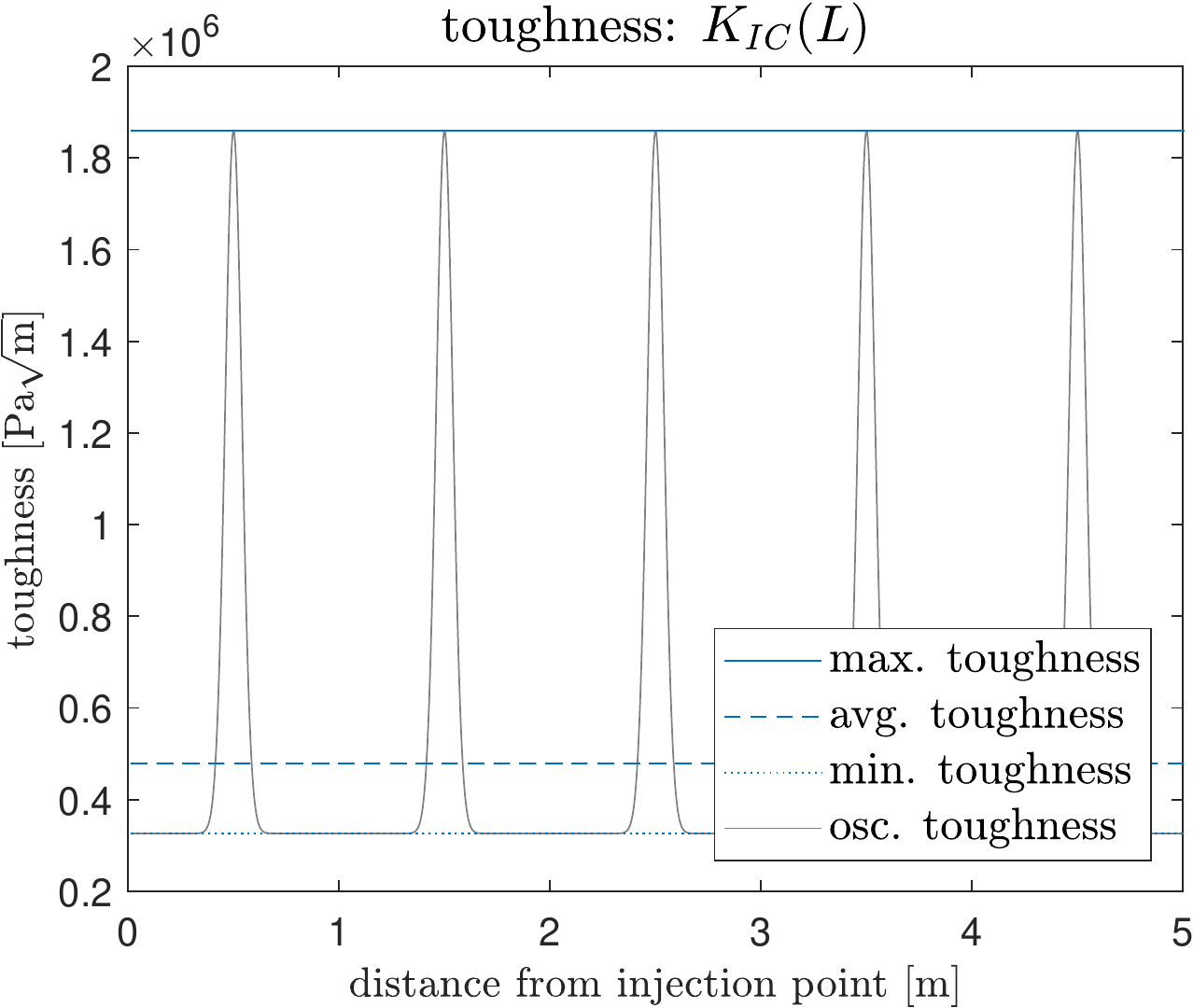}
\put(-225,155) {{\bf (c)}}
\hspace{12mm}
\includegraphics[width=0.45\textwidth]{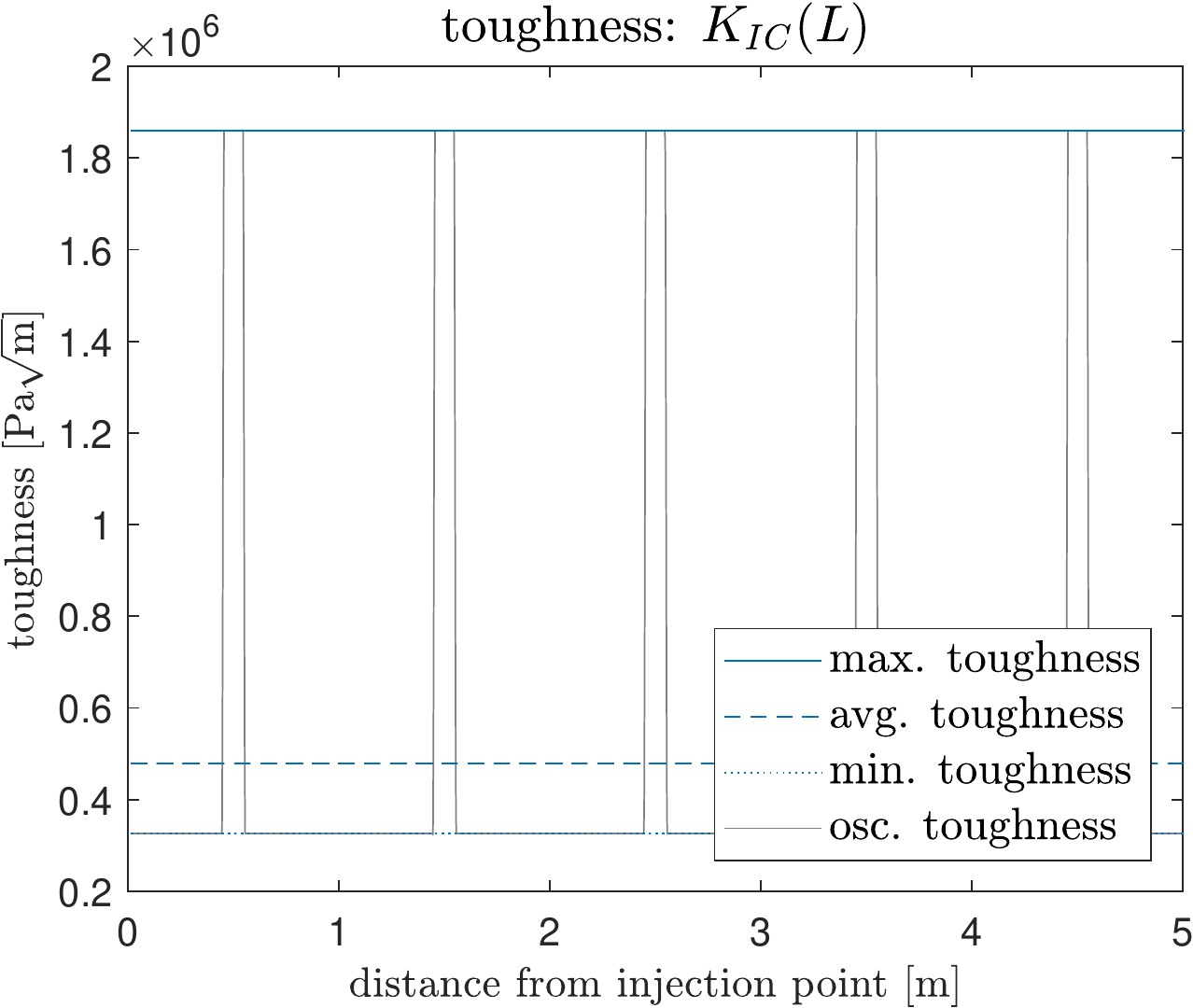}
\put(-225,155) {{\bf (d)}}
\caption{The material toughness distribution over space in the case of {\bf (a)}, {\bf (b)} balanced layering ($h=0.5$, see \eqref{defh}), and {\bf (c)}, {\bf (d)} unbalanced layering with $h=0.1$. The distributions are: {\bf (a)}, {\bf (c)} sinusoidal; {\bf (b)}, {\bf (d)} step-wise.}
 \label{Fig:ToughDist}
\end{figure}

All together, for each pair $K_{max}$ and $K_{min}$, four cases are considered: %
(i) balanced layering, step-wise distribution; %
(ii) balanced layering, sinusoidal distribution; %
(iii) unbalanced layering, step-wise distribution; %
(iv) unbalanced layering, sinusoidal distribution. 
Examples of these differing toughness distributions are provided in Fig.~\ref{Fig:ToughDist}. For simplicity, in all figures in this paper the sinusoidal distribution is displayed on the left-hand side, while the step-wise distribution is shown on the right-hand side.

Throughout all simulations used in this paper, the period of the toughness distribution can be assumed to be $X=1$ without loss of generality, as the solution over any other period can be recreated from this with the proper space-time rescaling. Meanwhile, the particular values of the maximum and minimum toughness are chosen to allow investigation of particular fracture regimes, as outlined in the next section. All remaining material constants are assumed to have been homogenised prior to this analysis, and as such do not vary between material layers. The values of these constants are taken identically in all simulations, and are stated in Table.~\ref{Table:parameters}.

\begin{table}[h]
	\centering
	\begin{tabular}{c|c|c|c}
		$E$ & $\nu$ & $\mu$ & $q_0 $ \\
		\hline \hline
		&&&\\
		$2.81 \times 10^{10}$ \, [Pa] & $0.25$ & $1 \times 10^{-3}$ \, [Pa s] & $6.62 \times 10^{-2}$ \, [m$^3$ / s] \\
		&&&\\
		\hline \hline
	\end{tabular}
	\caption{Problem parameters used in simulations. Note that $q_0 (t)$ is taken to be constant.}
	\label{Table:parameters}
\end{table}

\subsection{Parameterising the fracture regime}\label{Sect:delta}

Noting that the effectiveness of differing homogenisation strategies for the fracture toughness was shown to be dependent on the fracture regime \cite{Gaspare2022}, it will be useful to have a measure which allows us to, roughly, determine the regime being experienced at the crack tip at a given point in time. Towards this end, we adopt a method of parameterisation similar to that utilized in the aforementioned paper.

We note that the volume of fluid stored within the fracture is given by:
\begin{equation}
V(t) = 2\pi \int_0^{L(t)} r w(t,r) \, dr .
\end{equation}
Recall from equation \eqref{elasticity} that the aperture, $w$, can be represented as a sum of terms representing the contribution of the (viscous) fluid pressure, $w_1$, and the impact of the material toughness, $w_2$. As such, we can decompose the volume into that resulting from the viscosity and toughness dominated terms:
$$
V(t) = V_v (t) + V_T (t),
$$
with
$$
V_v (t) = 2\pi \int_0^{L(t)} r w_1 (t,r) \, dr , \quad V_T (t) = 2\pi \int_0^{L(t)} r w_2 (t,r) \, dr  .
$$
As a result, the ratio between the two terms
\begin{equation} \label{delta1}
\delta (t) = \frac{V_T (t)}{V_v (t)} ,
\end{equation}
will provide a measure of the extent to which the fracture evolution is being governed by the fluid viscosity or the material toughness, providing a rough parameterisation of the fracture regime. 

\begin{figure}[t!]
\centering
\includegraphics[width=0.45\textwidth]{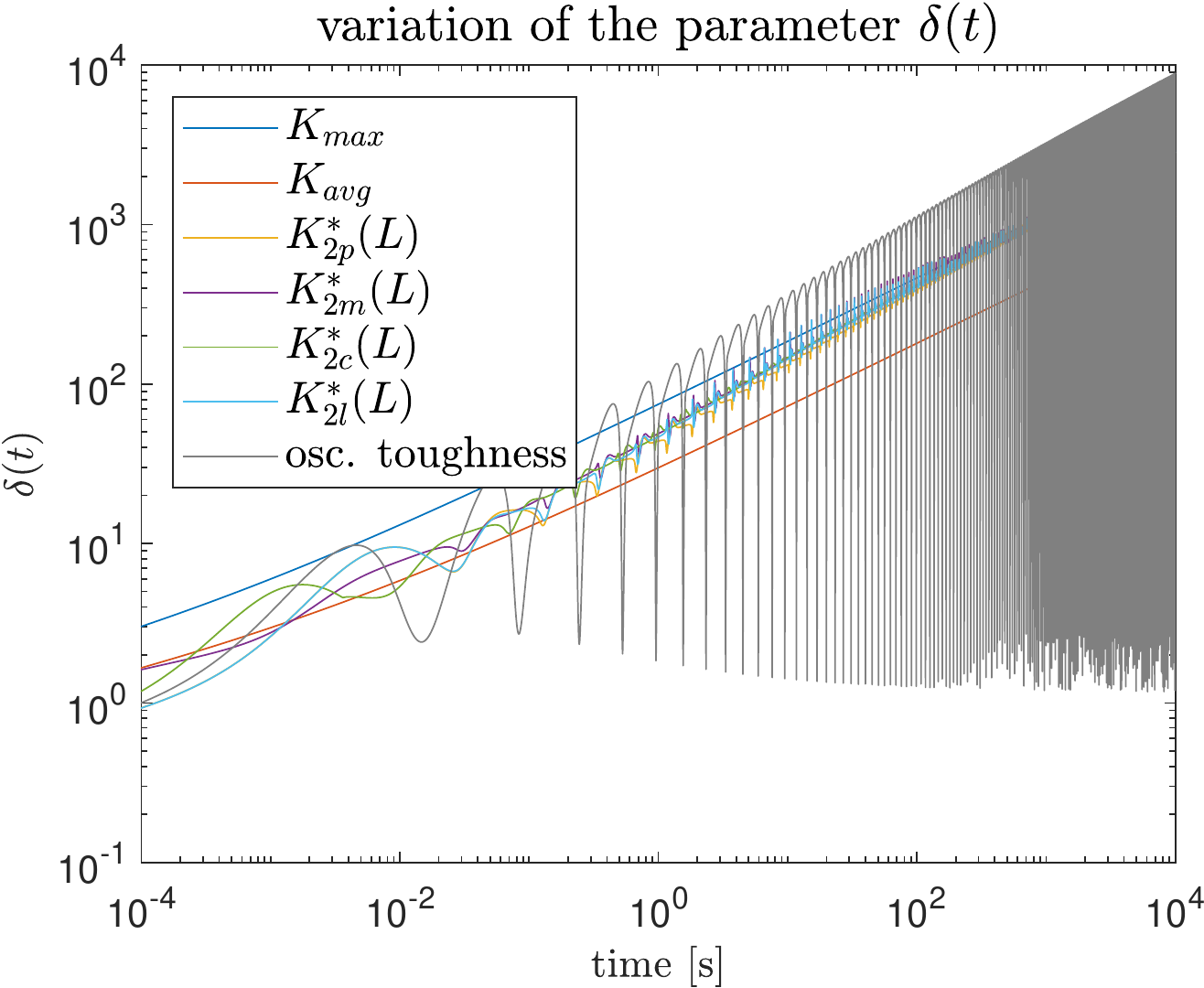}
\put(-225,155) {{\bf (a)}}
\hspace{12mm}
\includegraphics[width=0.45\textwidth]{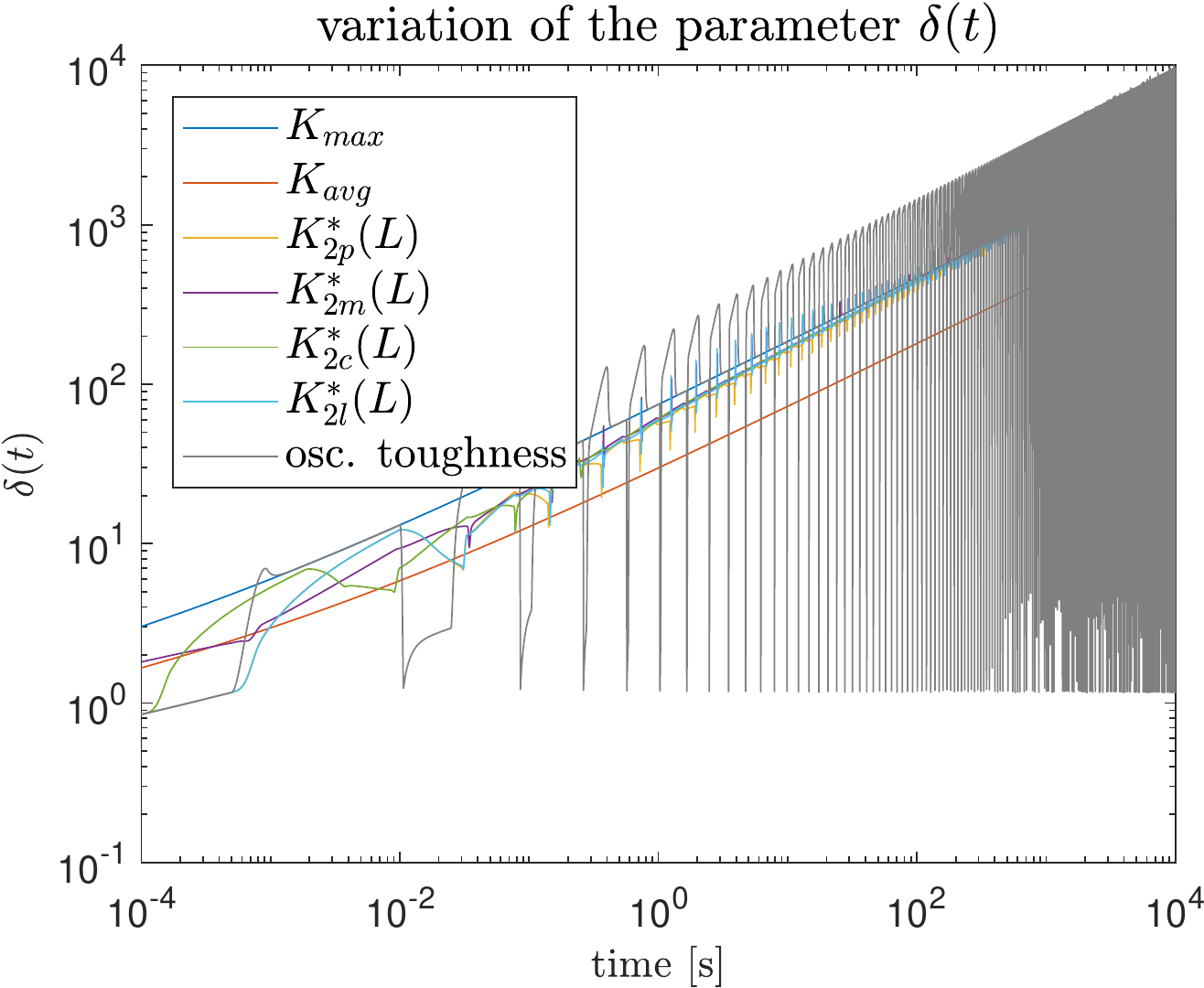}
\put(-225,155) {{\bf (b)}}
\\
\includegraphics[width=0.45\textwidth]{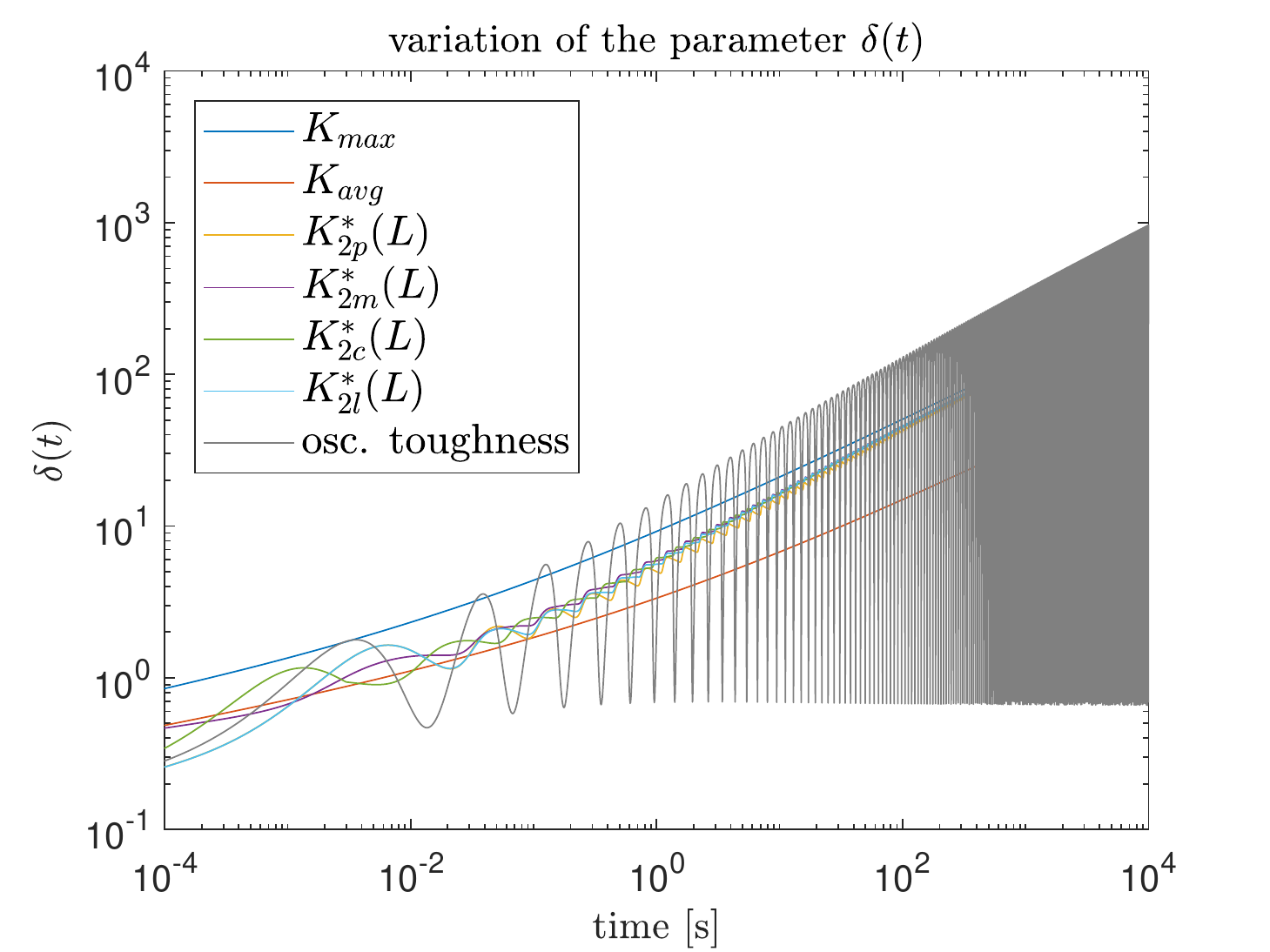}
\put(-225,155) {{\bf (c)}}
\hspace{12mm}
\includegraphics[width=0.45\textwidth]{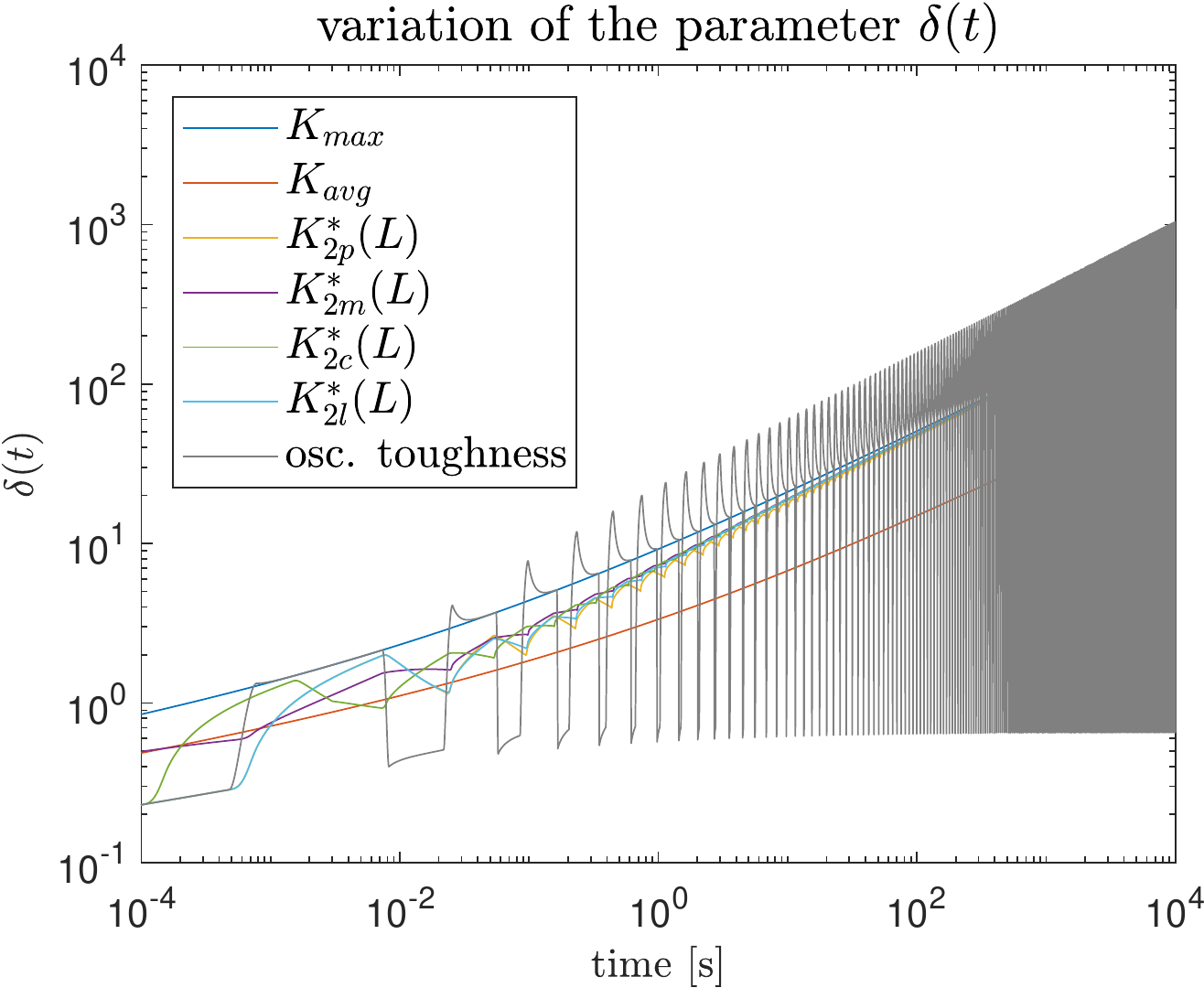}
\put(-225,155) {{\bf (d)}}
\\
\includegraphics[width=0.45\textwidth]{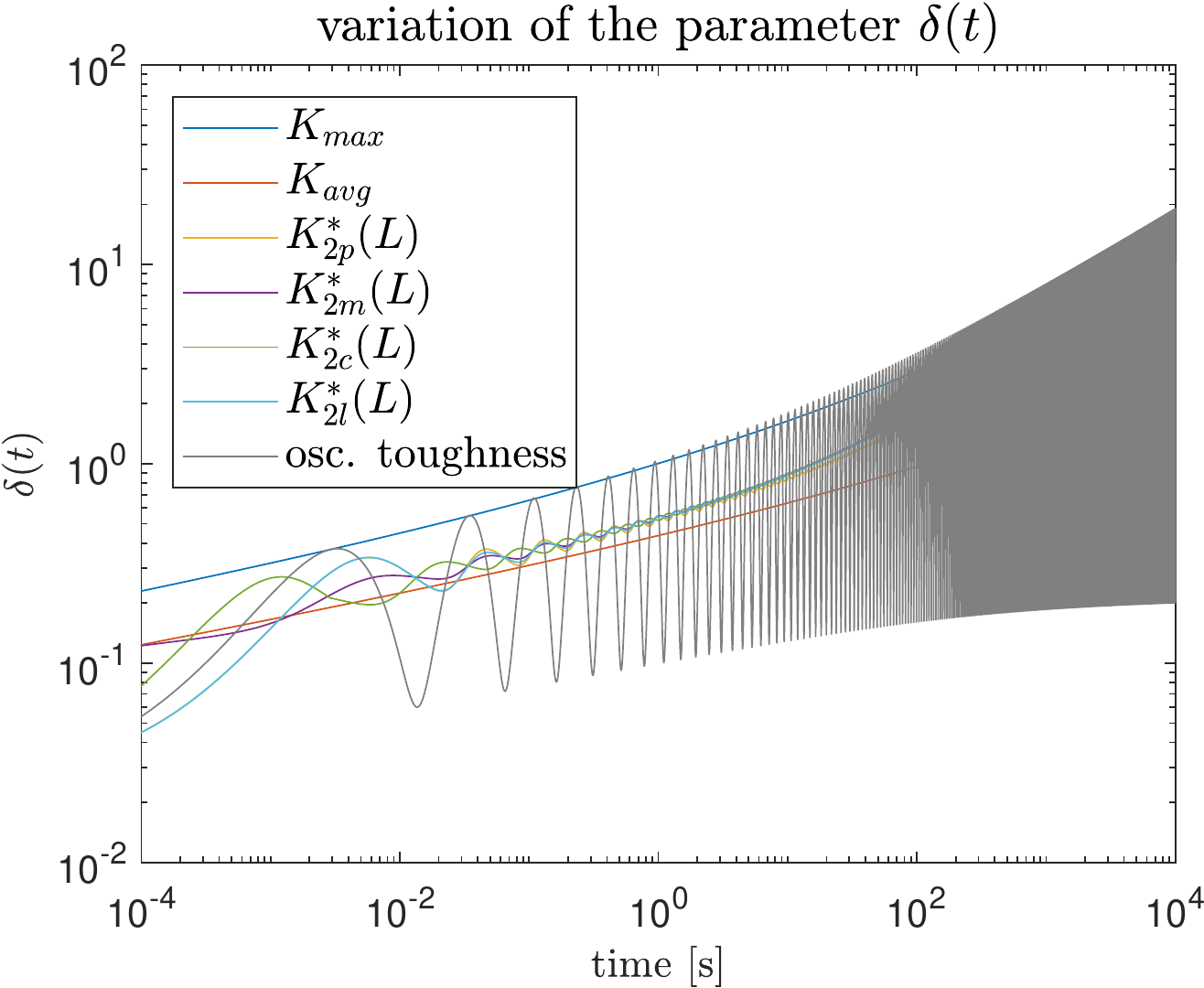}
\put(-225,155) {{\bf (e)}}
\hspace{12mm}
\includegraphics[width=0.45\textwidth]{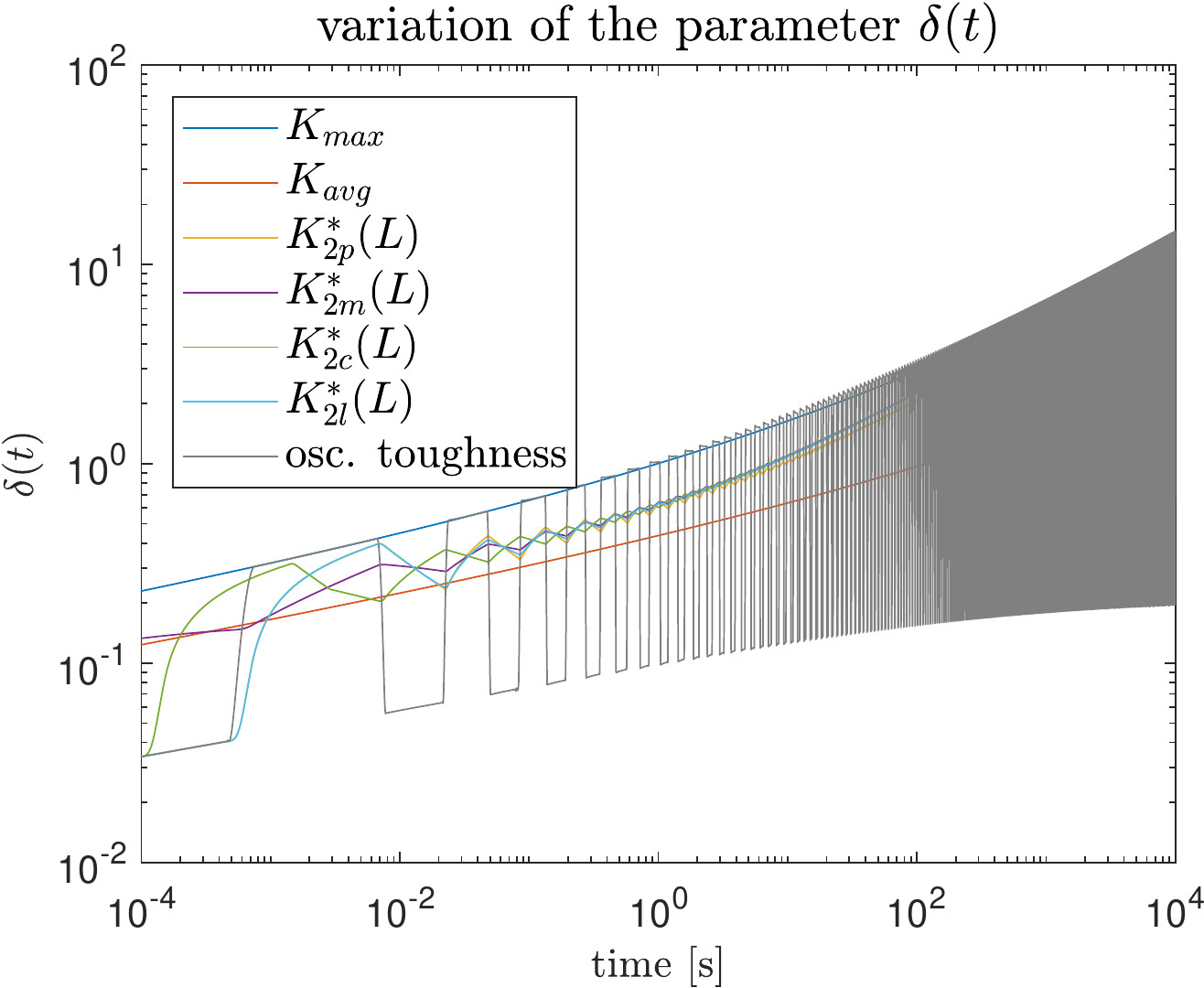}
\put(-225,155) {{\bf (f)}}
\caption{ Ratio between toughness and viscosity volumes $\delta(t)$.
The material toughness distribution, with balanced layering, are (see Table.~\ref{Table:toughness})  {\bf (a)}, {\bf (b)} Case 1; {\bf (c)}, {\bf (d)} Case 2; {\bf (e)}, {\bf (f)} Case 3. The toughness distributions are: {\bf (a)}, {\bf (c)}, {\bf (e)} sinusoidal; {\bf (b)}, {\bf (d)}, {\bf (f)} step-wise. Here $K_{2p}$ - $K_{2l}$ are different evaluations of the temporal-average of the fracture toughness \eqref{MeasureK2r}-\eqref{MeasureK2p}, outlined later in Sect.~\ref{Sect:Stratagems}.}
\label{Fig:Delta}
\end{figure}

There is one crucial difference between the parameter $\delta$ for the KGD and radial models, and that is that the value of $\delta (t)$ will inevitably change over time for the latter. This simply reflects the well-known transition of the fracture between the viscosity and toughness regimes as the fracture grows\footnote{Due to the absence of leak-off, the fracture will remain within the storage dominated regime, and we need only be concerned with the ${\cal M}$ -- ${\cal K}$ transition, see e.g. \cite{Peirce2008}}. The value of $\delta(t)$ computed for a range of different (periodic) toughness distributions is provided in Fig.~\ref{Fig:Delta}. Meanwhile, the rate at which the value of $\delta$ will change can be estimated by comparing with the natural scalings for the viscosity, ${\cal M}$, and toughness, ${\cal K}$, regimes. Using the natural scalings (see e.g. \cite{Garagash2000,GaragashSummary,Peirce2008,Savitski2002})
we obtain:
\begin{eqnarray}
	\text{Viscosity dominated:}\quad & \delta \sim 0.9642 \left(\frac{K_{Ic}^{18} (1-\nu^2)^{13} t^2}{Q_0^3 \mu^5 E^{13}}\right)^{\frac{1}{18}},\quad \delta \ll 1,  
	\label{eq:st}\\[2mm]
	\text{Toughness dominated:}\quad  & \delta \sim 2.5283   \left(\frac{K_{Ic}^{18} (1-\nu^2)^{13} t^2}{Q_0^3 \mu^5 E^{13}}\right)^{\frac{1}{5}},\quad \delta\gg 1.
	\label{eq:em}
\end{eqnarray}

The values of the ratio $\delta$ obtained for a fixed material toughness $K_{Ic}$ at various points in time are provided in Table.~\ref{Table:delta}. These values of the ratio were computed for the case of a material \emph{with homogeneous toughness}, while all other material parameters are as stated in Table.~\ref{Table:parameters}. The behaviour of $\delta$ was utilized to inform the choice of $K_{max}$ and $K_{min}$, which are used to define the toughness distribution throughout the investigation. These two values are taken to ensure that the material configuration when $t=1$ s corresponds to a desired pair of regimes, for instance having both materials start in the toughness dominated regime, or having one layer in the transient regime. The choice of values used in simulations, and the regimes they correspond to, are outlined in Table.~\ref{Table:toughness}.

\begin{table}[h!]
	\centering
	\begin{tabular}{c||c|c|c|c}
		$K_{Ic}$ [Pa $\sqrt{\text{m}}$] & $3.26 \times 10^{5}$ & $1.86 \times 10^{6}$ & $4.70 \times 10^{6}$ & $8.76\times 10^{6}$  \\
		\hline \hline
		${\delta(10^{-4})} $ & $0.0339$ & $0.2298$ & $0.8488$ & $3.018$ \\
		${\delta(10^0)} $ & $0.1005$ & $1.004$ & $9.224$ & $74.68$ \\
		${\delta(10^4)} $ & $0.3387$ & $12.49$ & $310.9$ & $2911$ \\
		\hline \hline
	\end{tabular}
	\caption{The values of the ratio $\delta (t)$ \eqref{delta1} corresponding to given values of the material toughness $K_{Ic}$ (to $3$ s.f.) at three moments in time, for a rock whose other material parameters are as stated in Table.~\ref{Table:parameters}.}
	\label{Table:delta}
\end{table}

\begin{table}[h]
	\centering
	\begin{tabular}{r|c|c}
		& $K_{max}$ [Pa $\sqrt{\text{m}}$] & $K_{min}$ [Pa $\sqrt{\text{m}}$] \\
		\hline \hline
		{\bf Case 1} & $8.76\times 10^{6}$ & $4.70 \times 10^{6}$ \\
		\hline
		{\bf Case 2} & $4.70 \times 10^{6}$ & $1.86 \times 10^{6}$ \\
		\hline
		{\bf Case 3} & $1.86 \times 10^{6}$ & $3.26 \times 10^{5}$ \\
		\hline \hline
	\end{tabular}
	\caption{The values of the maximum and minimum toughness $K_{Ic}$ used in simulations. Three cases are considered, to investigate the different fracture regimes, such that for $t=1$ s they correspond to: {\bf Case 1} Toughness-toughness layering; {\bf Case 2} Toughness-transient layering; {\bf Case 3} Transient-viscosity layering. Note that the regimes will change over time due to the fracture evolution (see Sect.~\ref{Sect:delta}, Table.~\ref{Table:delta}).}
	\label{Table:toughness}
\end{table}

\subsection{Numerical algorithm and behaviour of the key parameters}\label{Sect:Behaviour}

The solution of the system of equations \eqref{continuity1} -- \eqref{Kernel1}, where the toughness distribution is as outlined in Sect.~\ref{Sect:MaterTough} -- \ref{Sect:delta}, is obtained using a highly effective in-house-built solver based on the ``universal algorithm'' approach first outlined for the PKN/KGD models in \cite{Wrobel2015}, and first extended to the radial model in \cite{Peck2018a}. The algorithm has since been improved to be adaptive in both the spacial and temporal dimensions, allowing for the exceptionally high level of solution accuracy that we utilize in the following analysis. Full details of the algorithms construction are provided in \cite{Gaspare2020}.

\begin{figure}[t!]
\centering
\includegraphics[width=0.45\textwidth]{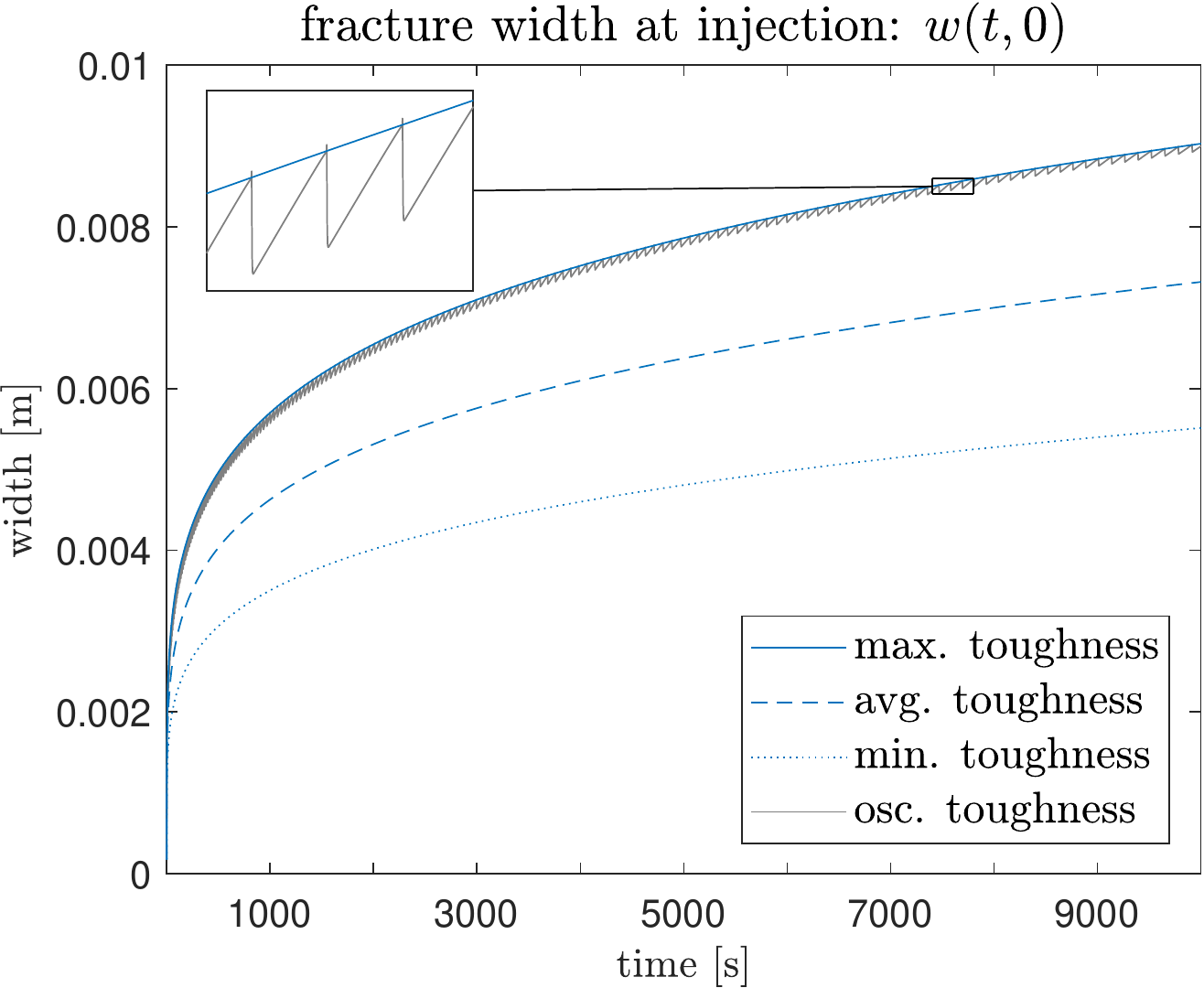}
\put(-225,155) {{\bf (a)}}
\hspace{12mm}
\includegraphics[width=0.45\textwidth]{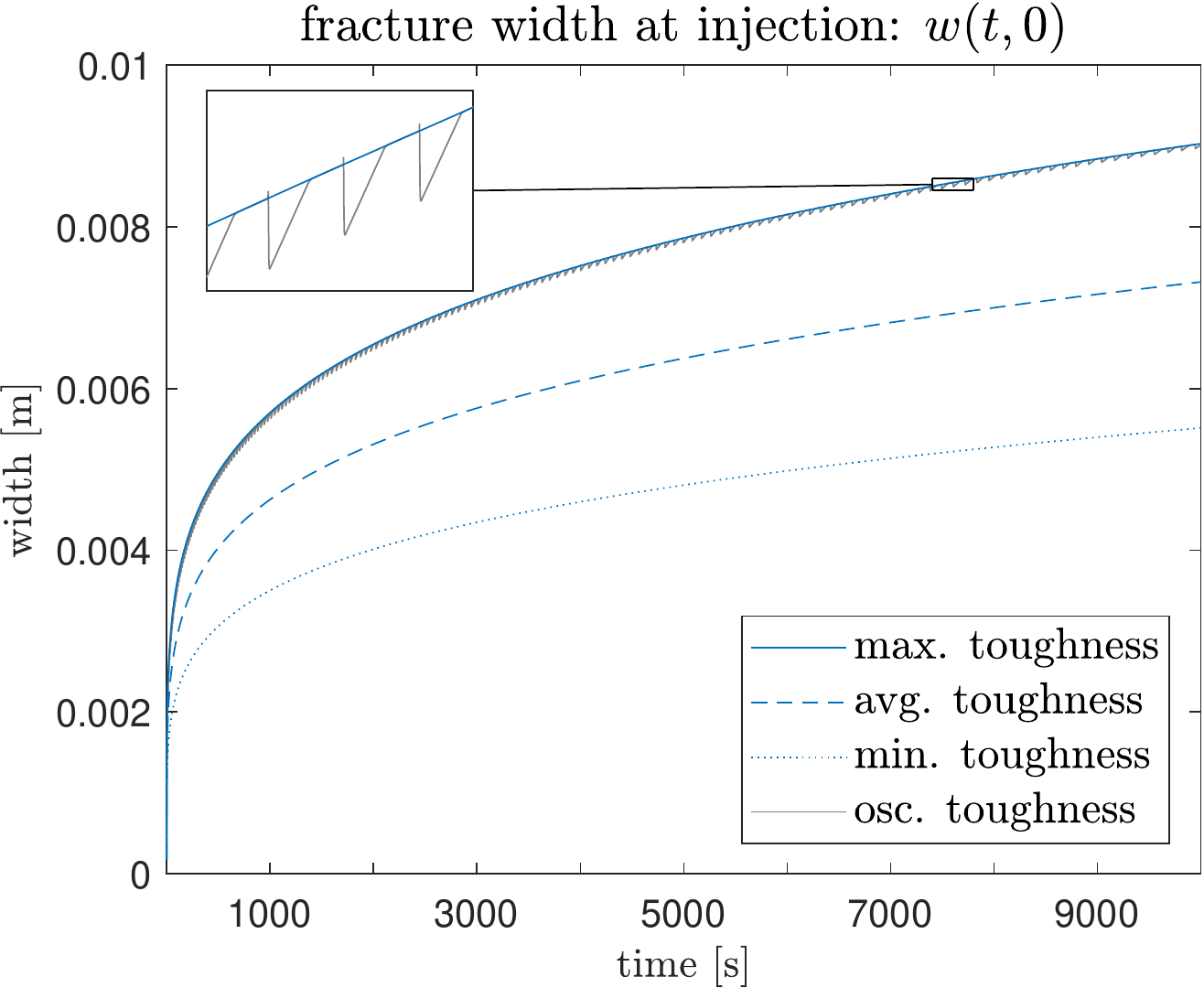}
\put(-225,155) {{\bf (b)}}
\\
\includegraphics[width=0.45\textwidth]{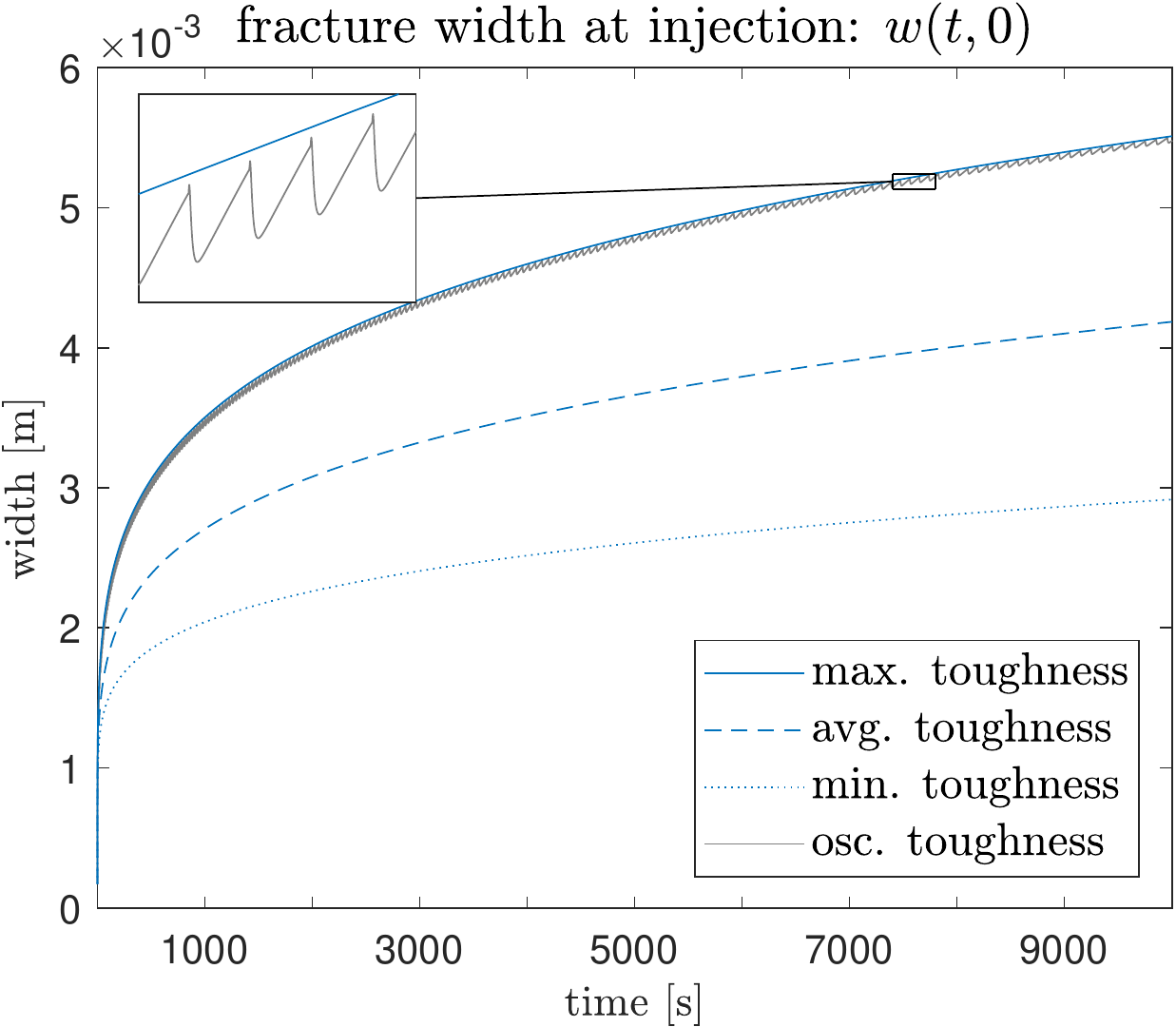}
\put(-225,155) {{\bf (c)}}
\hspace{12mm}
\includegraphics[width=0.45\textwidth]{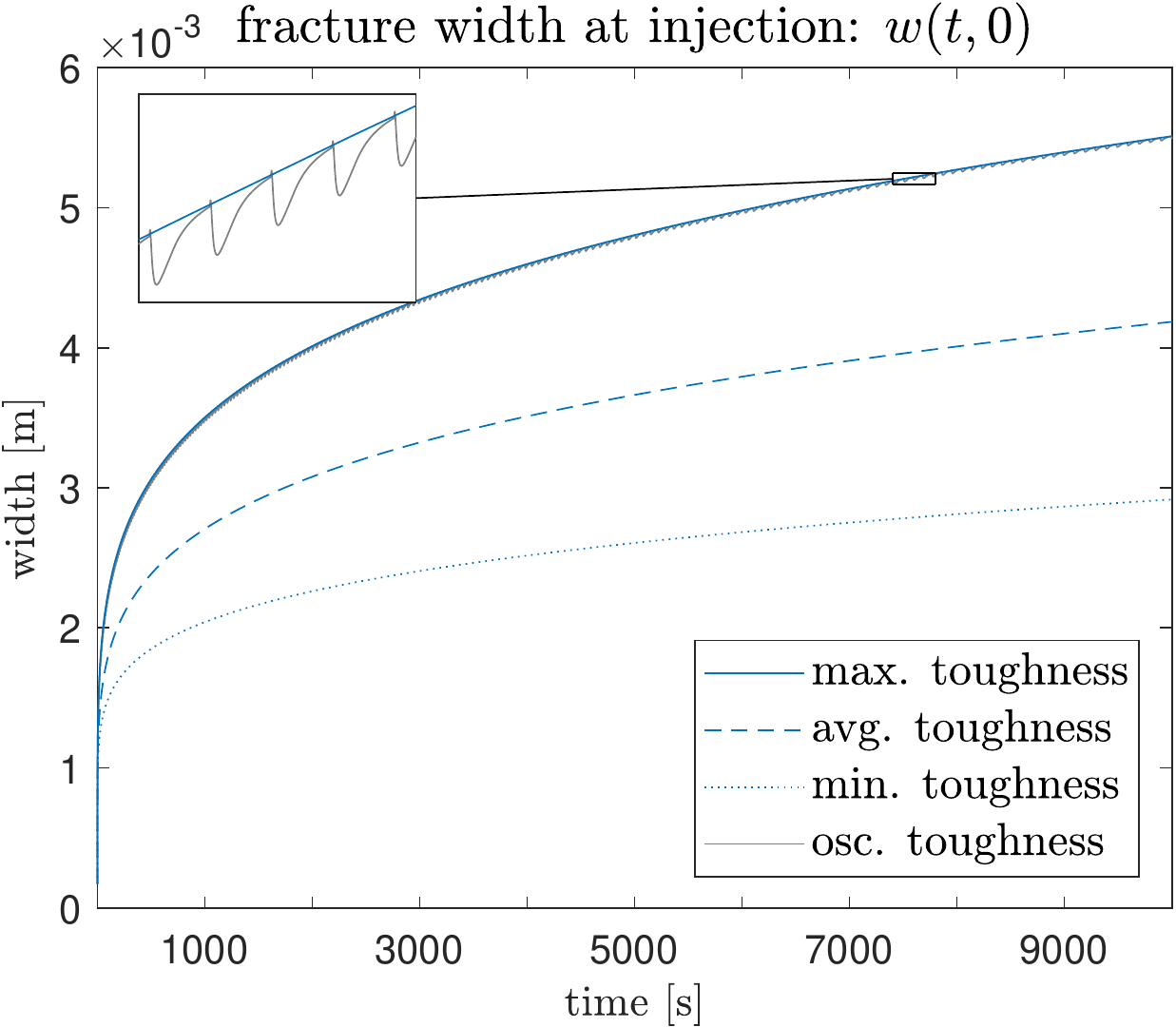}
\put(-225,155) {{\bf (d)}}
\\
\includegraphics[width=0.45\textwidth]{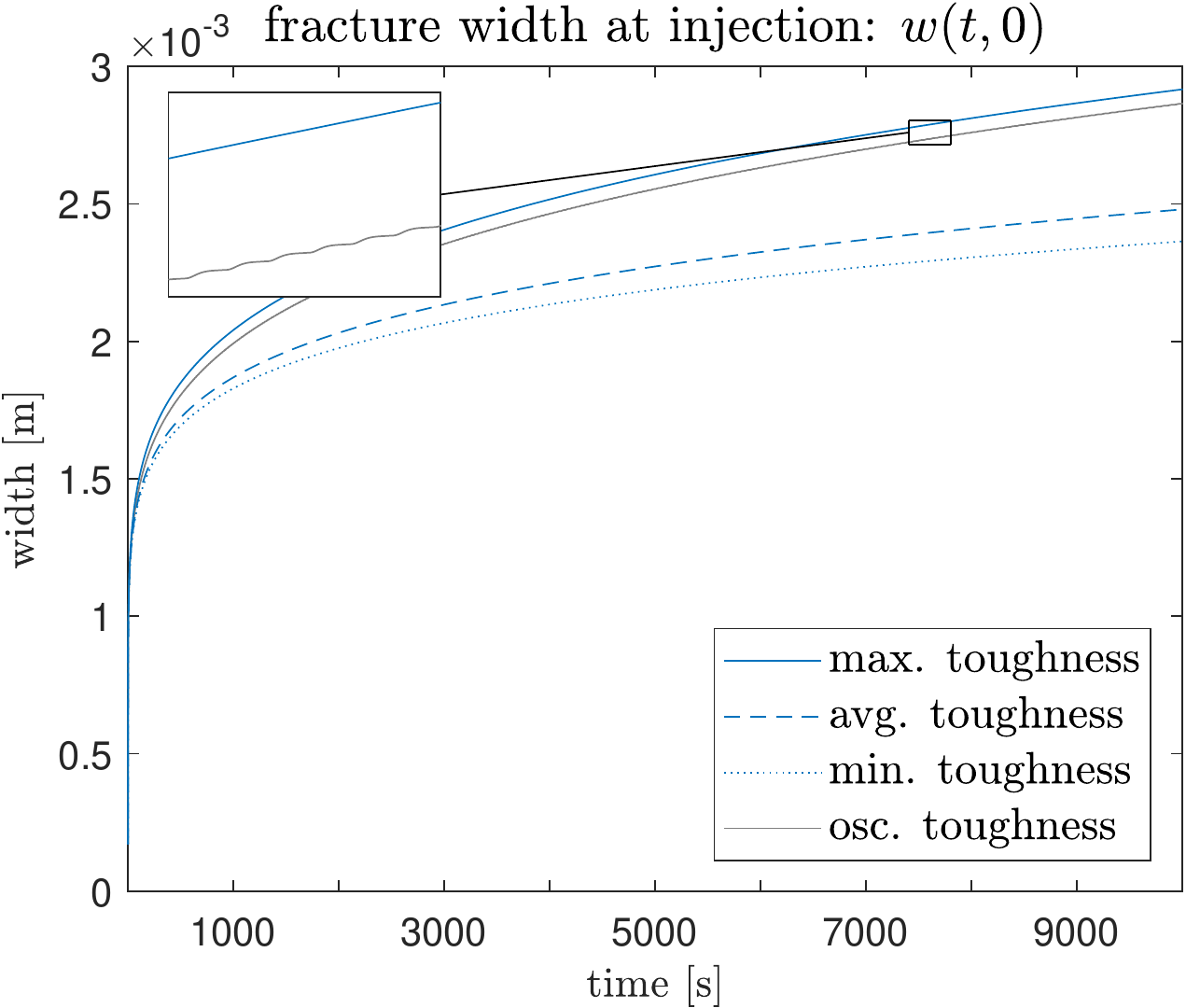}
\put(-225,155) {{\bf (e)}}
\hspace{12mm}
\includegraphics[width=0.45\textwidth]{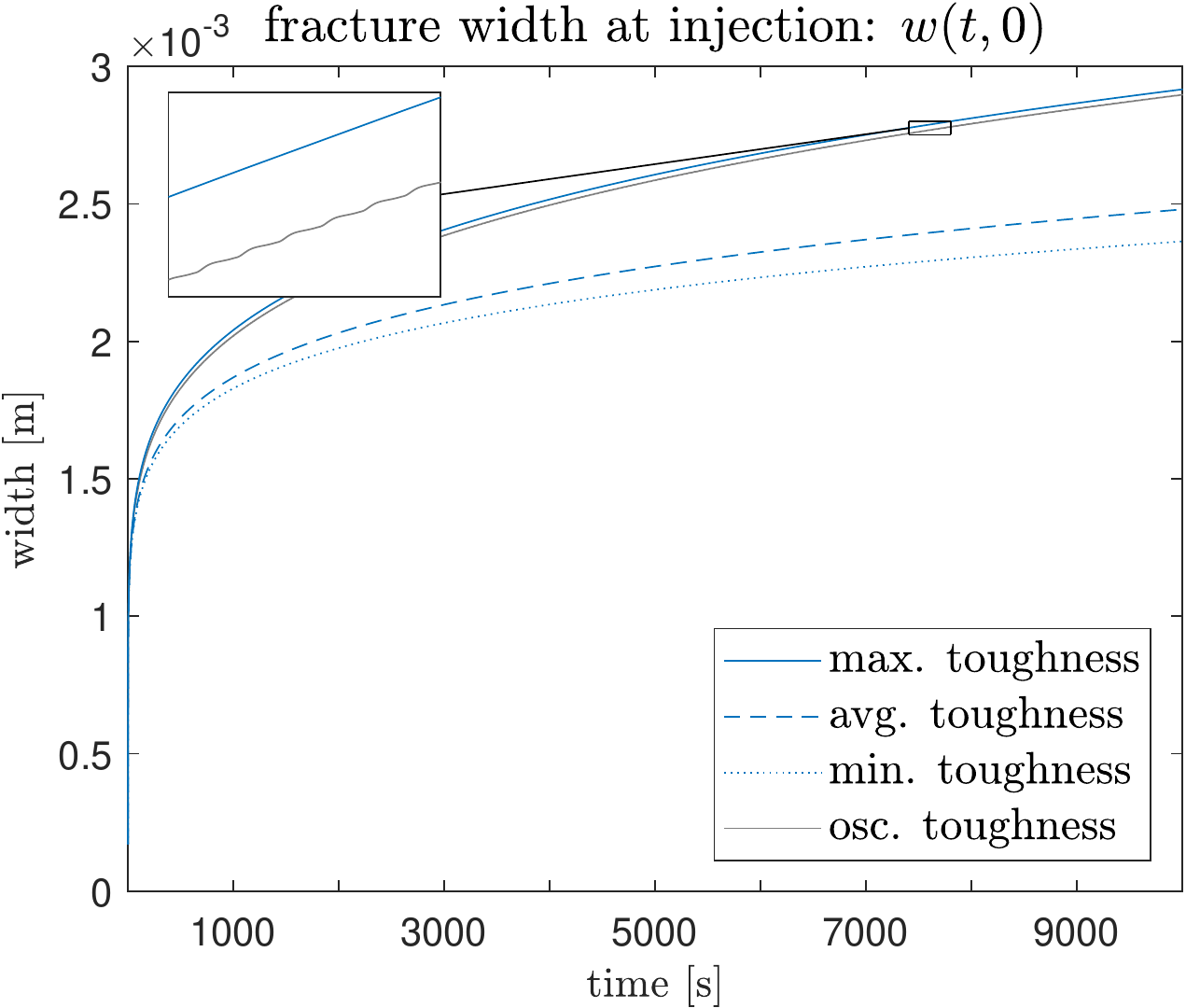}
\put(-225,155) {{\bf (f)}}
\caption{The fracture aperture at the crack opening $w(t,0)$ over time, for various material toughness distributions with balanced layering. The material toughness distribution, with balanced layering, are (see Table.~\ref{Table:toughness})  {\bf (a)}, {\bf (b)} Case 1; {\bf (c)}, {\bf (d)} Case 2; {\bf (e)}, {\bf (f)} Case 3. The toughness distributions are: {\bf (a)}, {\bf (c)}, {\bf (e)} sinusoidal; {\bf (b)}, {\bf (d)}, {\bf (f)} step-wise.}
\label{Fig:Aperture1}
\end{figure}

\begin{figure}[t!]
\centering
\includegraphics[width=0.45\textwidth]{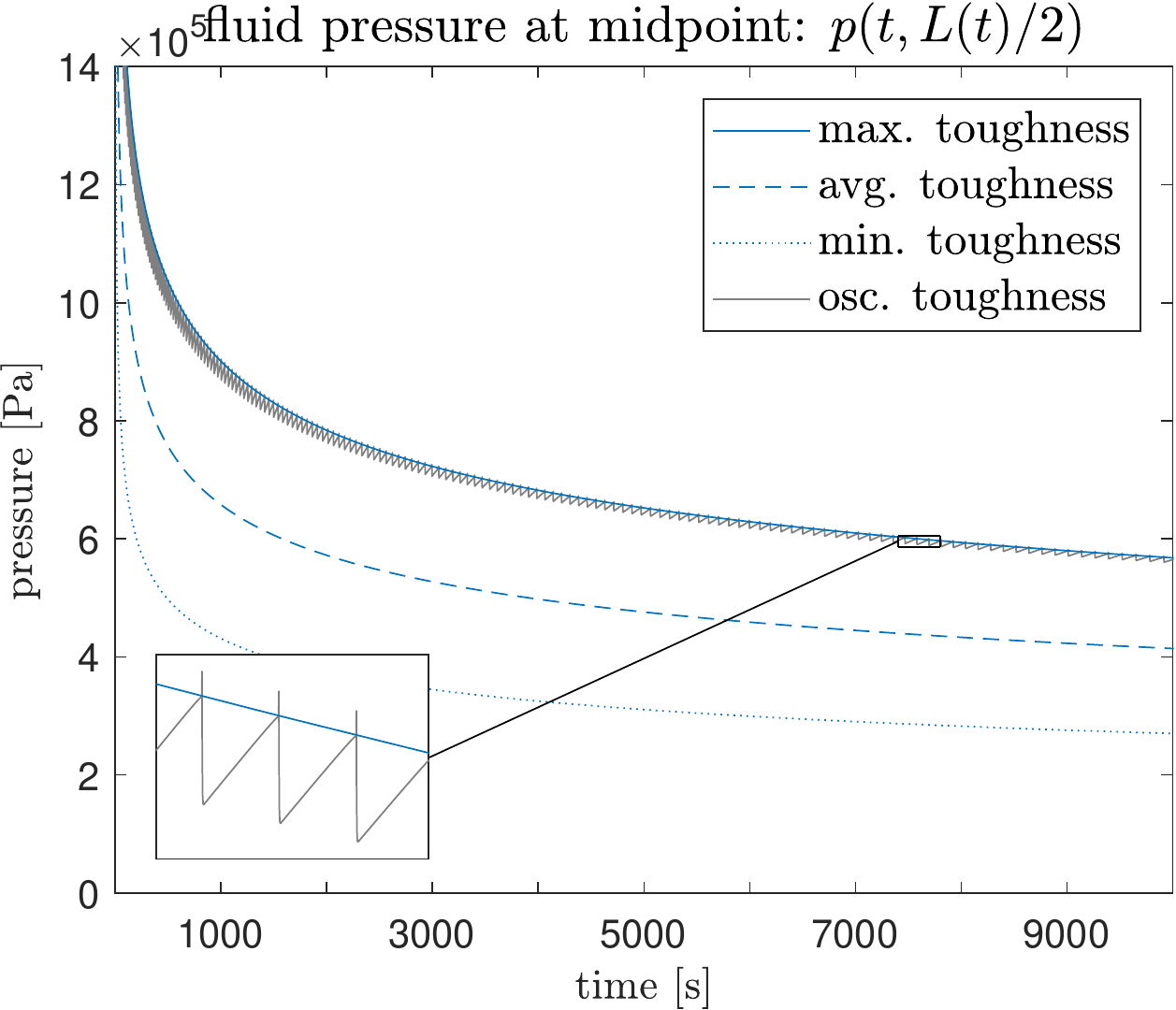}
\put(-225,155) {{\bf (a)}}
\hspace{12mm}
\includegraphics[width=0.45\textwidth]{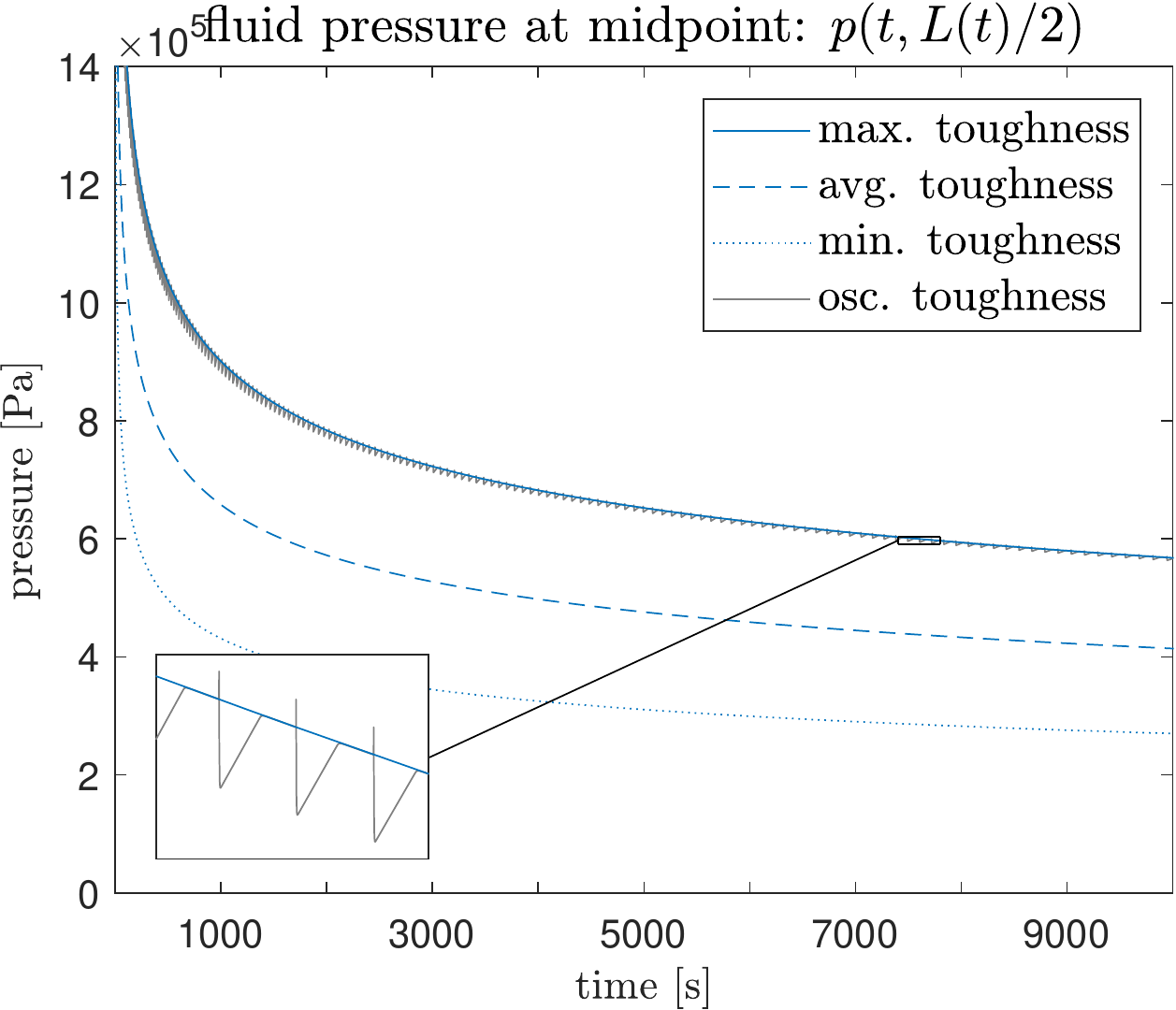}
\put(-225,155) {{\bf (b)}}
\\
\includegraphics[width=0.45\textwidth]{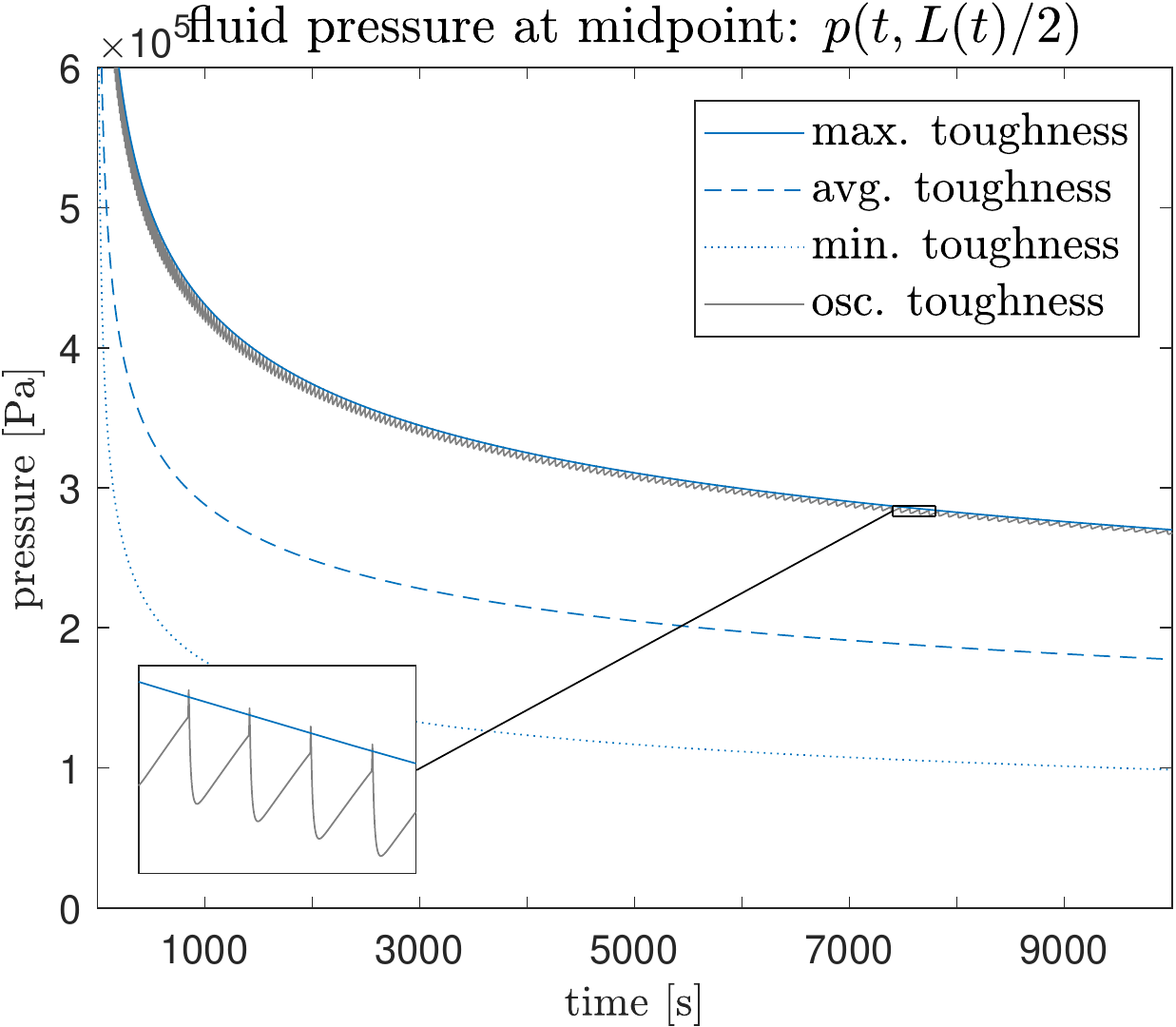}
\put(-225,155) {{\bf (c)}}
\hspace{12mm}
\includegraphics[width=0.45\textwidth]{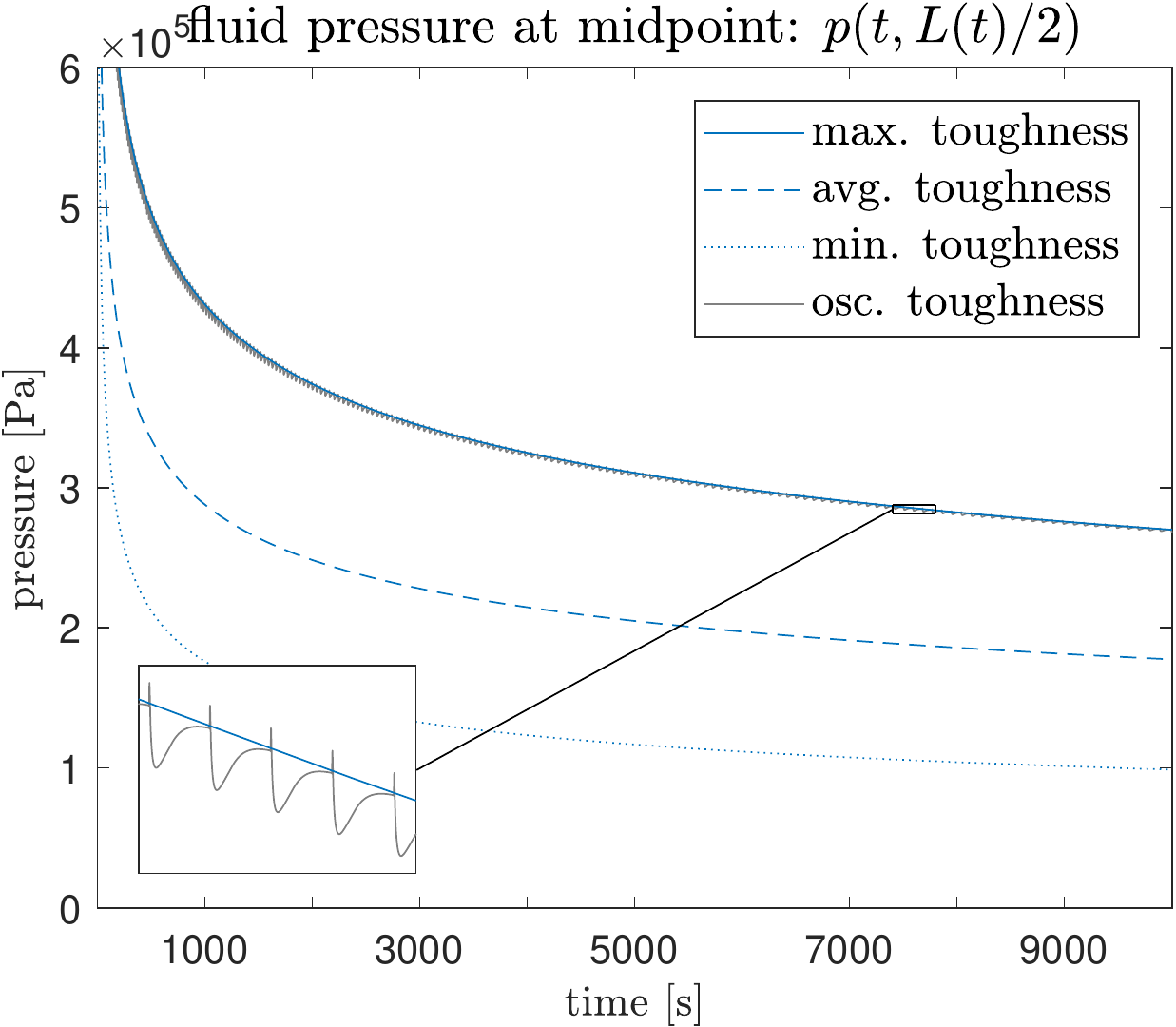}
\put(-225,155) {{\bf (d)}}
\\
\includegraphics[width=0.45\textwidth]{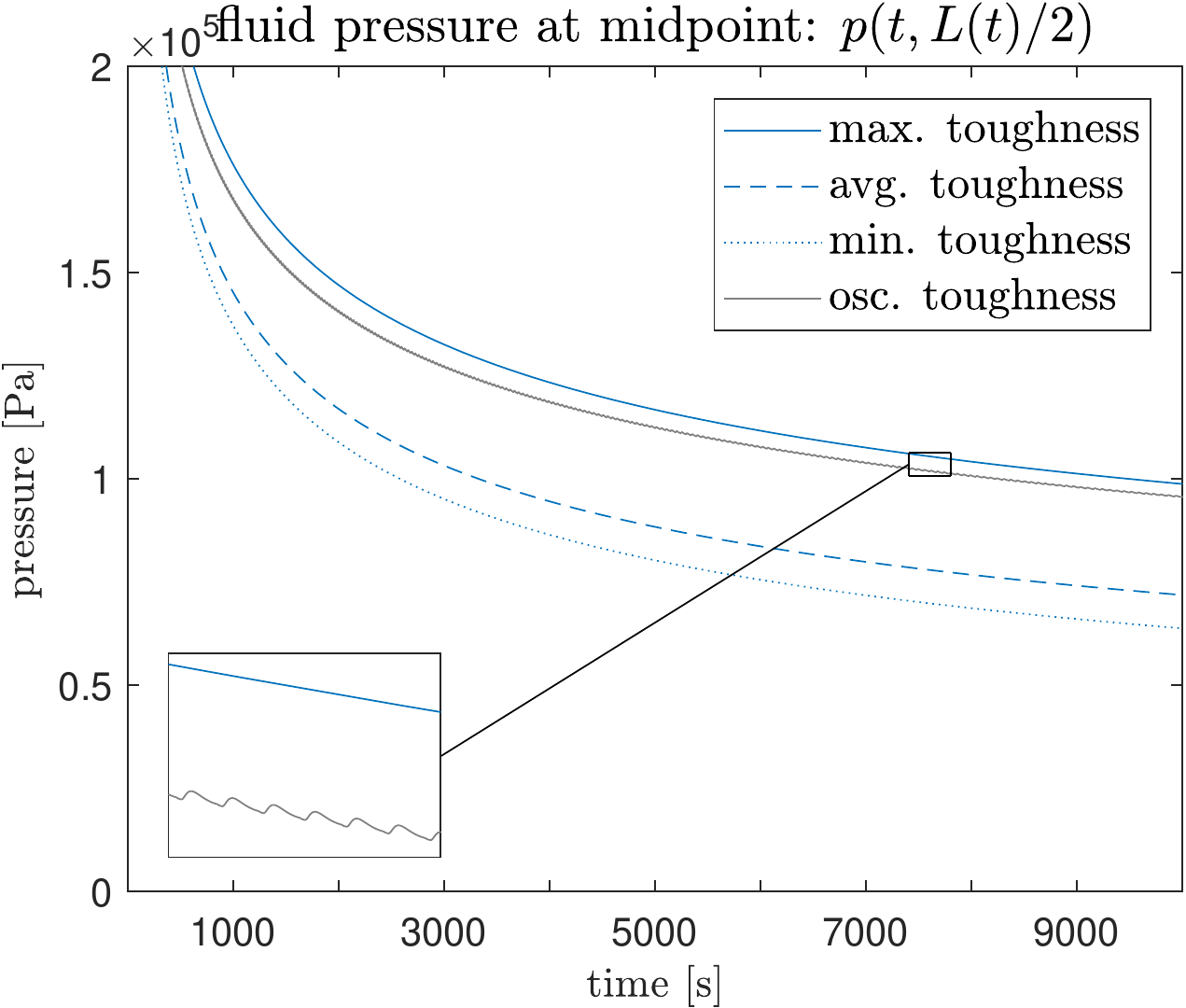}
\put(-225,155) {{\bf (e)}}
\hspace{12mm}
\includegraphics[width=0.45\textwidth]{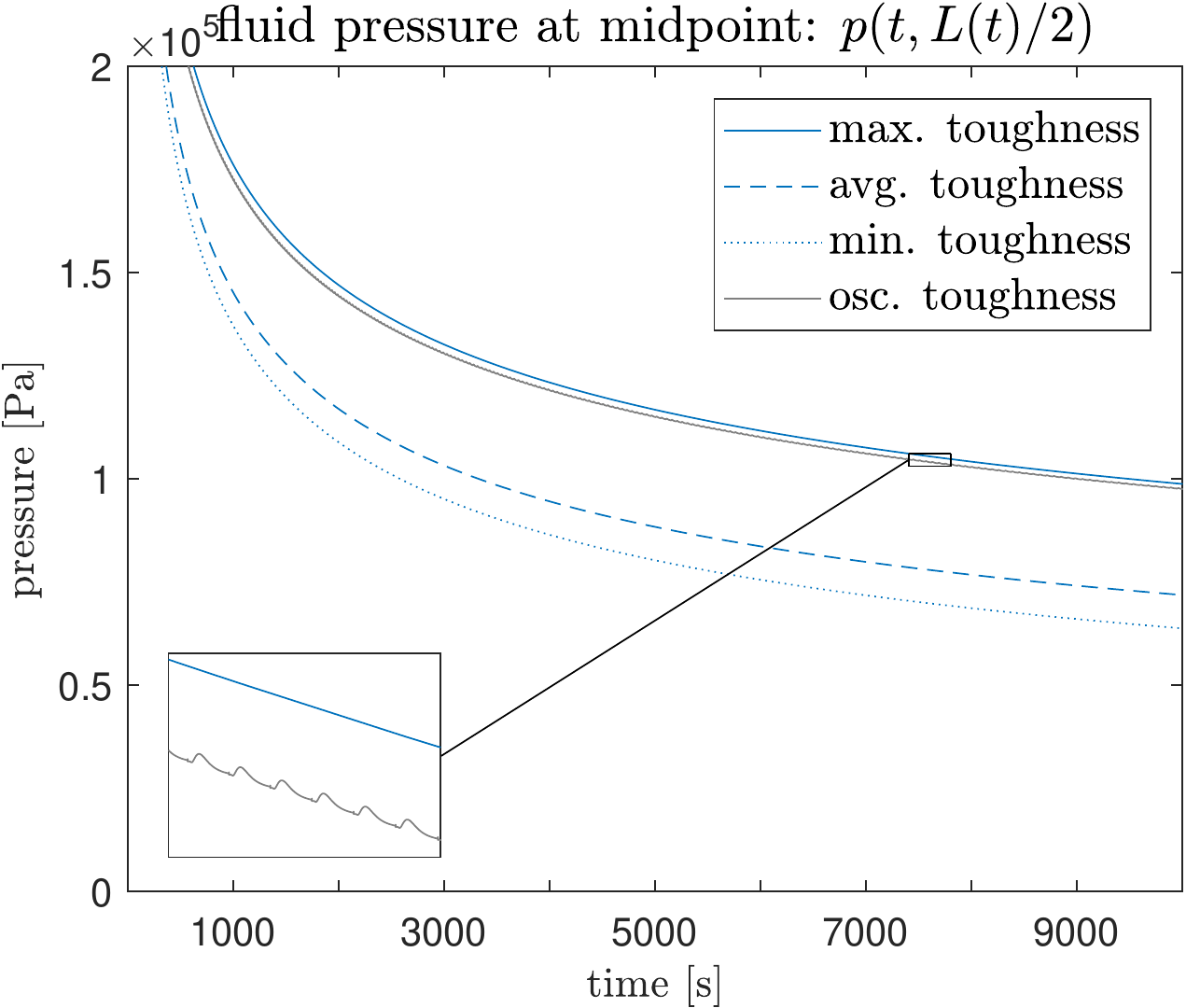}
\put(-225,155) {{\bf (f)}}
\caption{The fracture pressure at the mid-length $p(t,L(t)/2)$ over time, for various material toughness distributions with balanced layering. The material toughness distribution, with balanced layering, are (see Table.~\ref{Table:toughness})  {\bf (a)}, {\bf (b)} Case 1; {\bf (c)}, {\bf (d)} Case 2; {\bf (e)}, {\bf (f)} Case 3. The toughness distributions are: {\bf (a)}, {\bf (c)}, {\bf (e)} sinusoidal; {\bf (b)}, {\bf (d)}, {\bf (f)} step-wise.}
\label{Fig:Pressure1}
\end{figure}

\begin{figure}[t!]
\centering
\includegraphics[width=0.45\textwidth]{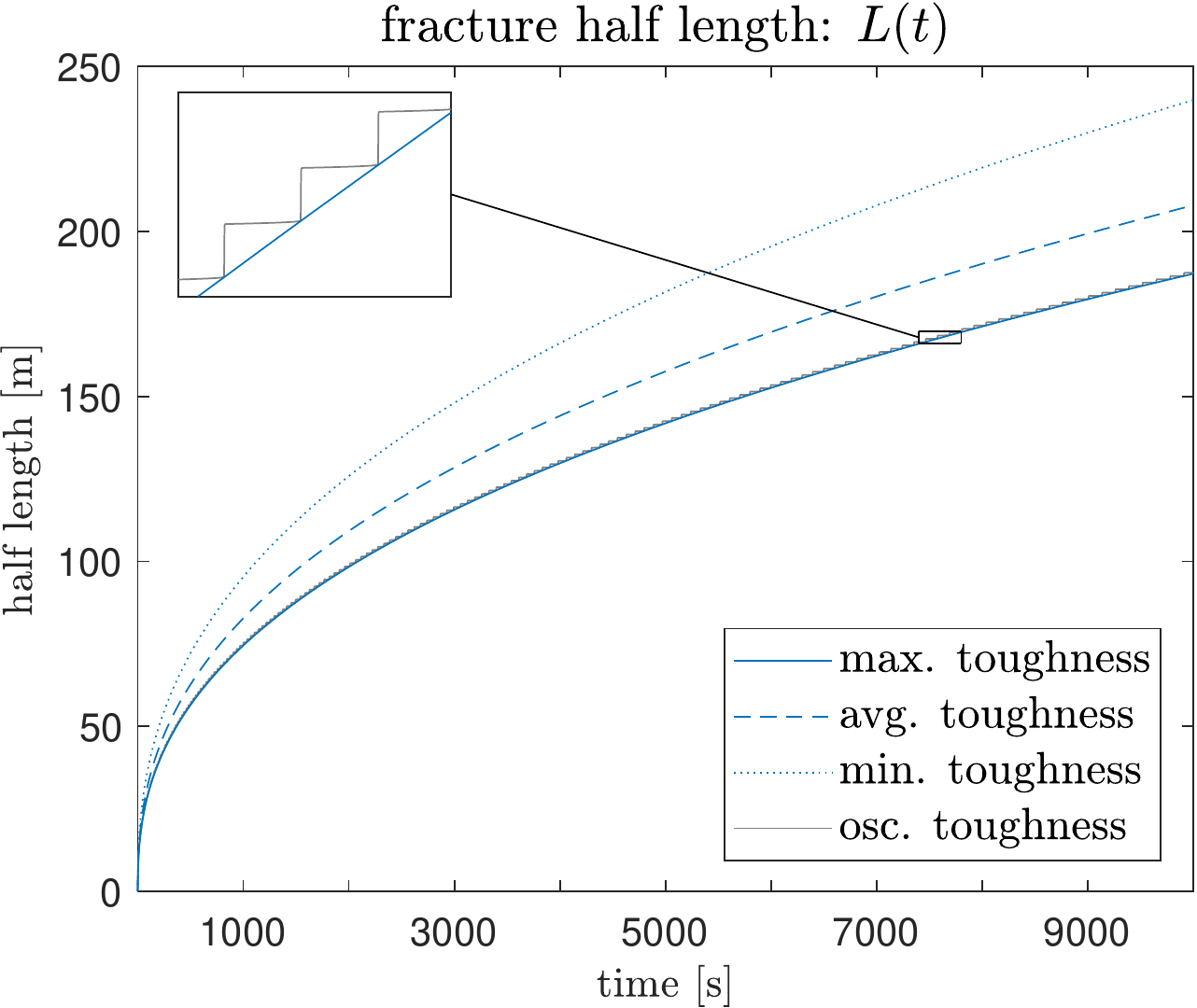}
\put(-225,155) {{\bf (a)}}
\hspace{12mm}
\includegraphics[width=0.45\textwidth]{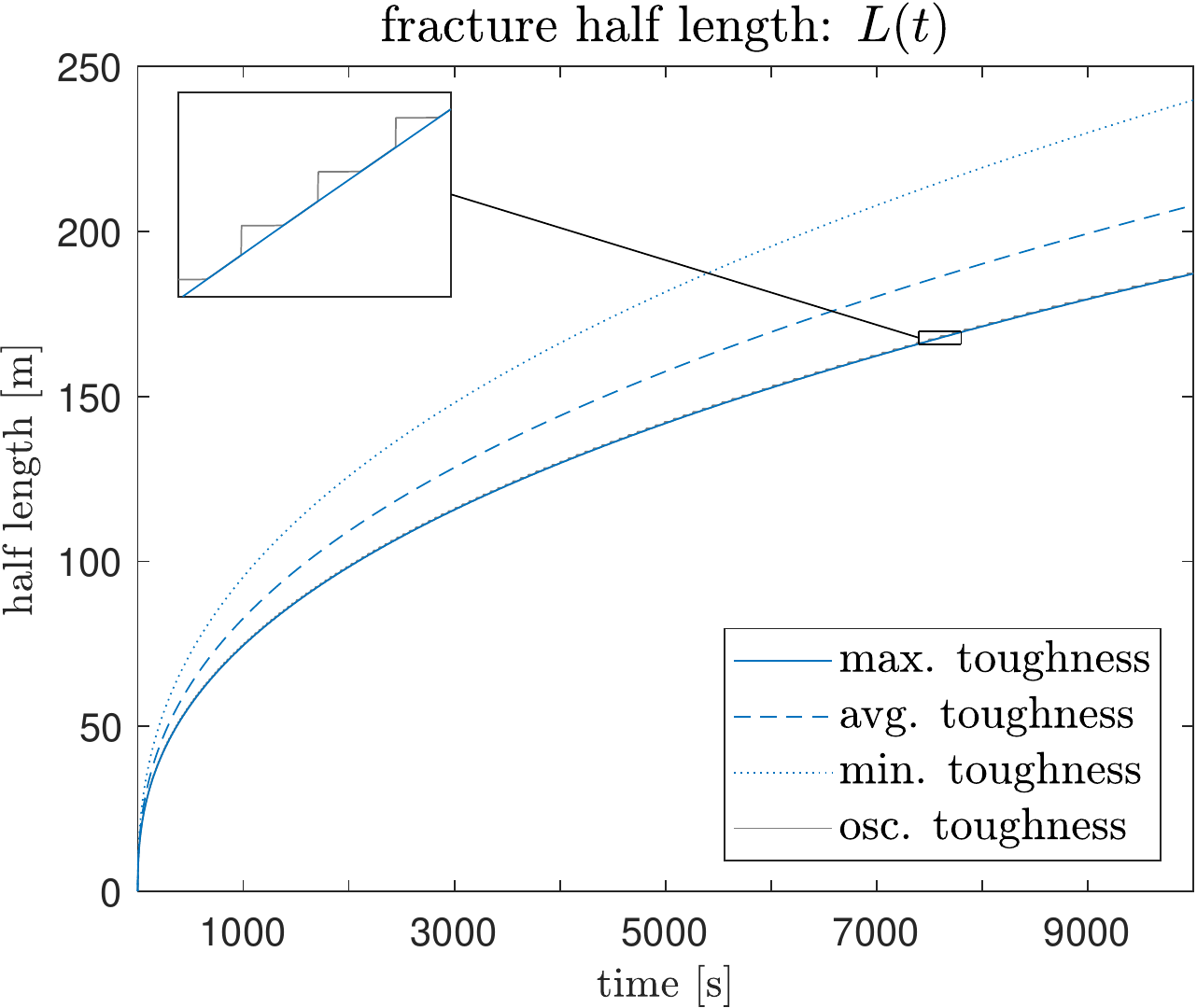}
\put(-225,155) {{\bf (b)}}
\\
\includegraphics[width=0.45\textwidth]{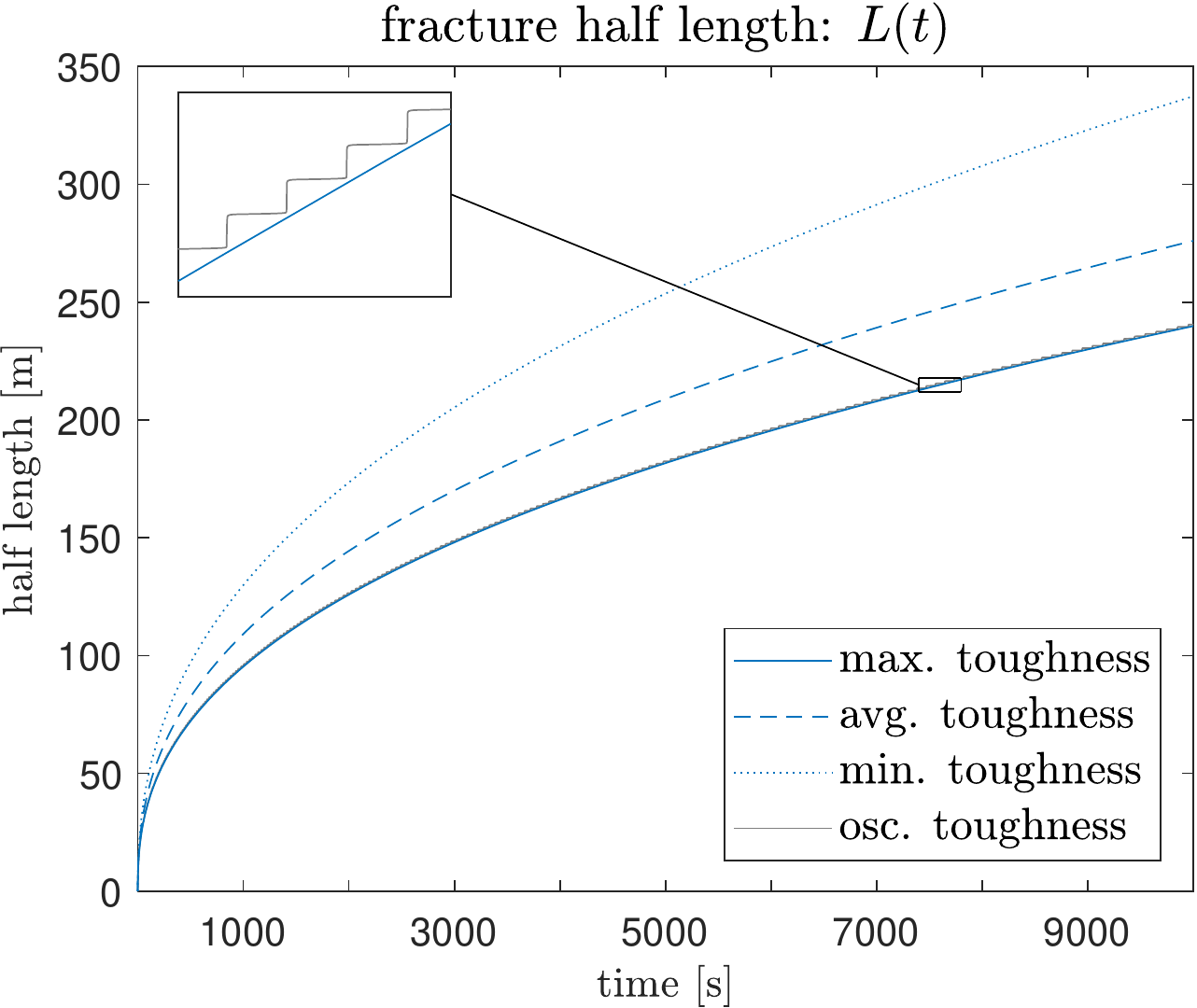}
\put(-225,155) {{\bf (c)}}
\hspace{12mm}
\includegraphics[width=0.45\textwidth]{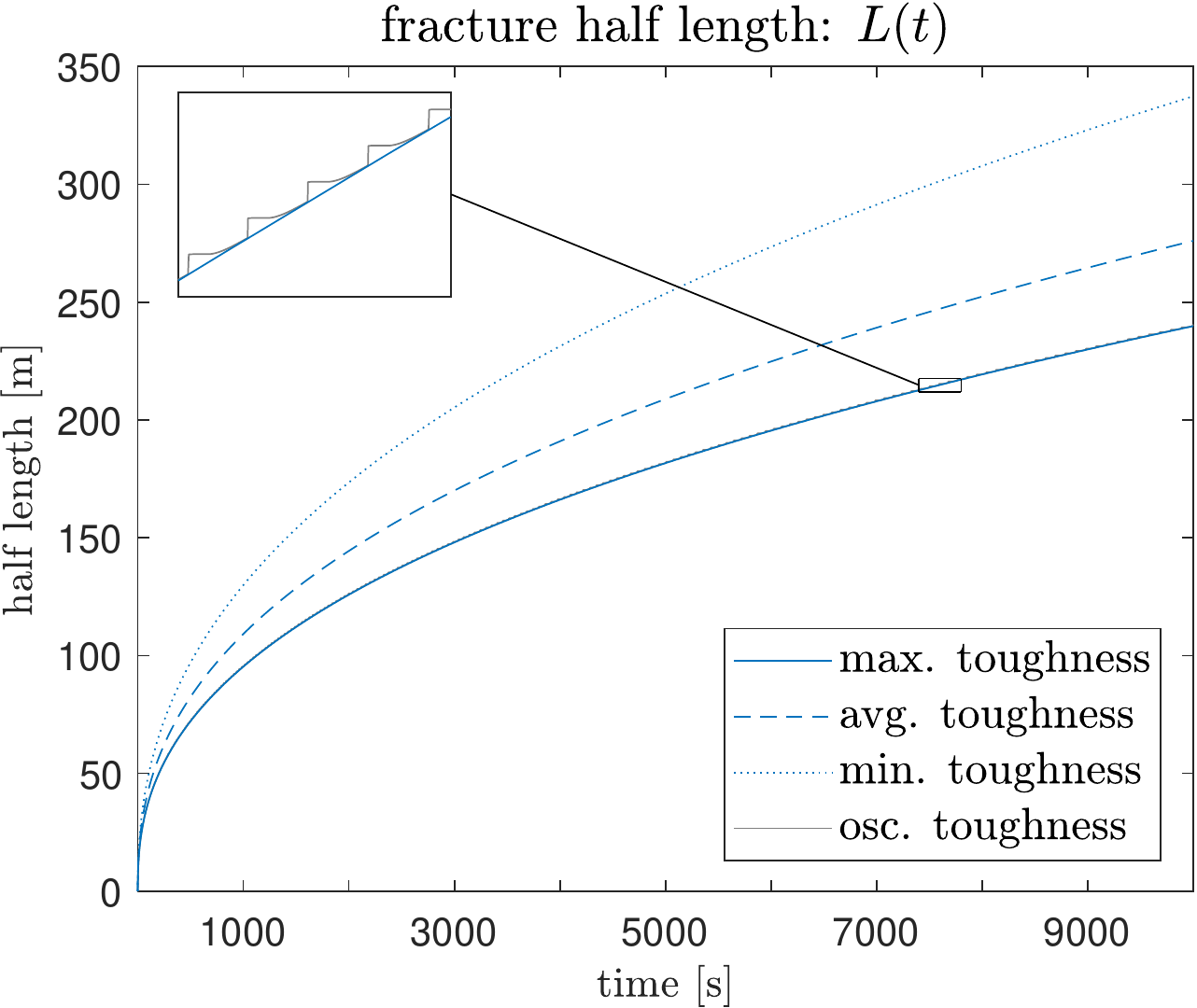}
\put(-225,155) {{\bf (d)}}
\\
\includegraphics[width=0.45\textwidth]{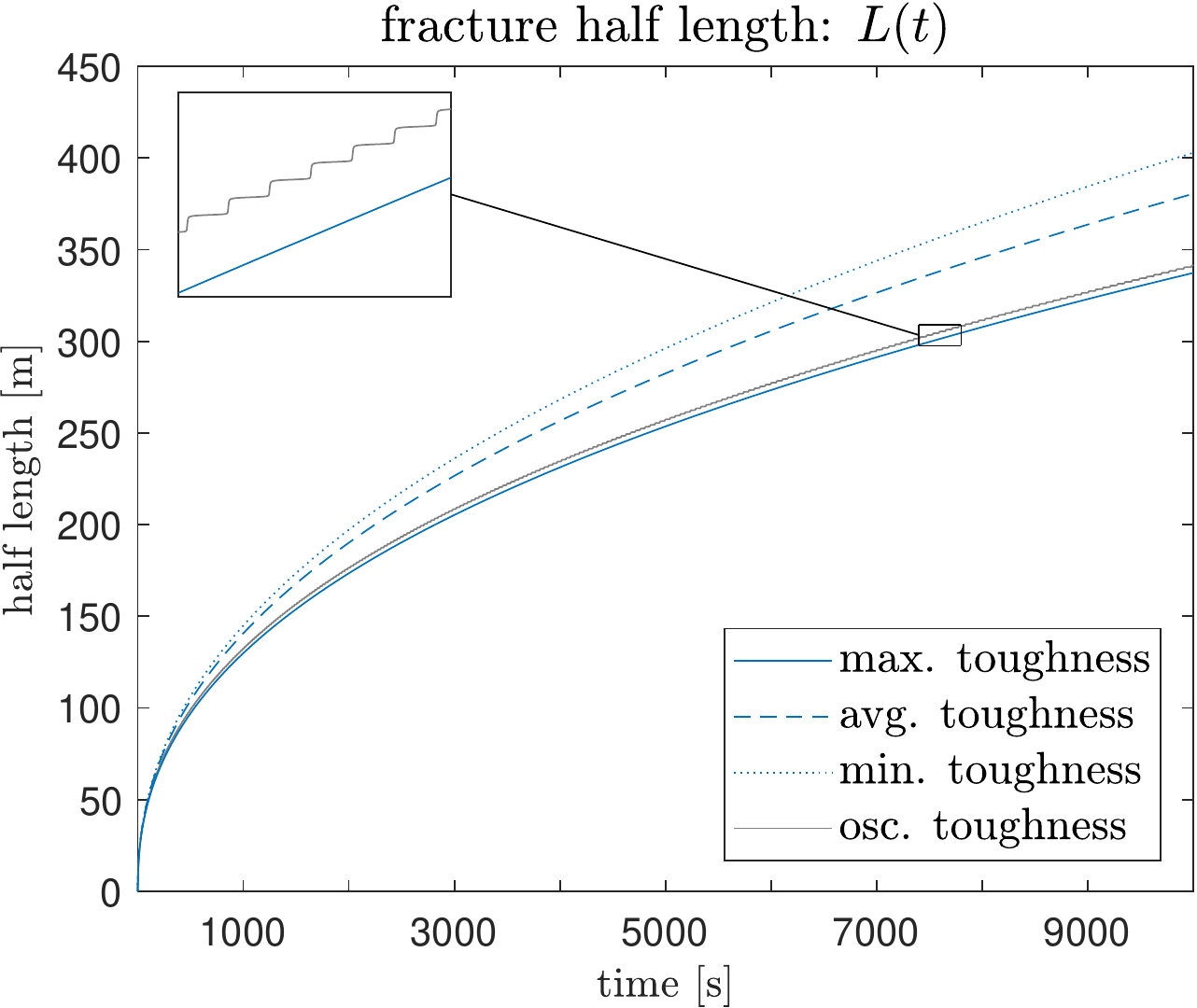}
\put(-225,155) {{\bf (e)}}
\hspace{12mm}
\includegraphics[width=0.45\textwidth]{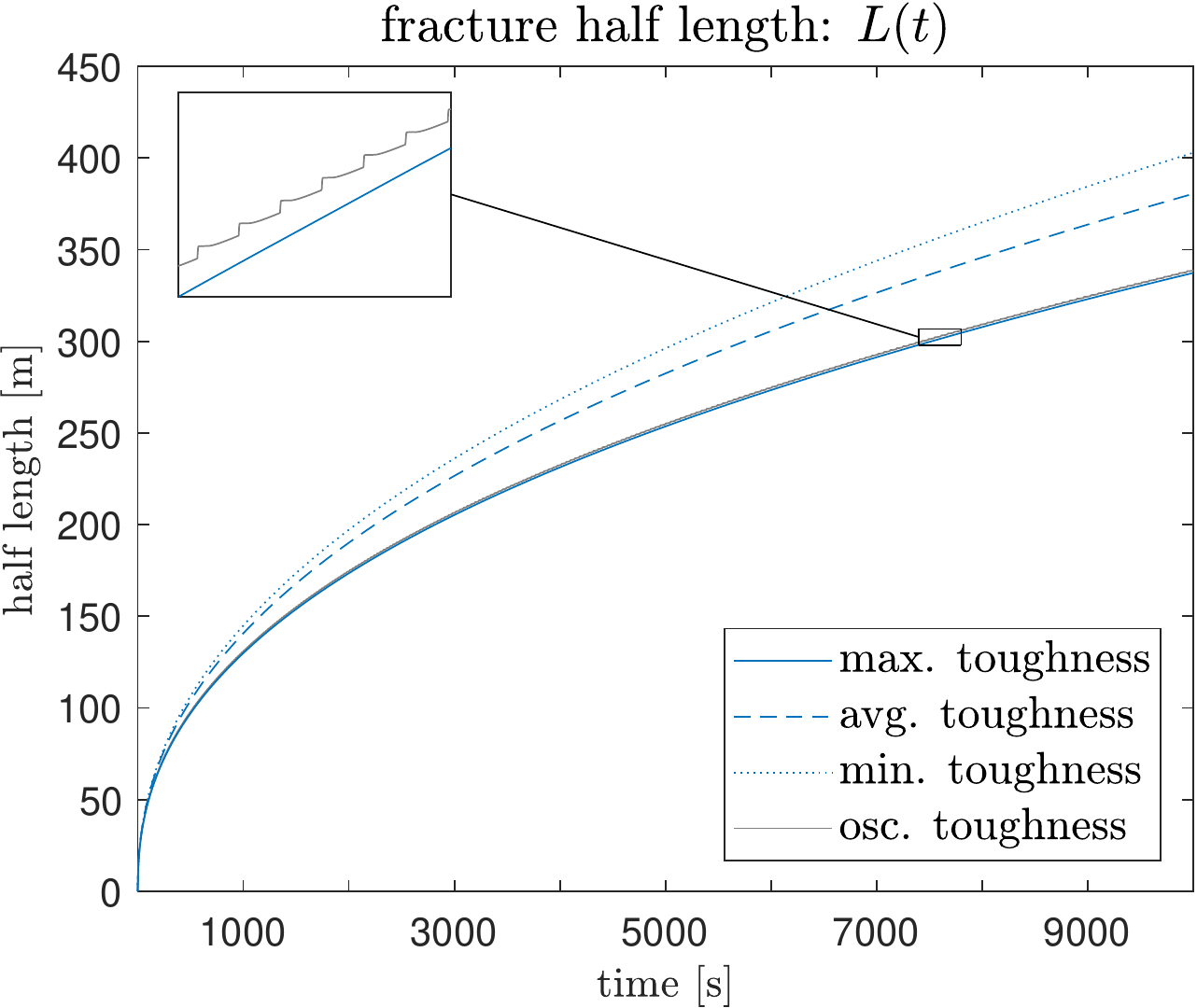}
\put(-225,155) {{\bf (f)}}
\caption{The fracture length $l(t)$ over time, for various material toughness distributions with balanced layering. The material toughness distribution, with balanced layering, are (see Table.~\ref{Table:toughness})  {\bf (a)}, {\bf (b)} Case 1; {\bf (c)}, {\bf (d)} Case 2; {\bf (e)}, {\bf (f)} Case 3. The toughness distributions are: {\bf (a)}, {\bf (c)}, {\bf (e)} sinusoidal; {\bf (b)}, {\bf (d)}, {\bf (f)} step-wise.}
\label{Fig:Length1}
\end{figure}

\begin{figure}[t!]
\centering
\includegraphics[width=0.45\textwidth]{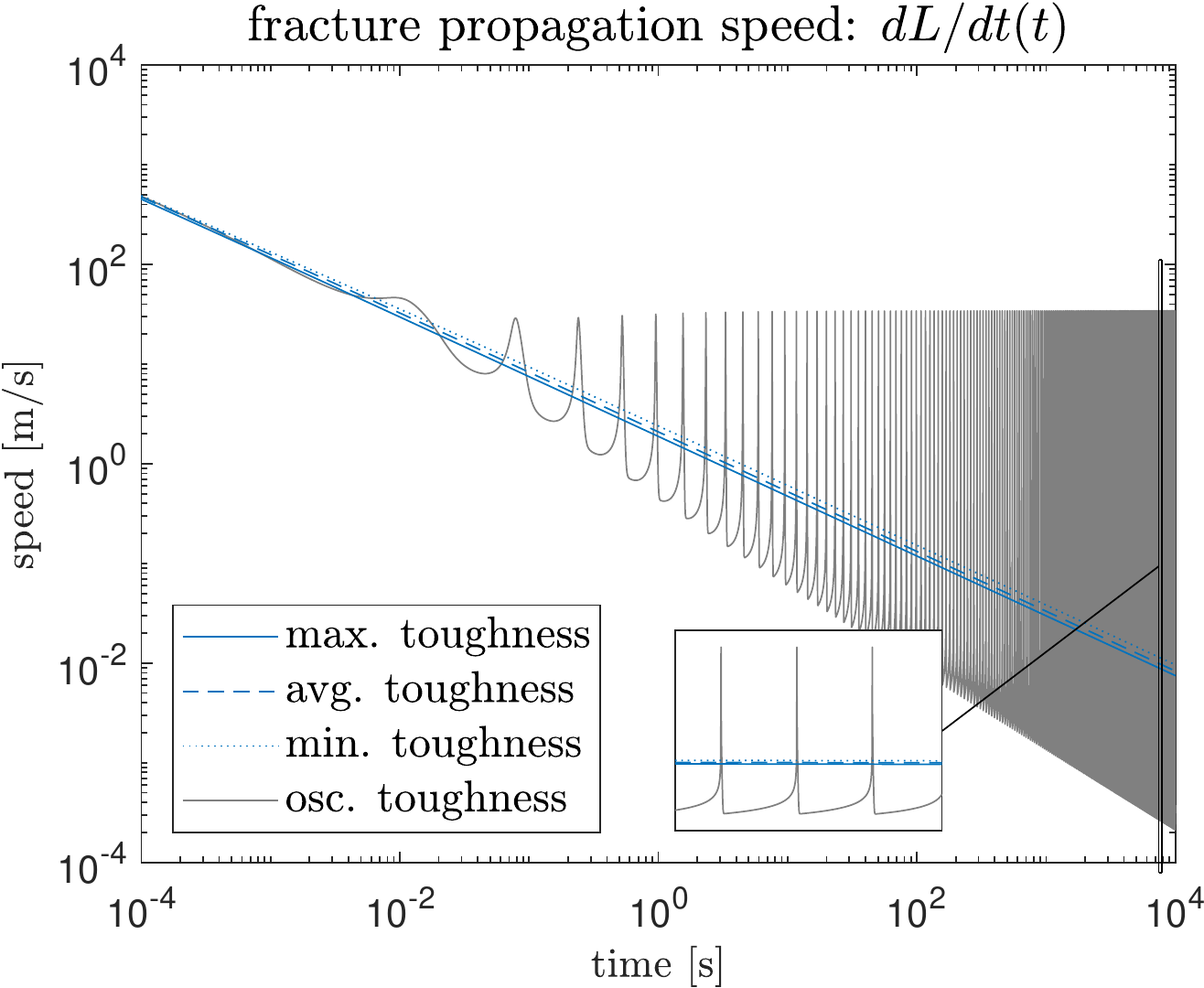}
\put(-225,155) {{\bf (a)}}
\hspace{12mm}
\includegraphics[width=0.45\textwidth]{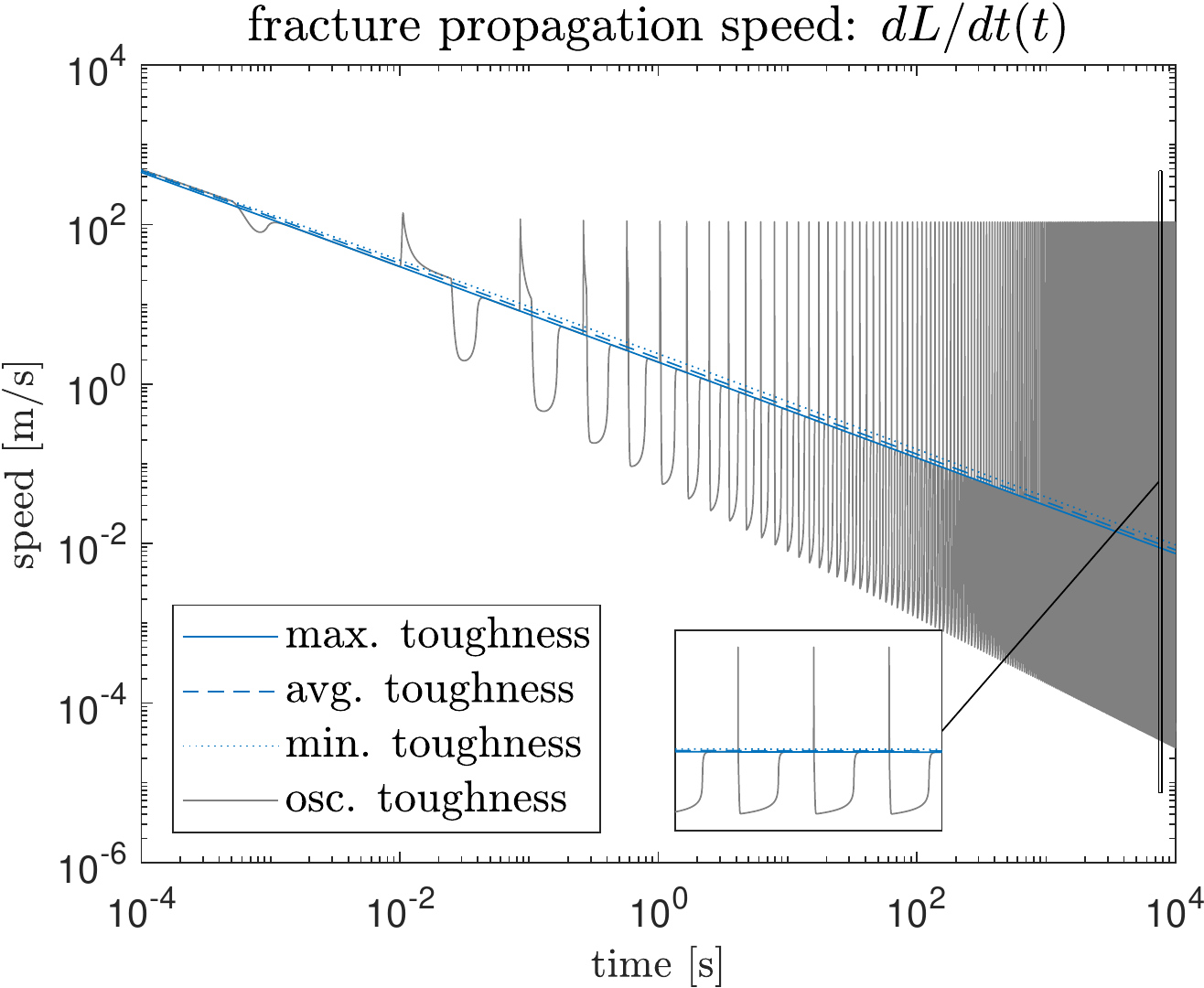}
\put(-225,155) {{\bf (b)}}
\\
\includegraphics[width=0.45\textwidth]{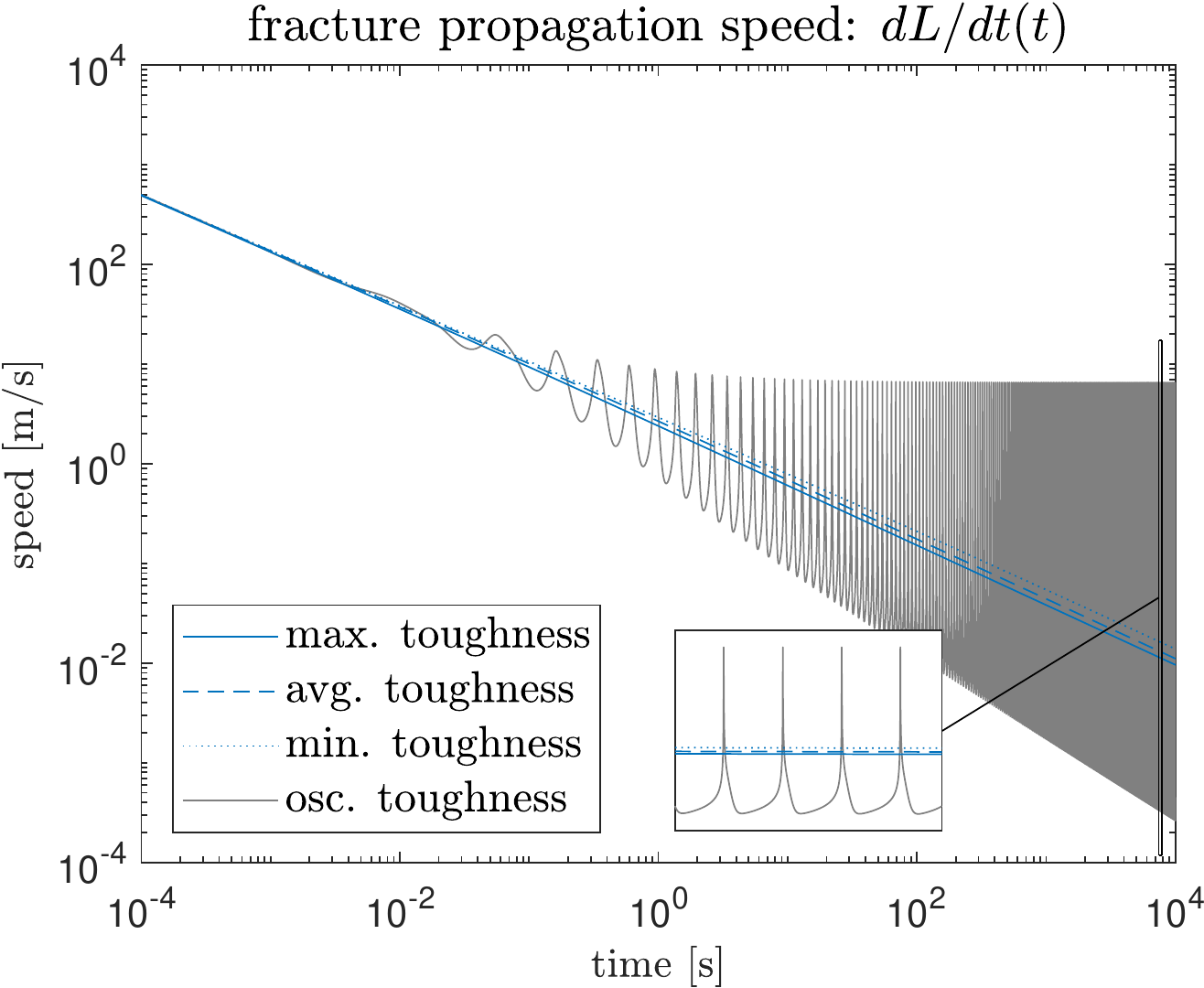}
\put(-225,155) {{\bf (c)}}
\hspace{12mm}
\includegraphics[width=0.45\textwidth]{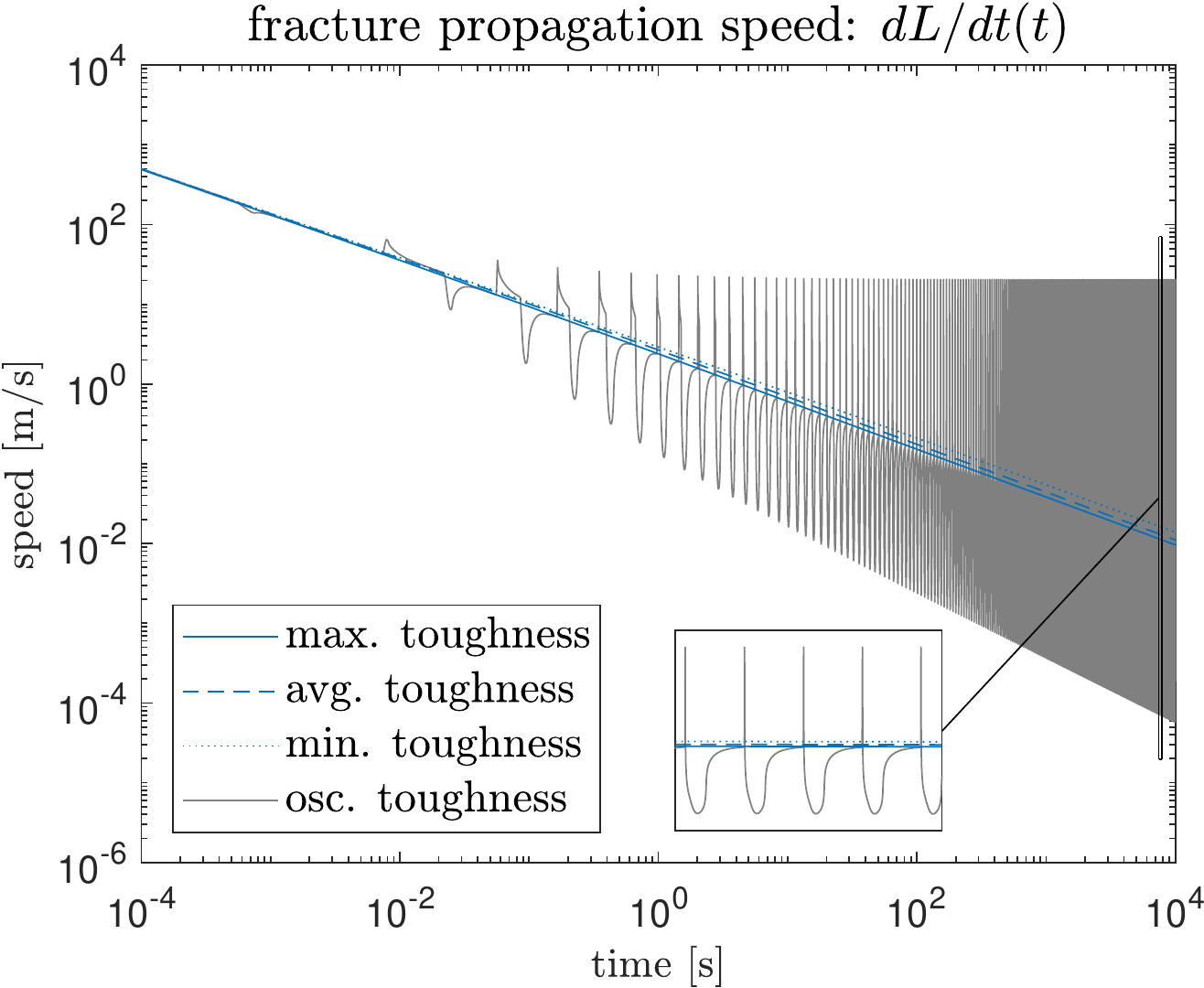}
\put(-225,155) {{\bf (d)}}
\\
\includegraphics[width=0.45\textwidth]{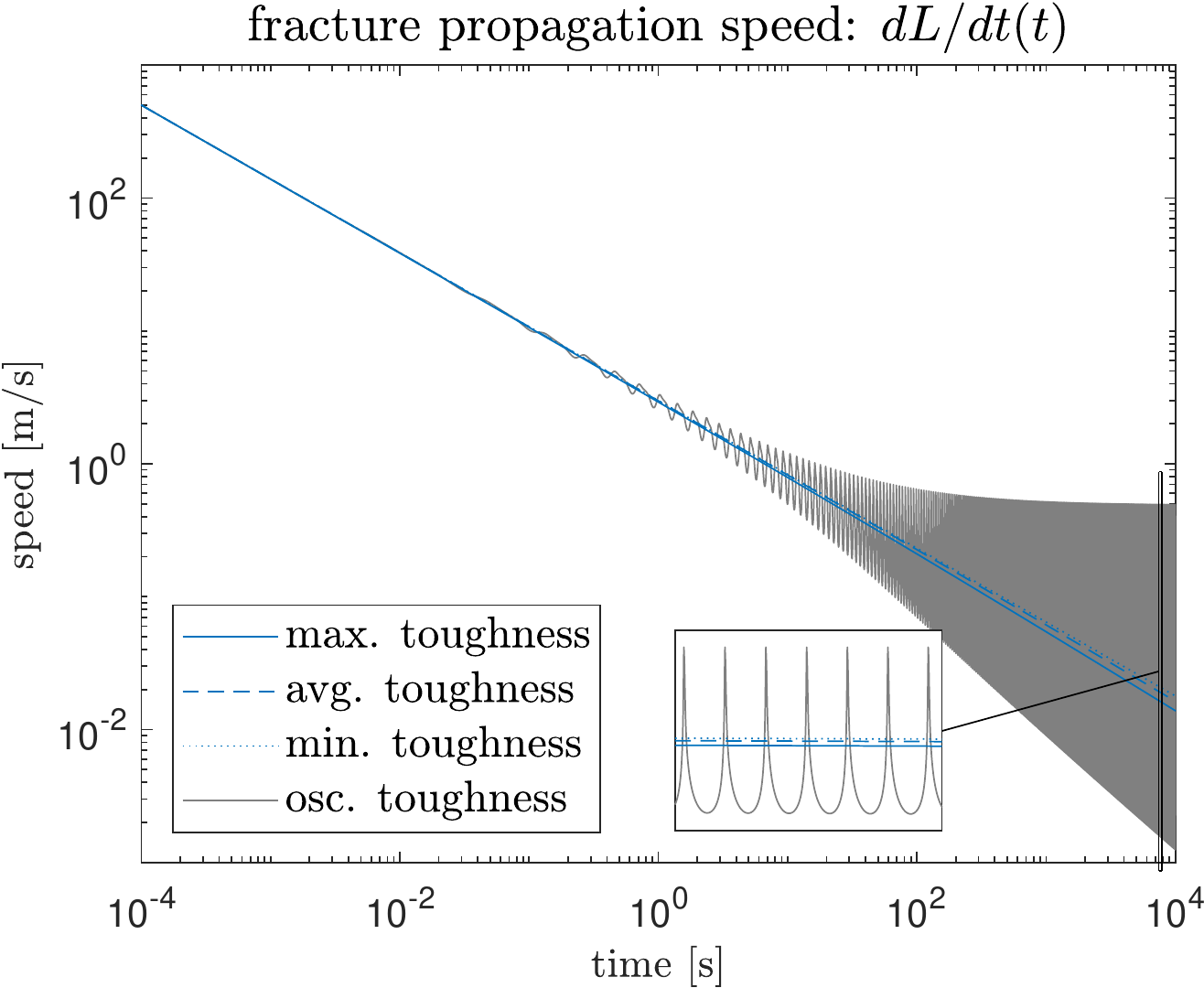}
\put(-225,155) {{\bf (e)}}
\hspace{12mm}
\includegraphics[width=0.45\textwidth]{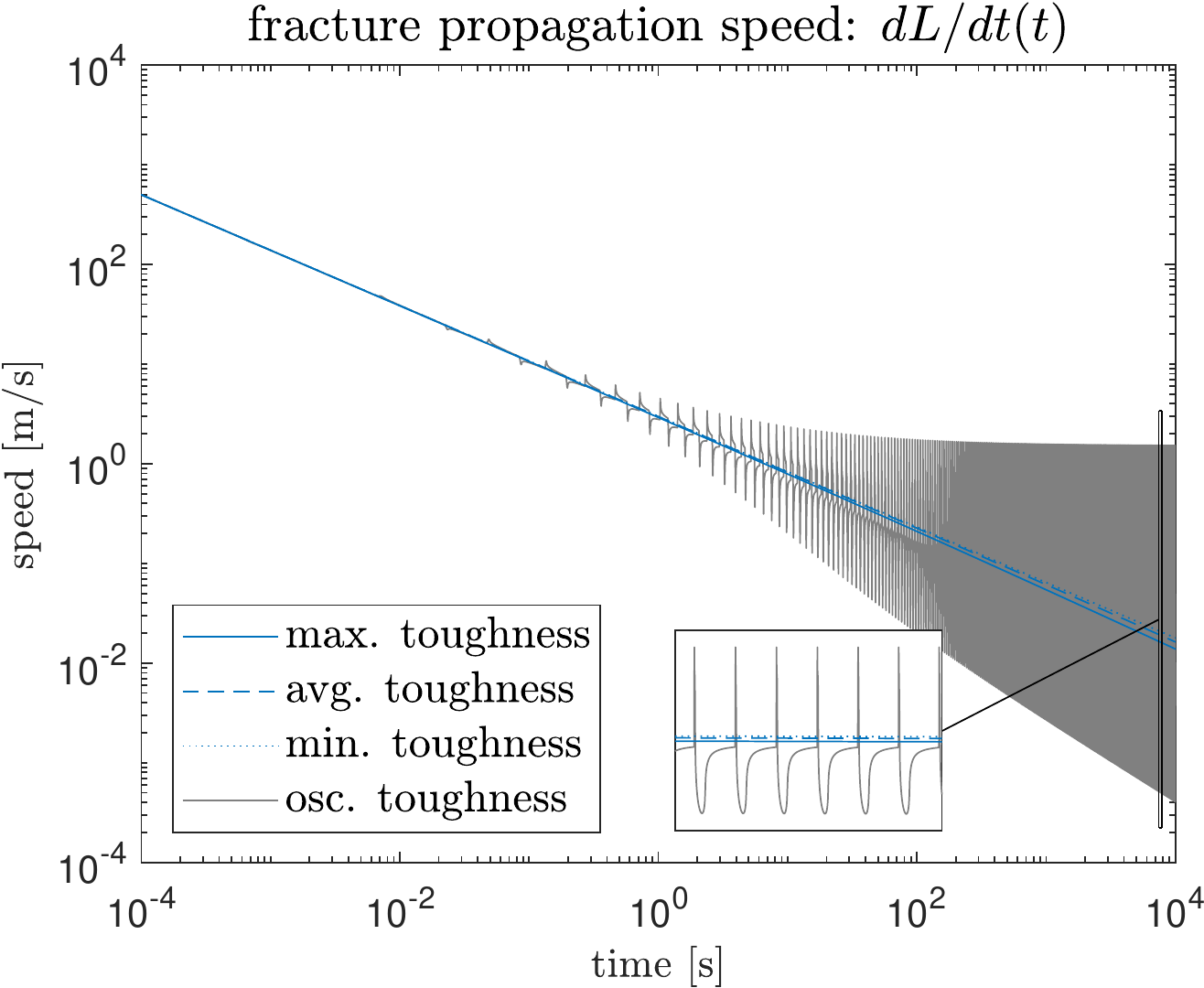}
\put(-225,155) {{\bf (f)}}
\caption{The tip velocity $dl/dt$ over time, for various material toughness distributions with balanced layering. The material toughness distribution, with balanced layering, are (see Table.~\ref{Table:toughness})  {\bf (a)}, {\bf (b)} Case 1; {\bf (c)}, {\bf (d)} Case 2; {\bf (e)}, {\bf (f)} Case 3. The toughness distributions are: {\bf (a)}, {\bf (c)}, {\bf (e)} sinusoidal; {\bf (b)}, {\bf (d)}, {\bf (f)} step-wise.}
\label{Fig:Speed1}
\end{figure}

\begin{figure}[t!]
\centering
\includegraphics[width=0.45\textwidth]{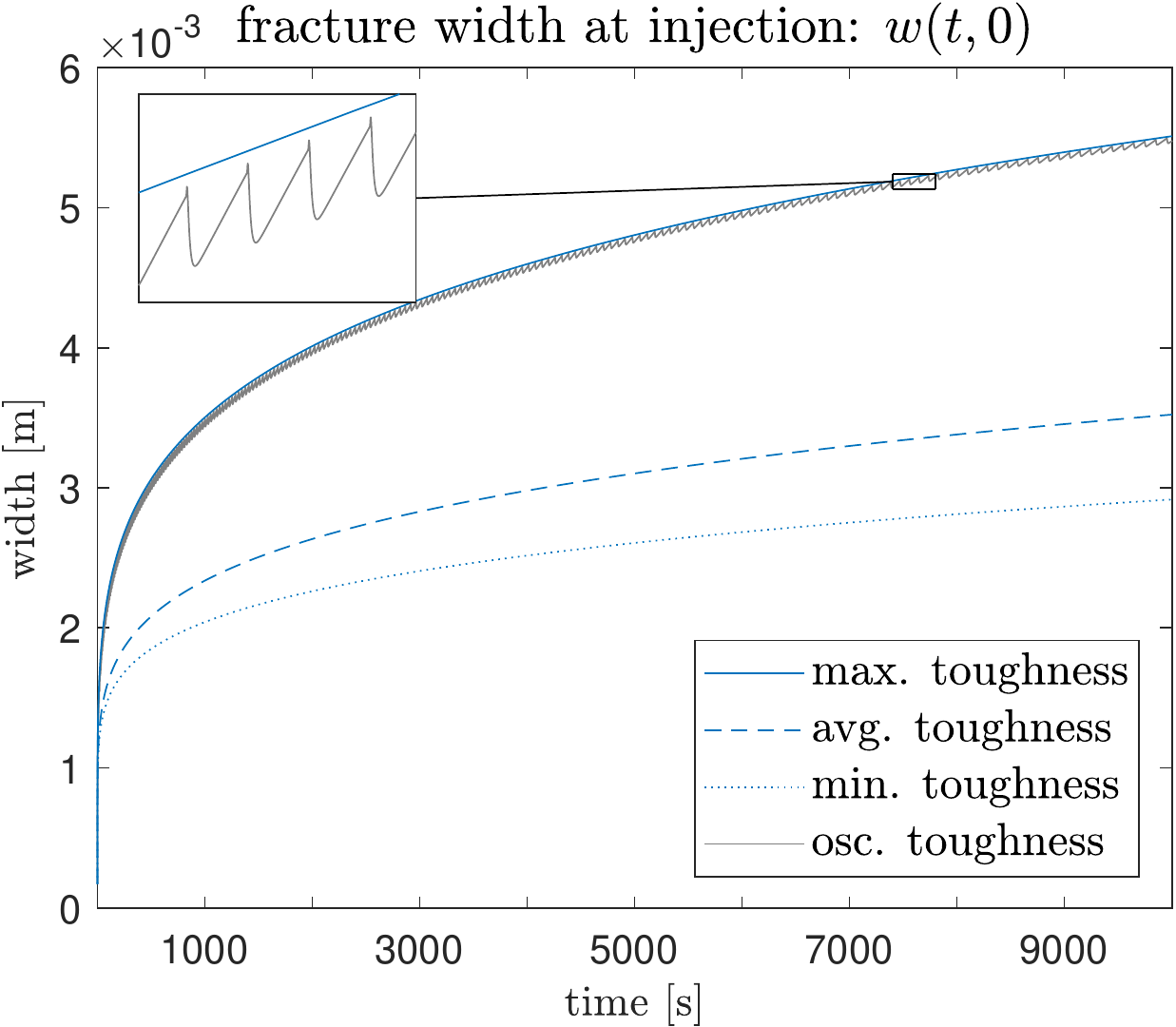}
\put(-225,155) {{\bf (a)}}
\hspace{12mm}
\includegraphics[width=0.45\textwidth]{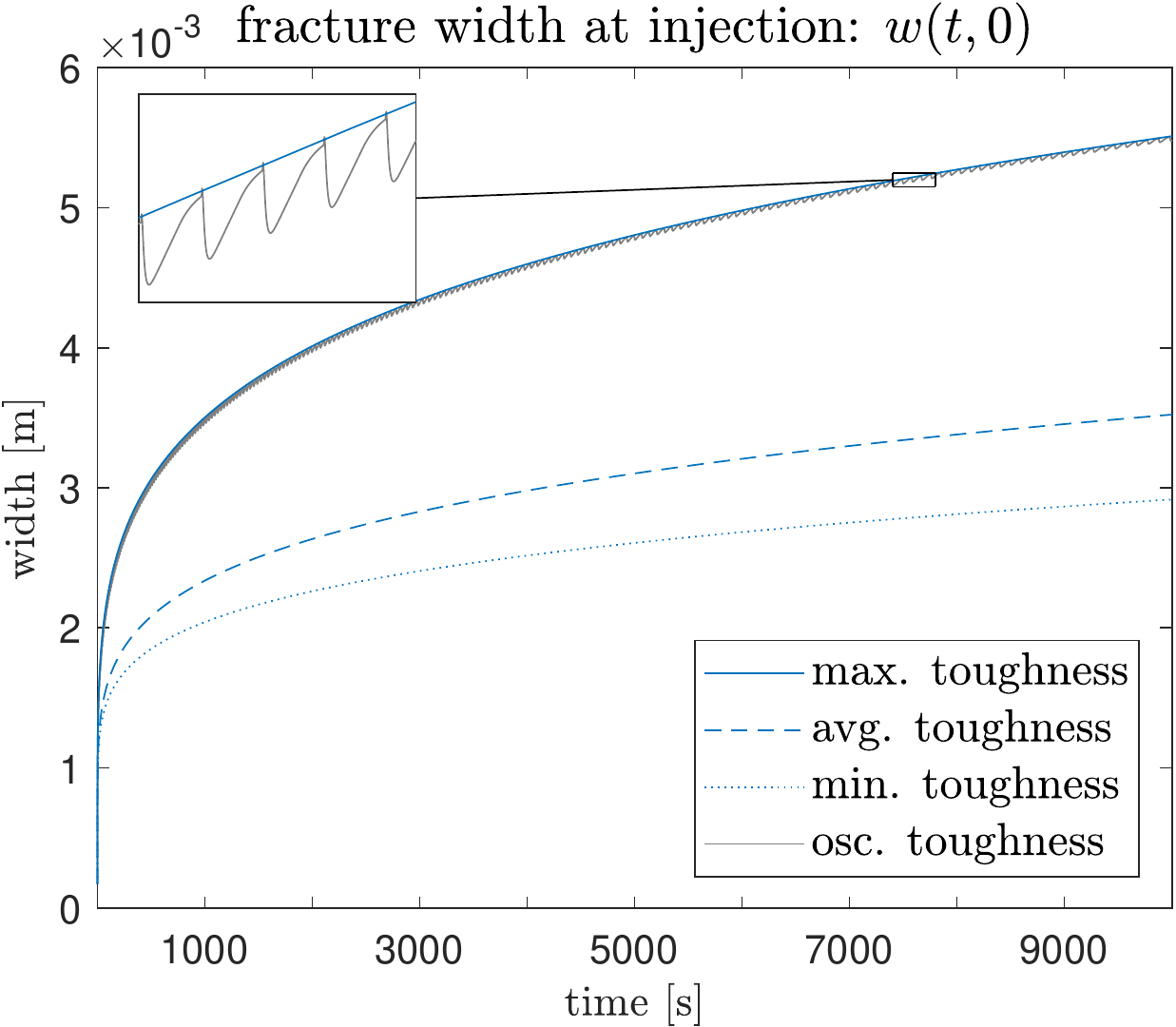}
\put(-225,155) {{\bf (b)}}
\\
\includegraphics[width=0.45\textwidth]{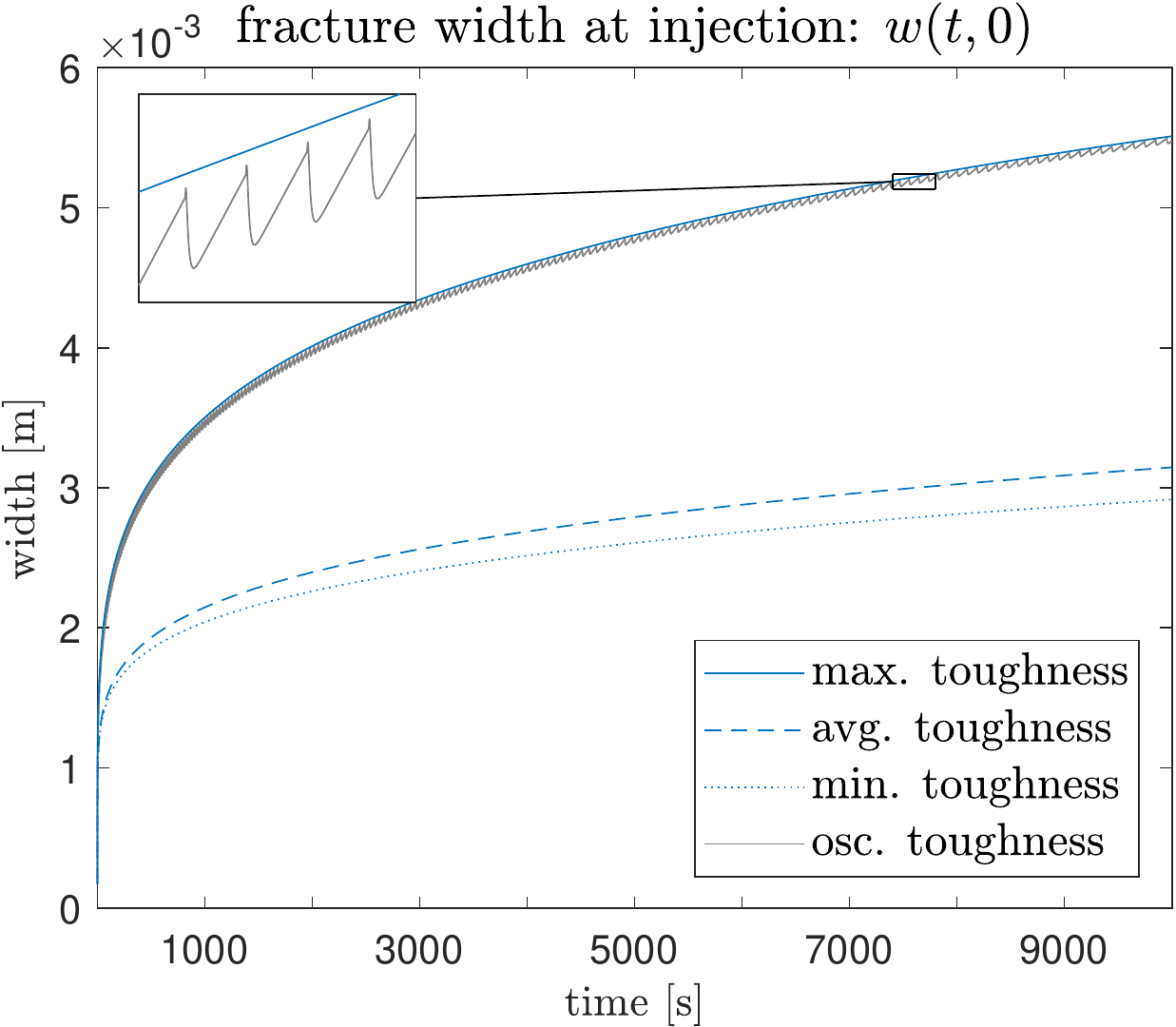}
\put(-225,155) {{\bf (c)}}
\hspace{12mm}
\includegraphics[width=0.45\textwidth]{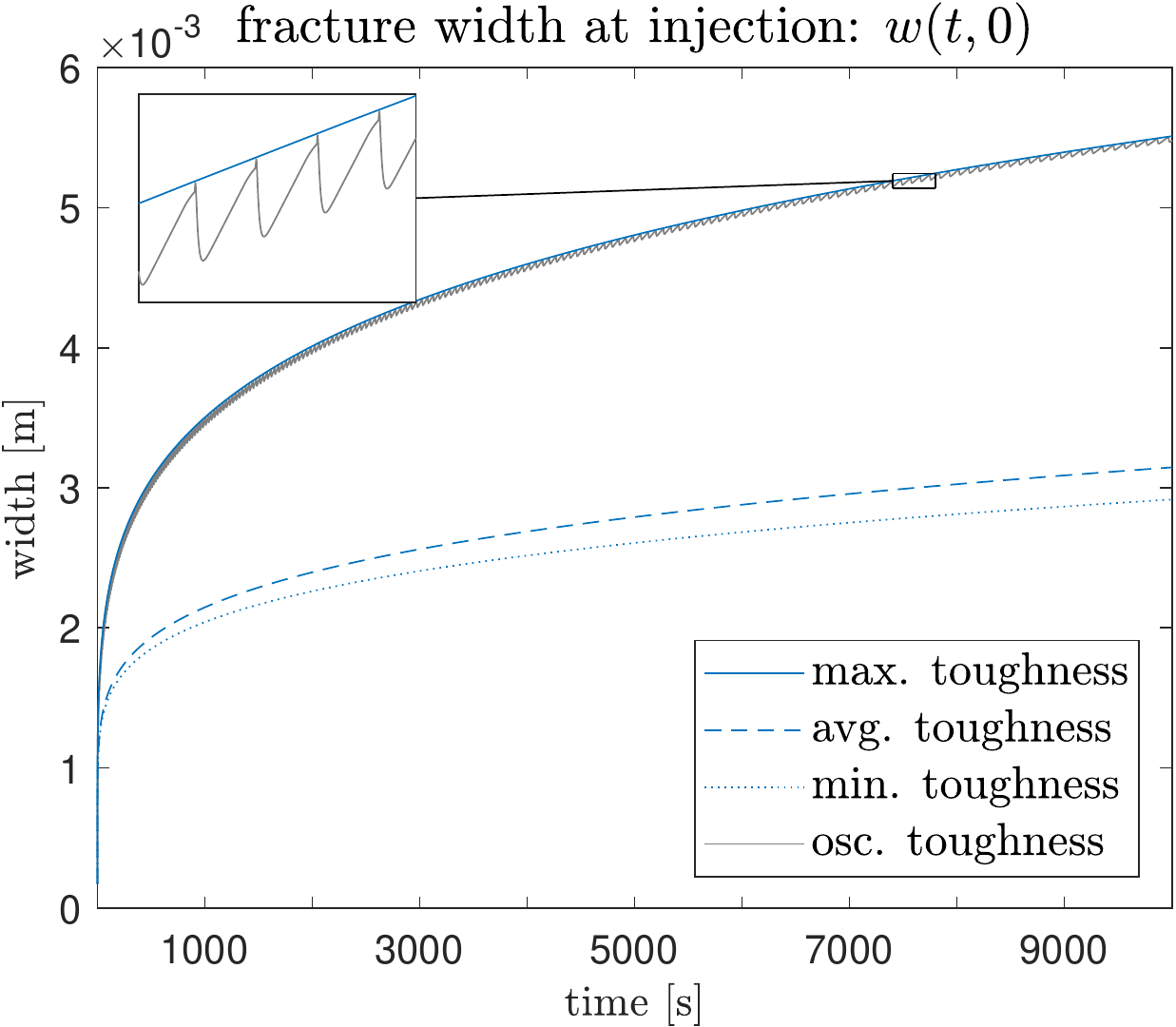}
\put(-225,155) {{\bf (d)}}
\\
\includegraphics[width=0.45\textwidth]{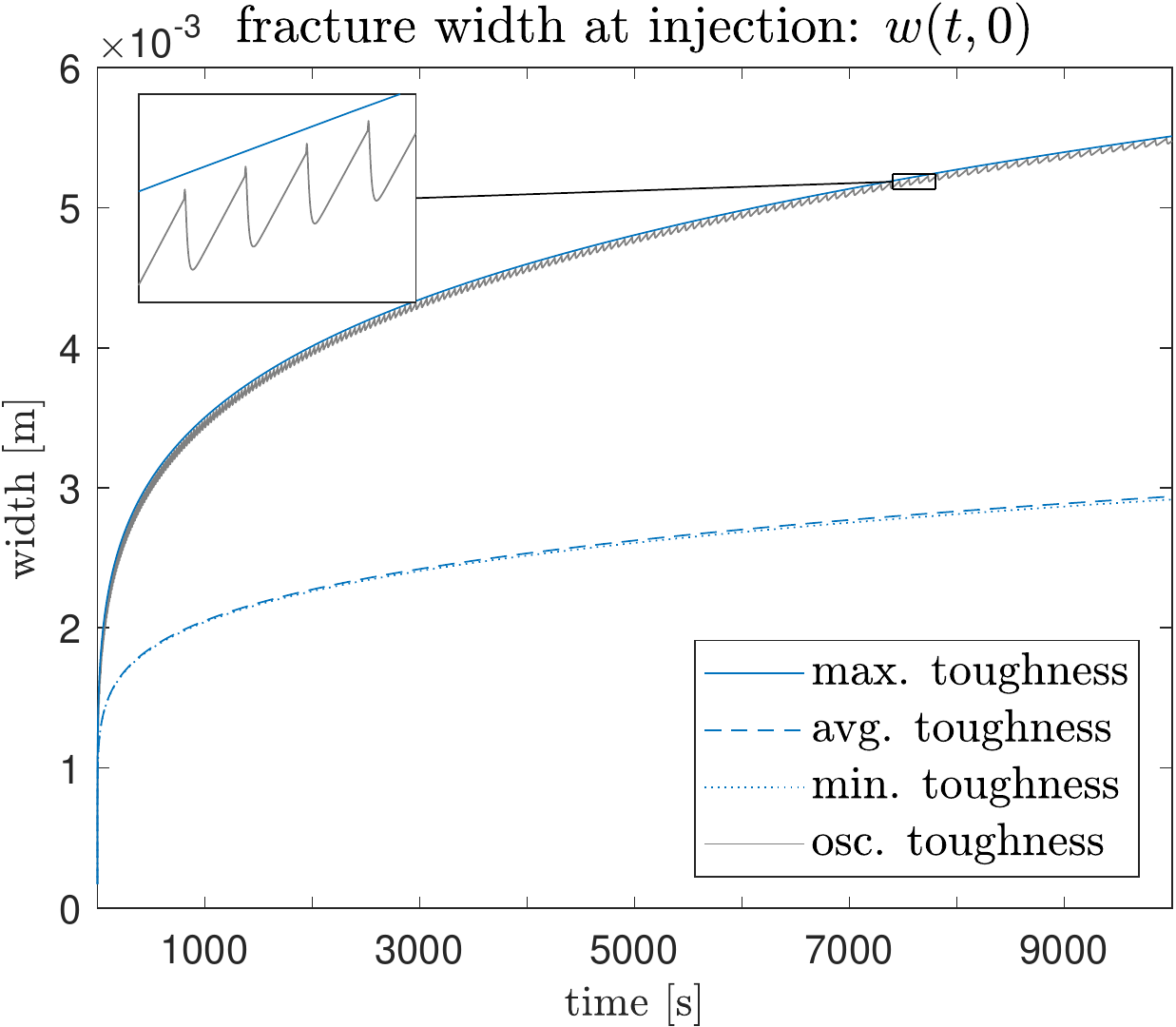}
\put(-225,155) {{\bf (e)}}
\hspace{12mm}
\includegraphics[width=0.45\textwidth]{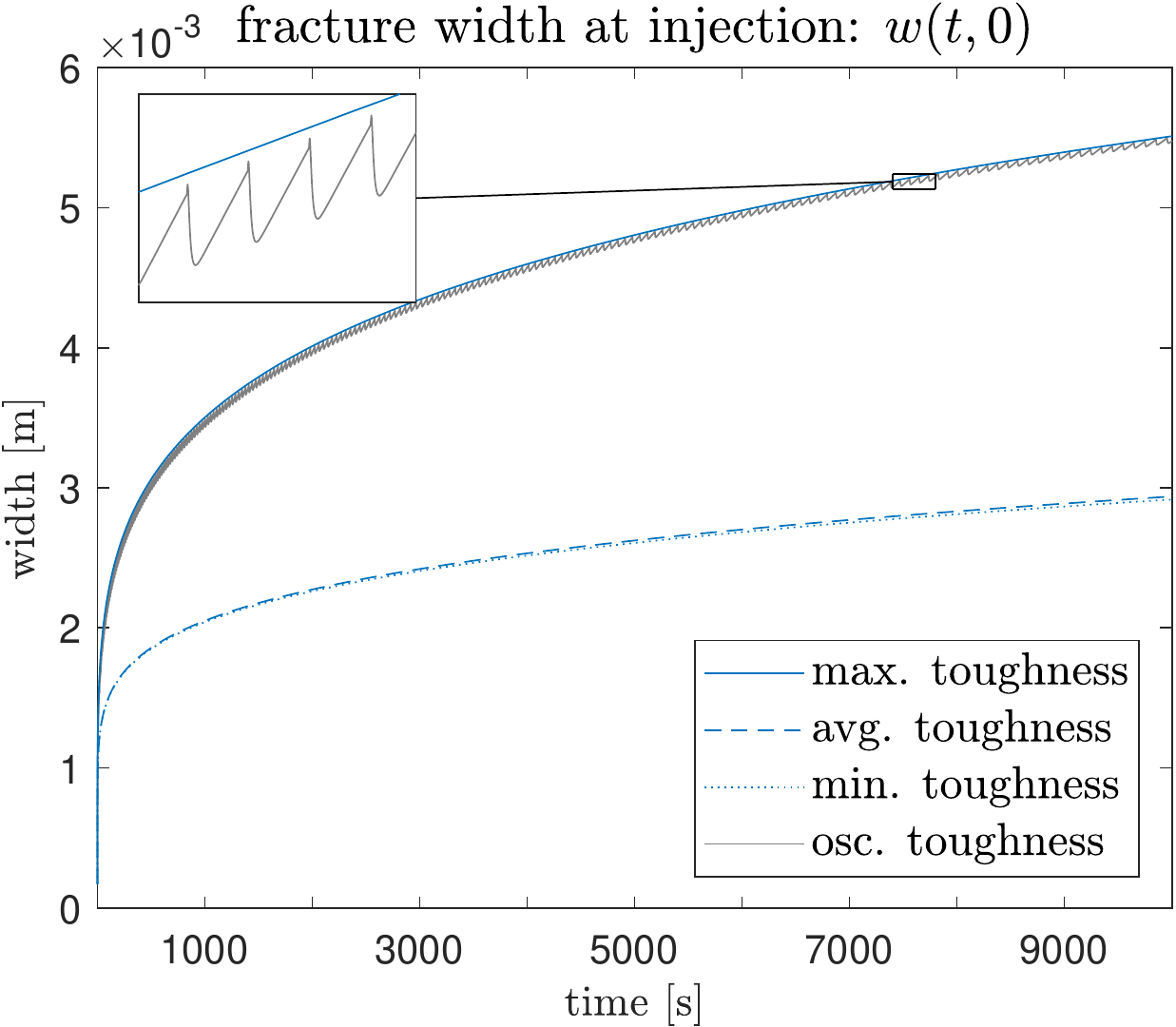}
\put(-225,155) {{\bf (f)}}
\caption{The fracture aperture at the crack opening $w(t,0)$ over time, for a material toughness in Case 2 (toughness-transient, see Table.~\ref{Table:toughness}), with unbalanced layering. The unbalanced layering constant \eqref{defh} is taken as  {\bf (a)}, {\bf (b)} $h=0.25$; {\bf (c)}, {\bf (d)} $h=0.1$; {\bf (e)}, {\bf (f)} $h=0.01$. The toughness distributions are: {\bf (a)}, {\bf (c)}, {\bf (e)} sinusoidal; {\bf (b)}, {\bf (d)}, {\bf (f)} step-wise.}
\label{Fig:Aperture2}
\end{figure}

\begin{figure}[t!]
\centering
\includegraphics[width=0.45\textwidth]{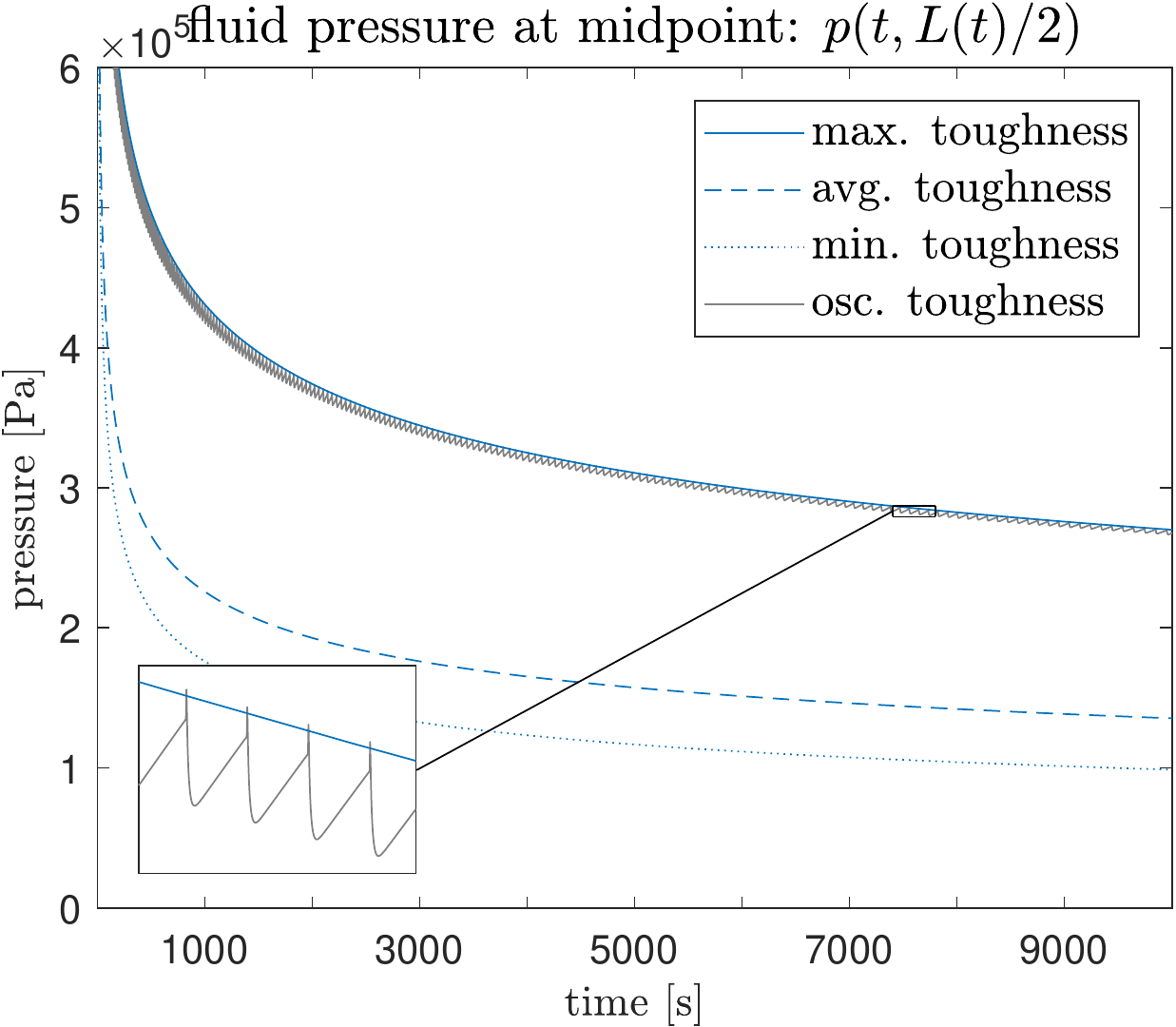}
\put(-225,155) {{\bf (a)}}
\hspace{12mm}
\includegraphics[width=0.45\textwidth]{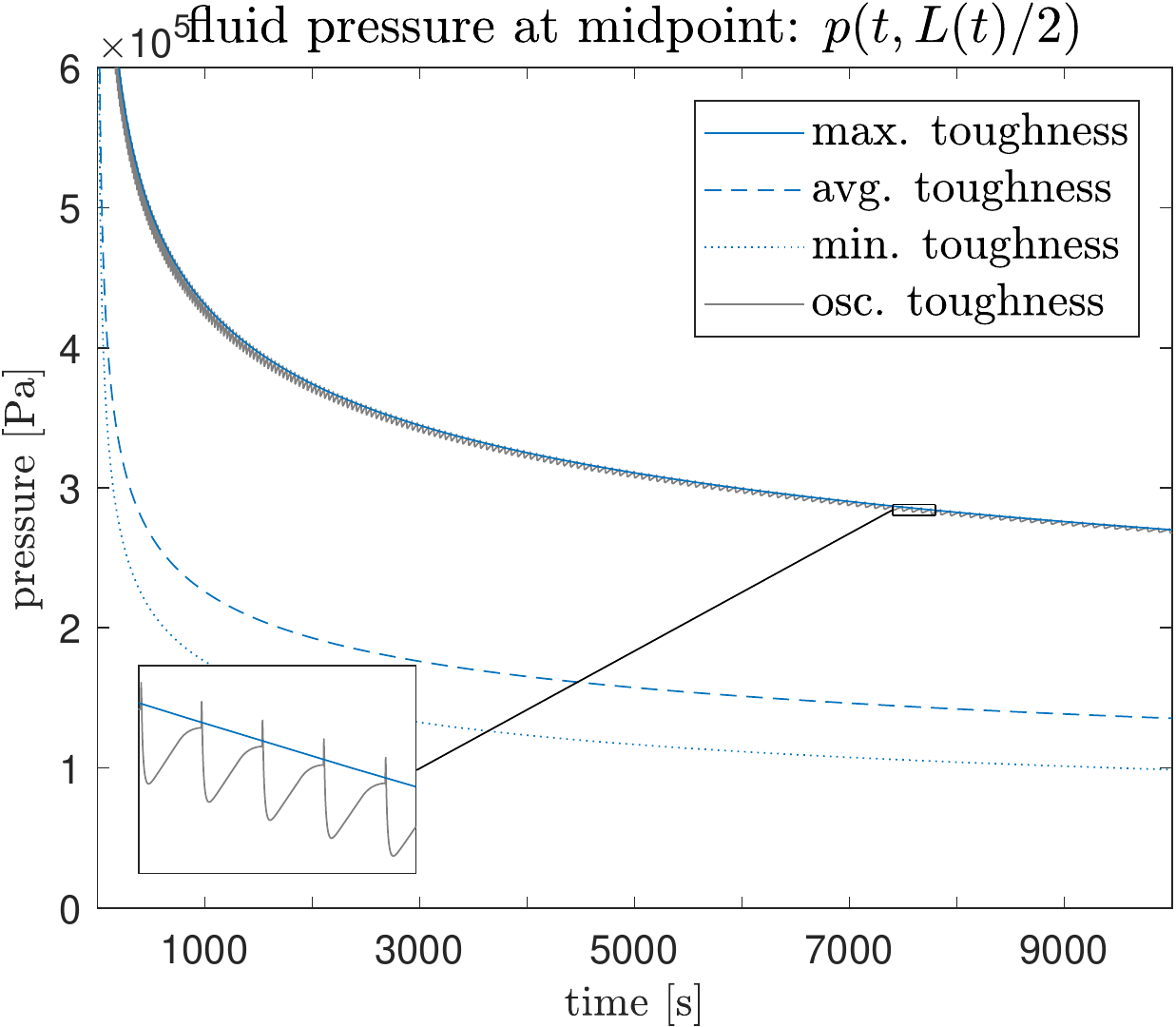}
\put(-225,155) {{\bf (b)}}
\\
\includegraphics[width=0.45\textwidth]{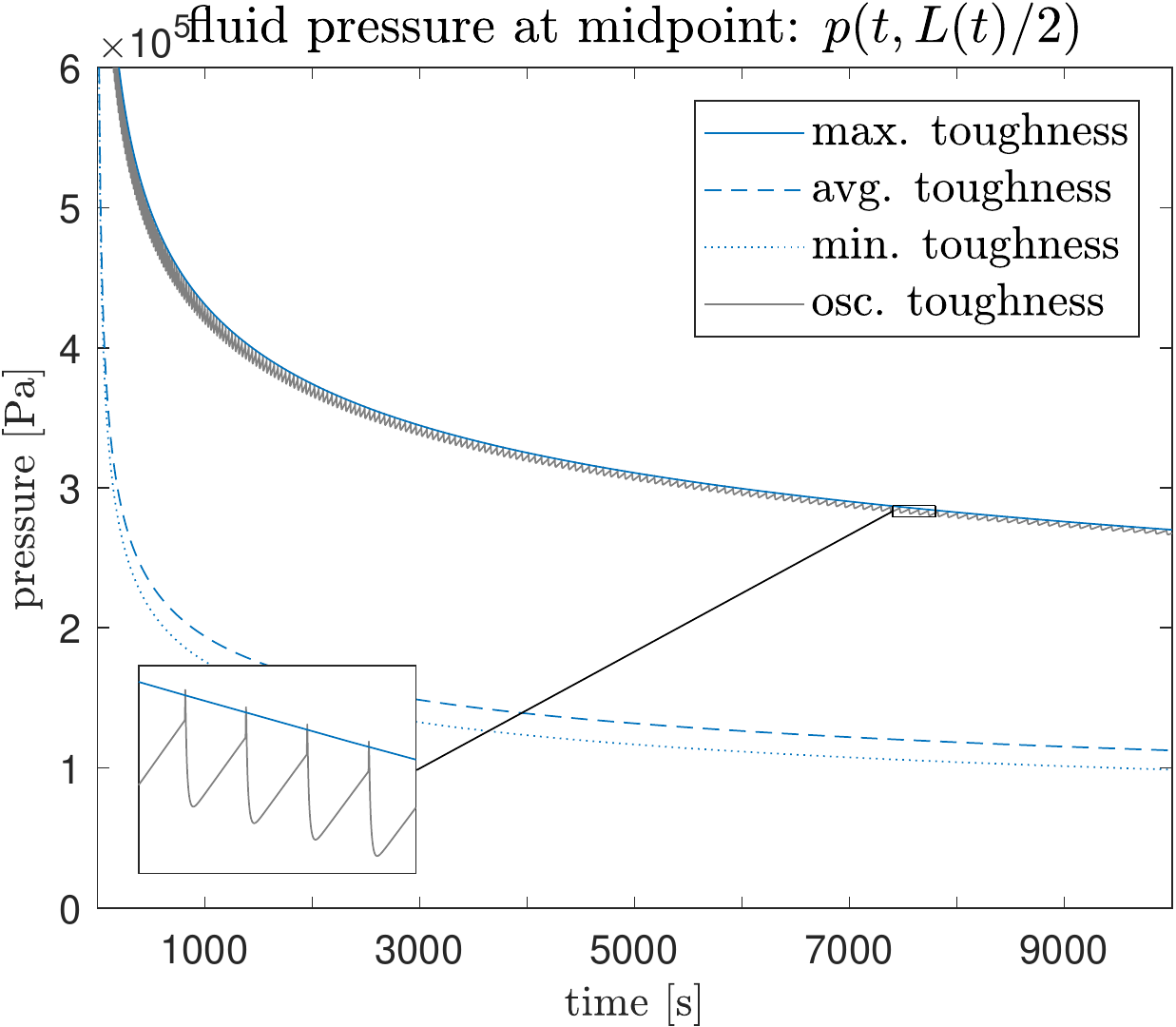}
\put(-225,155) {{\bf (c)}}
\hspace{12mm}
\includegraphics[width=0.45\textwidth]{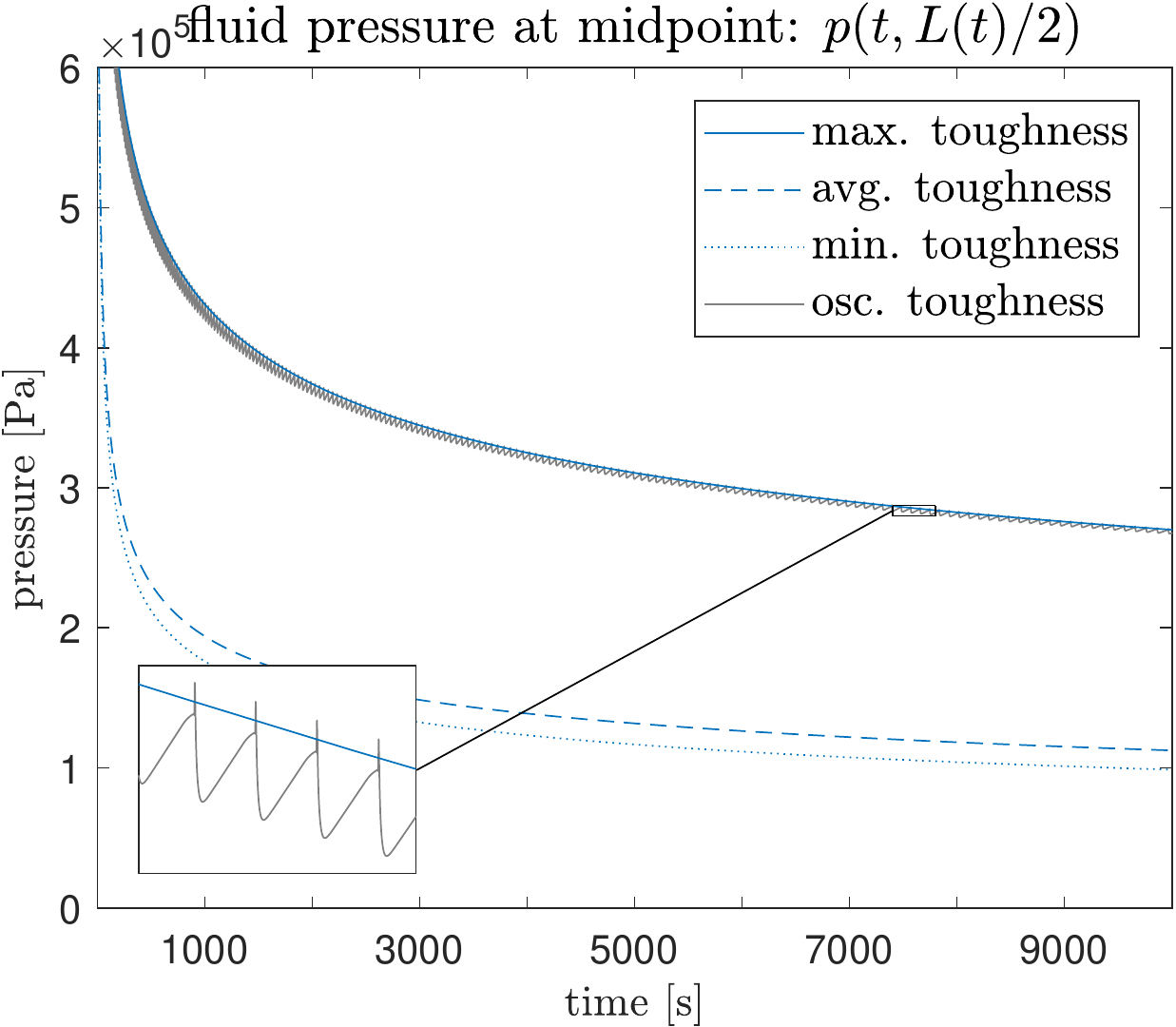}
\put(-225,155) {{\bf (d)}}
\\
\includegraphics[width=0.45\textwidth]{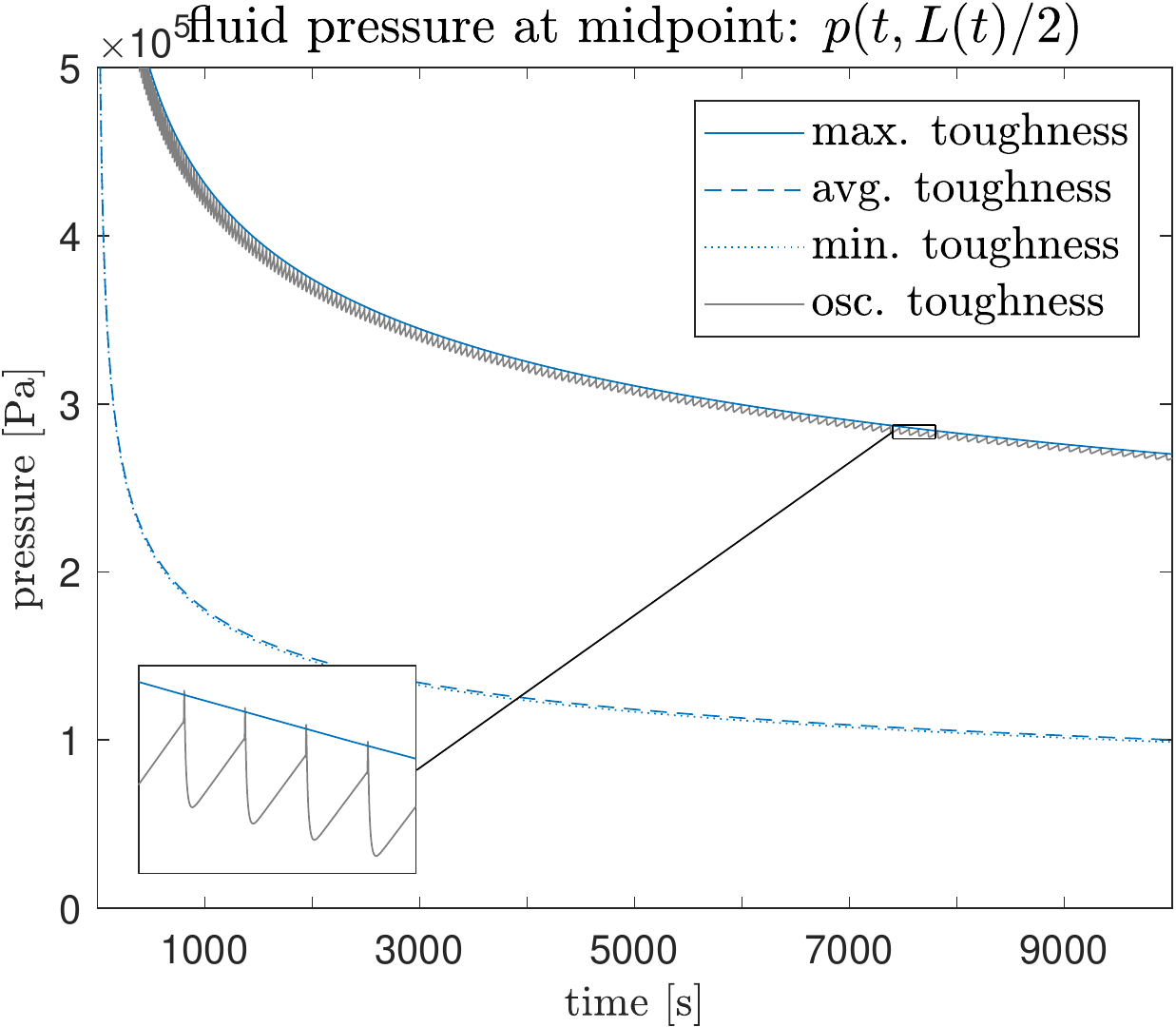}
\put(-225,155) {{\bf (e)}}
\hspace{12mm}
\includegraphics[width=0.45\textwidth]{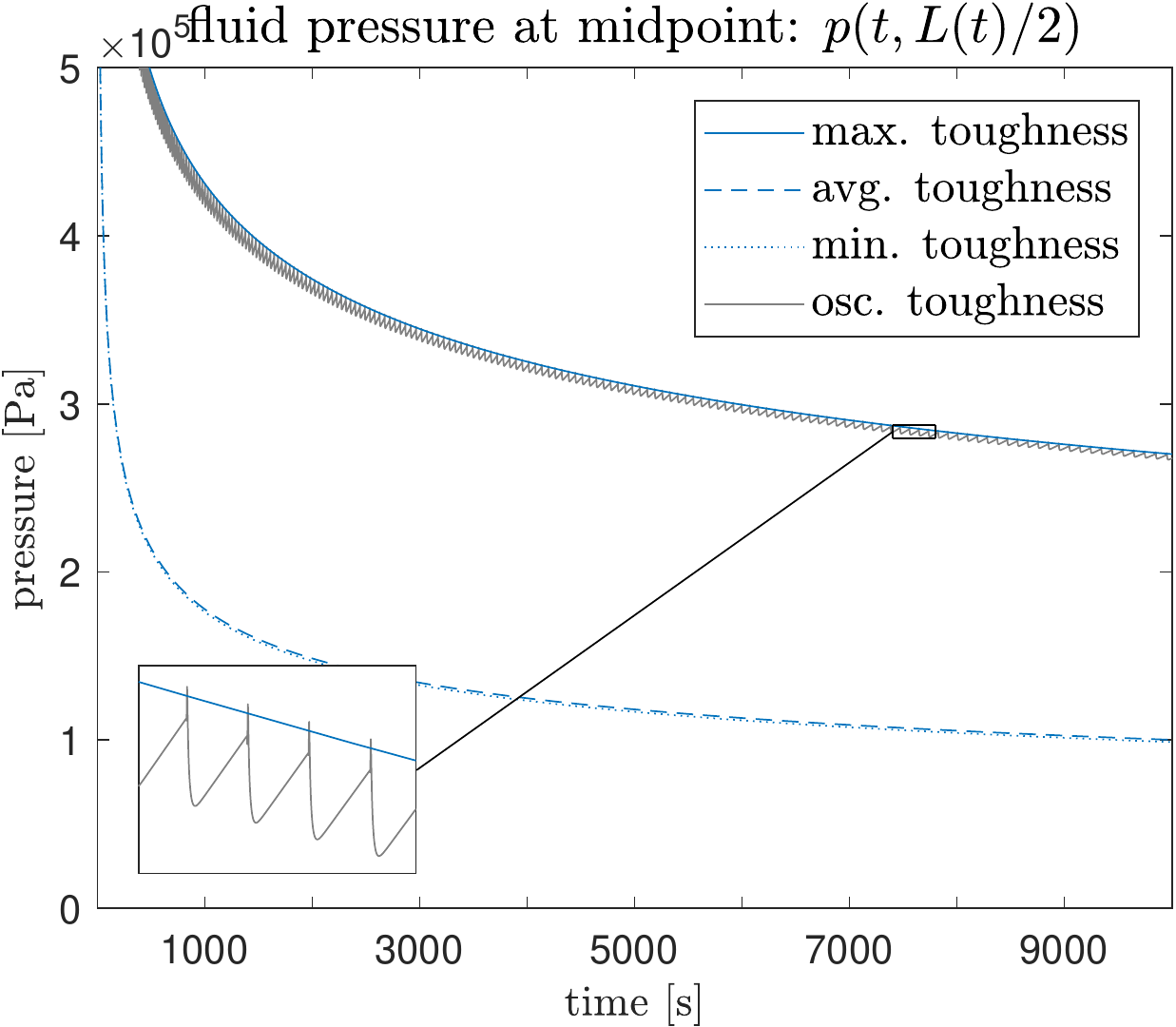}
\put(-225,155) {{\bf (f)}}
\caption{The fracture pressure at the mid-length $p(t,L(t)/2)$ over time, for a material toughness in Case 2 (toughness-transient, see Table.~\ref{Table:toughness}), with unbalanced layering. The unbalanced layering constant \eqref{defh} is taken as  {\bf (a)}, {\bf (b)} $h=0.25$; {\bf (c)}, {\bf (d)} $h=0.1$; {\bf (e)}, {\bf (f)} $h=0.01$. The toughness distributions are: {\bf (a)}, {\bf (c)}, {\bf (e)} sinusoidal; {\bf (b)}, {\bf (d)}, {\bf (f)} step-wise.}
\label{Fig:Pressure2}
\end{figure}

\begin{figure}[t!]
\centering
\includegraphics[width=0.45\textwidth]{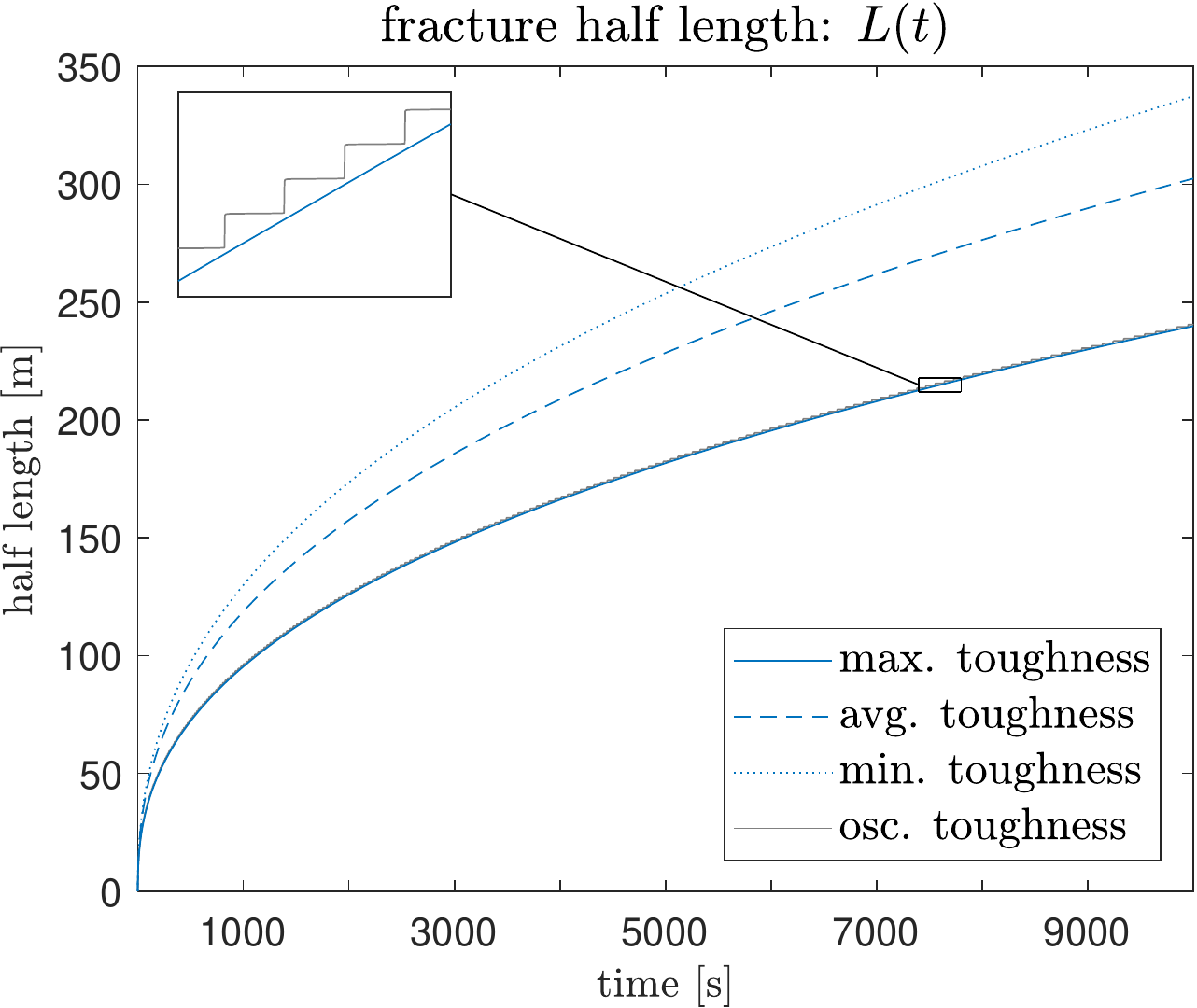}
\put(-225,155) {{\bf (a)}}
\hspace{12mm}
\includegraphics[width=0.45\textwidth]{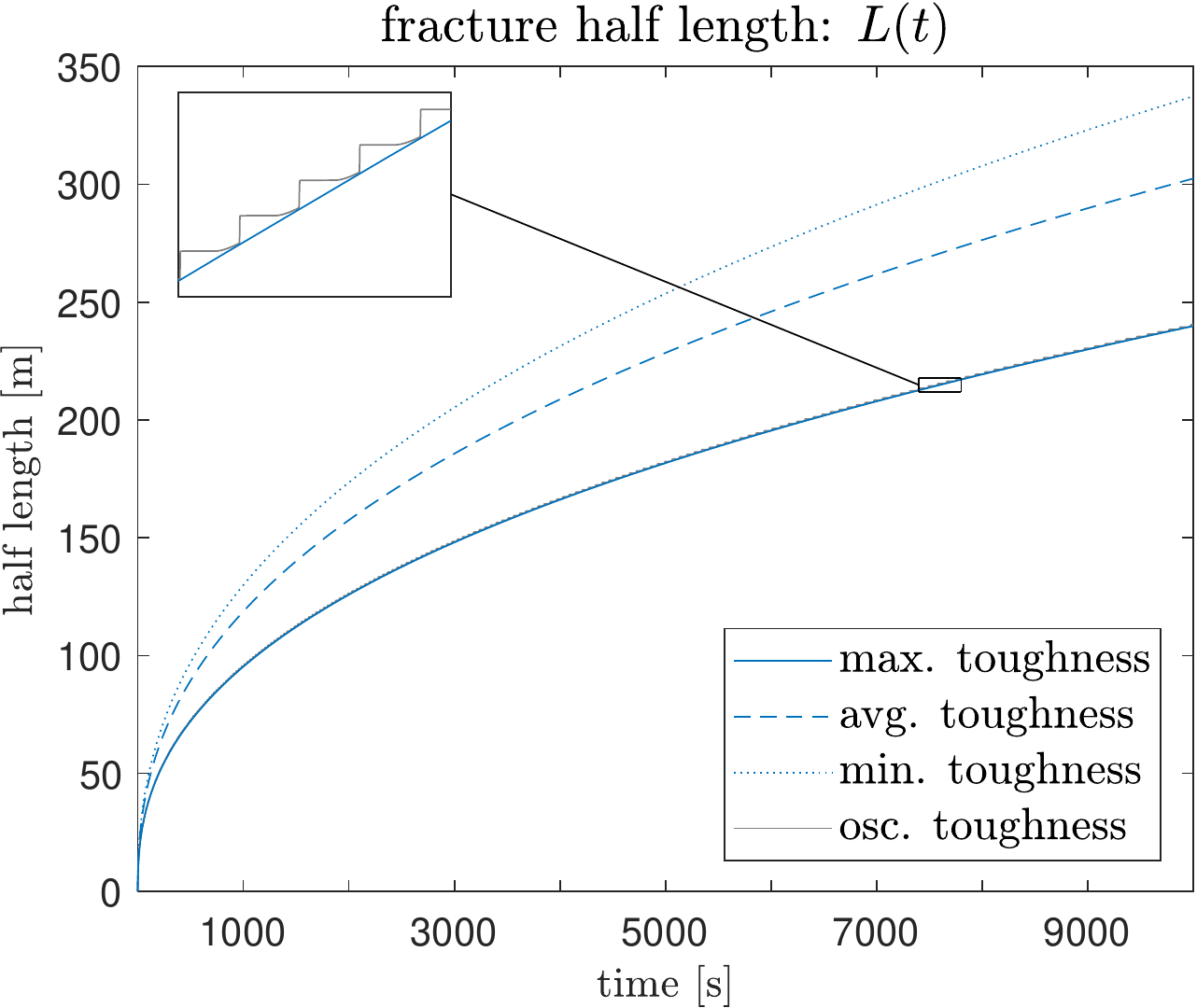}
\put(-225,155) {{\bf (b)}}
\\
\includegraphics[width=0.45\textwidth]{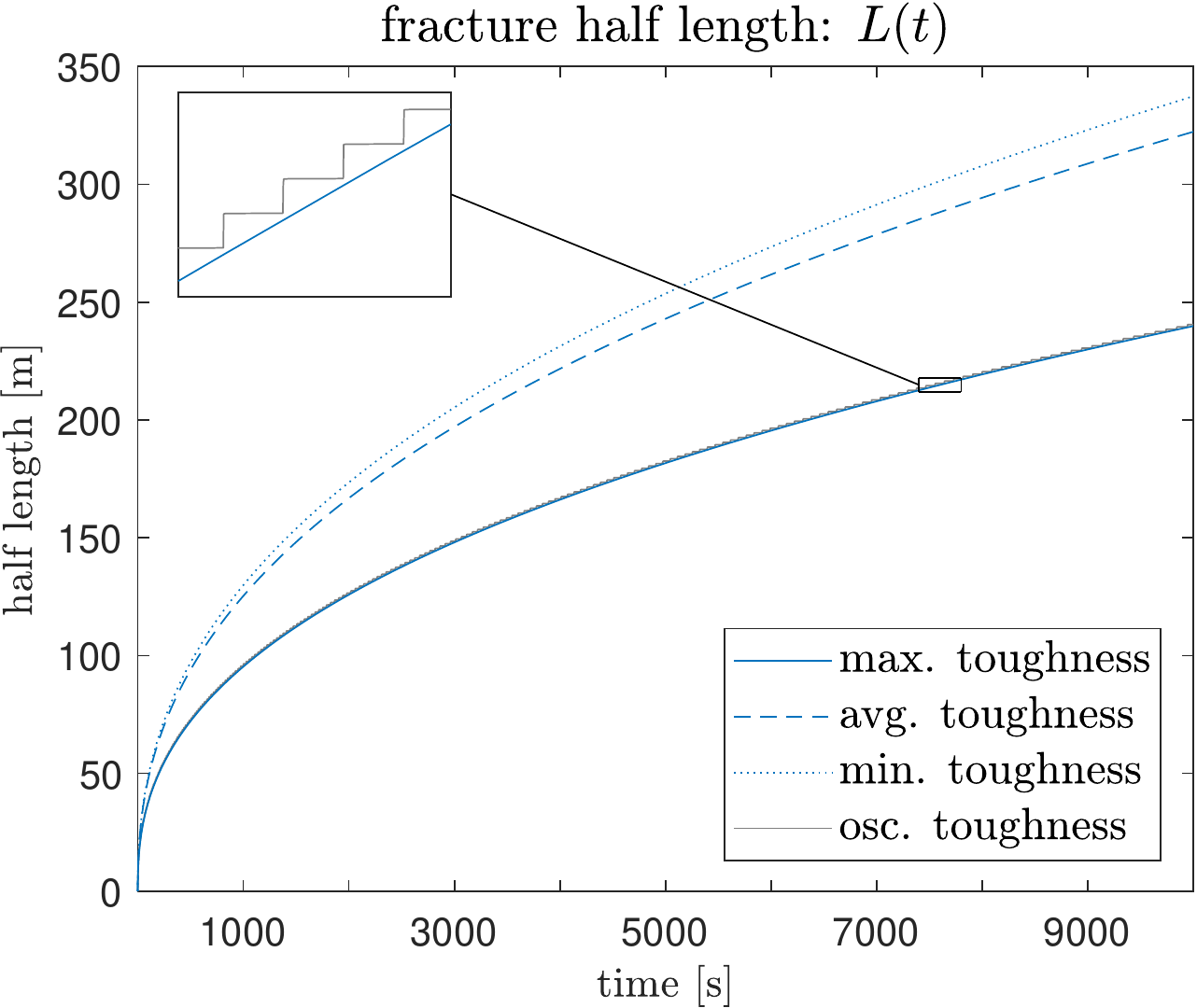}
\put(-225,155) {{\bf (c)}}
\hspace{12mm}
\includegraphics[width=0.45\textwidth]{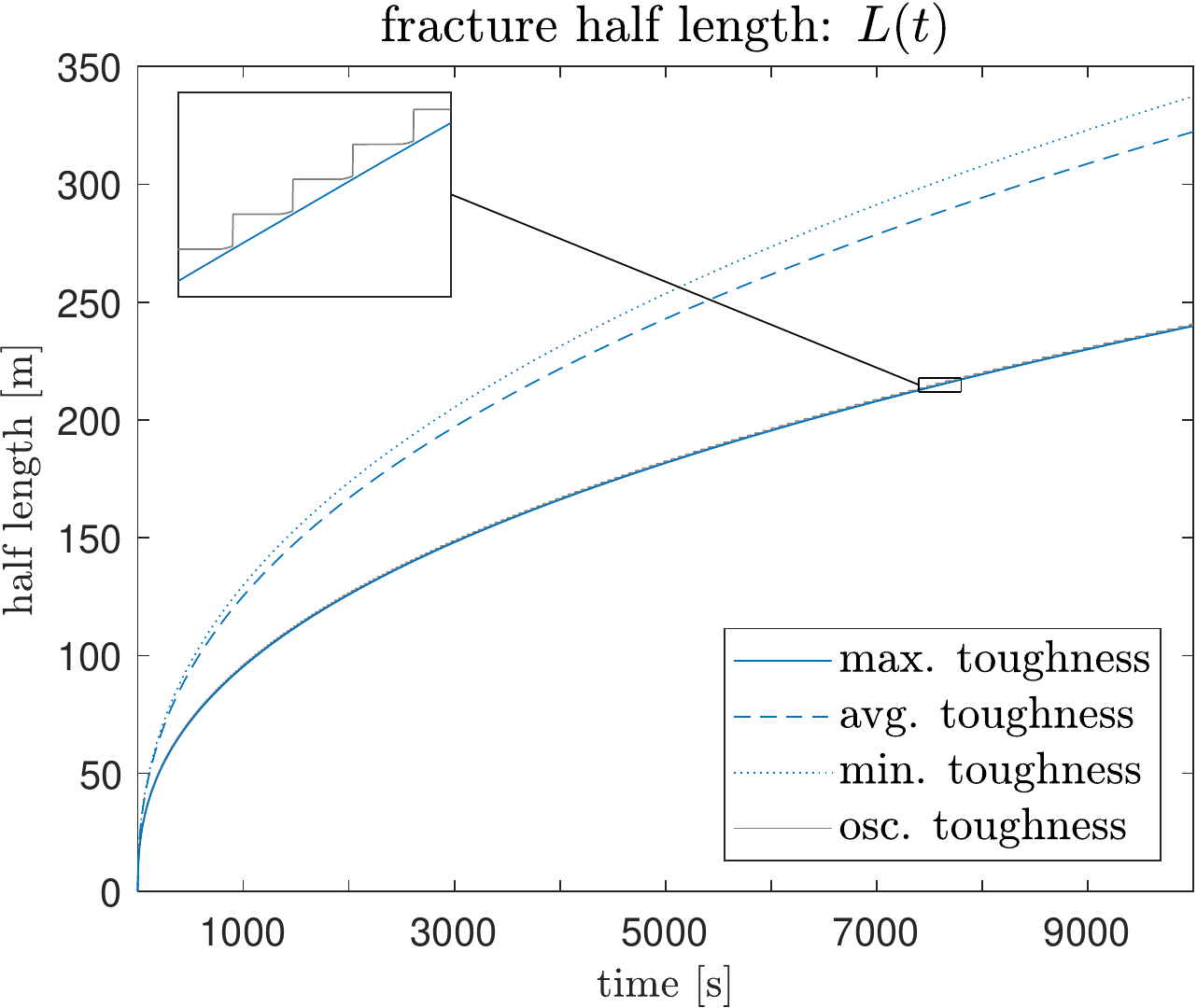}
\put(-225,155) {{\bf (d)}}
\\
\includegraphics[width=0.45\textwidth]{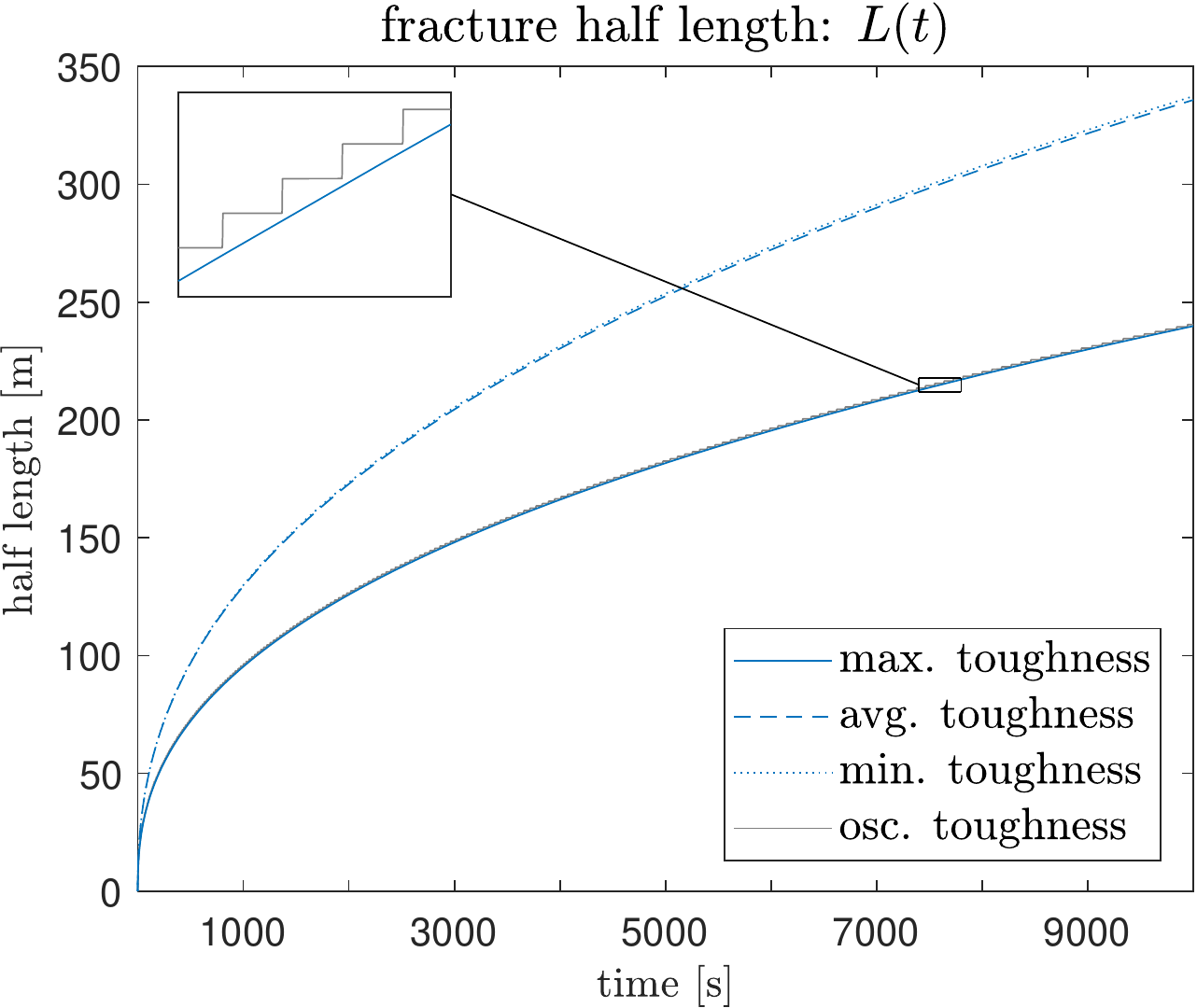}
\put(-225,155) {{\bf (e)}}
\hspace{12mm}
\includegraphics[width=0.45\textwidth]{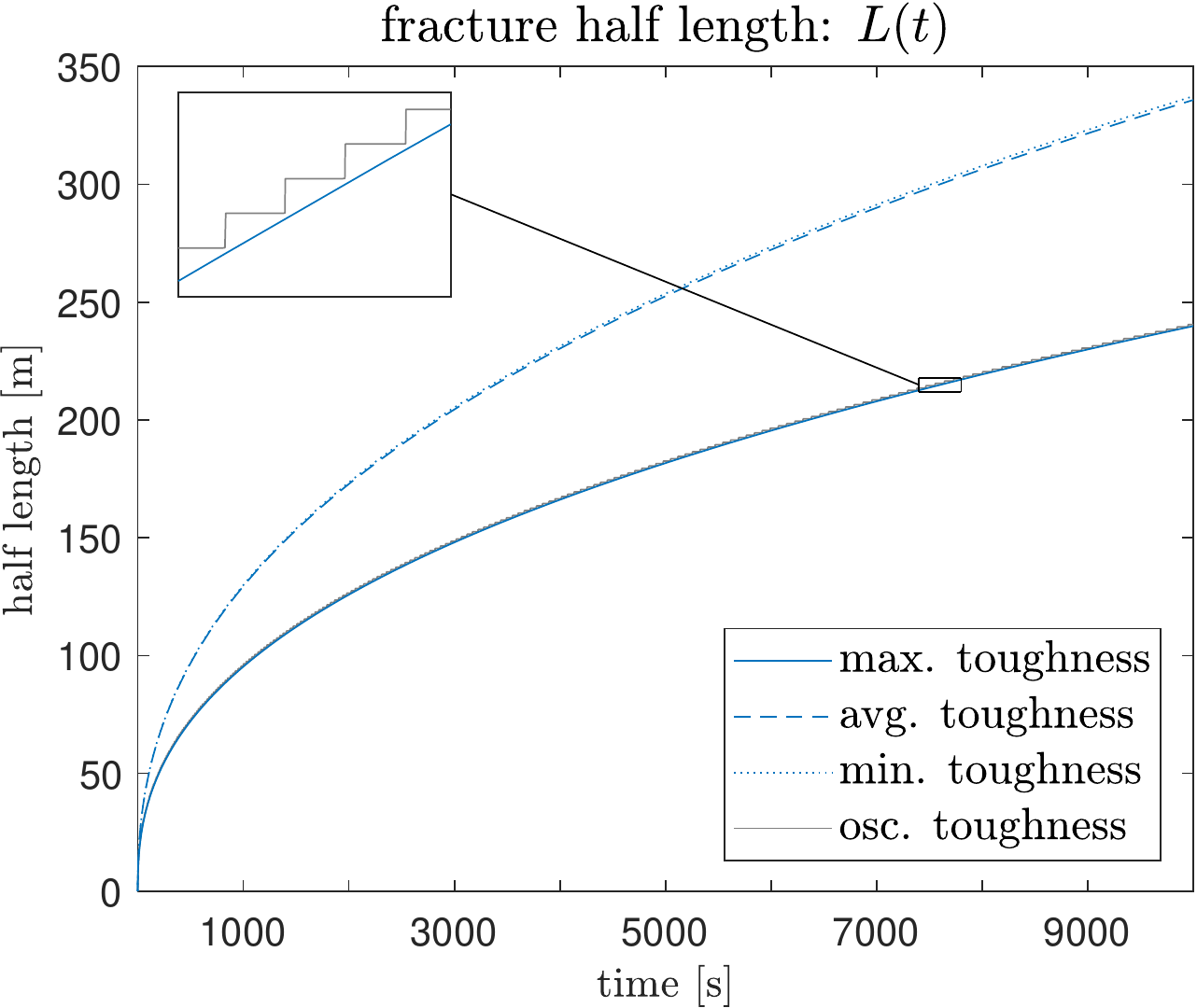}
\put(-225,155) {{\bf (f)}}
\caption{The fracture length $l(t)$ over time, for a material toughness in Case 2 (toughness-transient, see Table.~\ref{Table:toughness}), with unbalanced layering. The unbalanced layering constant \eqref{defh} is taken as  {\bf (a)}, {\bf (b)} $h=0.25$; {\bf (c)}, {\bf (d)} $h=0.1$; {\bf (e)}, {\bf (f)} $h=0.01$. The toughness distributions are: {\bf (a)}, {\bf (c)}, {\bf (e)} sinusoidal; {\bf (b)}, {\bf (d)}, {\bf (f)} step-wise.}
\label{Fig:Length2}
\end{figure}

\begin{figure}[t!]
\centering
\includegraphics[width=0.45\textwidth]{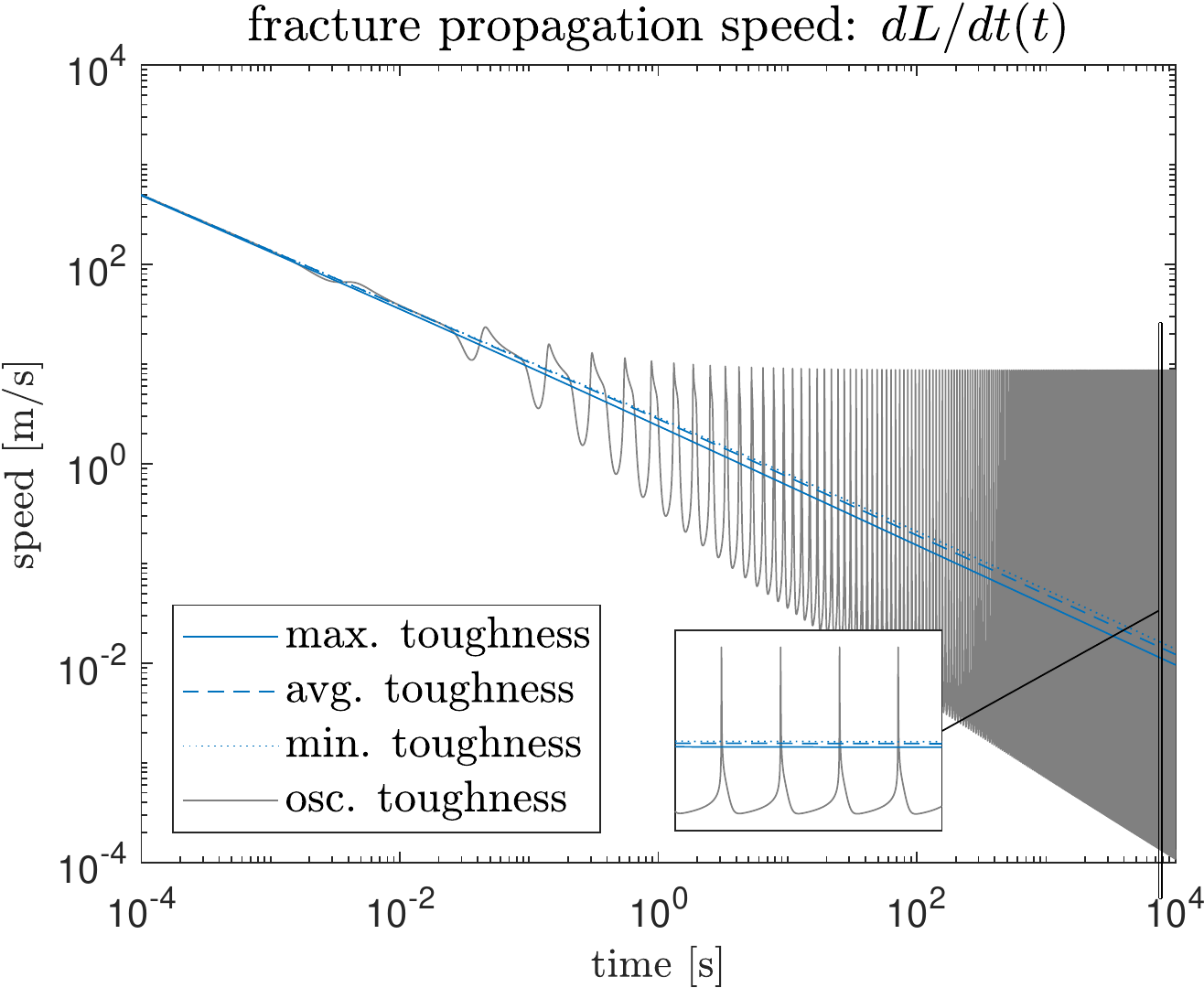}
\put(-225,155) {{\bf (a)}}
\hspace{12mm}
\includegraphics[width=0.45\textwidth]{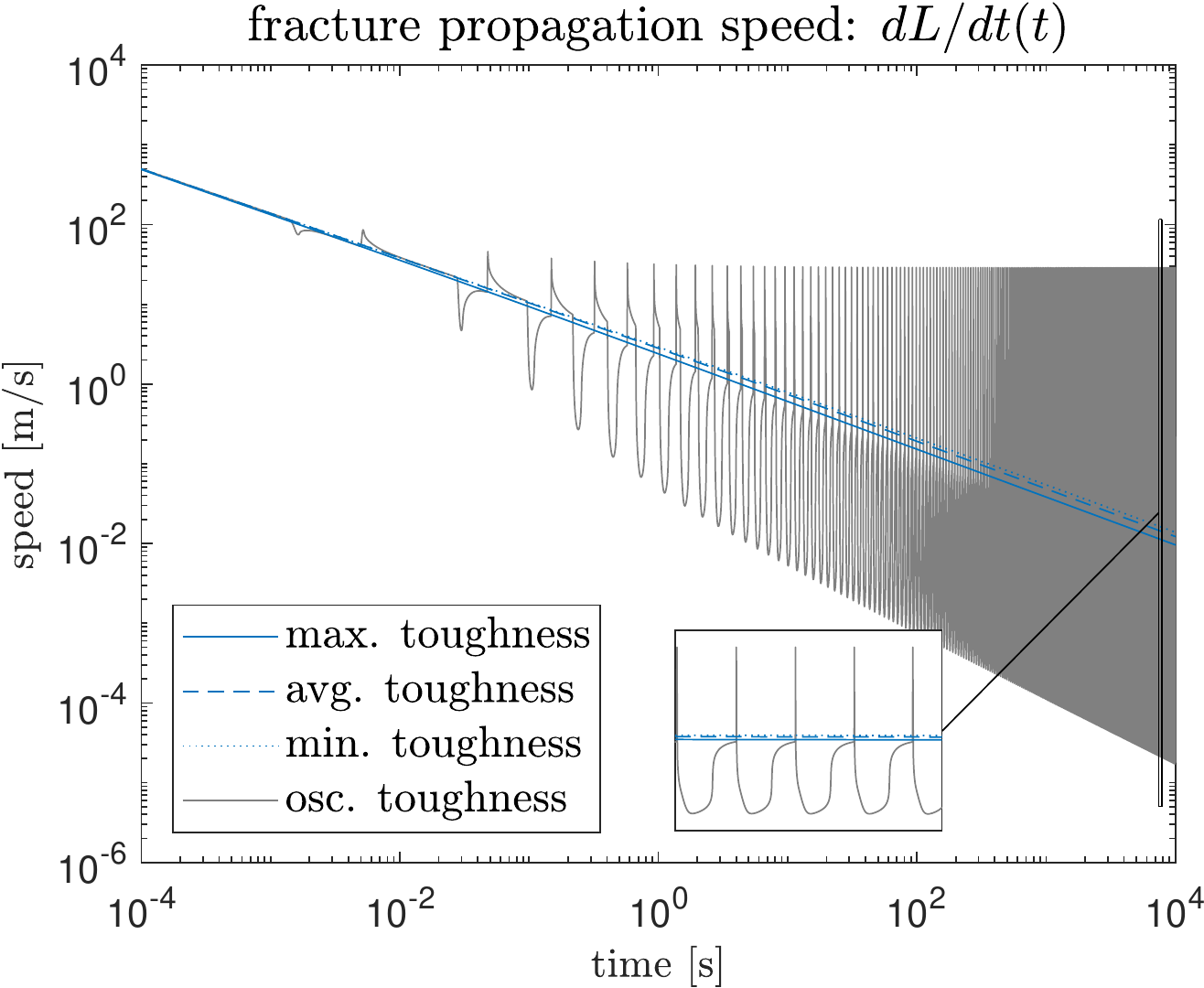}
\put(-225,155) {{\bf (b)}}
\\
\includegraphics[width=0.45\textwidth]{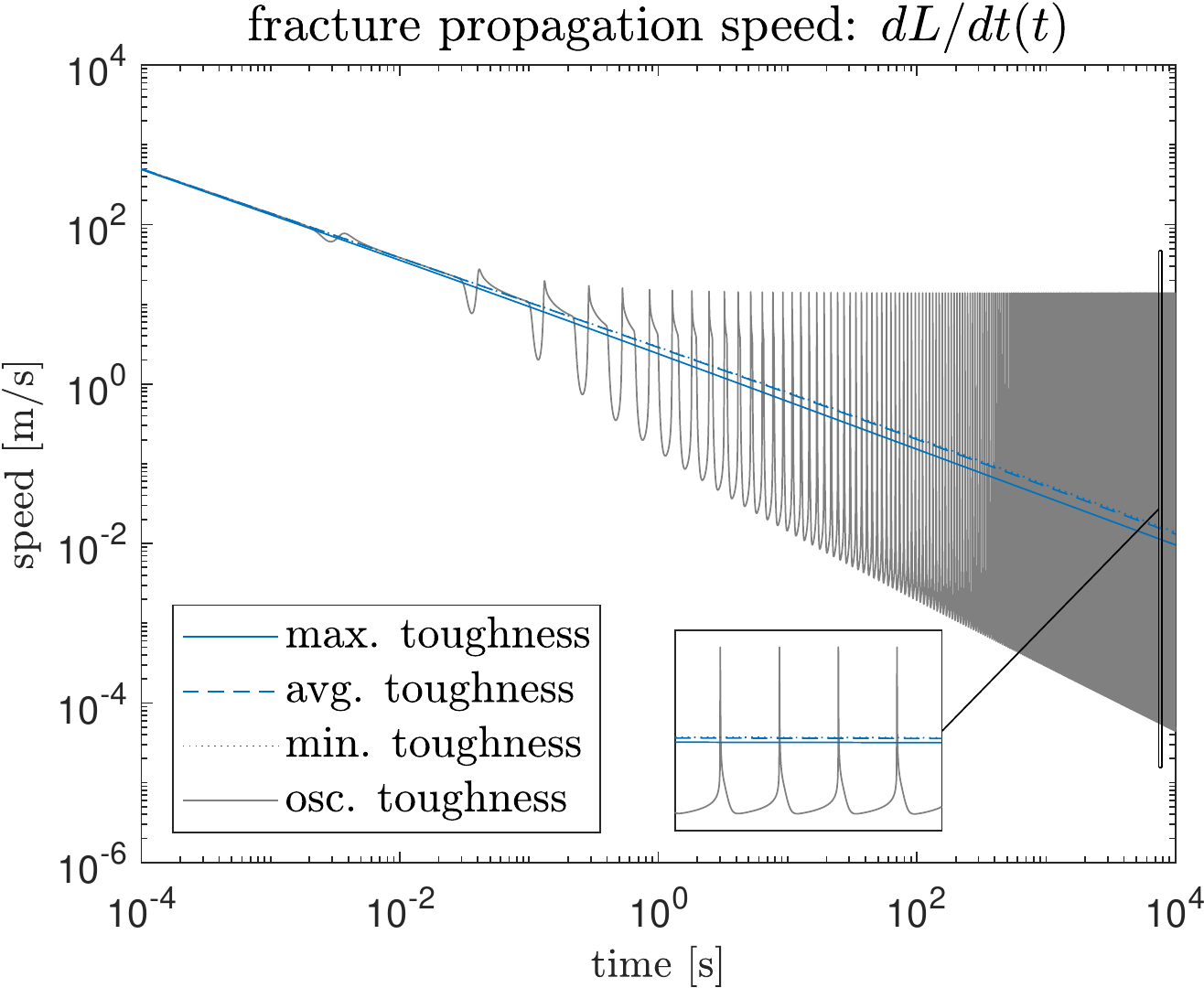}
\put(-225,155) {{\bf (c)}}
\hspace{12mm}
\includegraphics[width=0.45\textwidth]{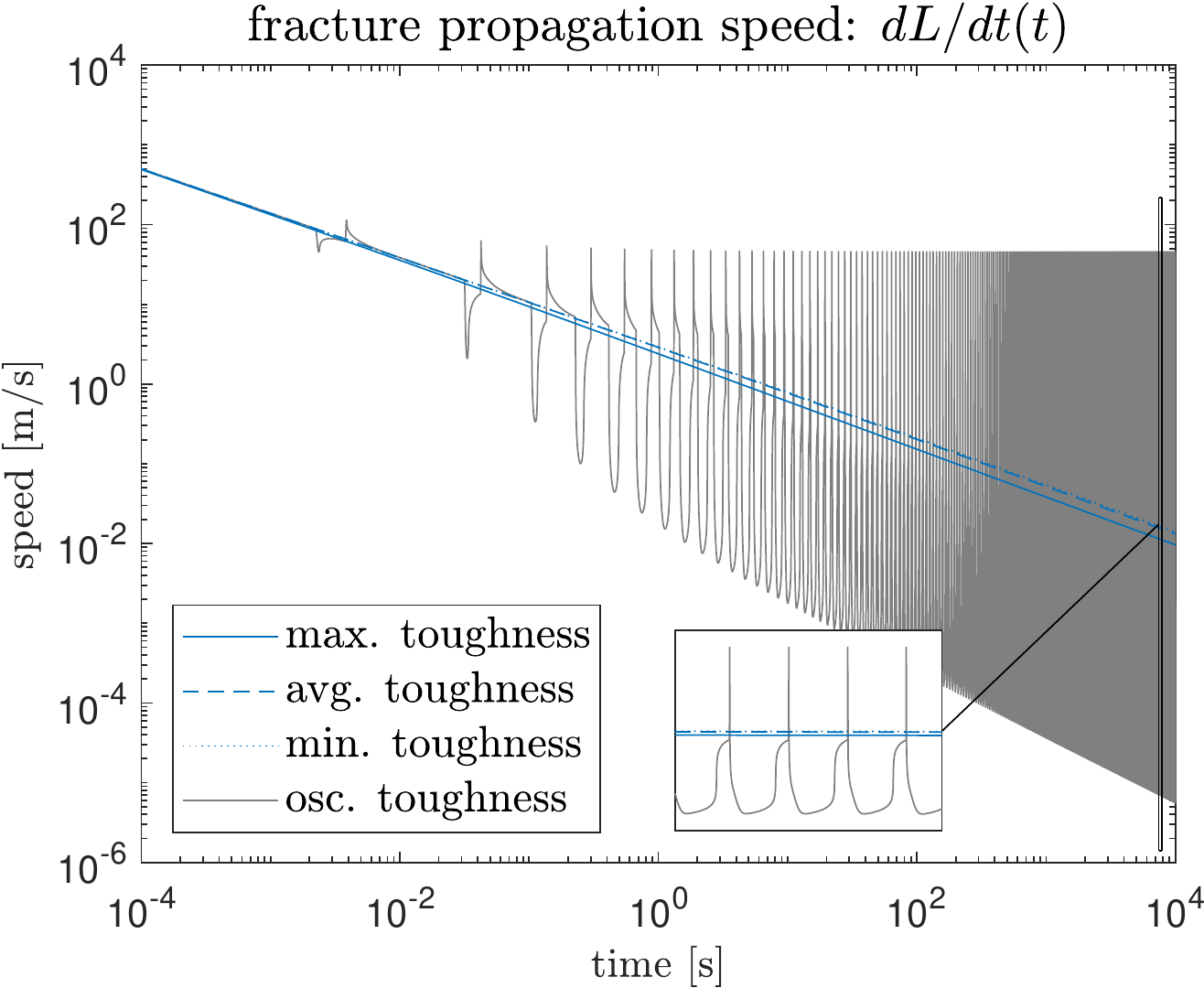}
\put(-225,155) {{\bf (d)}}
\\
\includegraphics[width=0.45\textwidth]{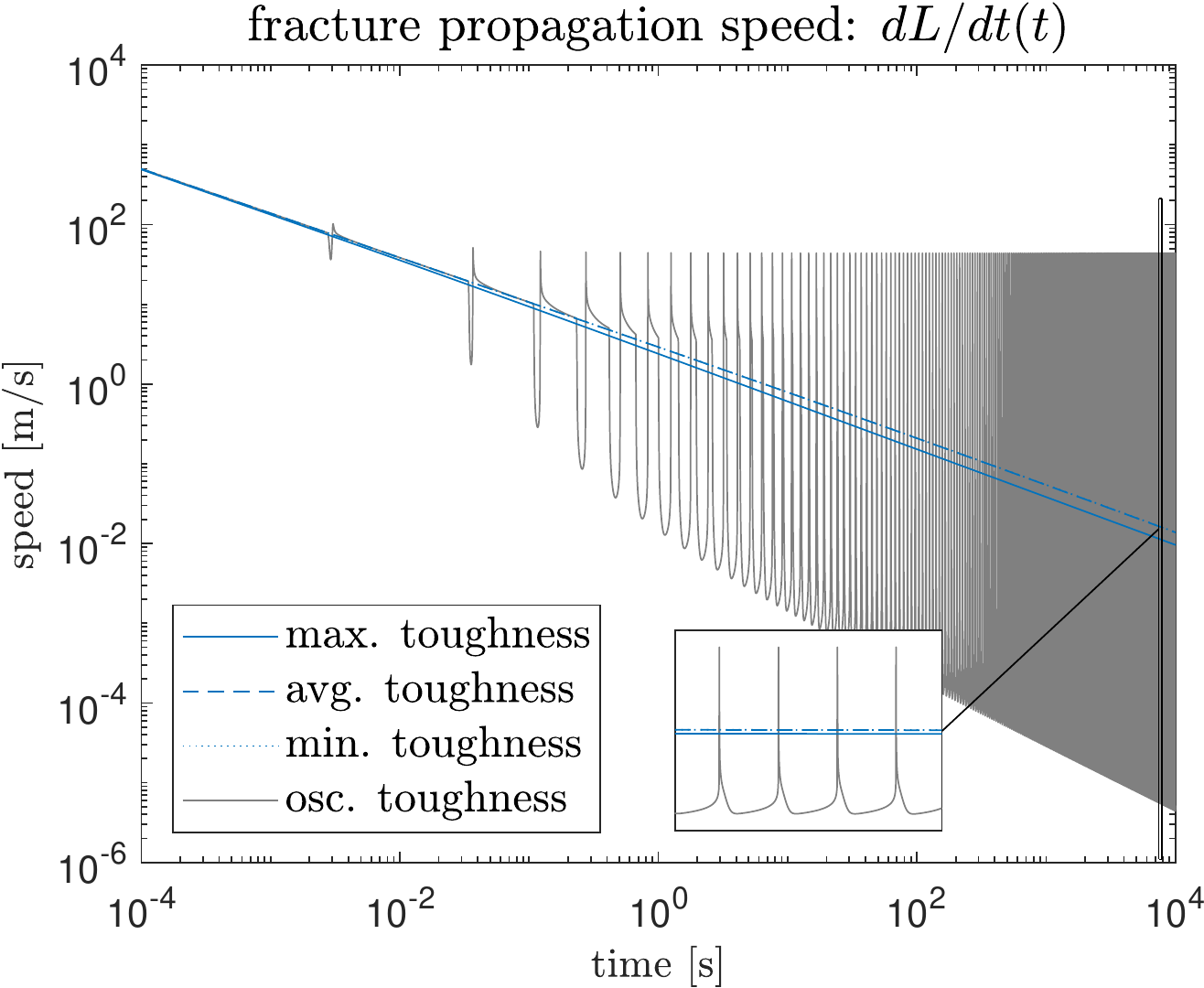}
\put(-225,155) {{\bf (e)}}
\hspace{12mm}
\includegraphics[width=0.45\textwidth]{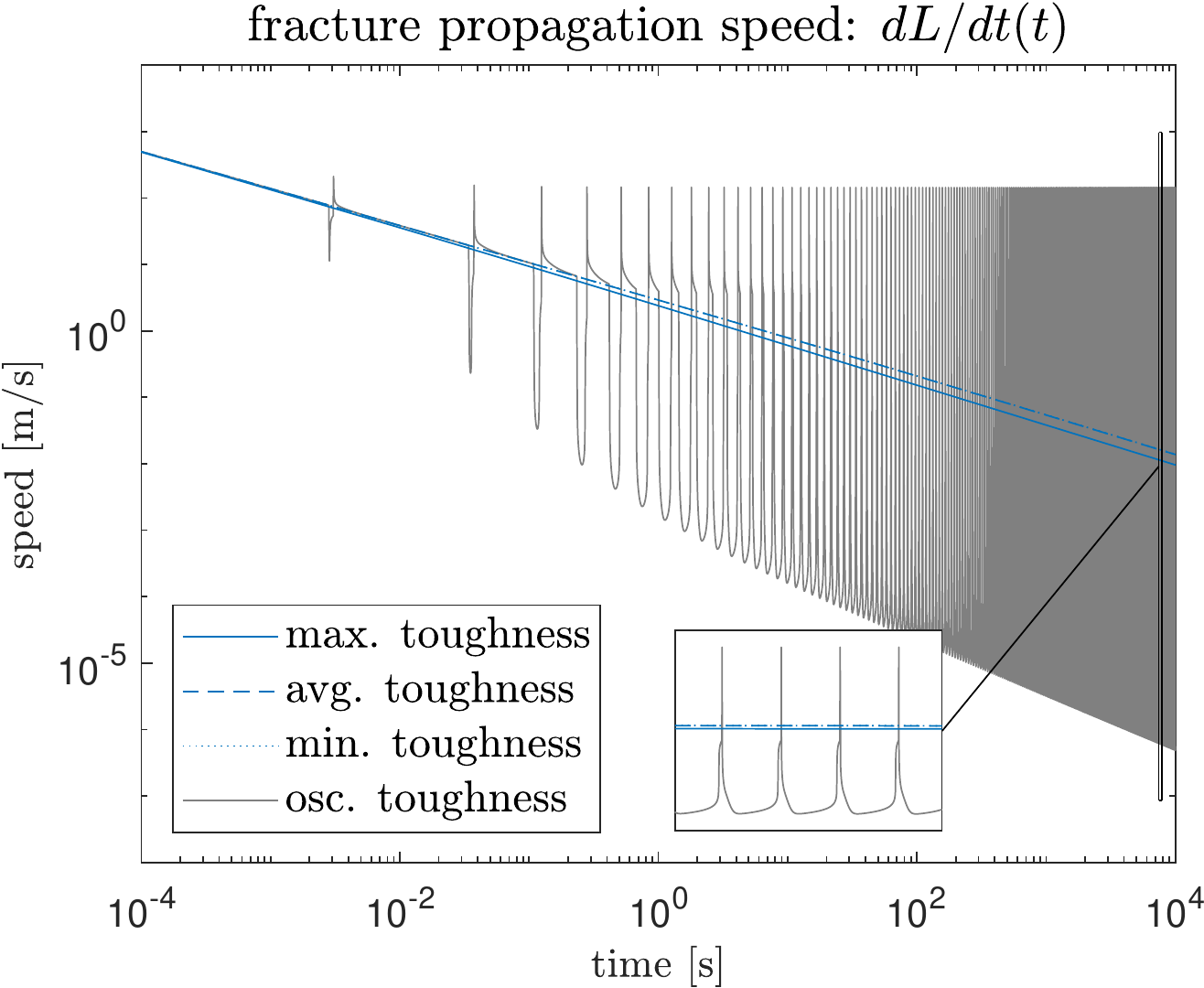}
\put(-225,155) {{\bf (f)}}
\caption{The tip velocity $dL/dt$ over time, for a material toughness in Case 2 (toughness-transient, see Table.~\ref{Table:toughness}), with unbalanced layering. The unbalanced layering constant \eqref{defh} is taken as  {\bf (a)}, {\bf (b)} $h=0.25$; {\bf (c)}, {\bf (d)} $h=0.1$; {\bf (e)}, {\bf (f)} $h=0.01$. The toughness distributions are: {\bf (a)}, {\bf (c)}, {\bf (e)} sinusoidal; {\bf (b)}, {\bf (d)}, {\bf (f)} step-wise.}
\label{Fig:Speed2}
\end{figure}

The solutions obtained for four key process parameters, namely the crack opening $w(t,0)$, the mid-length fluid pressure $p(t,L(t)/2)$, the fracture (half-)length $L(t)$ and the tip velocity $dL/dt$ are provided in Figs.~\ref{Fig:Aperture1} - \ref{Fig:Speed2} for a variety of material toughness distributions. In all simulations the material constants are taken as listed in Table.~\ref{Table:parameters}.

In Figs.~\ref{Fig:Aperture1} - \ref{Fig:Speed1} the solution behaviour for the balanced toughness configuration in a variety of regimes (toughness-toughness,  toughness-transient and transient-viscosity) are provided. The results for the fracture width, fluid pressure and length all display an interesting trend. The effect of the oscillating toughness on the smoothness of the solution is more pronounced for larger values of the (maximum) toughness (compare (a), (b) and (e), (f) for each figure), however these solutions remain close to the solution obtained using the maximum toughness homogenisation. Meanwhile, the solution for the transient-viscosity configuration (parts (e), (f) in all figures) is visibly further from the solution obtained by the maximum toughness homogenisation strategy, even through the solution for periodic toughness shows little to no oscillation. This is evidently due to the periodic nature of the toughness effecting the transition from the viscosity to the toughness dominated regime, which the maximum toughness homogenisation does not account for, but which does not play a significant role for fractures already in the toughness dominated regime.

Next, the results in Figs.~\ref{Fig:Aperture2} - \ref{Fig:Speed2} give the behaviour of the same four key process parameters ($w(t,0)$, $p(t,L(t)/2)$, $L(t)$, $dL/dt$) for unbalanced toughness in the toughness-transient case. The three cases considered, $h=0.25$, $h=0.1$ and $h=0.01$ represent a minor imbalance in the toughness distribution, a significant imbalance, and a case where the maximum toughness layer makes up only a seemingly insignificant portion of the body. Intriguingly, the effect of the unbalanced nature of the toughness distribution is almost negligible (compare (a), (b) and (e), (f) for each figure). This can be explained by noting the physical mechanism by which the periodic toughness effects the solution, as previously outlined in\linebreak \cite{Gaspare2022} (and noted previously by others, see that paper for details). Namely, considering the step-wise case, when the fracture is transitioning from the less tough material to the higher toughness material, the fluid is under-pressure for fracturing the material ahead of the tip. As a result, the fracture pauses while this pressure increases, until it is capable of continuing the crack extension (as can be seen in the figures for fracture length and fluid pressure, Figs.~\ref{Fig:Pressure1} - \ref{Fig:Length1} and Fig.~\ref{Fig:Pressure2} - \ref{Fig:Length2}). Conversely, when transitioning from the higher toughness material to a lower toughness one, the crack is over-pressurised for the material is is fracturing, causing the crack to extend rapidly. Note that these mechanisms are not affected by the width of the higher toughness layer, nor is the fluid pressure required to start the initial fracture in each layer. 

It should be stated however that the assumption that $h<0.5$ does play a role in this analysis, and we would not necessarily expect the same result if instead considering the case where $h$ is close to $1$ (as the `under-pressure' effect would become less significant). However, somewhat surprisingly, we can conclude that only the maximum and minimum toughness strongly effect the fracture evolution when $h<0.5$, and the width of each layer only has a minimal effect (for the periodic distribution considered here).





$\quad$
\newpage


\newpage
$\quad$
\newpage
$\quad$
\newpage
$\quad$
\newpage

\section{Comparison of homogenisation strategies}

With it now possible to obtain highly accurate solutions for a penny-shaped fracture growing within a domain with (axisymmetric) periodic toughness, and an initial examination of the key process parameters undertaken, we are now in a position to investigate the effectiveness of various homogenisaion techniques for the material toughness. We begin by introducing the proposed strategies.

\subsection{The maximum toughness and temporal averaging approaches}\label{Sect:Stratagems}

One of the simplest approaches to homogenising the material toughness is the maximum toughness strategy proposed in \cite{DONTSOV2021108144}. Here, the heterogeneous material is replaced with a homogeneous one, whose uniform toughness is taken as the maximum of that in the heterogeneous material. This approach is incredibly simple to implement, and was demonstrated by the authors to be effective for the KGD model in the viscosity dominated regime \cite{Gaspare2022}. The effectiveness of the approximation however decreases when dealing with the transient or toughness dominated regimes.

An alternative approach, previously proposed by the authors in \cite{Gaspare2022}, is to use a method based on the temporal averaging of the toughness. For the rational behind this strategy, we direct the reader to the stated paper, however it was shown to be effective for all considered regimes in the case of the KGD model with periodic toughness. 

%
%
%
%

\begin{figure}[t!]
\centering
\includegraphics[width=0.45\textwidth]{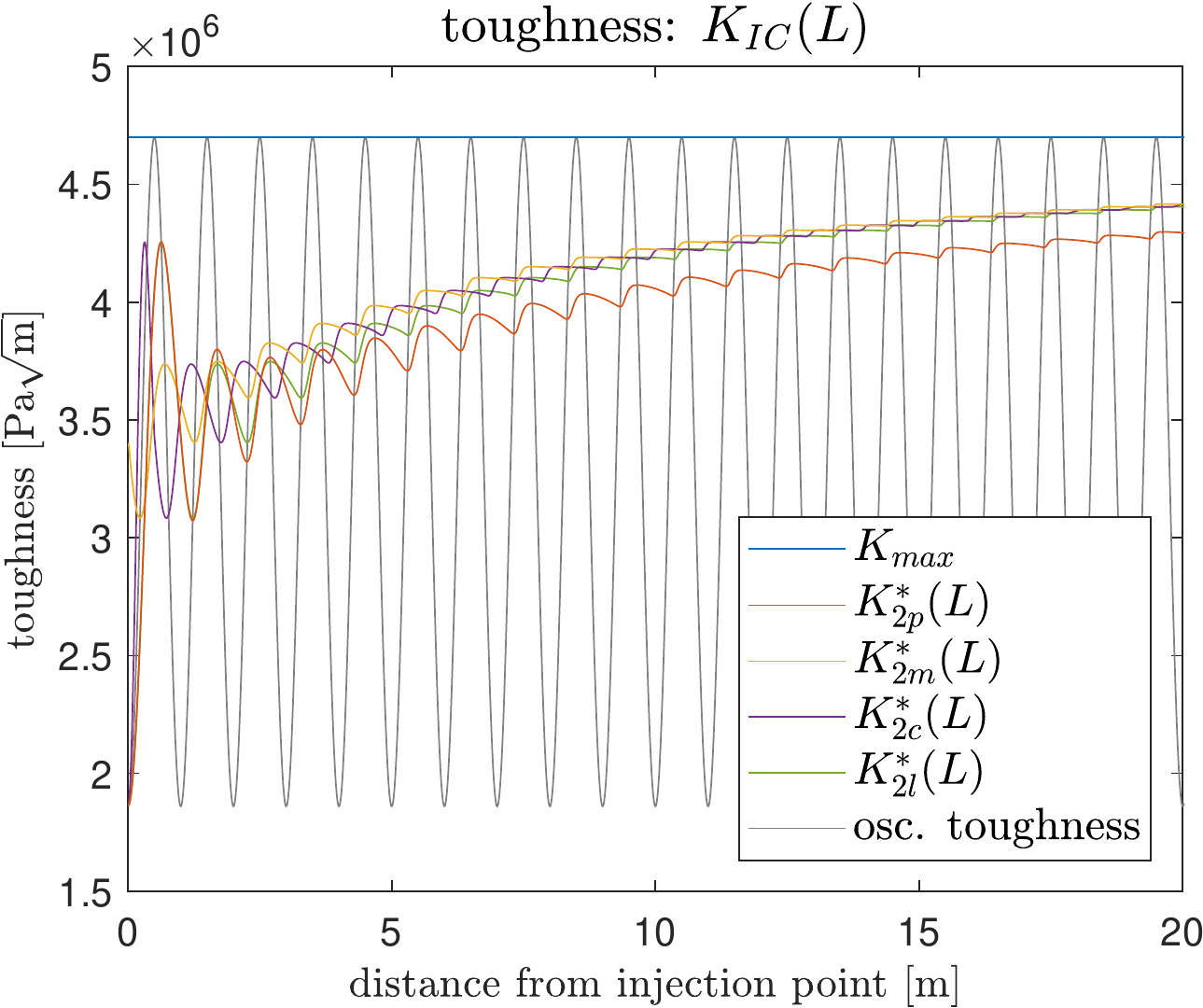}
\put(-225,155) {{\bf (a)}}
\hspace{12mm}
\includegraphics[width=0.45\textwidth]{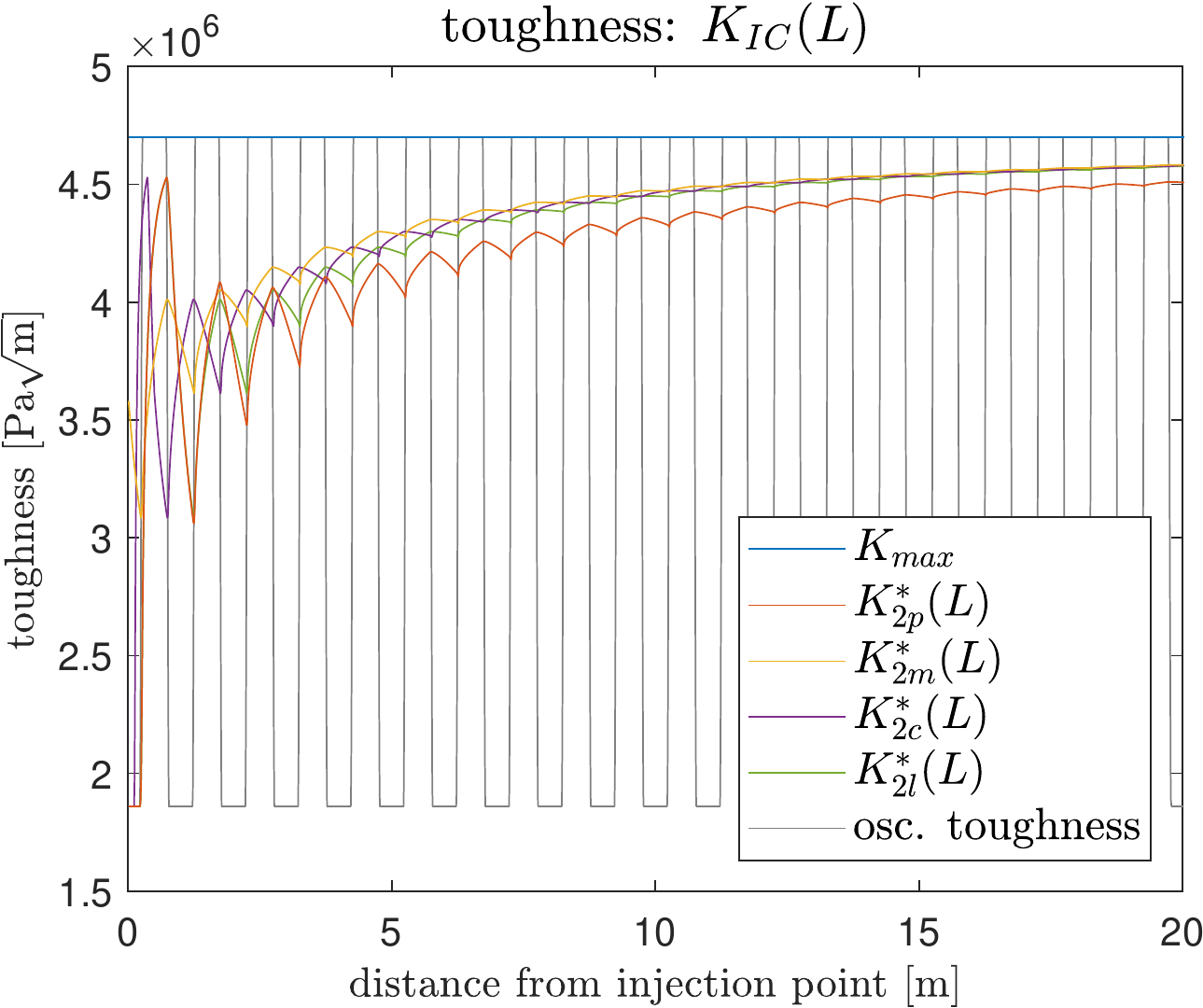}
\put(-225,155) {{\bf (b)}}
\\
\includegraphics[width=0.45\textwidth]{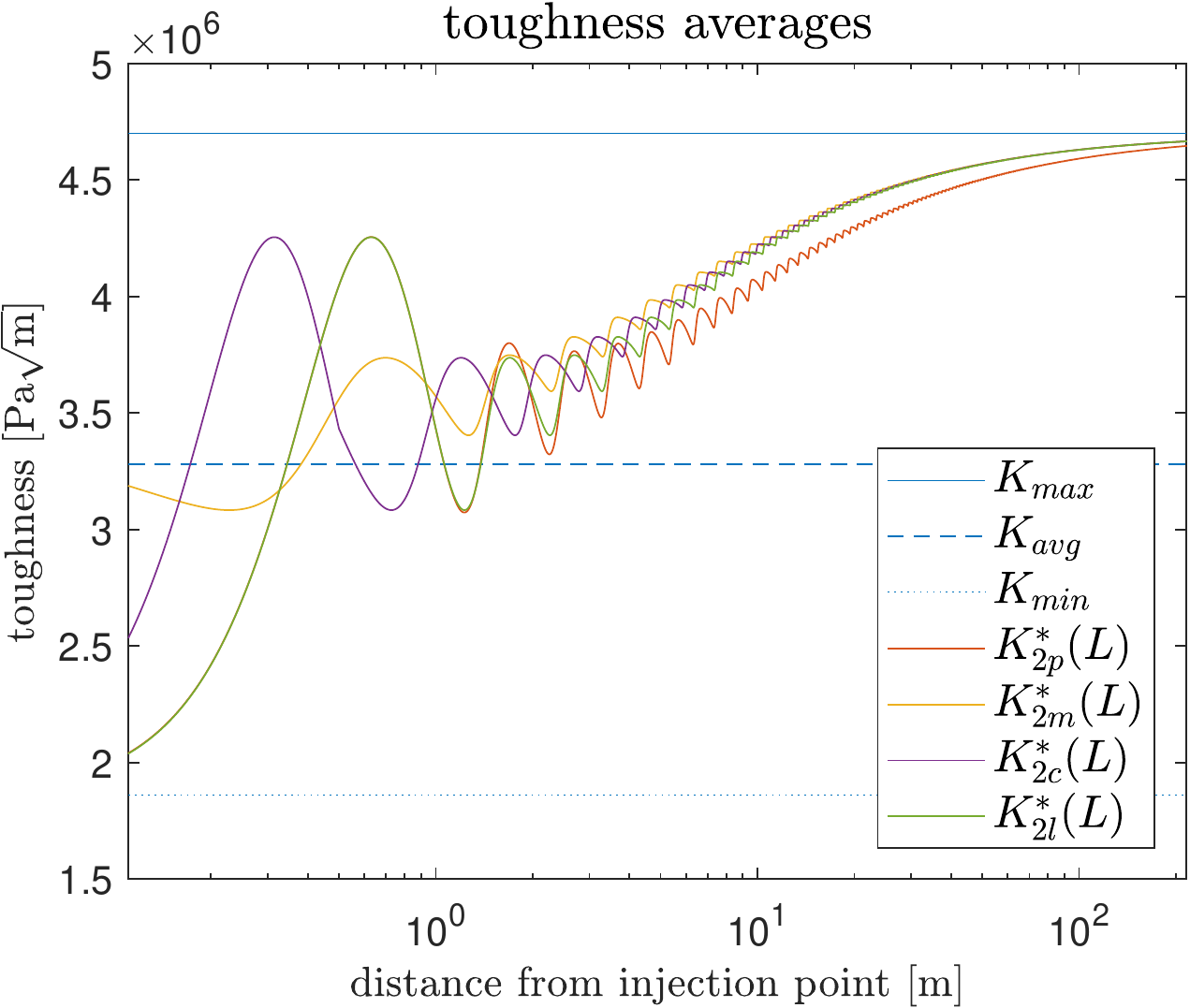}
\put(-225,155) {{\bf (c)}}
\hspace{12mm}
\includegraphics[width=0.45\textwidth]{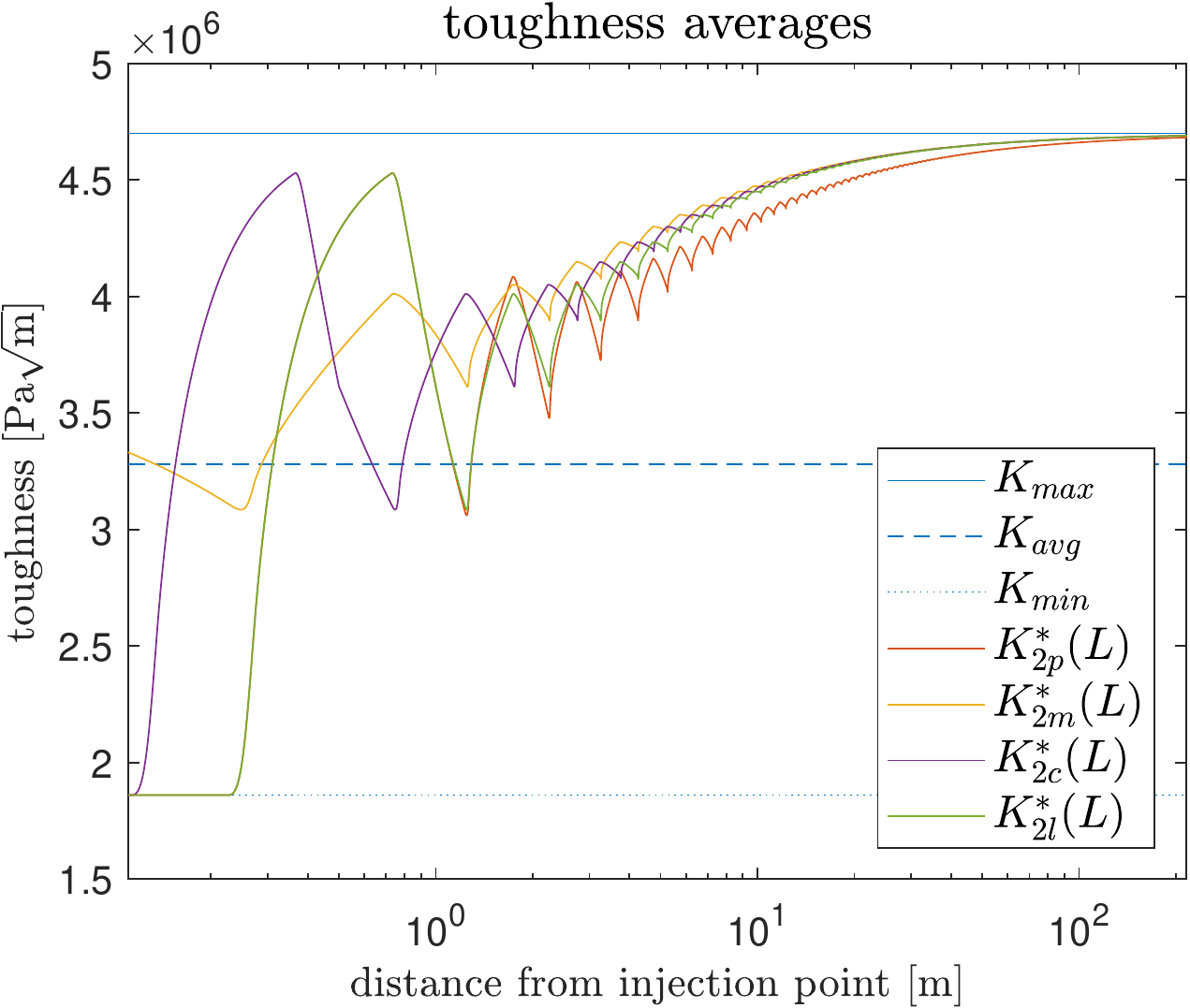}
\put(-225,155) {{\bf (d)}}
\caption{Results of the measures \eqref{MeasureK2r}-\eqref{MeasureK2p} for the periodic toughness distribution in Case 2 (toughness-transient regime) with balanced layering. Here we show {\bf (a)}, {\bf (b)} the initial behaviour; {\bf (c)}, {\bf (d)} the long-time behaviour. The toughness distributions are: {\bf (a)}, {\bf (c)} sinusoidal; {\bf (b)}, {\bf (d)} step-wise. }
\label{Fig:Measures1}
\end{figure}

We consider the formulation of the temporal-averaging approach based on evaluating the fracture energy. As this corresponds to the $L_2$-norm, it will be denoted with the subscript $2$. One key advantage of this approach is that it remains effective even when other material parameters are also considered to be heterogeneous (as pointed out in \cite{DONTSOV2021108144}). We consider three variants of the moving temporal average of $K_{Ic}$ over the period, denoted `left', `center' and `right' (subscript $l$, $c$, $r$), depending on which side the integration limits are based: 
\begin{equation} \label{MeasureK2r}
K_{2r}^* (L) = \sqrt{\frac{1}{t(L+dL)-t(L)}\int_{t(L)}^{t(L+dL)}K_{IC}^2\big(L(\xi)\big)d\xi}
\end{equation}
\begin{equation}
K_{2c}^* (L) = \begin{cases}
	\sqrt{\displaystyle\frac{1}{t(2L)-t(0)}\displaystyle\int_{t(0)}^{t(2L)}K_{IC}^2\big(L(\xi)\big)d\xi}, \quad 0 < L\leq dL/2 \\
	 \sqrt{\displaystyle\frac{1}{t(L+dL/2)-t(L-dL/2)}\displaystyle\int_{t(L-dL/2)}^{t(L+dL/2)}K_{IC}^2\big(L(\xi)\big)d\xi}, \quad L>dL/2
\end{cases}
\end{equation}
\begin{equation}\label{MeasureK2l}
K_{2l}^* (L) = \begin{cases}
	\sqrt{\displaystyle\frac{1}{t(L)-t(0)}\displaystyle\int_{t(0)}^{t(L)}K_{IC}^2\big(L(\xi)\big)d\xi}, \quad 0 < L\leq dL \\
	\sqrt{\displaystyle\frac{1}{t(L)-t(L-dL)}\displaystyle\int_{t(L-dL)}^{t(L)}K_{IC}^2\big(L(\xi)\big)d\xi}, \quad L>dL
\end{cases}
\end{equation}
For completeness, we also include the progressive measure, denoted with subscript $p$, which is given by
\begin{equation} \label{MeasureK2p}
K_{2p}^* (L) = \sqrt{\frac{1}{t(L+dL)}\int_{0}^{t(L+dL)}K_{IC}^2\big(L(\xi)\big)d\xi}
\end{equation}
Examples of the homogenised values of $K_{Ic}$ obtained by measures \eqref{MeasureK2r}-\eqref{MeasureK2p}, alongside the maximum toughness, are provided in Fig.~\ref{Fig:Measures1}.

\subsection{Results for balanced toughness distributions}

\begin{figure}[t!]
 \centering
 \includegraphics[width=0.45\textwidth]{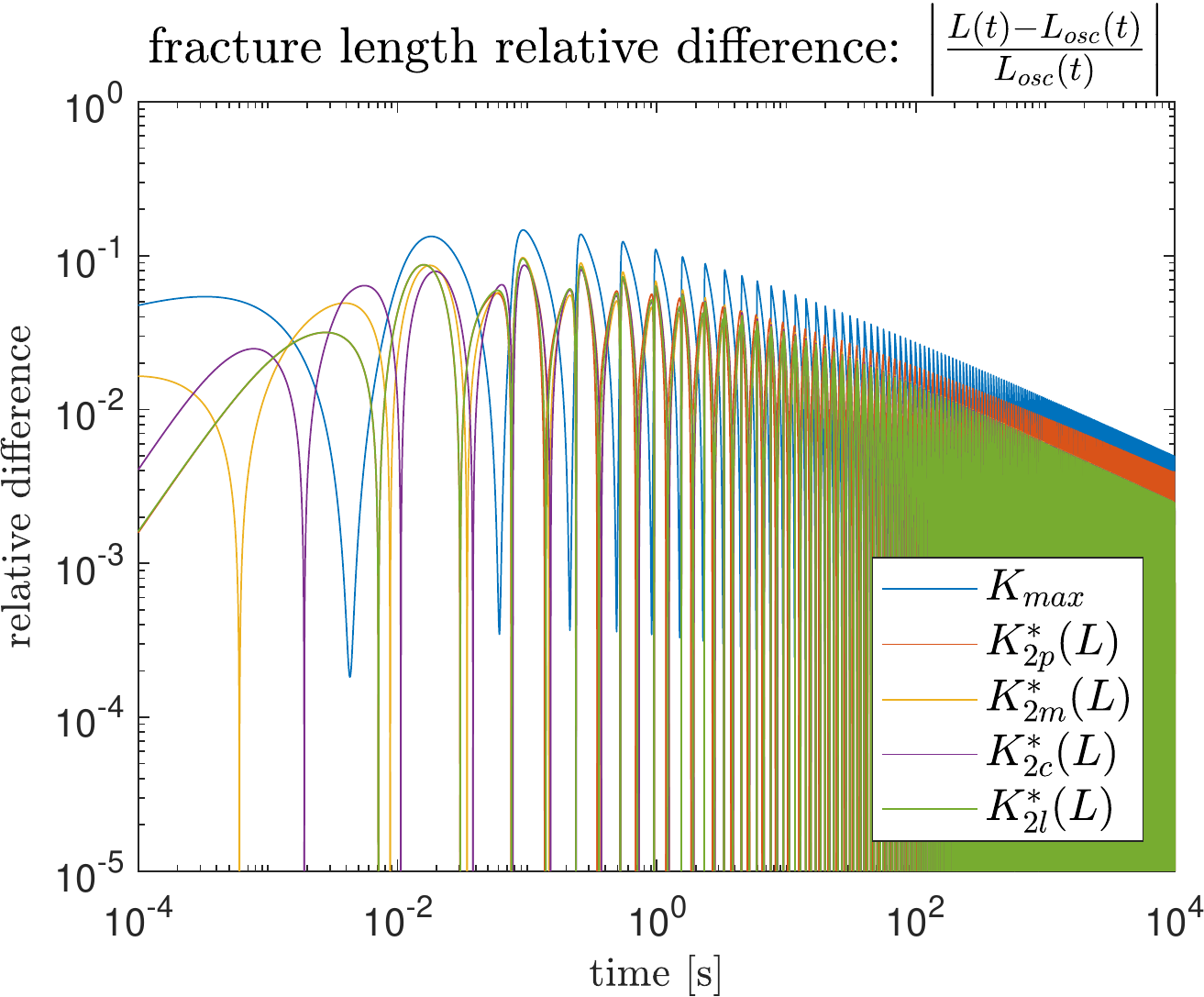}
\put(-225,155) {{\bf (a)}}
\hspace{12mm}
\includegraphics[width=0.45\textwidth]{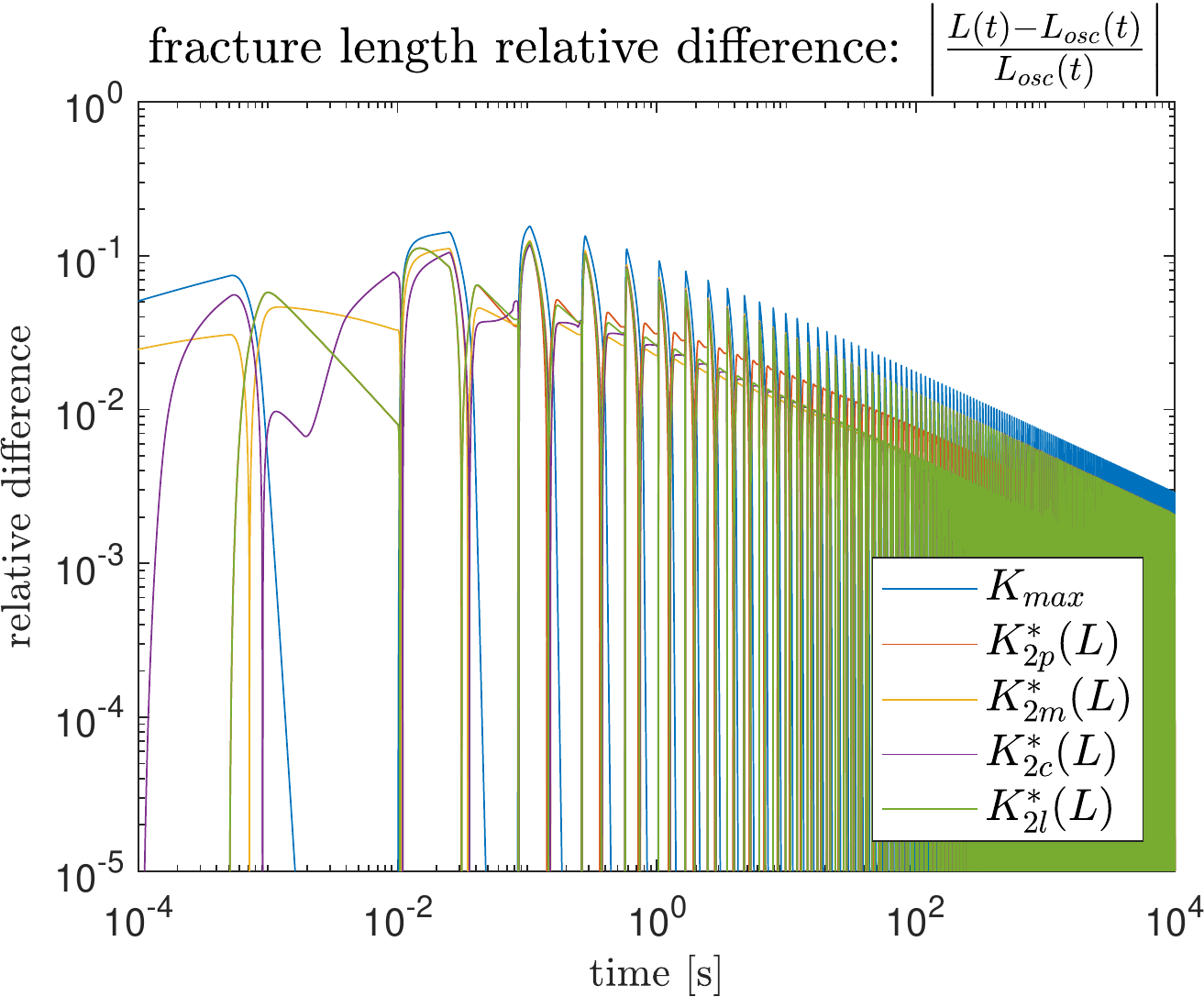}
\put(-225,155) {{\bf (b)}}
\\
\includegraphics[width=0.45\textwidth]{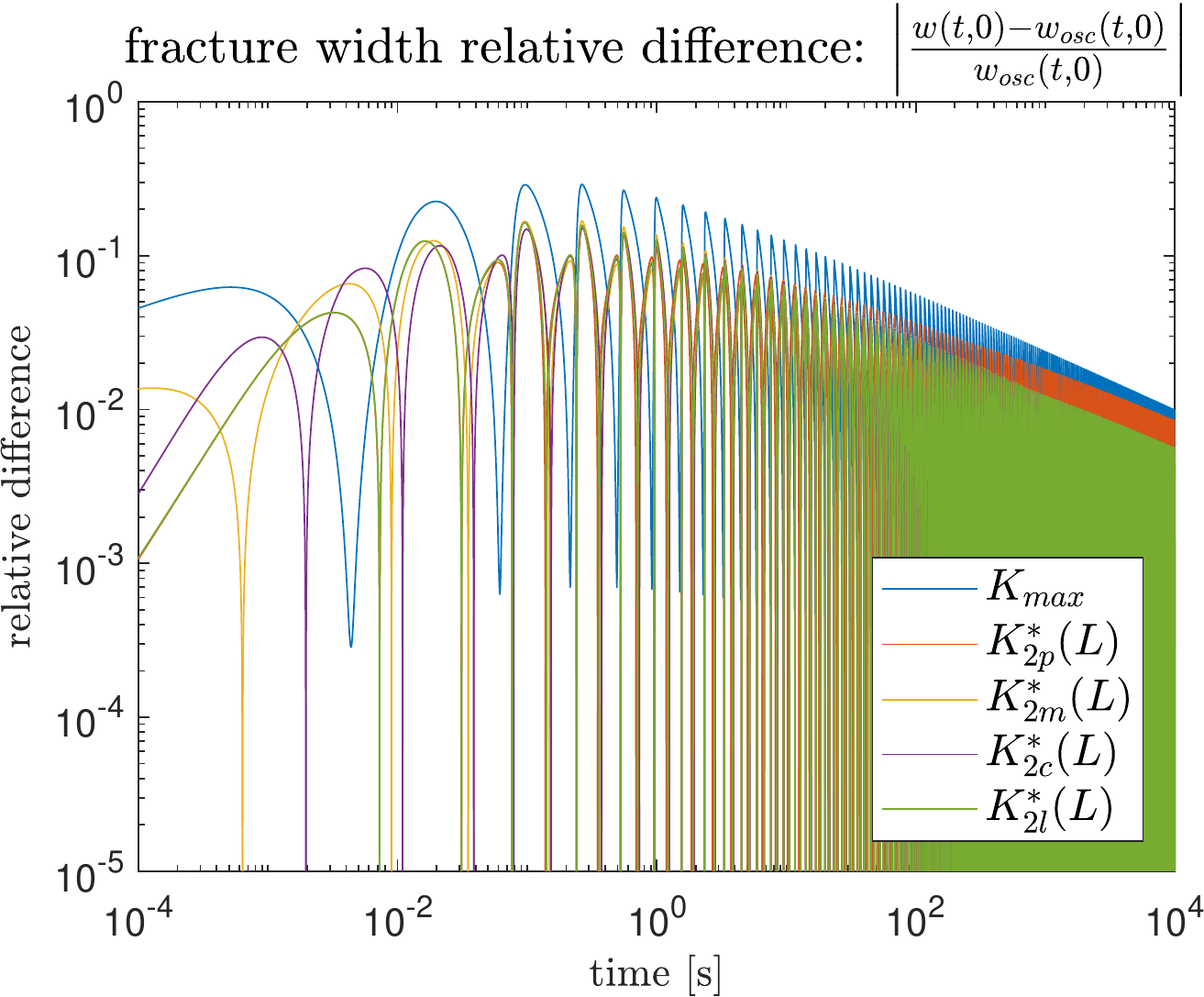}
\put(-225,155) {{\bf (c)}}
\hspace{12mm}
\includegraphics[width=0.45\textwidth]{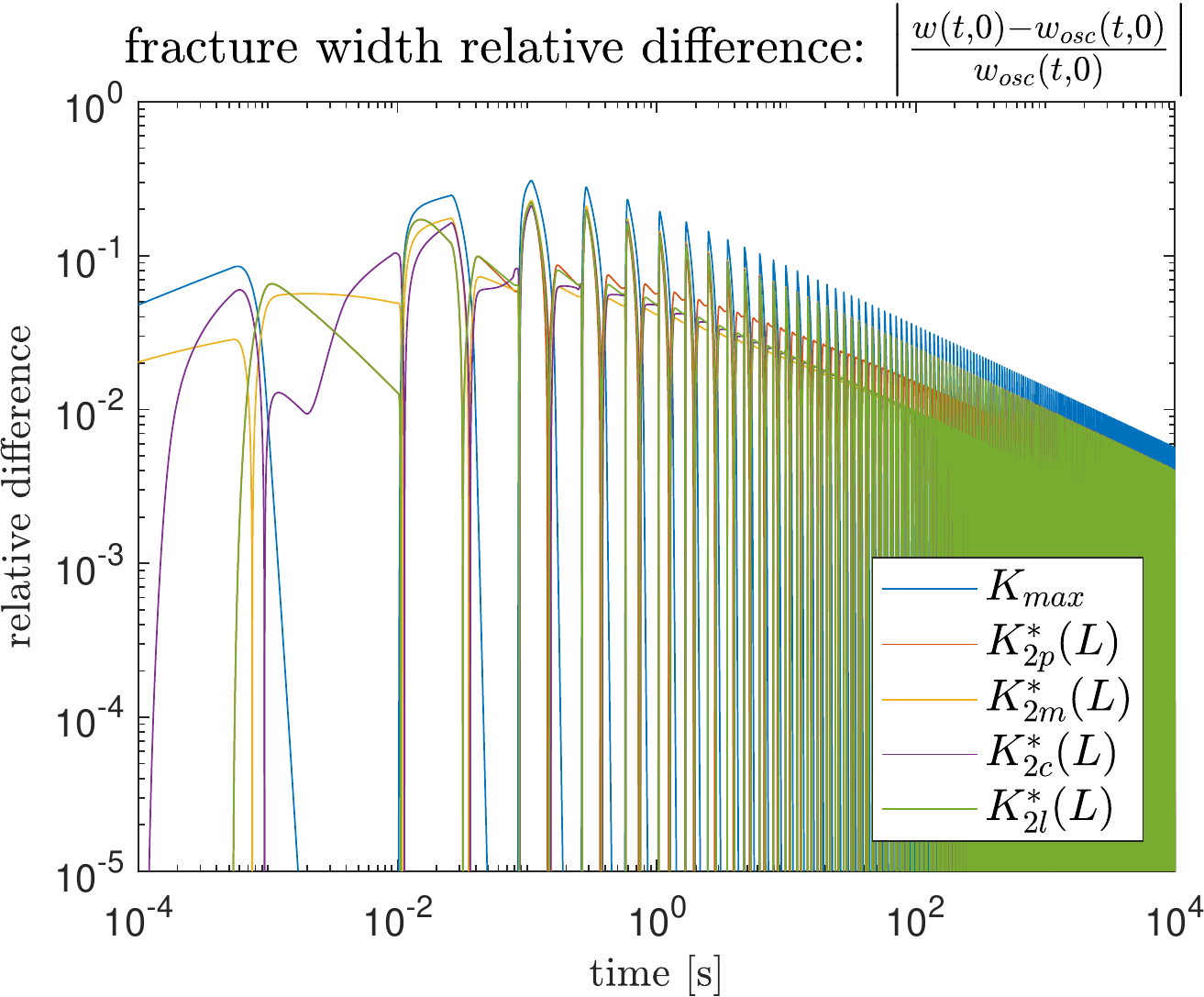}
\put(-225,155) {{\bf (d)}}
\\
\includegraphics[width=0.45\textwidth]{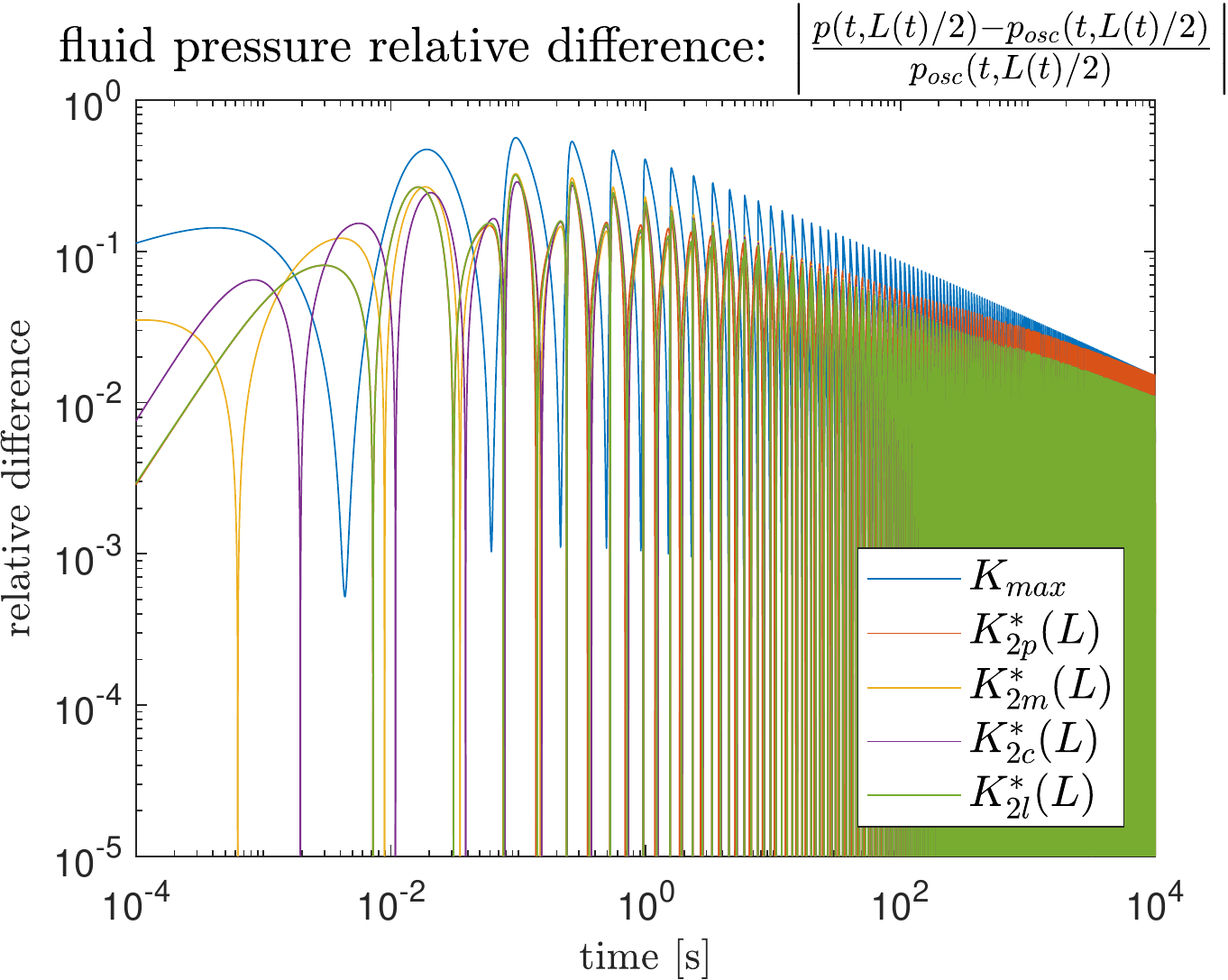}
\put(-225,155) {{\bf (e)}}
\hspace{12mm}
\includegraphics[width=0.45\textwidth]{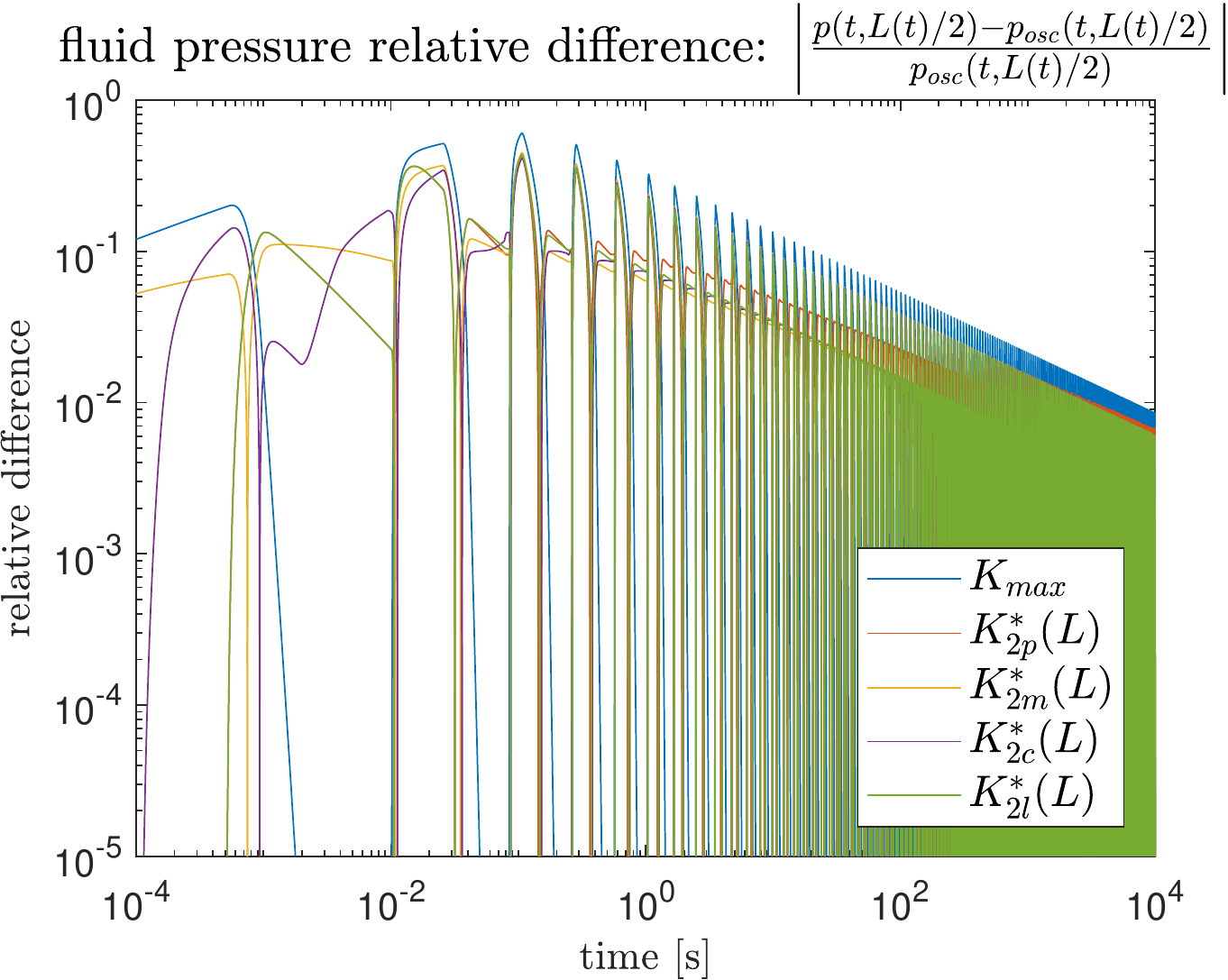}
\put(-225,155) {{\bf (f)}}
\caption{Results for the toughness-toughness distribution with balanced layering (Case 1, see Table.~\ref{Table:toughness}). Relative differences for the {\bf (a)}, {\bf (b)} fracture length $L(t)$, {\bf (c)}, {\bf (d)} fracture opening at the injection point $w(t,0)$, {\bf (e)}, {\bf (f)} fluid pressure at the mid-point $p(t, L(t)/2 )$.}
 \label{Even_Del100_10}
\end{figure}

\begin{figure}[t!]
 \centering
 \includegraphics[width=0.45\textwidth]{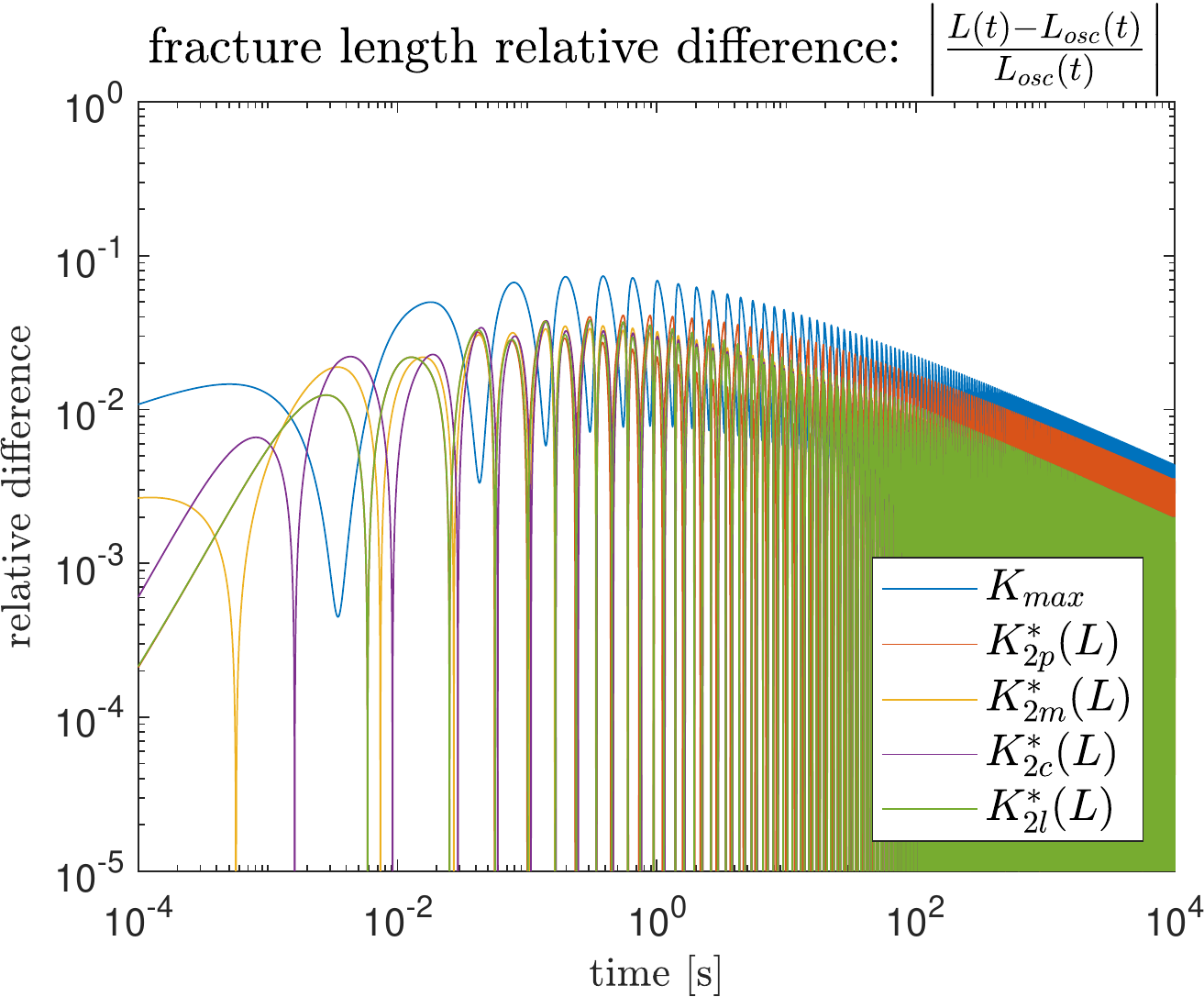}
\put(-225,155) {{\bf (a)}}
\hspace{12mm}
\includegraphics[width=0.45\textwidth]{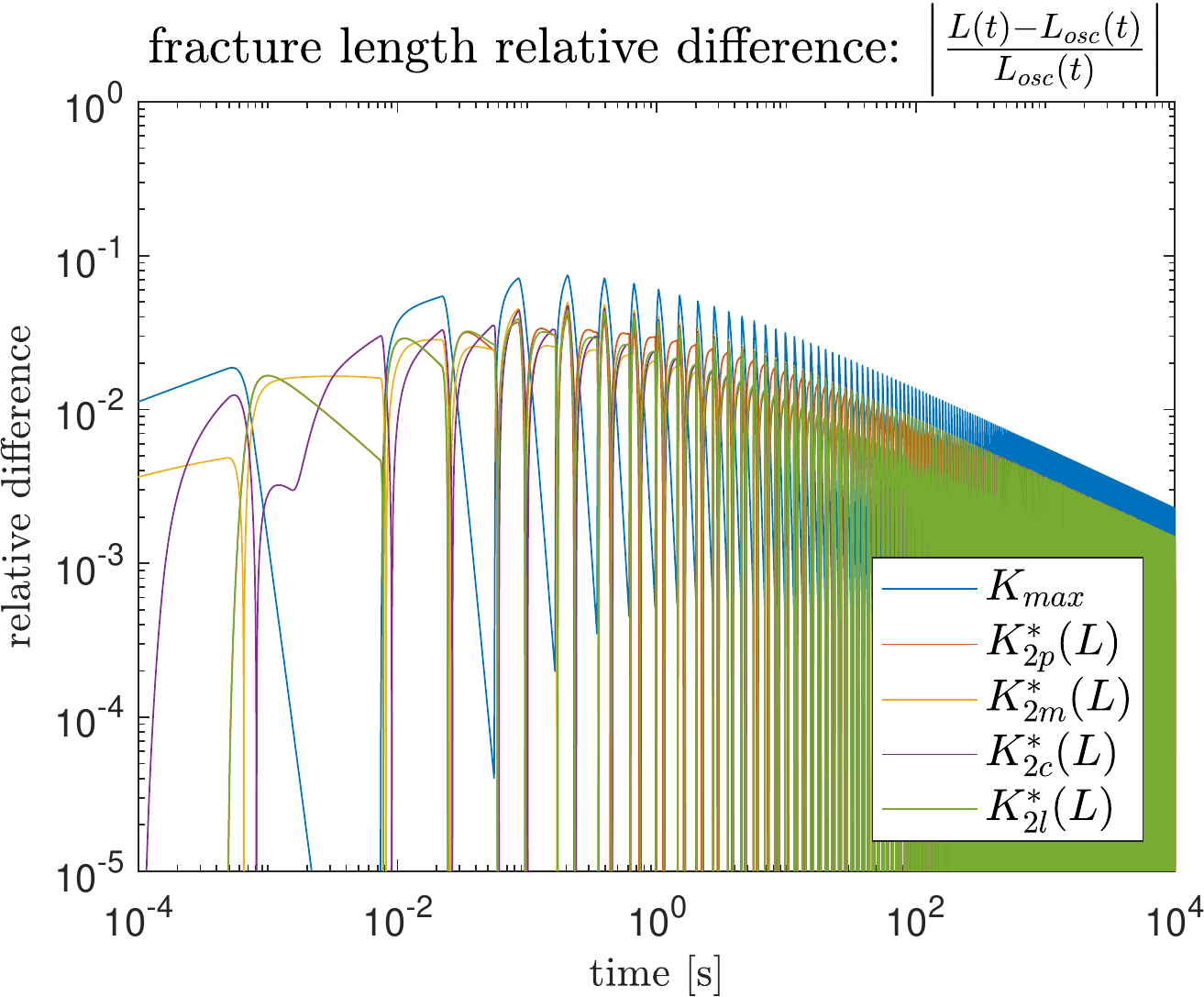}
\put(-225,155) {{\bf (b)}}
\\
\includegraphics[width=0.45\textwidth]{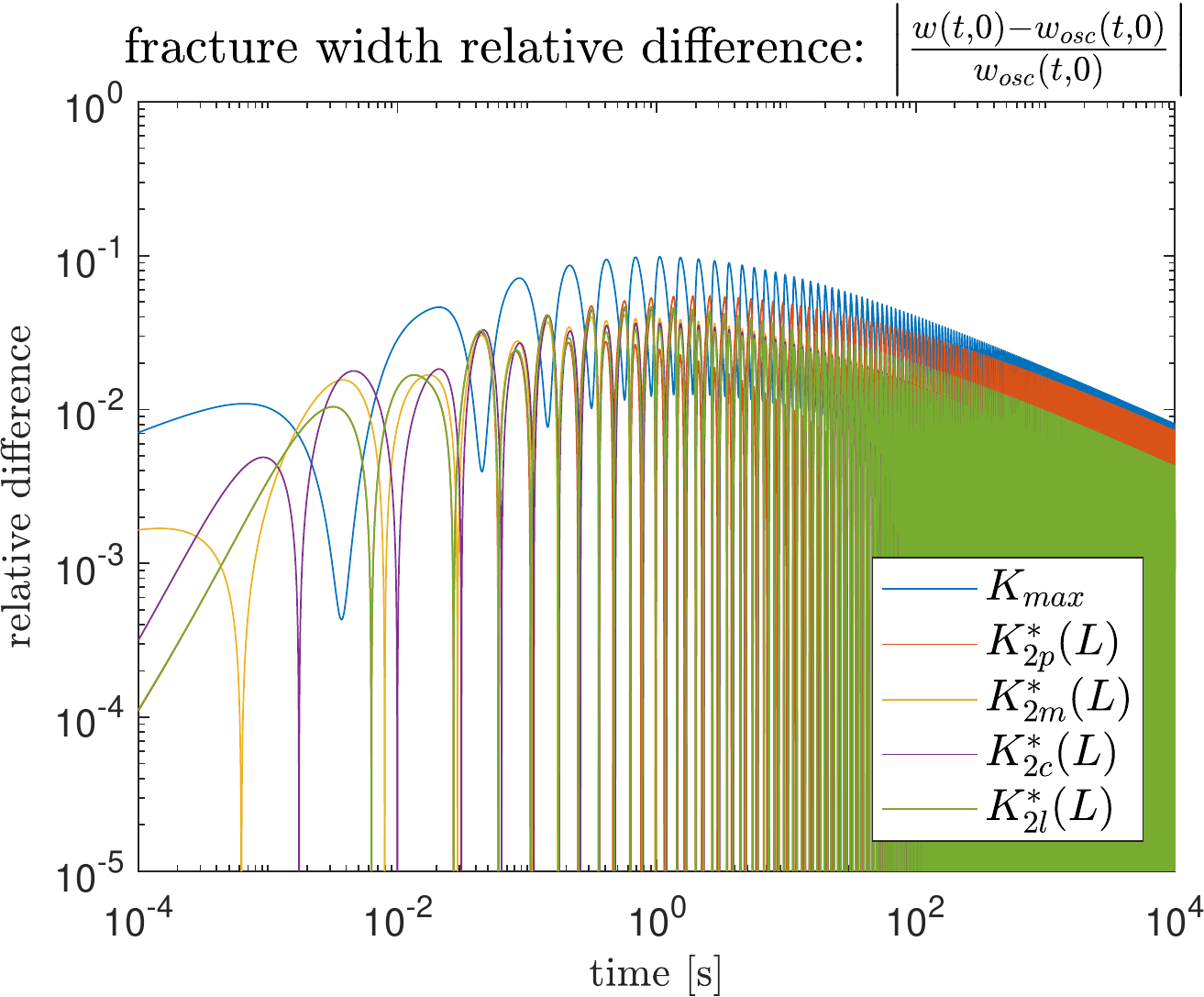}
\put(-225,155) {{\bf (c)}}
\hspace{12mm}
\includegraphics[width=0.45\textwidth]{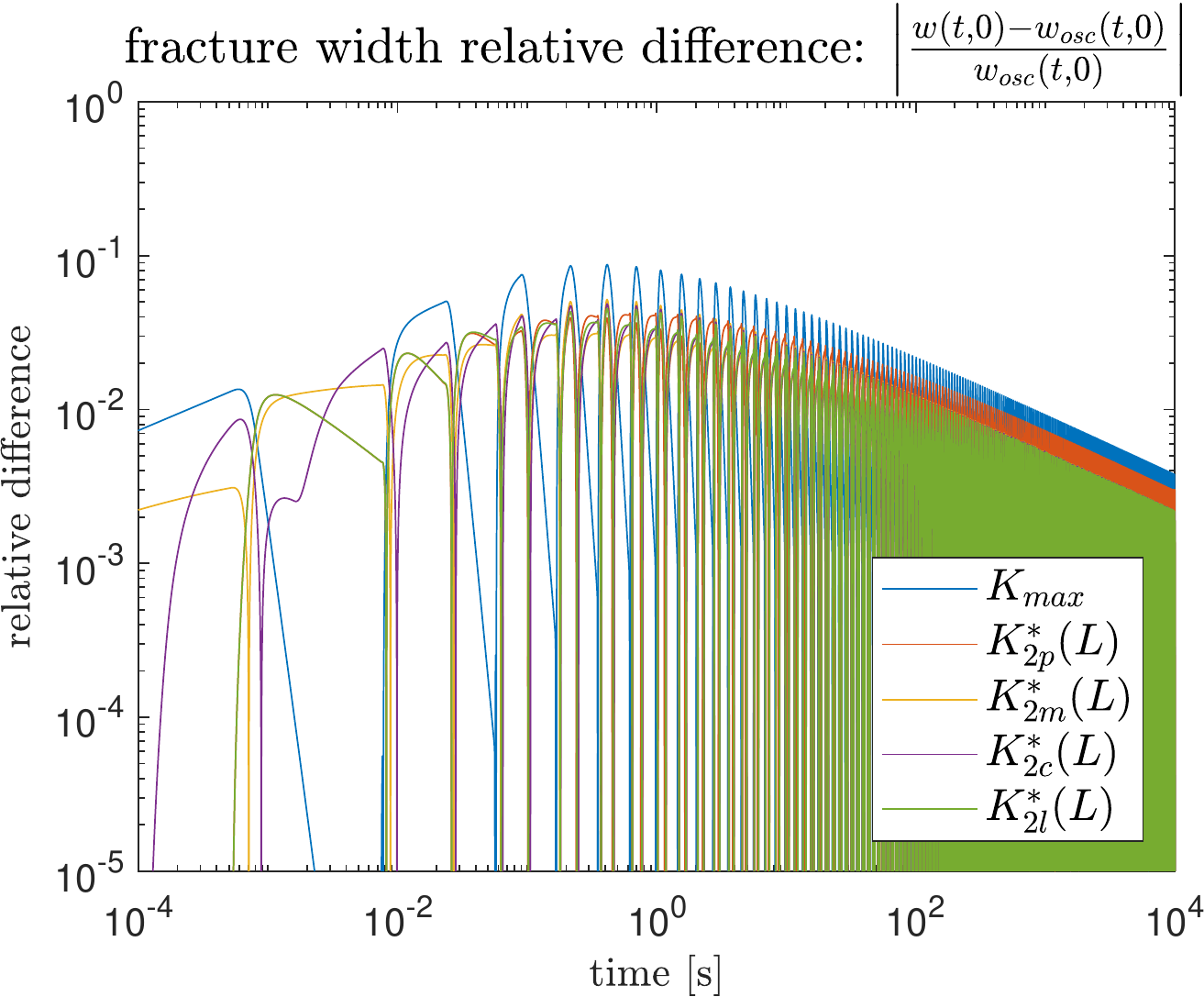}
\put(-225,155) {{\bf (d)}}
\\
\includegraphics[width=0.45\textwidth]{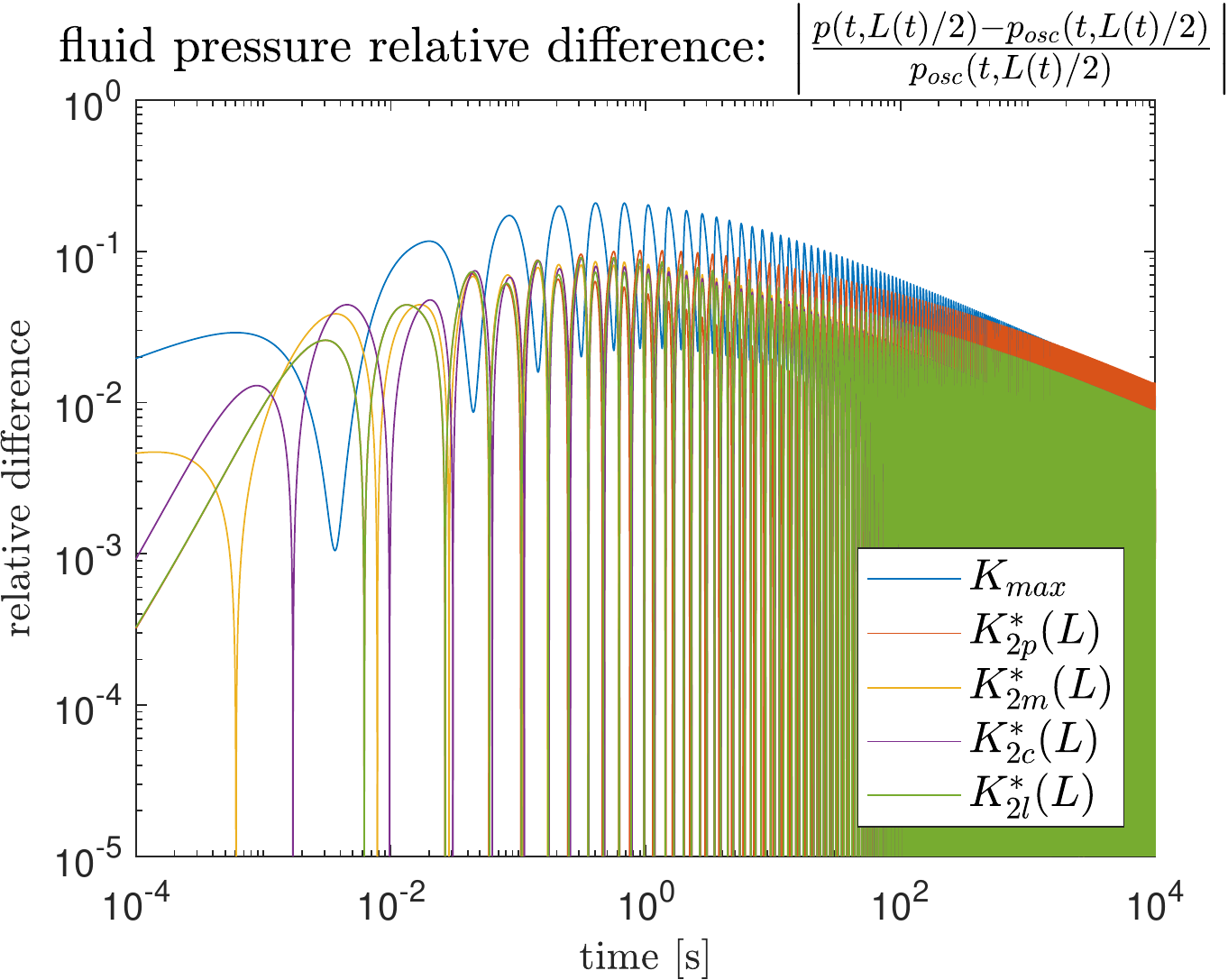}
\put(-225,155) {{\bf (e)}}
\hspace{12mm}
\includegraphics[width=0.45\textwidth]{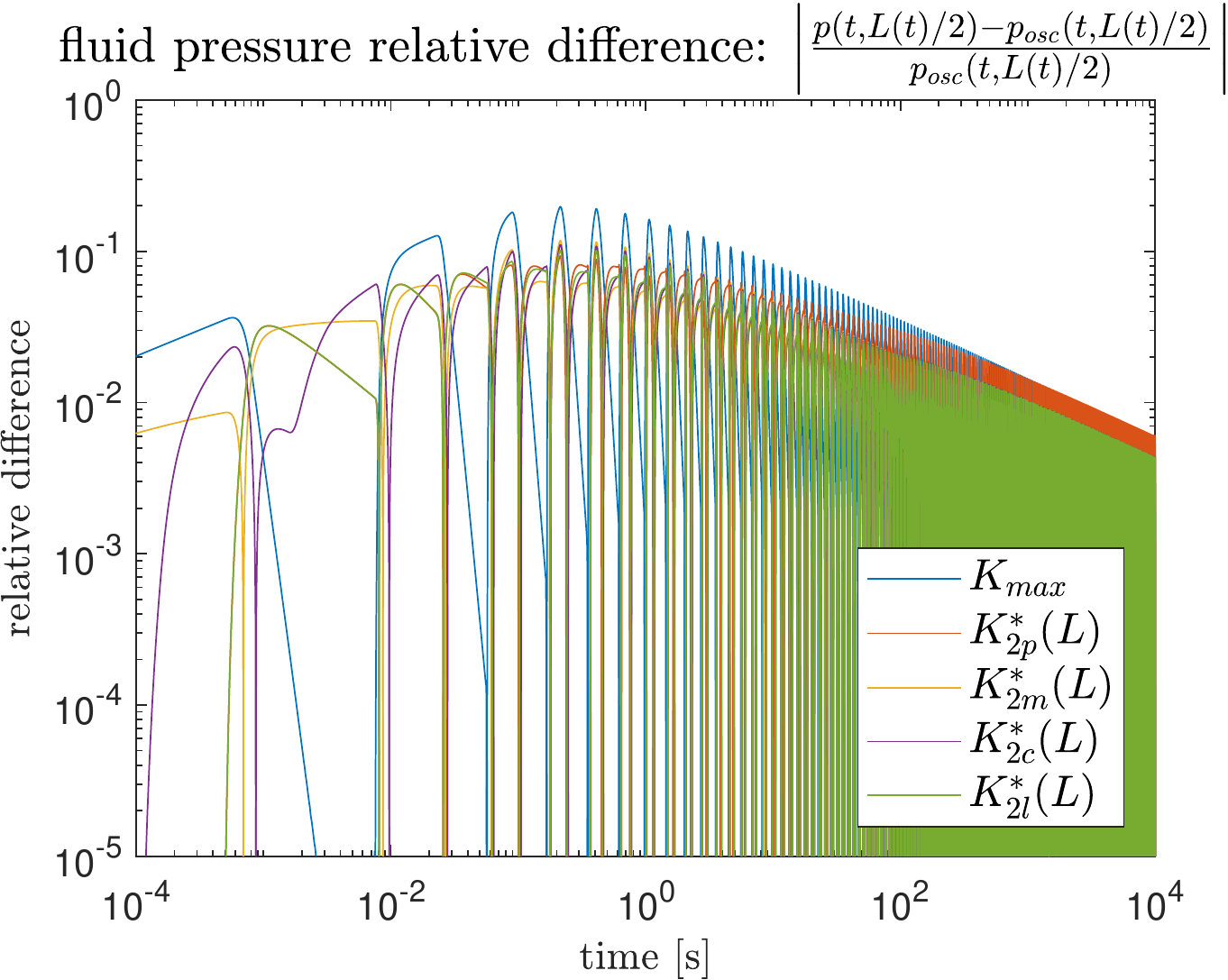}
\put(-225,155) {{\bf (f)}}
\caption{Results for the toughness-transient distribution with balanced layering (Case 2, see Table.~\ref{Table:toughness}). Relative differences for the {\bf (a)}, {\bf (b)} fracture length $L(t)$, {\bf (c)}, {\bf (d)} fracture opening at the injection point $w(t,0)$, {\bf (e)}, {\bf (f)} fluid pressure at the mid-point $p(t, L(t)/2 )$.}
 \label{Even_Del10_1}
\end{figure}

\begin{figure}[t!]
 \centering
 \includegraphics[width=0.45\textwidth]{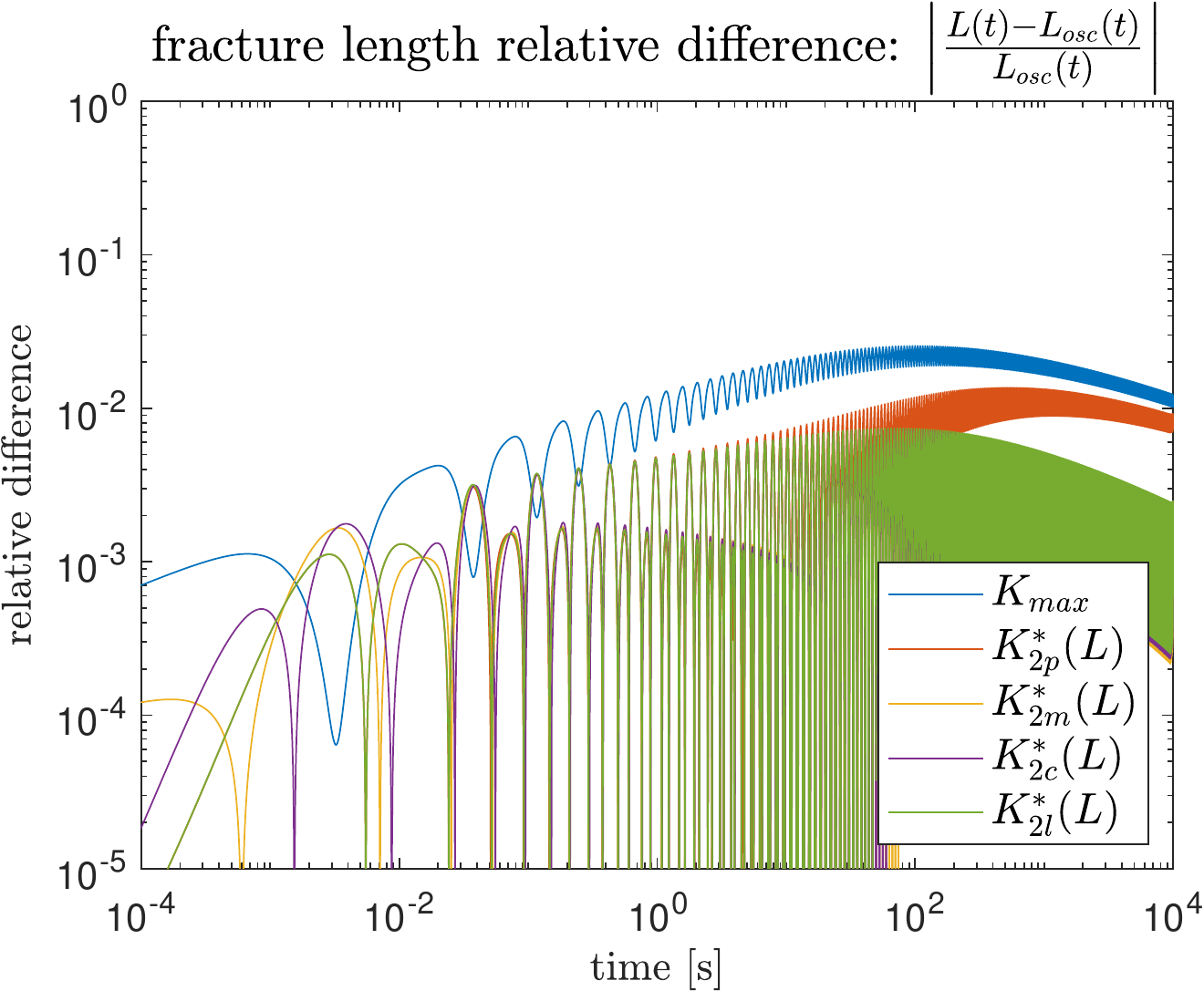}
\put(-225,155) {{\bf (a)}}
\hspace{12mm}
\includegraphics[width=0.45\textwidth]{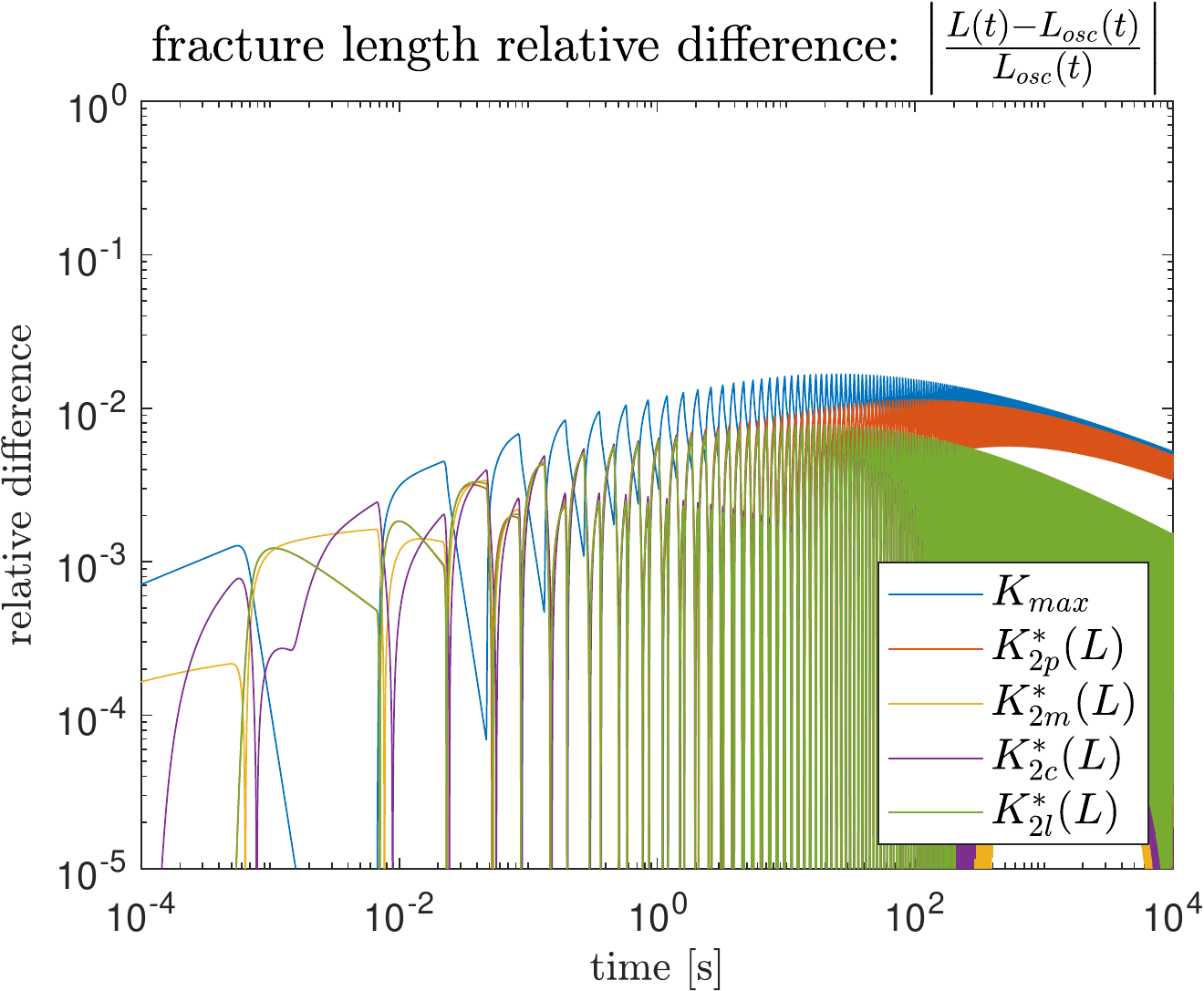}
\put(-225,155) {{\bf (b)}}
\\
\includegraphics[width=0.45\textwidth]{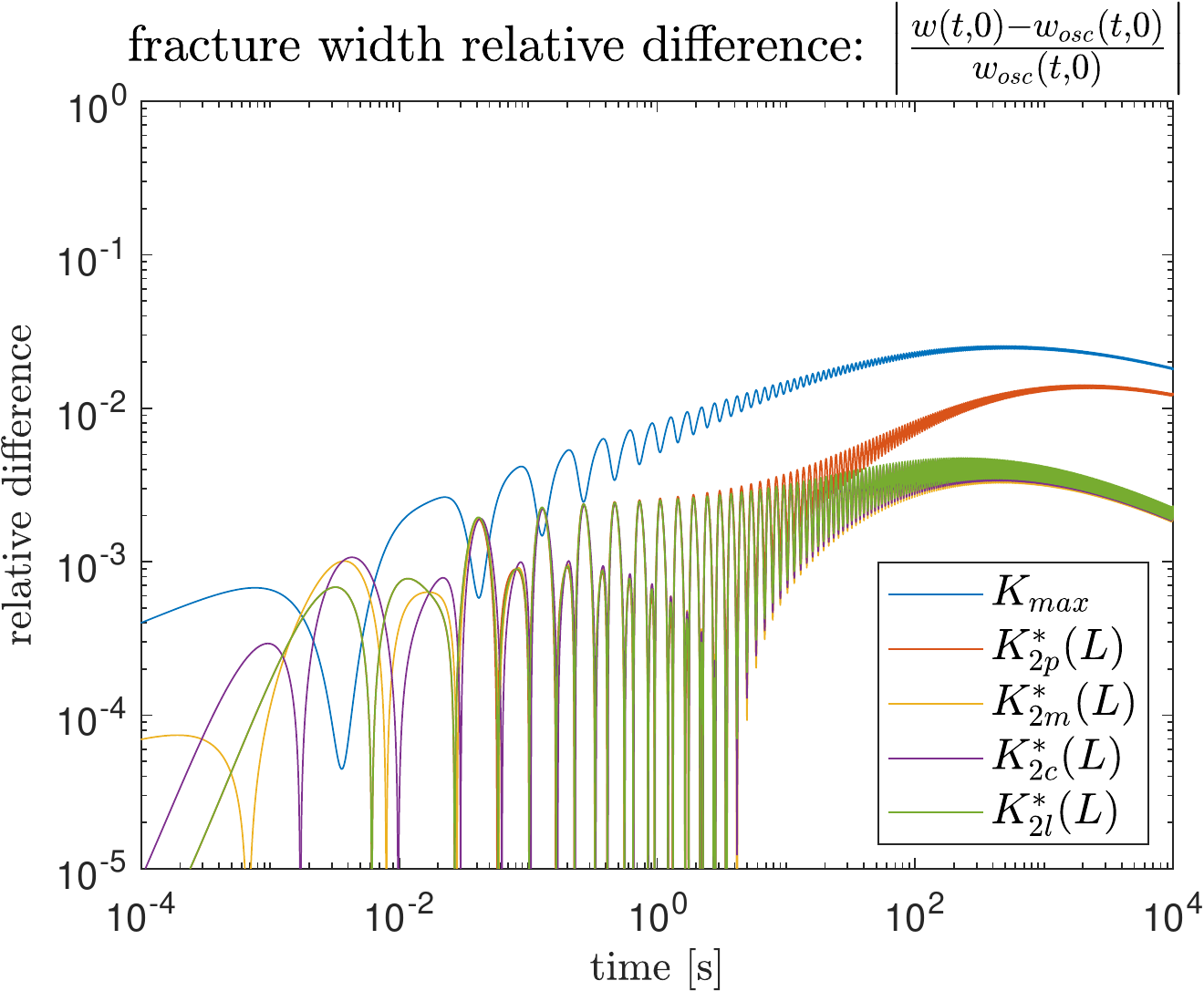}
\put(-225,155) {{\bf (c)}}
\hspace{12mm}
\includegraphics[width=0.45\textwidth]{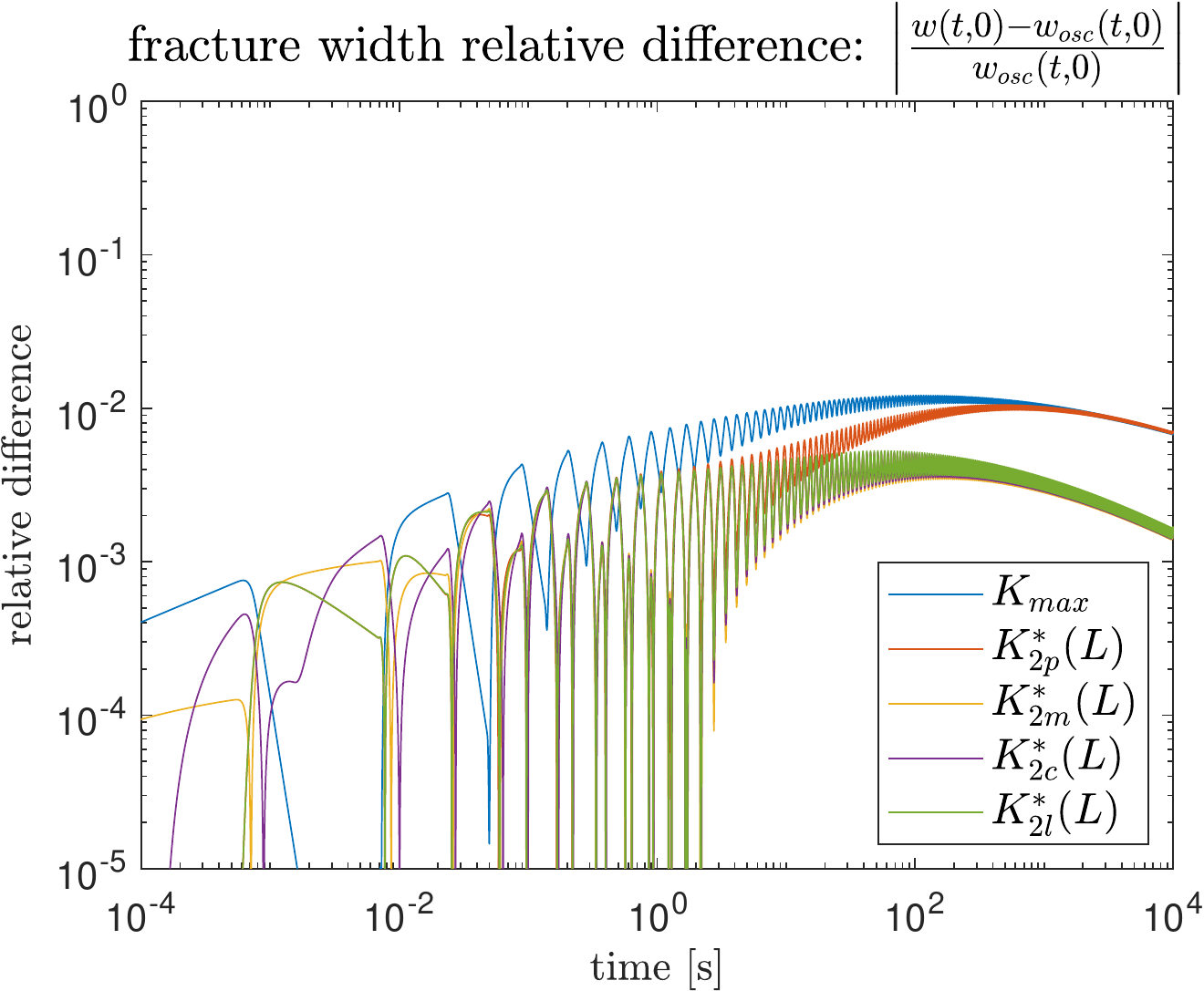}
\put(-225,155) {{\bf (d)}}
\\
\includegraphics[width=0.45\textwidth]{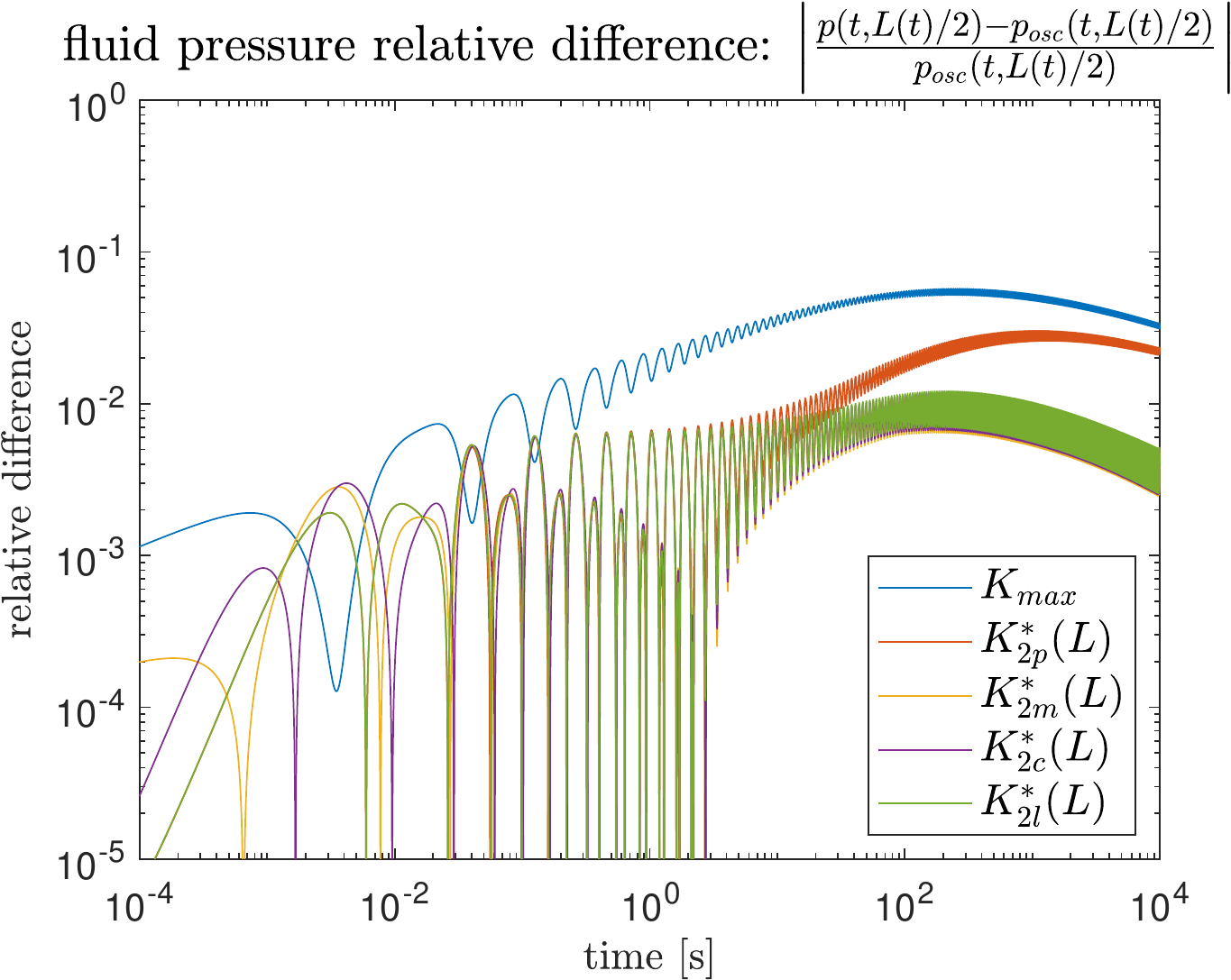}
\put(-225,155) {{\bf (e)}}
\hspace{12mm}
\includegraphics[width=0.45\textwidth]{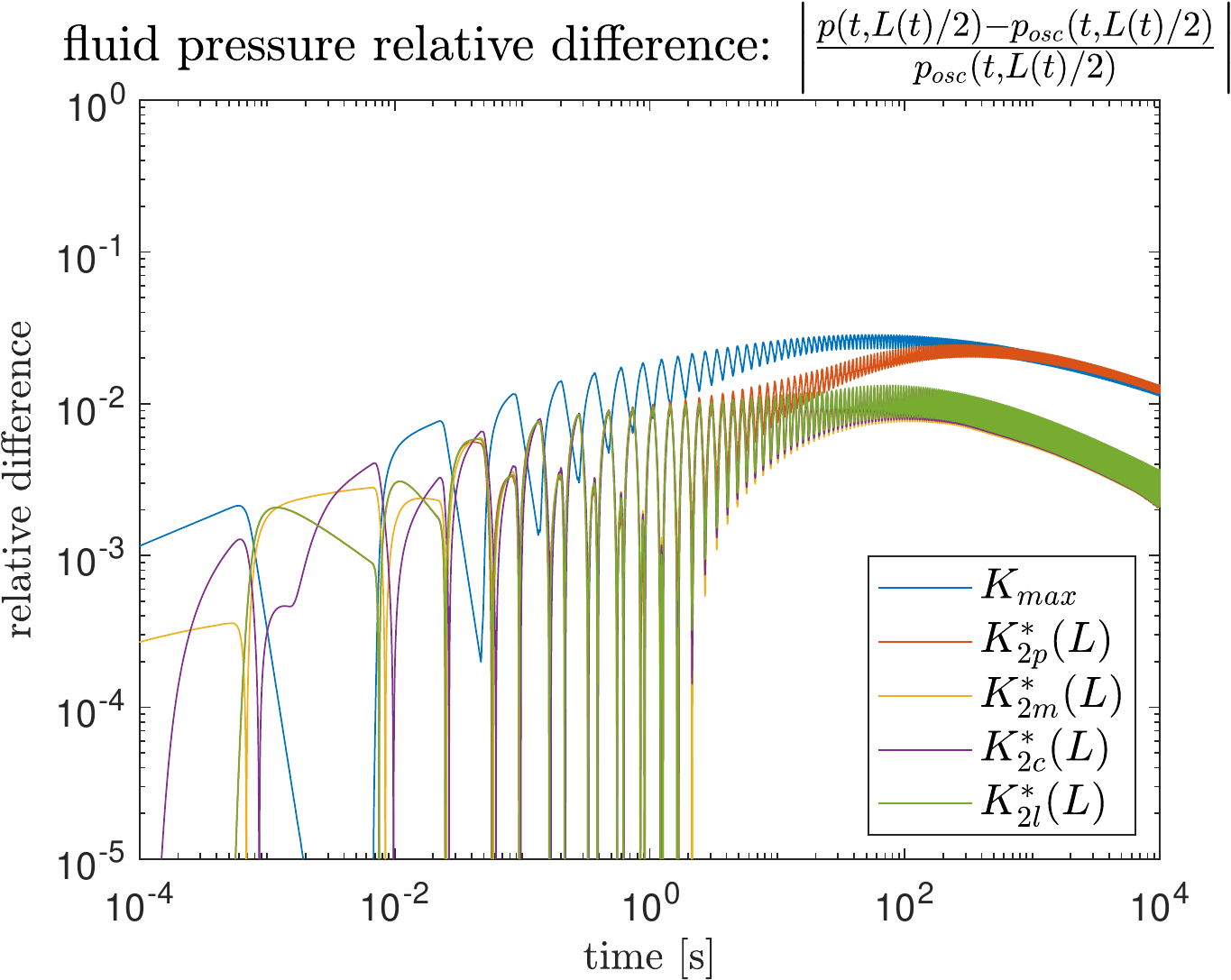}
\put(-225,155) {{\bf (f)}}
\caption{Results for the transient-viscosity distribution with balanced layering (Case 3, see Table.~\ref{Table:toughness}). Relative differences for the {\bf (a)}, {\bf (b)} fracture length $L(t)$, {\bf (c)}, {\bf (d)} fracture opening at the injection point $w(t,0)$, {\bf (e)}, {\bf (f)} fluid pressure at the mid-point $p(t, L(t)/2 )$.}
 \label{Even_Del1_01}
\end{figure}

With the model in place and the various measures established, we can now begin an investigation of the effectiveness of each homogenisation strategy for the fracture toughness. We start with the case of balanced toughness. Simulations were conducted for periodic toughness with each of the step-wise and sinusoidal distributions, for each of the regime parings outlined in Table.~\ref{Table:toughness}. These were compared with results obtained by simulating the same system, but utilizing the toughness distribution obtained by each of the homogenisation strategies: the maximum toughness strategy, and the temporal-averaging measures \eqref{MeasureK2r} - \eqref{MeasureK2p}. The relative difference obtained for the key process parameters, the fracture (half-)length, $l(t)$, the crack opening, $w(t,0)$, and the pressure at the midpoint, $p(t, l(t)/2)$, are provided in Figs.~\ref{Even_Del100_10} - \ref{Even_Del1_01}.

The first case, when the fracture propagates in the toughness-toughness regime for almost the entire duration, is shown in Fig.~\ref{Even_Del100_10}. Here, the relative error of the maximum toughness strategy exceeds $10$\% for all three parameters. This improves over time, particularly for the step-wise distribution, but requires more than $10^2$ seconds to decrease below $1$\% even for the fracture length, and does not achieve this until almost $10^4$ seconds for the fluid pressure. The temporal-averaging measures fair slightly better in all cases, with the maximum relative error remaining just below $10$\% for the crack length over the entire duration, but exceeding it for the other parameters. However, it should be noted that the results for measures \eqref{MeasureK2r}-\eqref{MeasureK2p} are almost as accurate as a homogenisation strategy can be in this case, as the inherent oscillations of the parameters over time will always yield some difference (see Fig.~\ref{Fig:Aperture1} - \ref{Fig:Speed2}), and the measures are already close to the `ideal' average.

Meanwhile, the toughness-transient case provided in Fig.~\ref{Even_Del10_1} sees a clear difference in relative error for the two forms of homogenisation strategy considered here. The maximum toughness strategy has a relative error close to or exceeding $10$\% for all the process parameters, while the temporal-averaging based approaches remain below this threshold. Note that over long-time (after $10^2$ seconds), the different strategies all yield similar results to those for the toughness-toughness case, as the regime changes over time.

Finally, Fig.~\ref{Even_Del1_01} provides the transient-viscosity case. In this instance is is clear that all of the homogenisation strategies are reasonably effective, particularly during the early stages, with the error only increasing as the fracture regime changes. It is clear however that the temporal-averaging homogenisation strategy is visibly more effective, with the relative error never exceeding $1$\% for almost all process parameters over the entire duration (only figure (e), for the fluid pressure under sinusoidal toughness distribution, briefly exceeds it). 

To summarise, the approaches based on temporal averaging remain consistently more effective than the maximum toughness strategy, although the difference is less than that seen for the KGD model. For the toughness-toughness case, and to some extent the toughness-transient case, the measures based on temporal averaging approach the ideal limit of what can be achieved by any homogenisation strategy, with the error largely resulting from the inherent oscillation of the system parameters. For cracks with one region in the viscosity dominated regime, the regime change over time for penny-shaped cracks does play a notable role, as can be seen clearly in Fig.~\ref{Even_Del1_01}. As the regime changes to the transient-toughness case (see Table.~\ref{Table:delta}), the effectiveness of the maximum toughness strategy decreases significantly, although the relative error for the crack length remains below $4$\%. The temporal averaging based approach is notably less affected by this. 

\newpage


$\quad$
\newpage


$\quad$
\newpage


$\quad$
\newpage

\subsection{Results for unbalanced layering}

With the effectiveness of the various homogenisation strategies evaluated for the case where the differing material layers are evenly distributed, now we move onto the case where the maximum toughness layer represents a smaller portion of the total material. This degree to which the material layering in `unbalanced' is dictated by the constant $0<h<1$, defined in \eqref{defh}, and represents the extent to which the average toughness of the material (over space) is closer to the maximum toughness (as $h\to 1$) or the minimum toughness (as $h\to 0$) of the heterogeneous material. We only consider the latter case, with $h<0.5$.

To obtain the fullest possible picture of the effect of the unbalanced layering on the effectiveness of the various homogenisation strategies, we split the investigation into two parts. First, we will analyse the effect of varying $h$ when keeping the maximum and minimum toughness fixed (the toughness-transient case). Next, we will keep $h$ fixed at an arbitrarily low value, $h=0.01$, and examine the impact of the regime within which the crack is initially propagating on the homogenisation strategies (varying $K_{max}$, $K_{min}$).




\subsubsection{Toughness-transient layering with unbalanced toughness}

\begin{figure}[t!]
 \centering
 \includegraphics[width=0.45\textwidth]{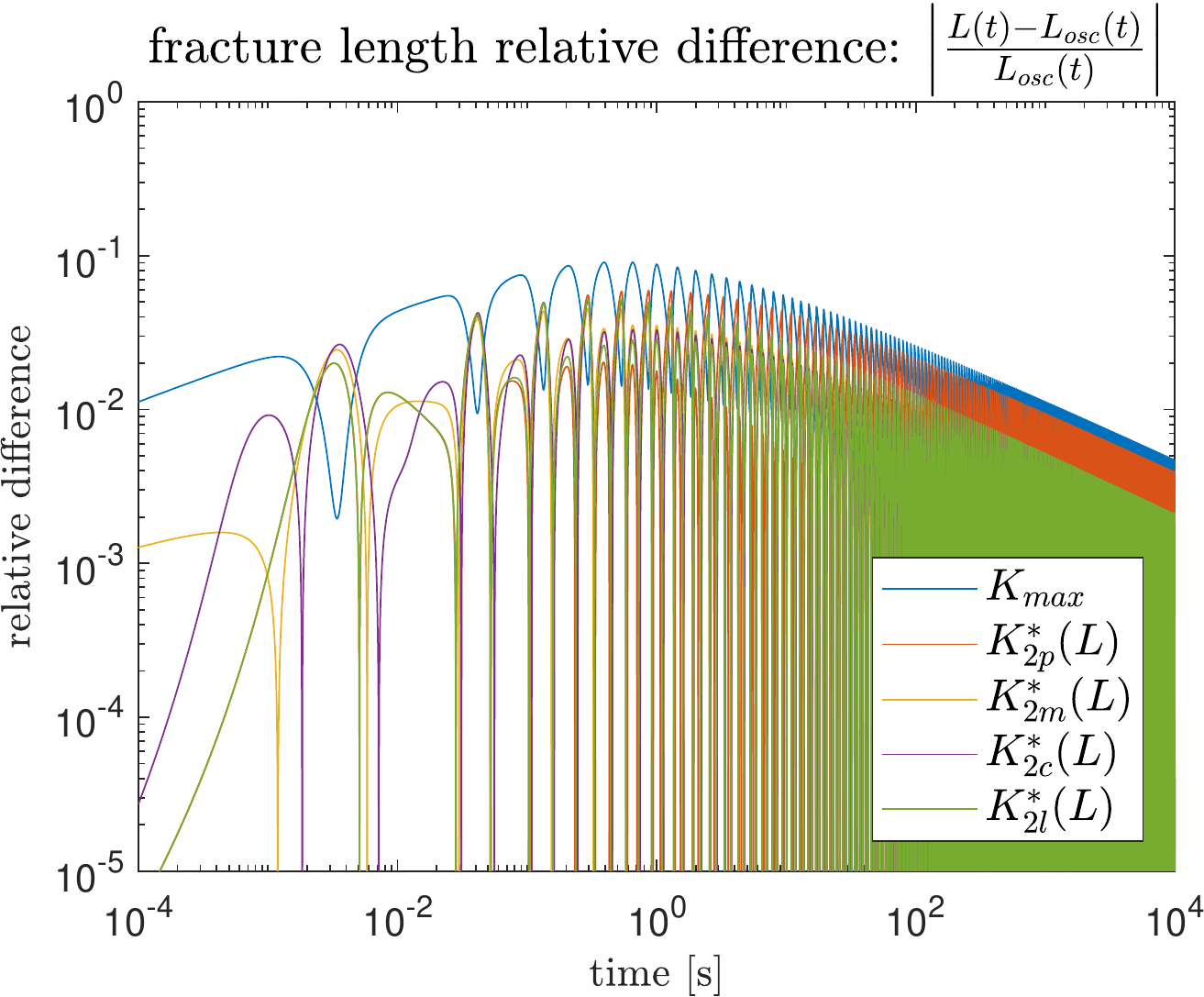}
\put(-225,155) {{\bf (a)}}
\hspace{12mm}
\includegraphics[width=0.45\textwidth]{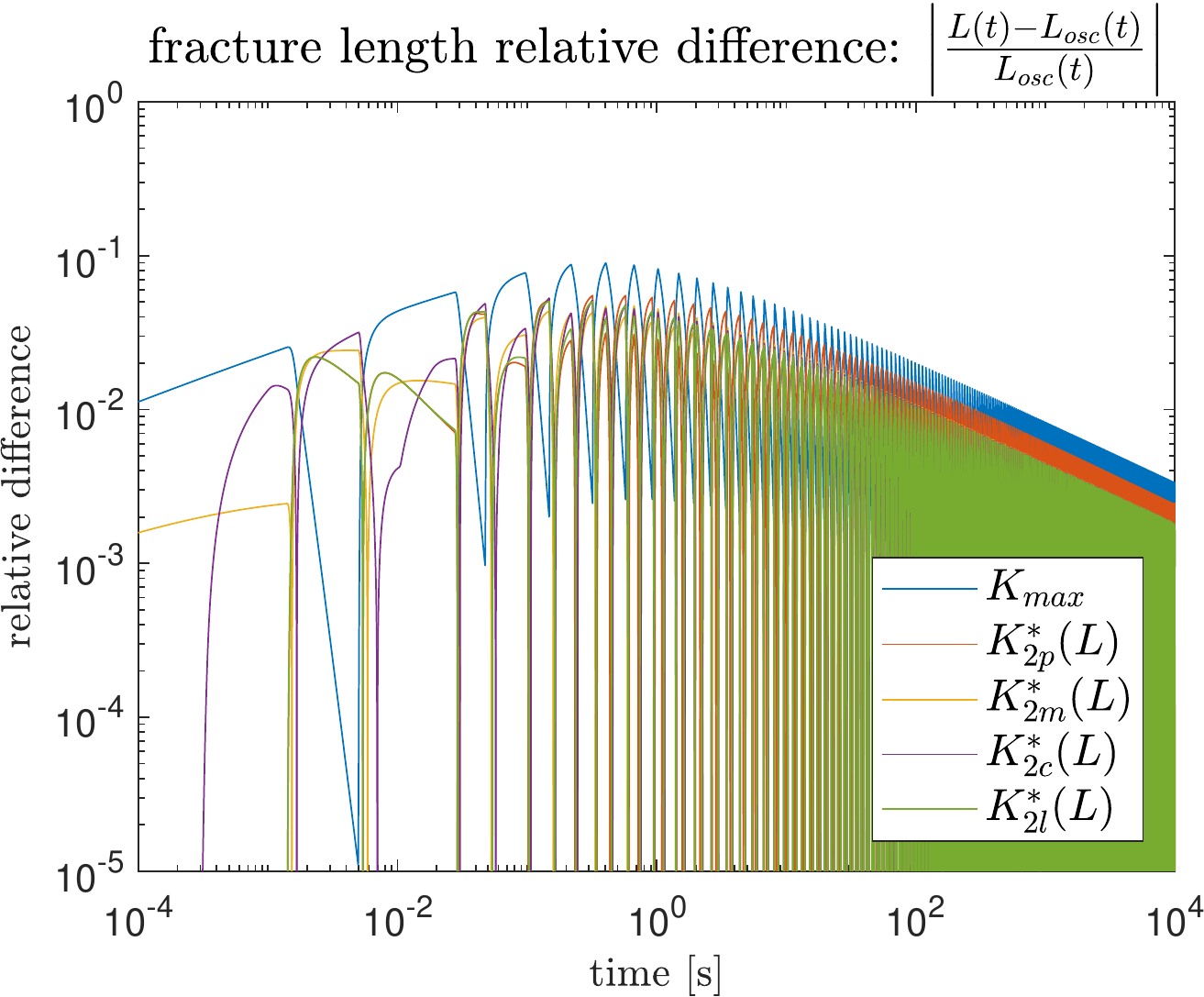}
\put(-225,155) {{\bf (b)}}
\\
 \includegraphics[width=0.45\textwidth]{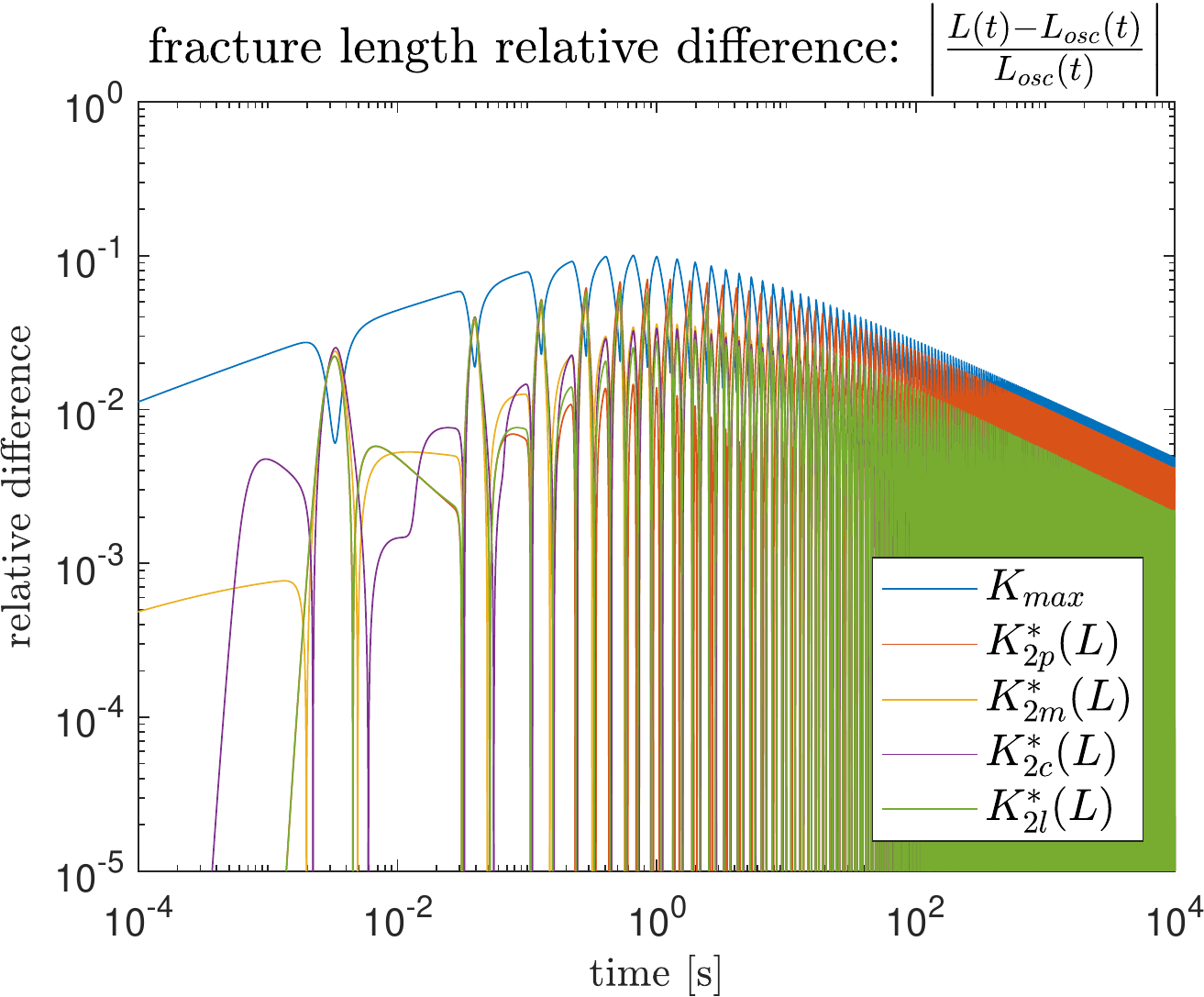}
\put(-225,155) {{\bf (c)}}
\hspace{12mm}
\includegraphics[width=0.45\textwidth]{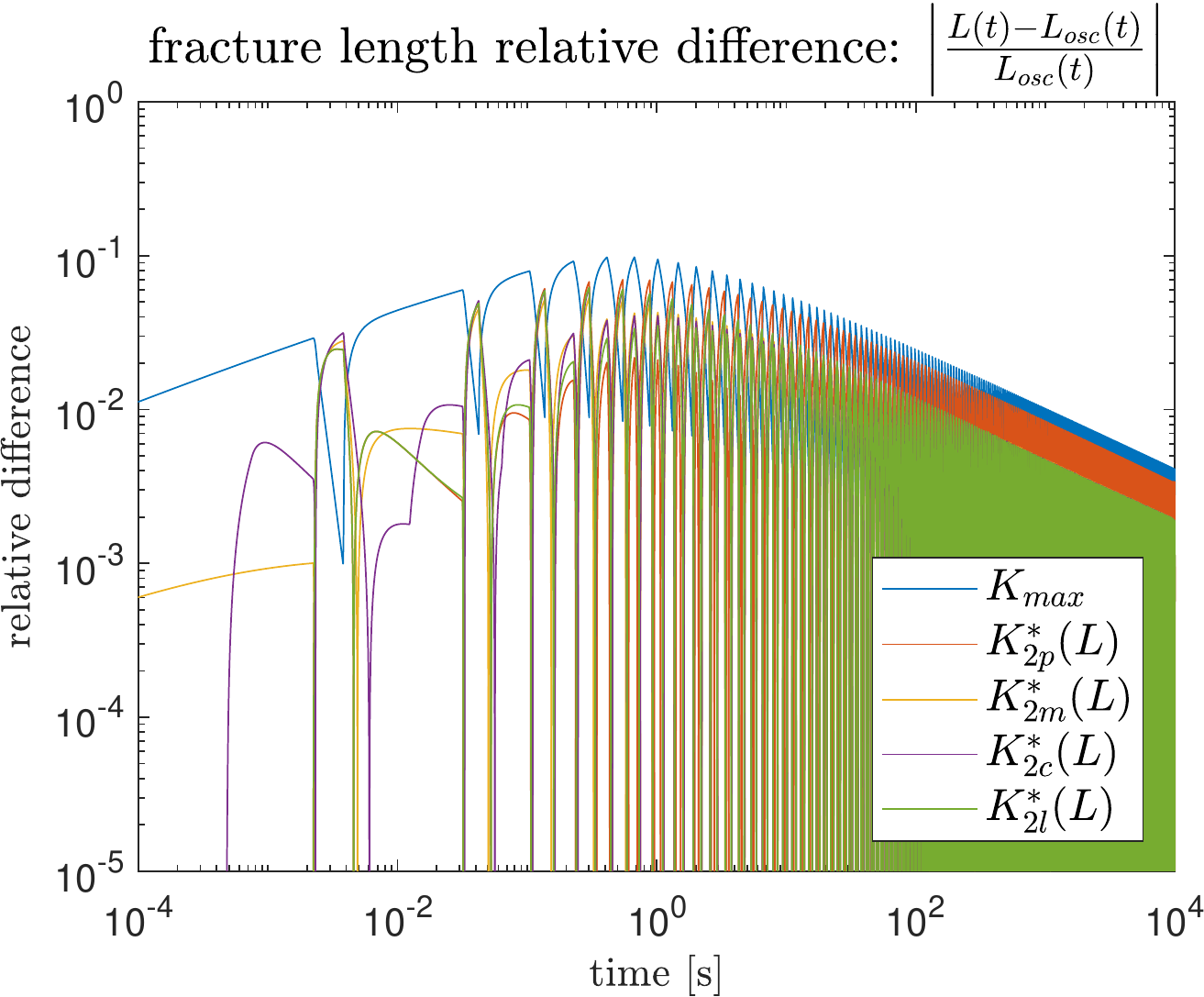}
\put(-225,155) {{\bf (d)}}
\\
 \includegraphics[width=0.45\textwidth]{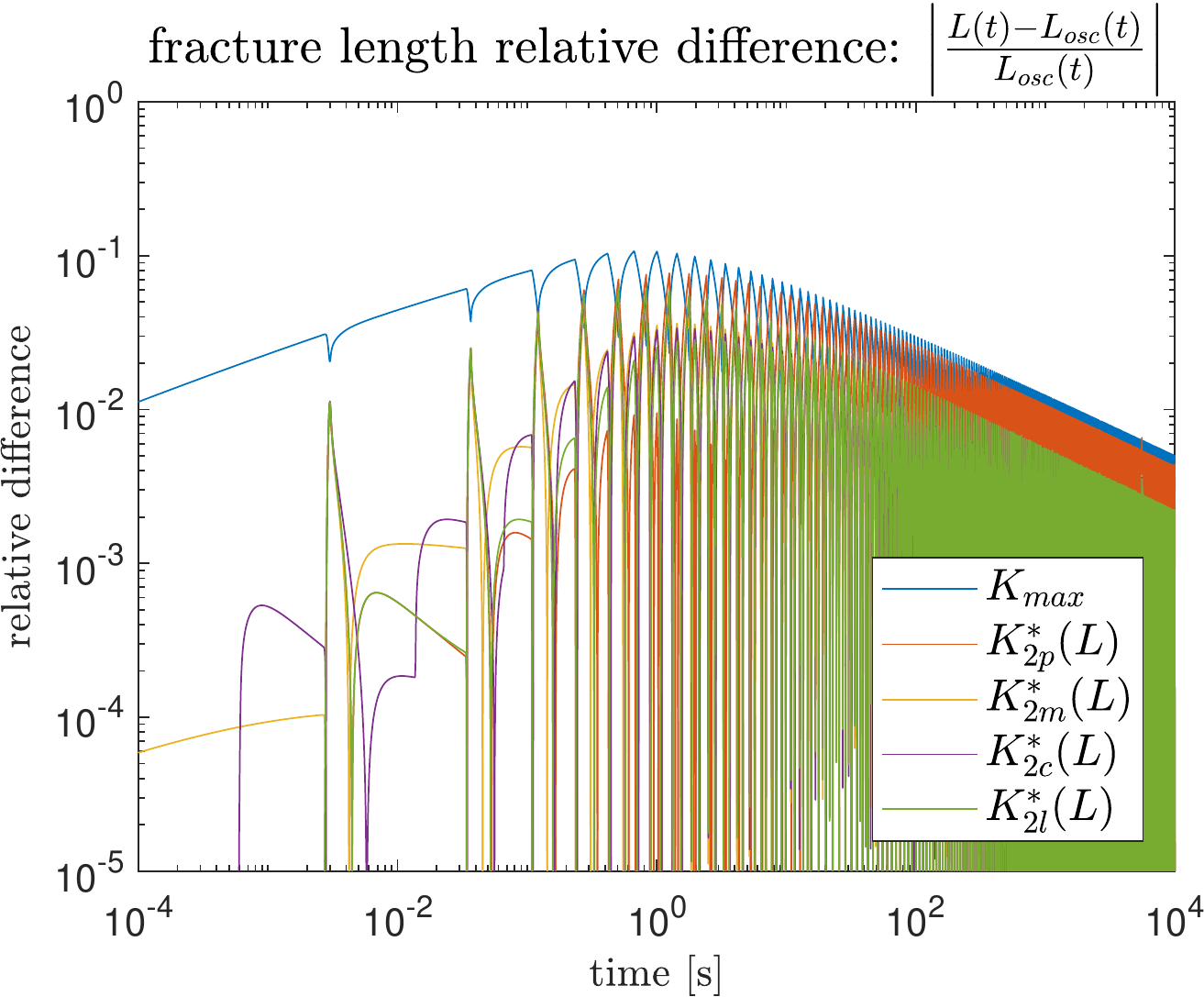}
\put(-225,155) {{\bf (e)}}
\hspace{12mm}
\includegraphics[width=0.45\textwidth]{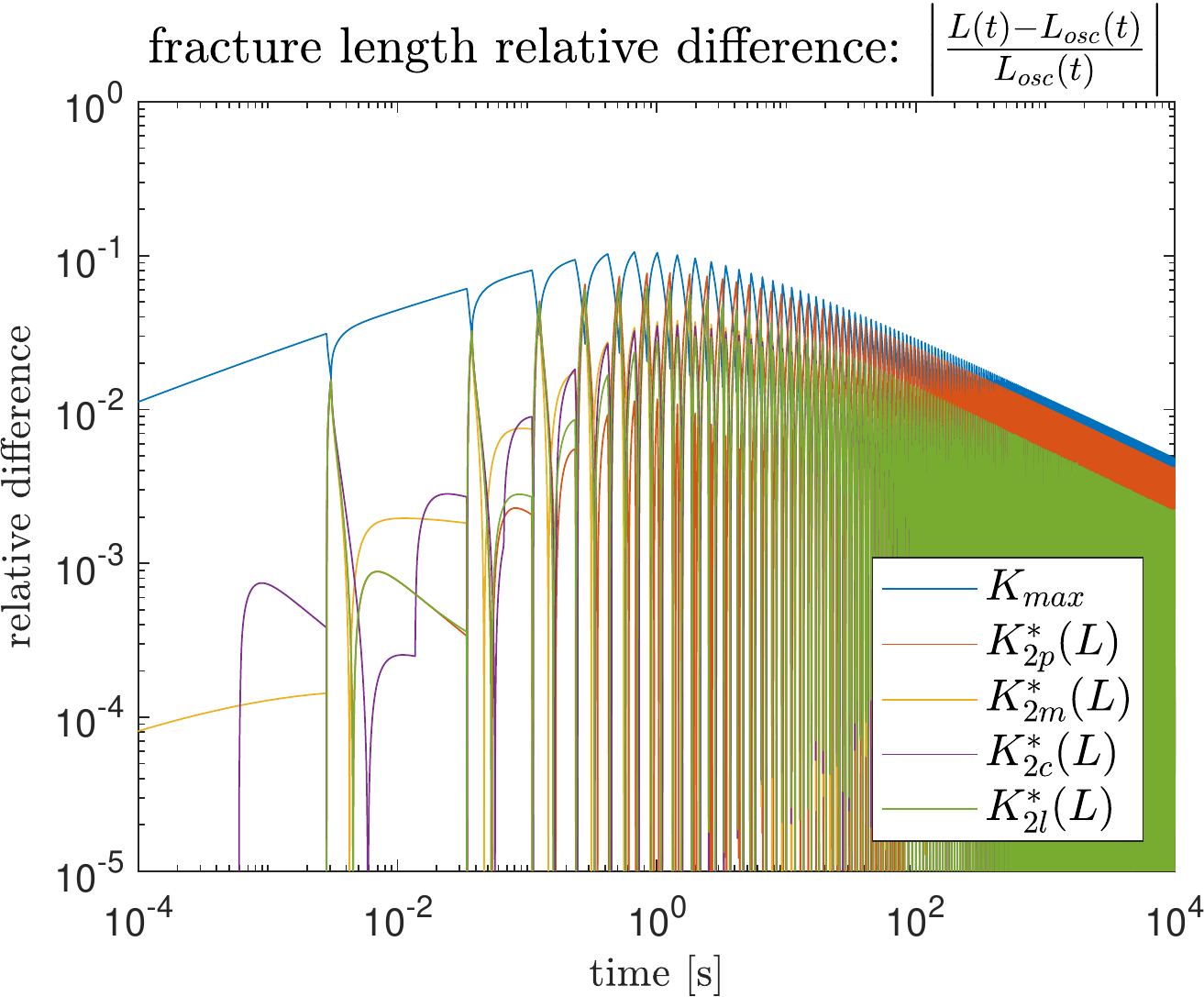}
\put(-225,155) {{\bf (f)}}
\caption{Relative difference in the fracture (half-)length $L(t)$ for the toughness-transient distribution with unbalanced layering (Case 2, see Table.~\ref{Table:toughness}). Relative differences for {\bf (a)}, {\bf (b)} $h=0.25$ (see \eqref{defh}, {\bf (c)}, {\bf (d)} $h=0.1$, {\bf (e)}, {\bf (f)} $h=0.01$.}
 \label{Unbalanced_Rel_L}
\end{figure}

\begin{figure}[t!]
 \centering
 \includegraphics[width=0.45\textwidth]{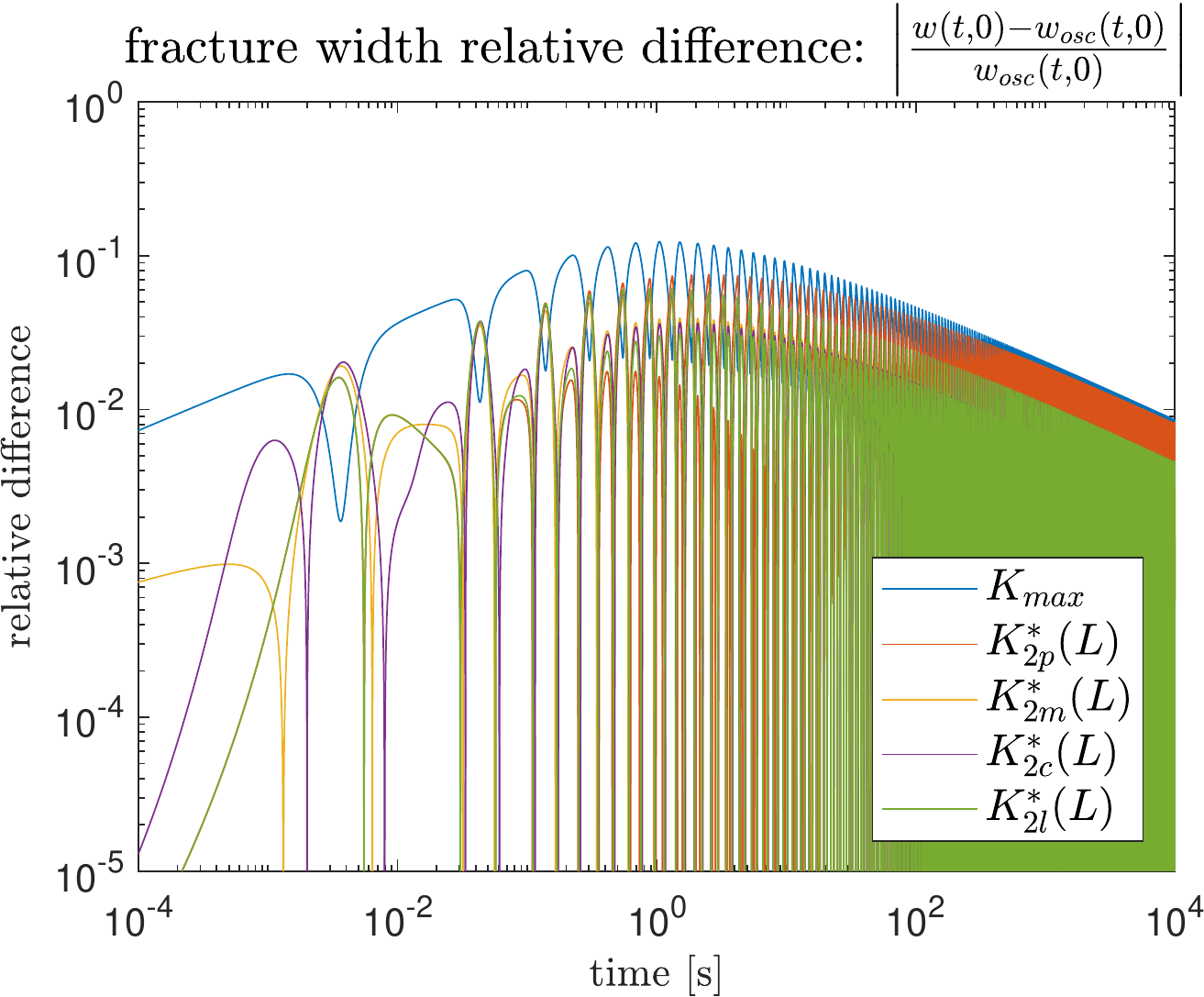}
\put(-225,155) {{\bf (a)}}
\hspace{12mm}
\includegraphics[width=0.45\textwidth]{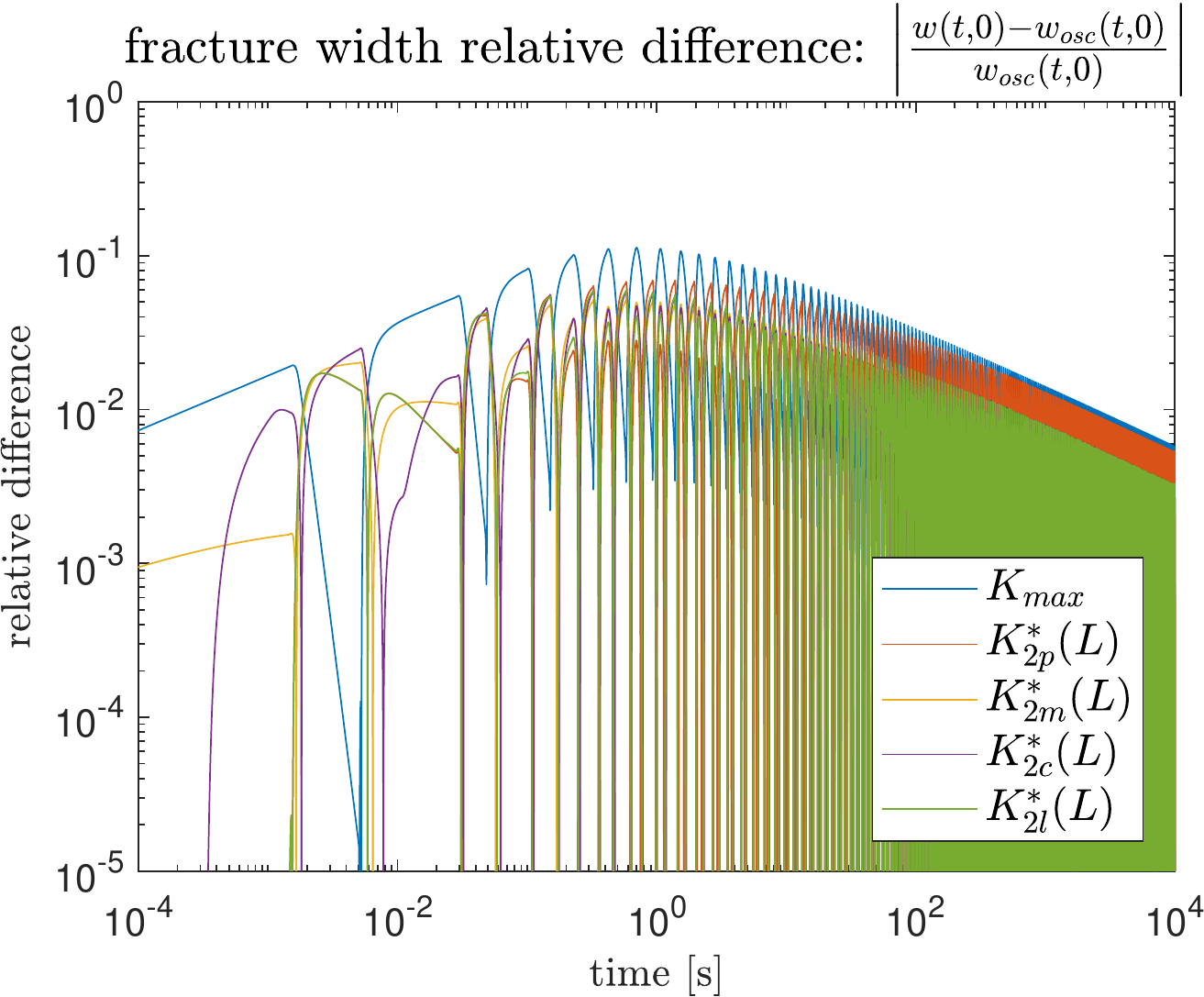}
\put(-225,155) {{\bf (b)}}
\\
 \includegraphics[width=0.45\textwidth]{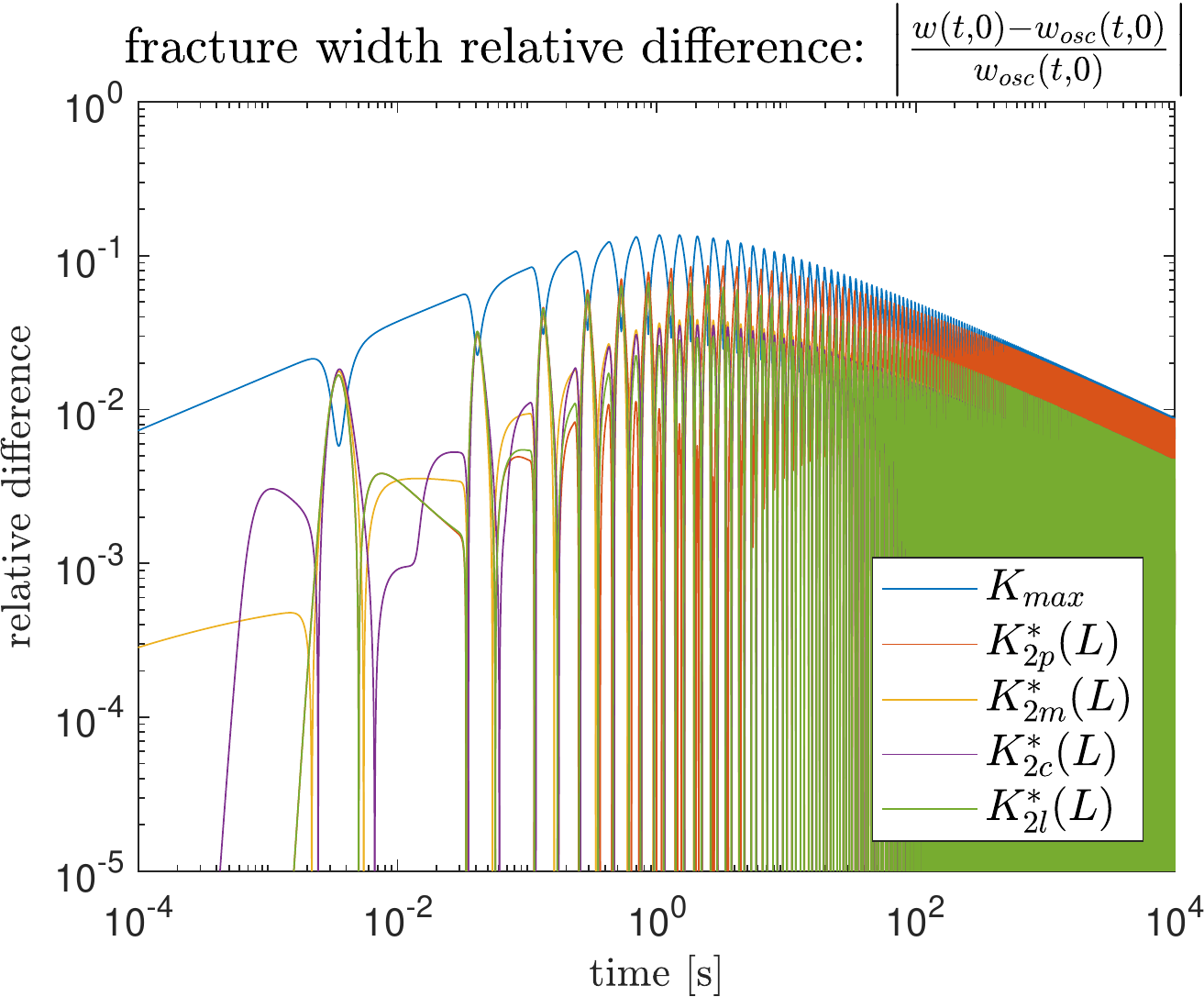}
\put(-225,155) {{\bf (c)}}
\hspace{12mm}
\includegraphics[width=0.45\textwidth]{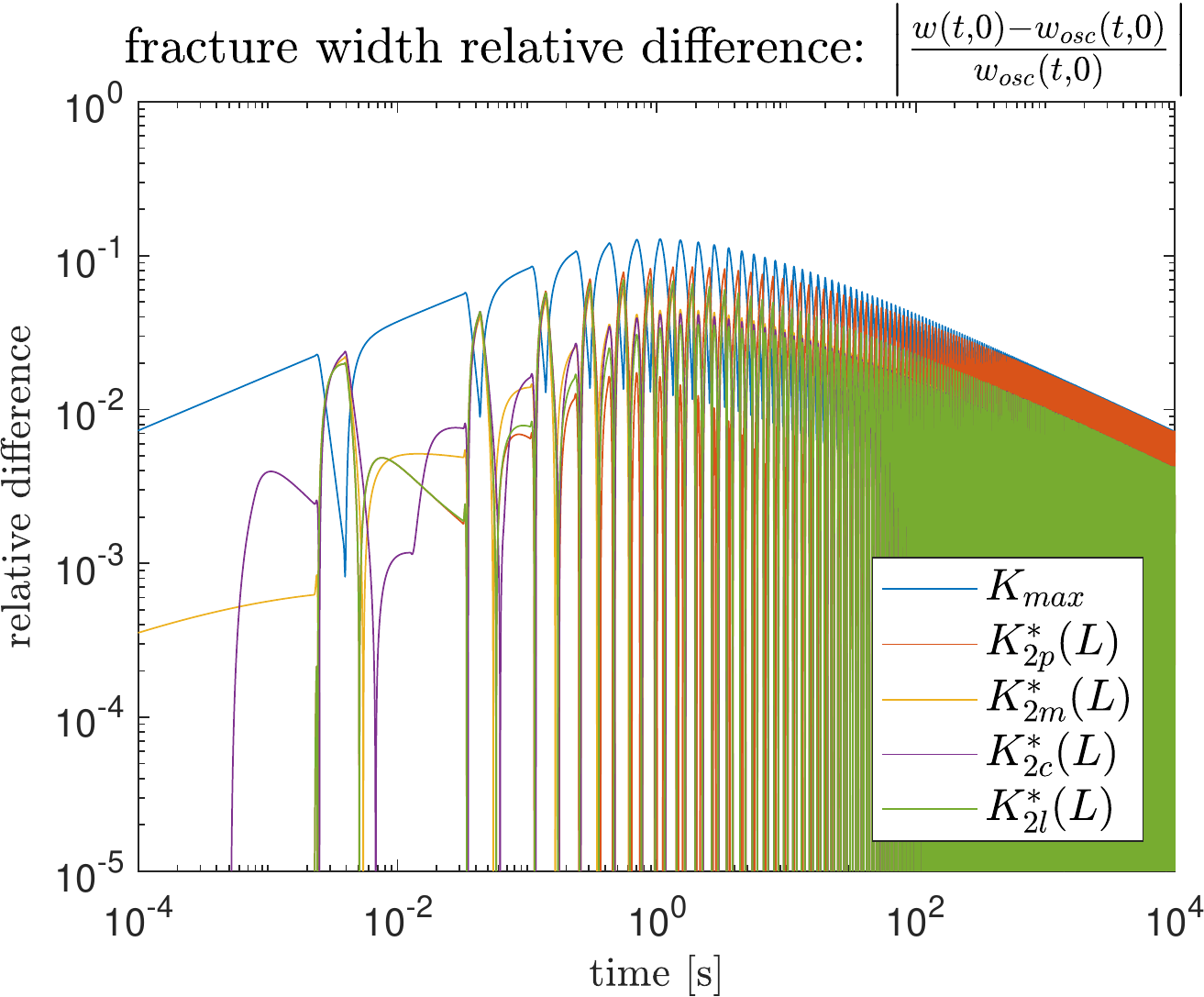}
\put(-225,155) {{\bf (d)}}
\\
 \includegraphics[width=0.45\textwidth]{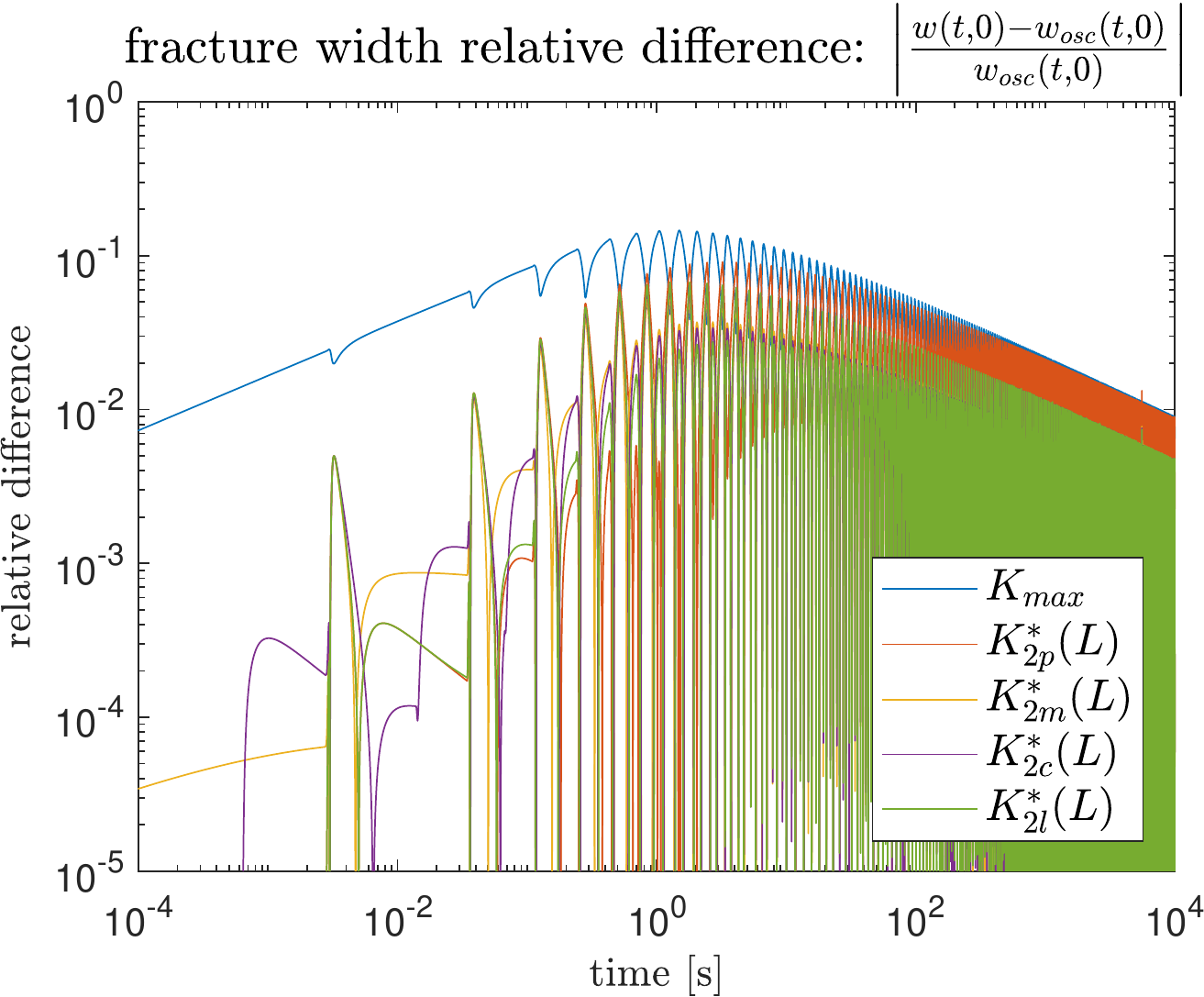}
\put(-225,155) {{\bf (e)}}
\hspace{12mm}
\includegraphics[width=0.45\textwidth]{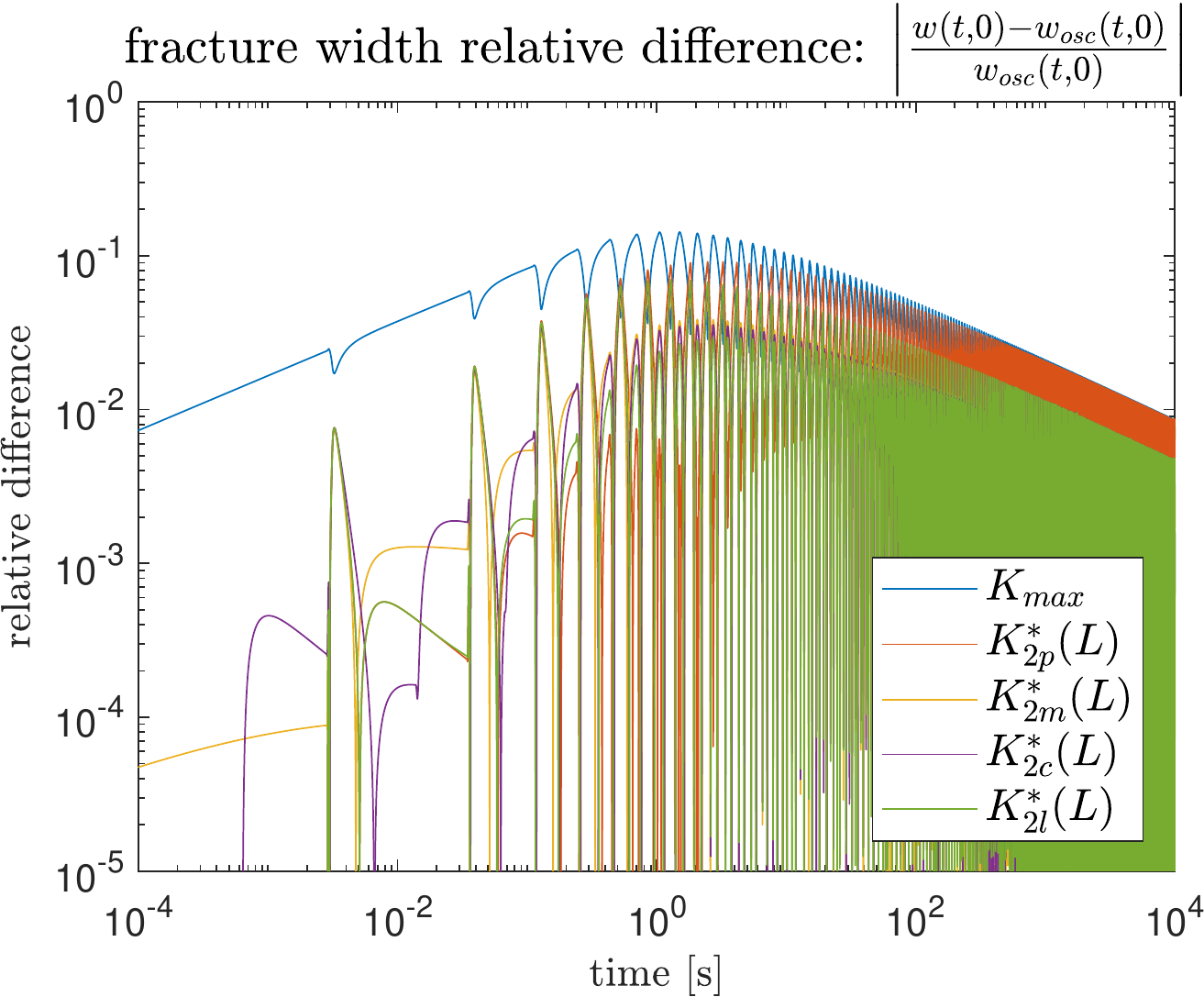}
\put(-225,155) {{\bf (f)}}
\caption{Relative difference in the crack opening $w(t,0)$ for the toughness-transient distribution with unbalanced layering (Case 2, see Table.~\ref{Table:toughness}). Relative differences for {\bf (a)}, {\bf (b)} $h=0.25$ (see \eqref{defh}, {\bf (c)}, {\bf (d)} $h=0.1$, {\bf (e)}, {\bf (f)} $h=0.01$.}
 \label{Unbalanced_Rel_w0}
\end{figure}

\begin{figure}[t!]
 \centering
 \includegraphics[width=0.45\textwidth]{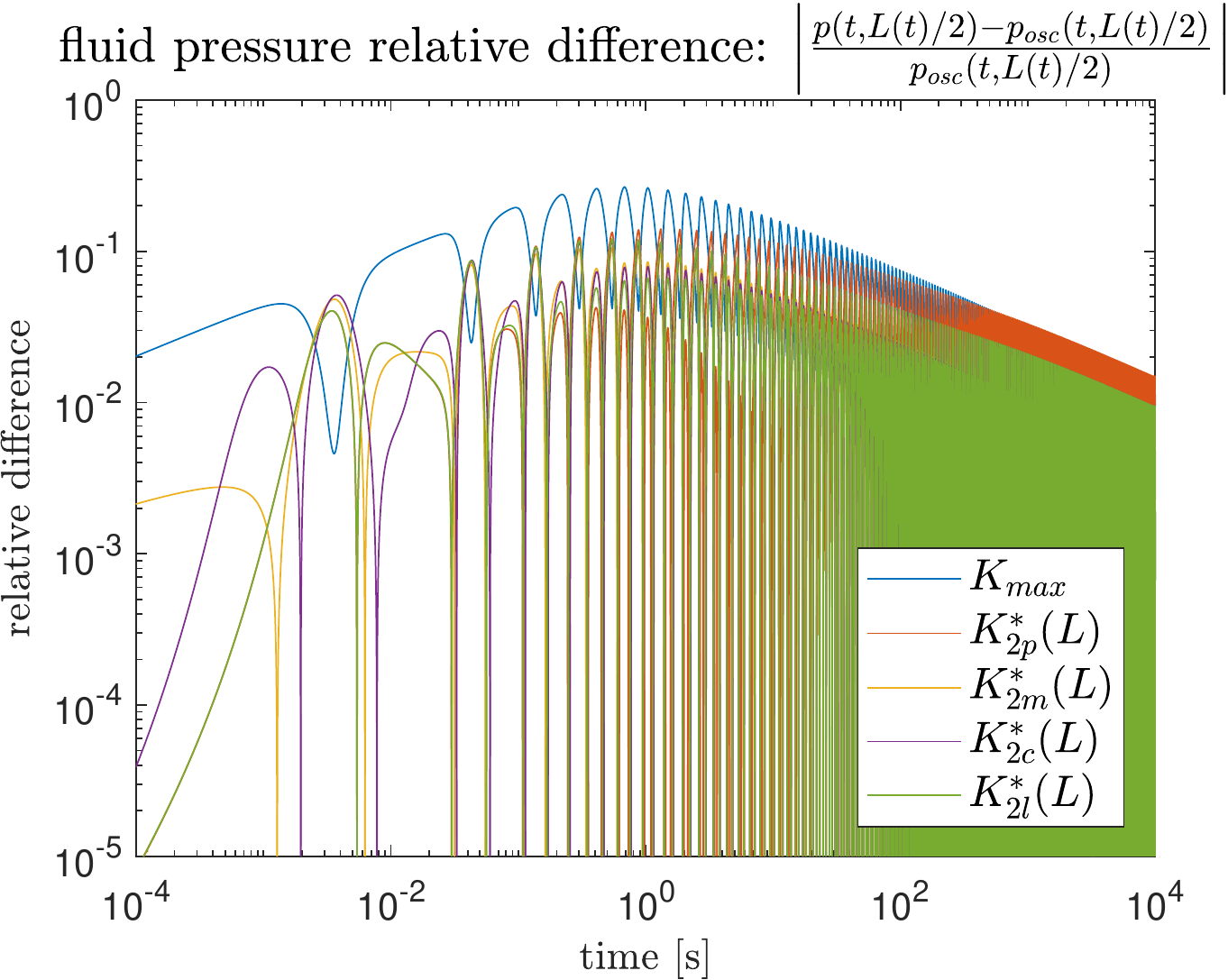}
\put(-225,155) {{\bf (a)}}
\hspace{12mm}
\includegraphics[width=0.45\textwidth]{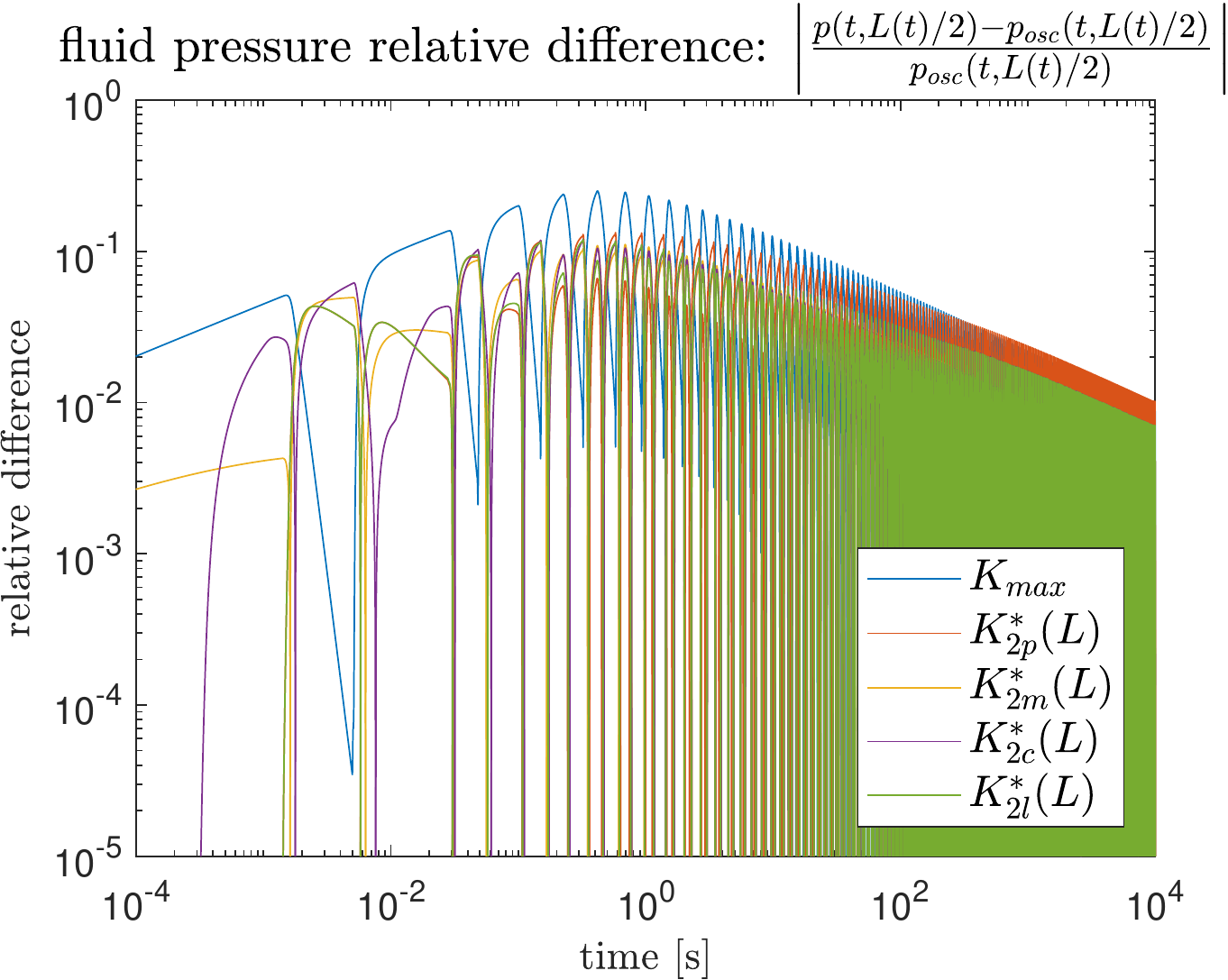}
\put(-225,155) {{\bf (b)}}
\\
 \includegraphics[width=0.45\textwidth]{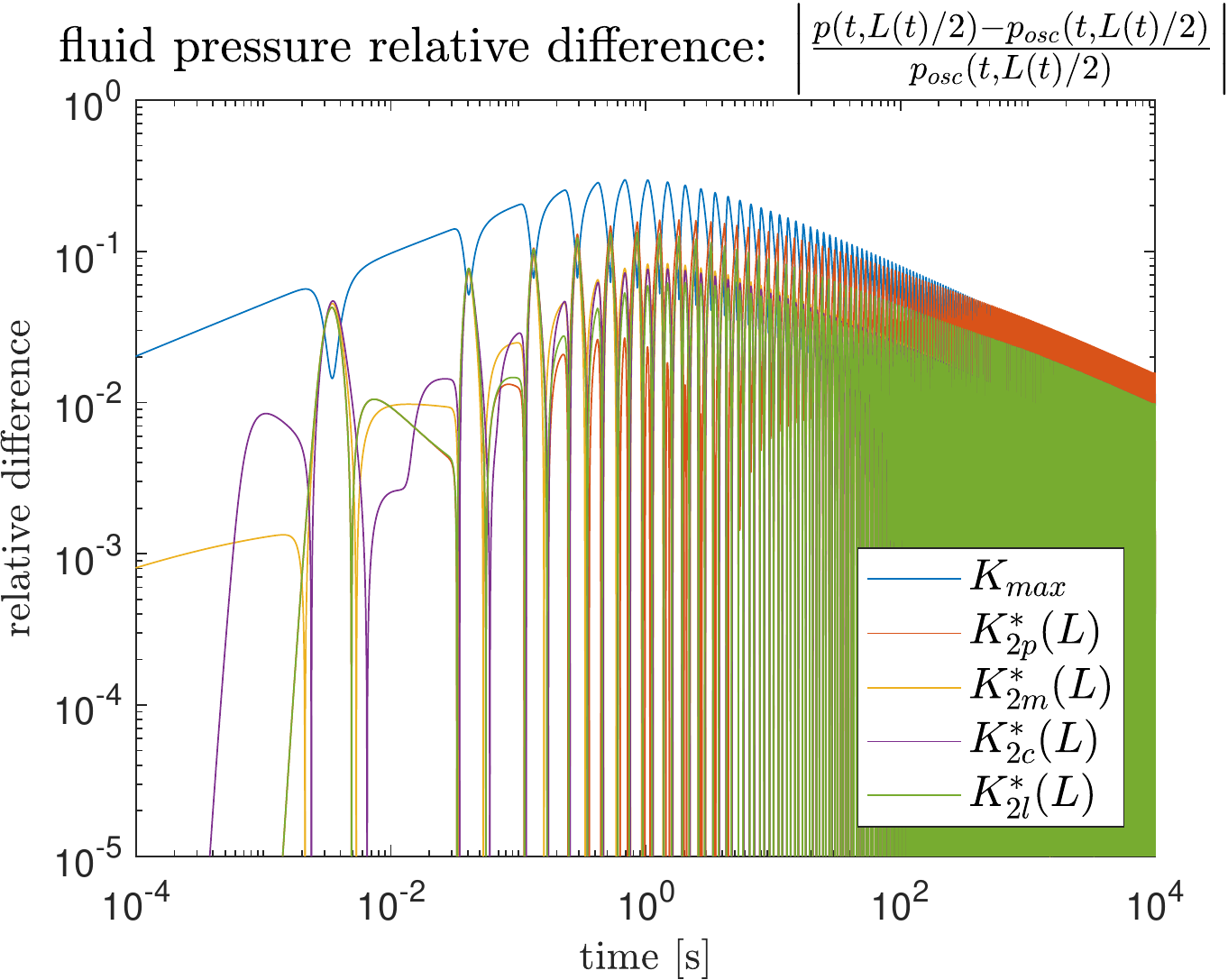}
\put(-225,155) {{\bf (c)}}
\hspace{12mm}
\includegraphics[width=0.45\textwidth]{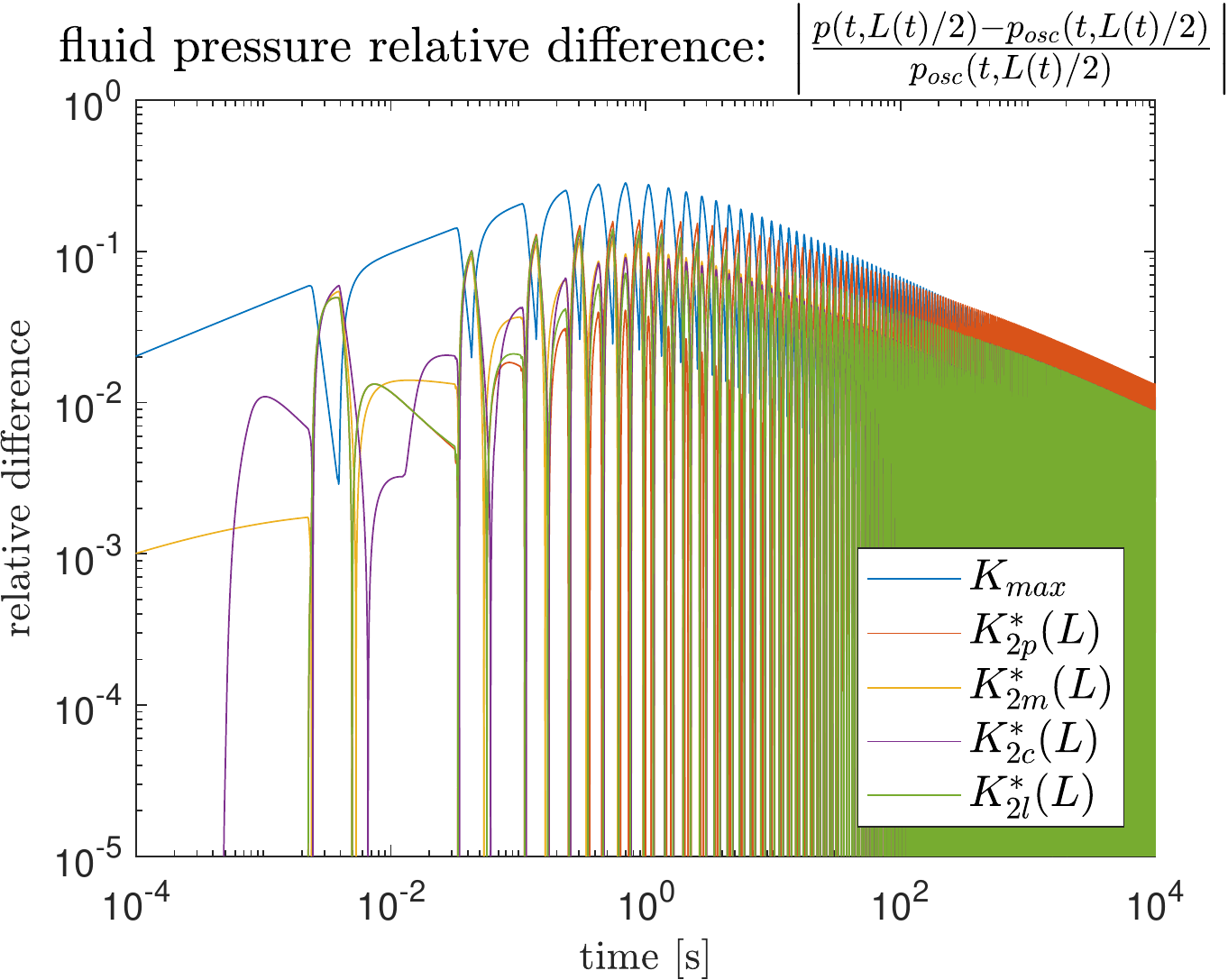}
\put(-225,155) {{\bf (d)}}
\\
 \includegraphics[width=0.45\textwidth]{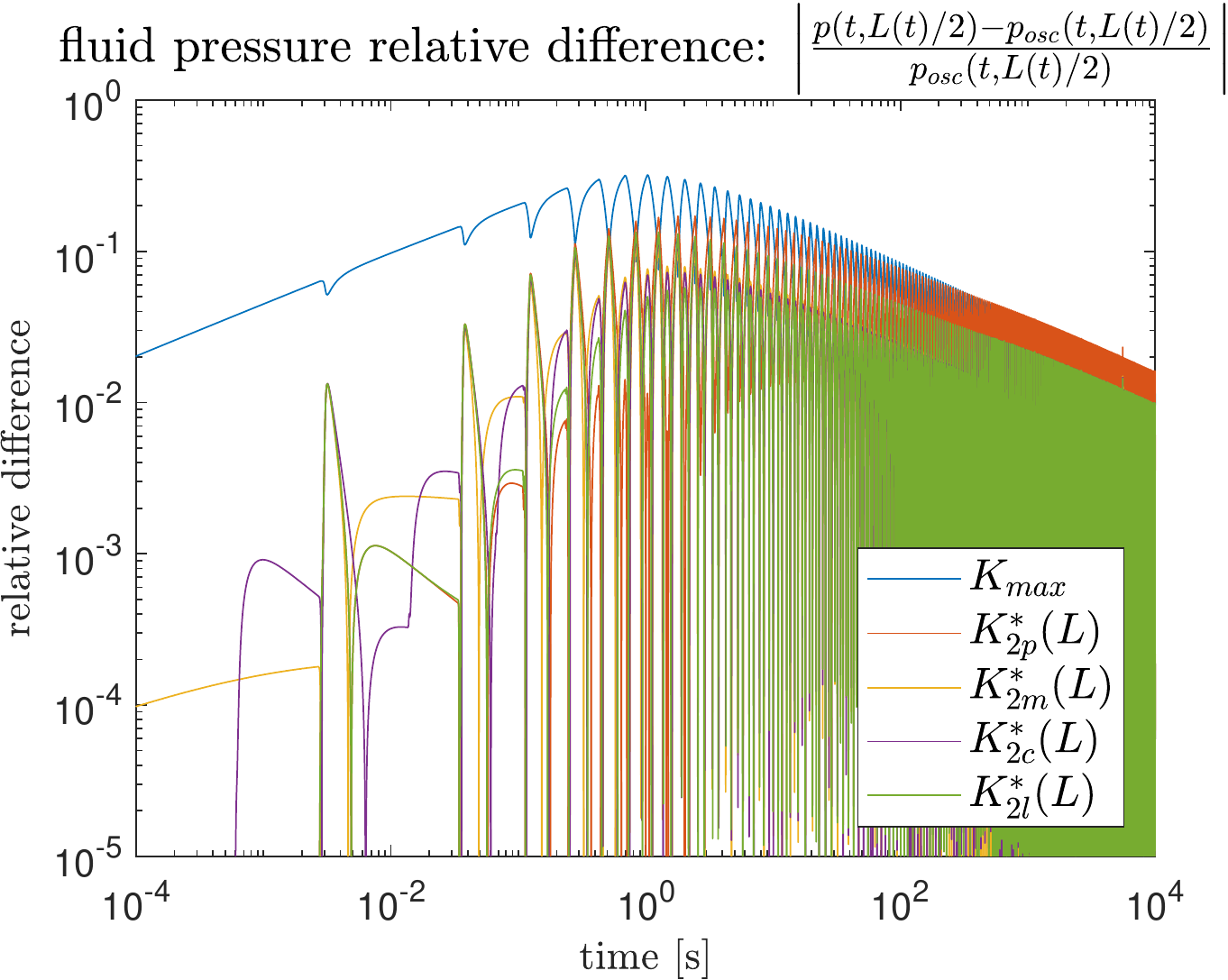}
\put(-225,155) {{\bf (e)}}
\hspace{12mm}
\includegraphics[width=0.45\textwidth]{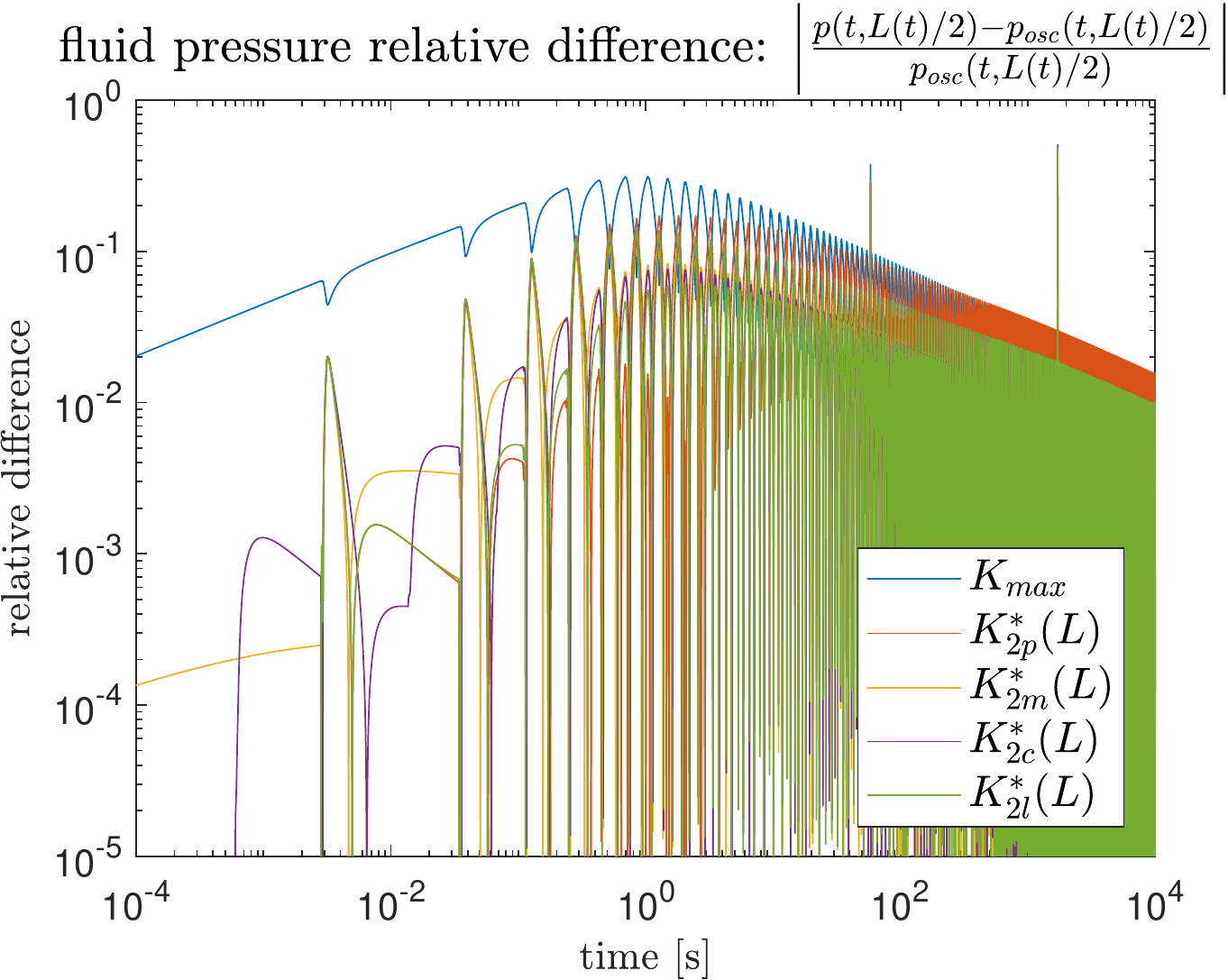}
\put(-225,155) {{\bf (f)}}
\caption{Relative difference in the fluid pressure at the midpoint $p(t,L(t)/2)$ for the toughness-transient distribution with unbalanced layering (Case 2, see Table.~\ref{Table:toughness}). Relative differences for {\bf (a)}, {\bf (b)} $h=0.25$ (see \eqref{defh}, {\bf (c)}, {\bf (d)} $h=0.1$, {\bf (e)}, {\bf (f)} $h=0.01$.}
 \label{Unbalanced_Rel_p0}
\end{figure}

We begin by examining the effect of varying degrees of unbalanced layering on the effectiveness of the different homogenisation strategies. The relative difference between the solution obtained for the periodic toughness distribution, taking $h=0.01$, $0.1$, $0.25$, and those obtained using the various homogenisation strategies (maximum toughness, temporal averaging), are provided in Figs.~\ref{Unbalanced_Rel_L} - \ref{Unbalanced_Rel_p0}.

From the figures, the effectiveness of the maximum toughness strategy does not appear to depend on the width of the layering, for either the sinusoidal or step-wise distributions, even in the case with a significant imbalance ($h=0.01$). This is in line with the previous observations about the fracture behaviour for the unbalanced case, discussed in Sect.~\ref{Sect:Behaviour}, and again we note that it may not hold when considering $h>0.5$ (in fact, it is likely that the maximum toughness strategy would become almost optimal as $h\to 1$).

The same appears to hold true for the temporal-averaging based approaches \eqref{MeasureK2r}-\eqref{MeasureK2p}, with the unbalanced layering having only a negligible impact on the effectiveness of the homogenisation strategy in the toughness-transient case.


\newpage


$\quad$
\newpage


$\quad$
\newpage

$\quad$
\newpage




\subsubsection{The case with a minimal width toughness layer}

\begin{figure}[t!]
 \centering
 \includegraphics[width=0.45\textwidth]{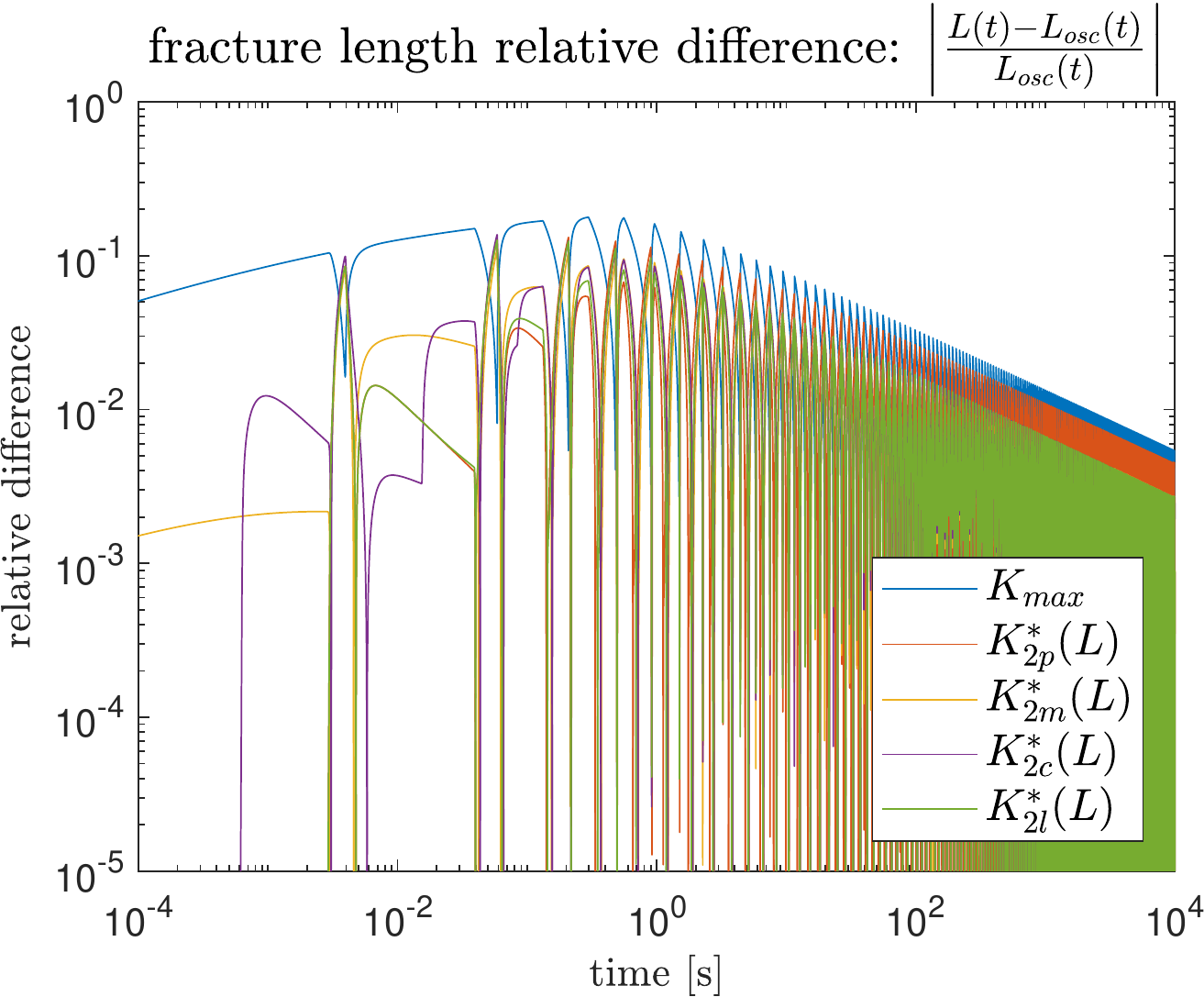}
\put(-225,155) {{\bf (a)}}
\hspace{12mm}
\includegraphics[width=0.45\textwidth]{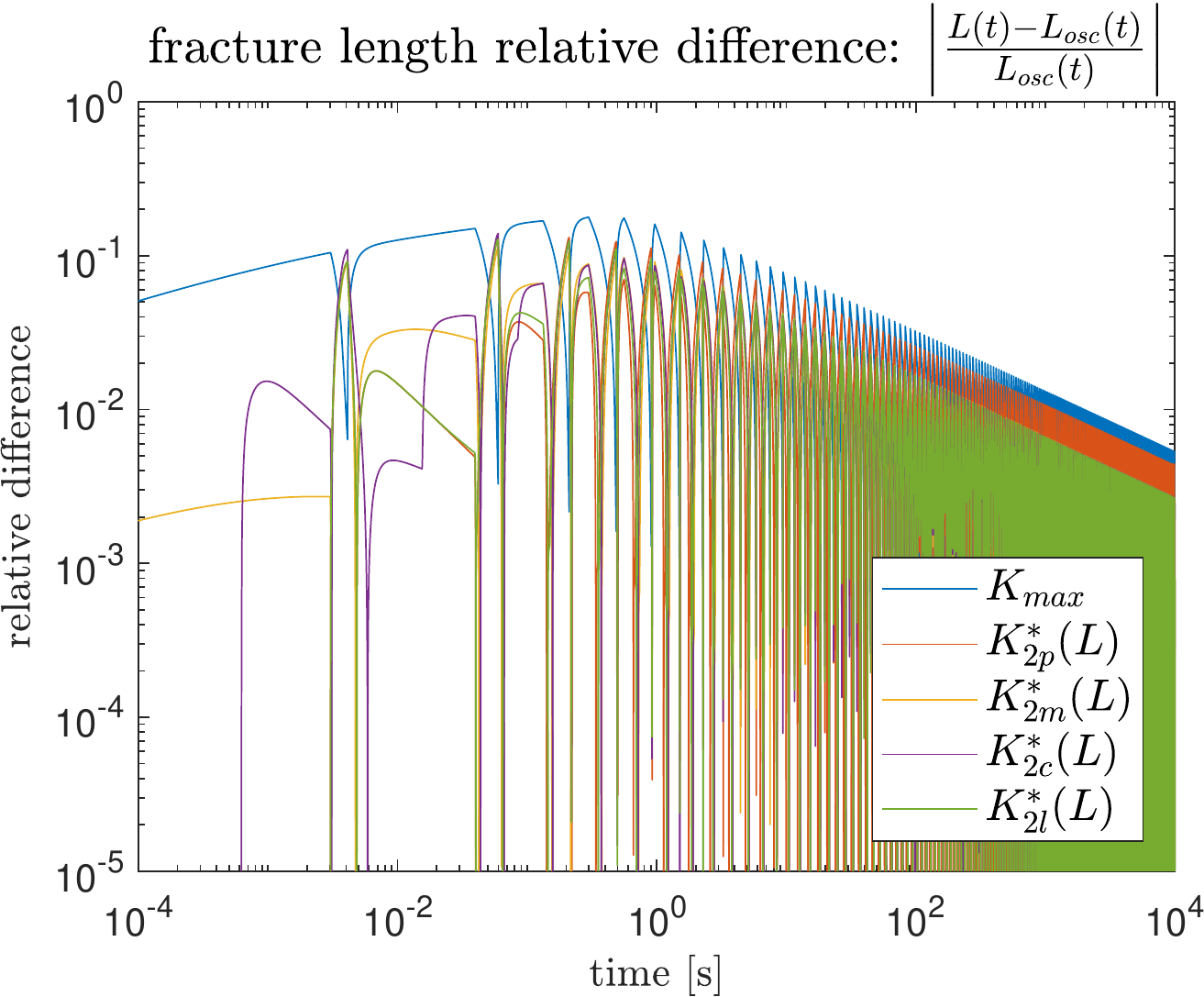}
\put(-225,155) {{\bf (b)}}
\\
\includegraphics[width=0.45\textwidth]{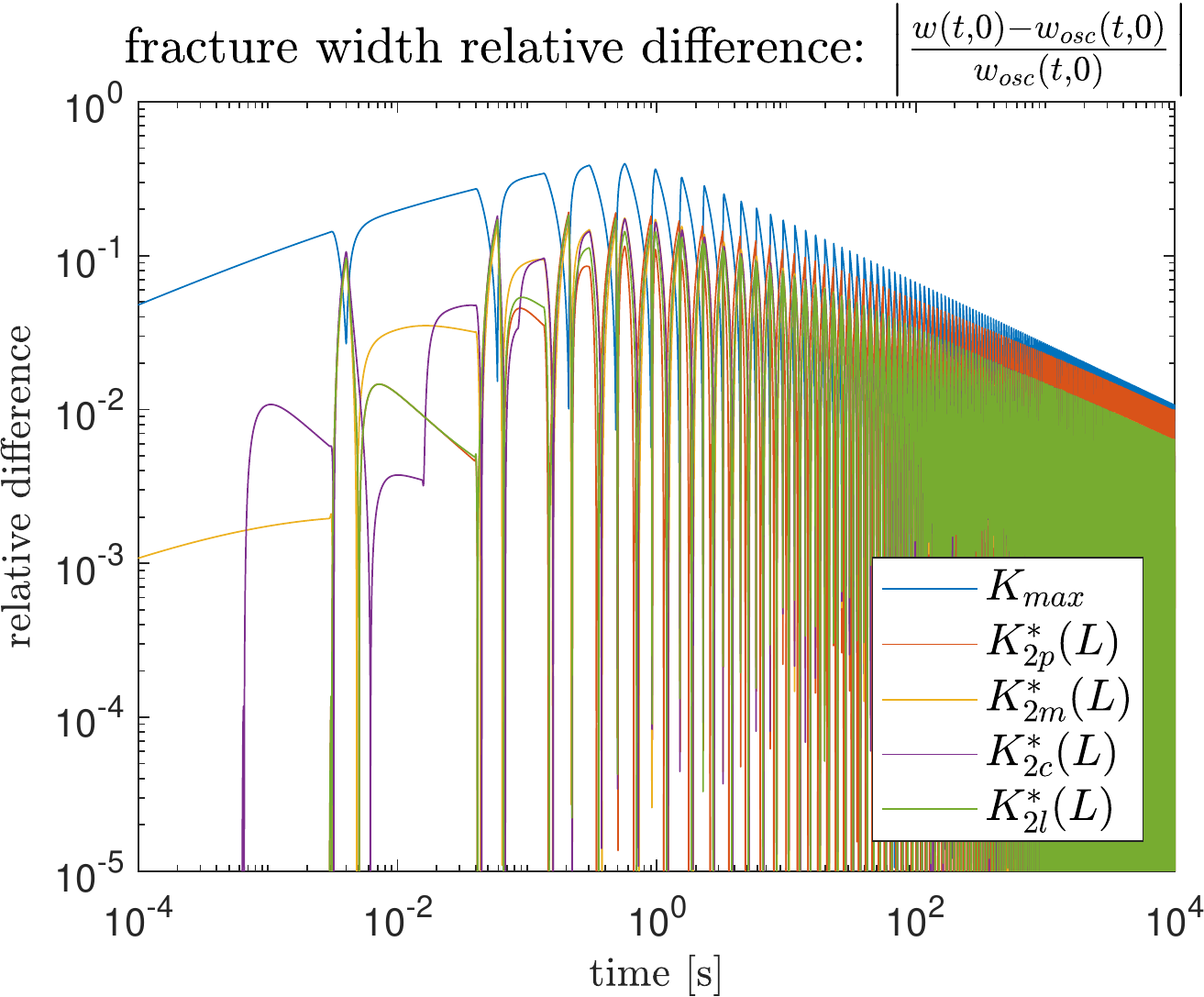}
\put(-225,155) {{\bf (c)}}
\hspace{12mm}
\includegraphics[width=0.45\textwidth]{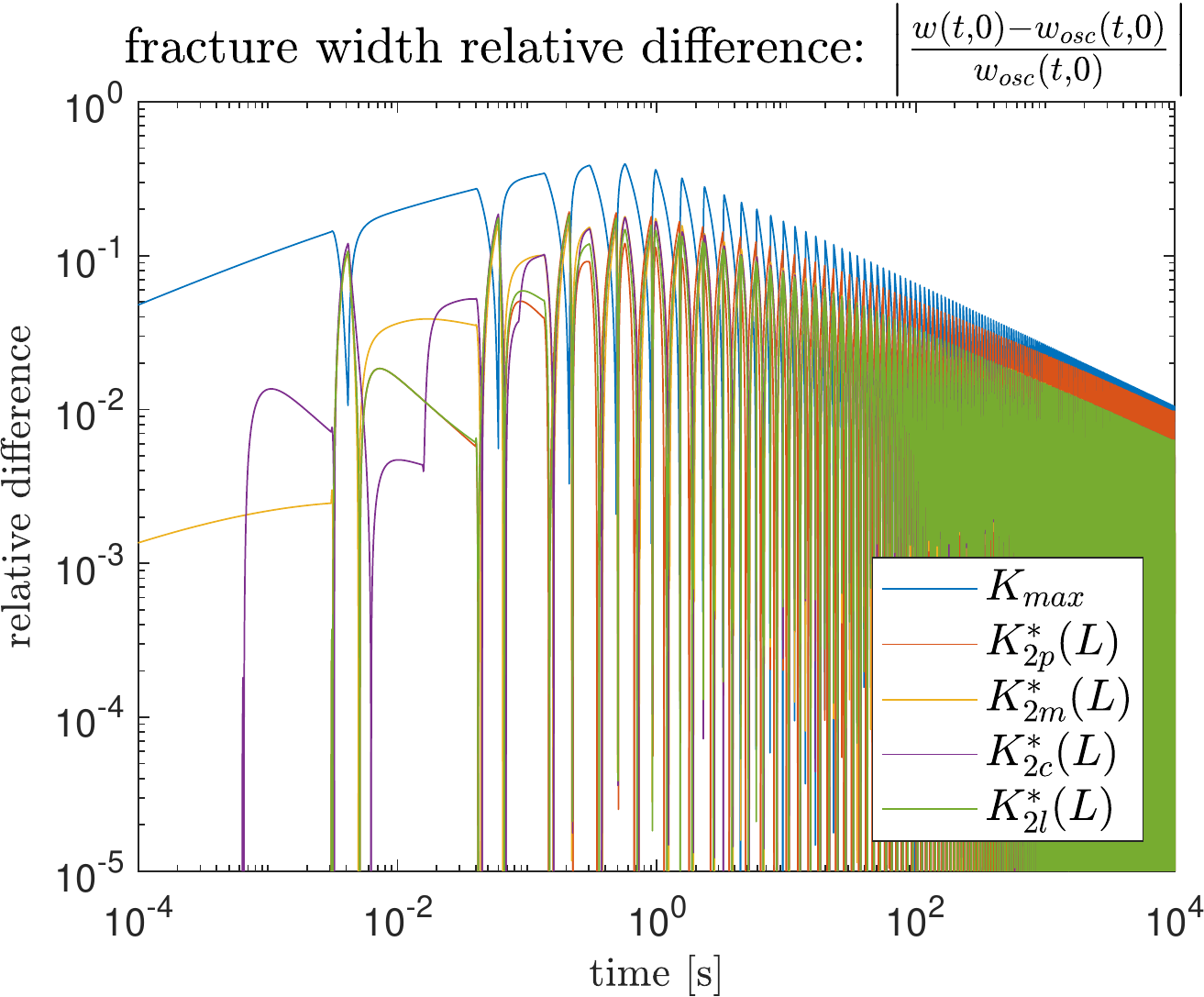}
\put(-225,155) {{\bf (d)}}
\\
\includegraphics[width=0.45\textwidth]{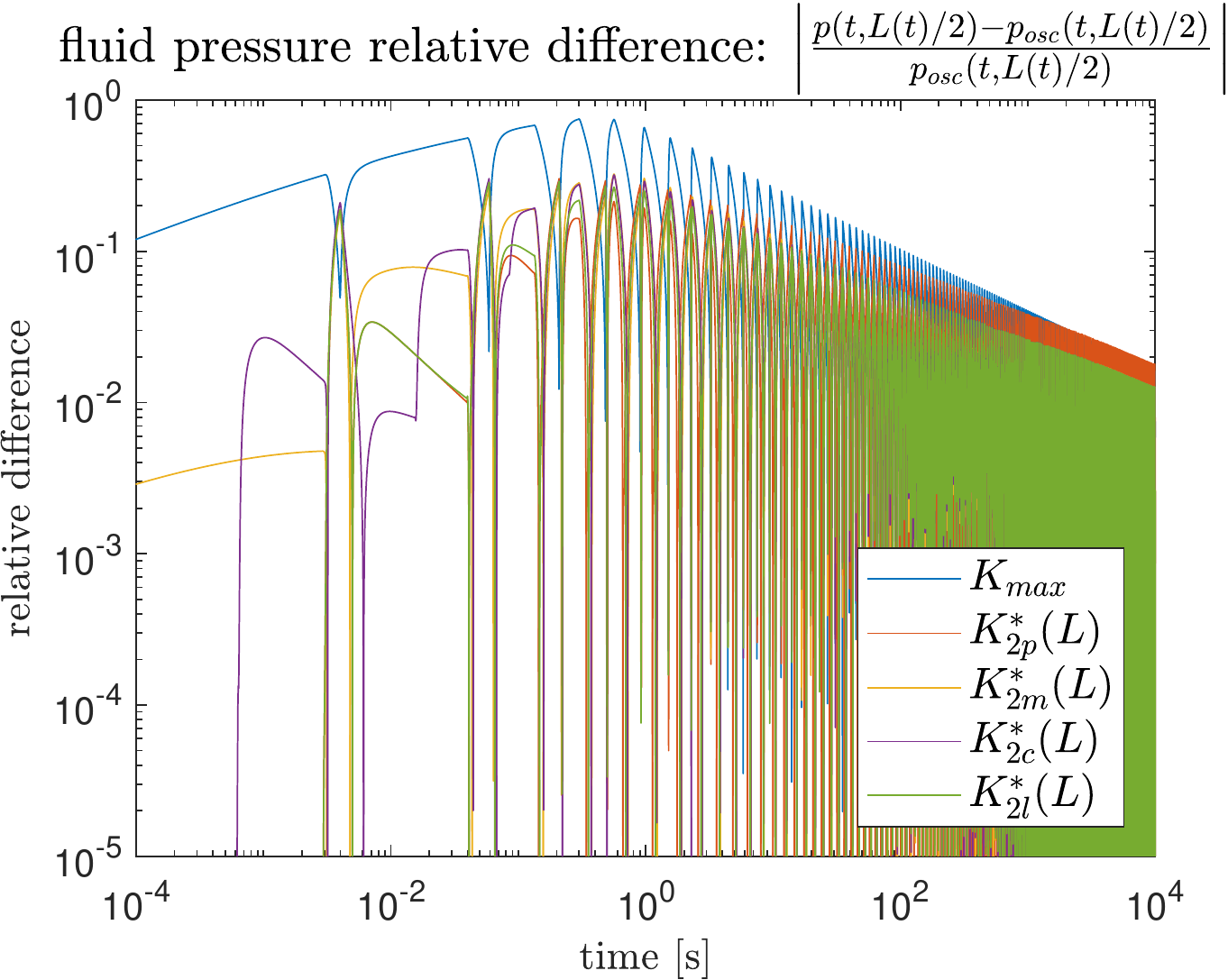}
\put(-225,155) {{\bf (e)}}
\hspace{12mm}
\includegraphics[width=0.45\textwidth]{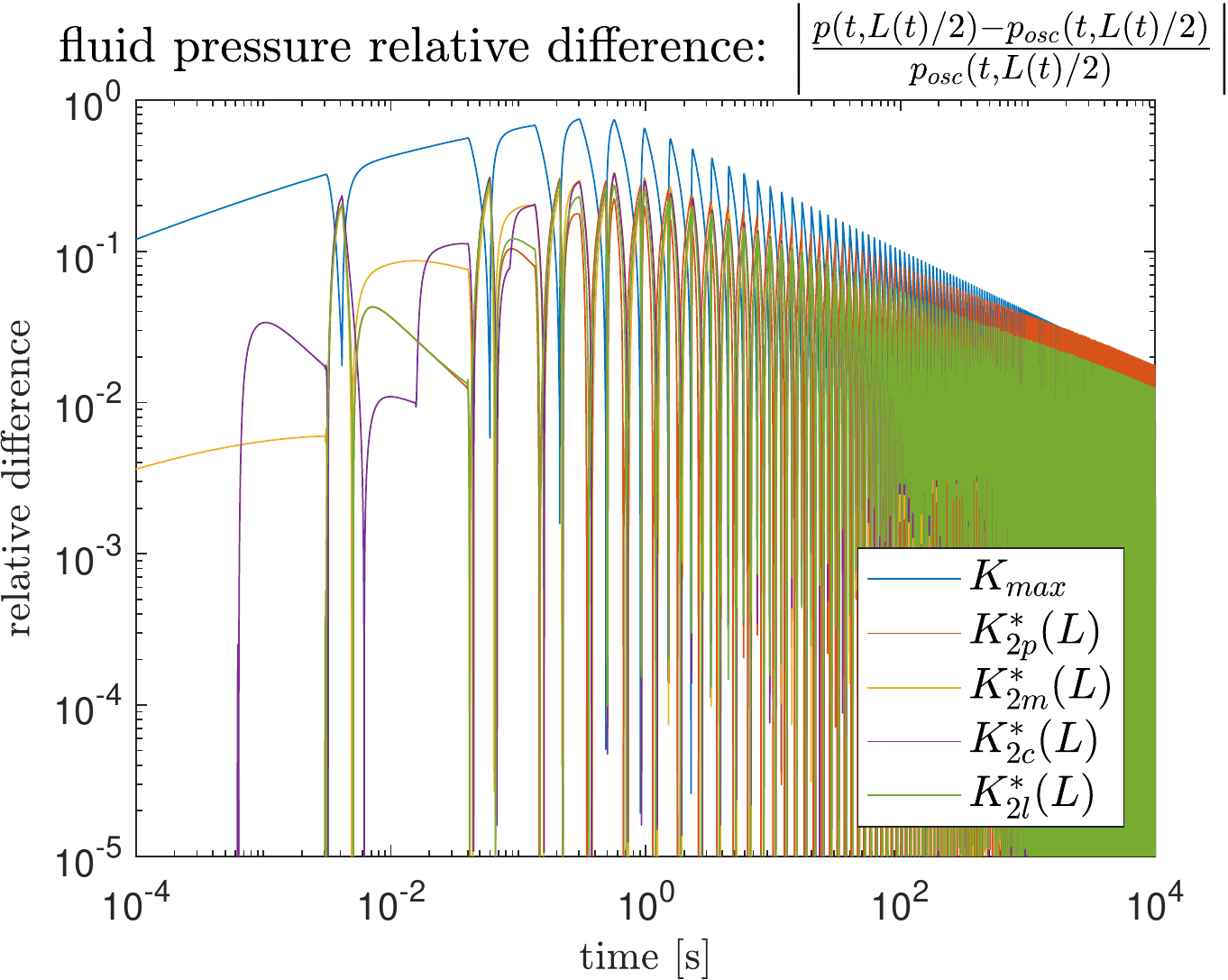}
\put(-225,155) {{\bf (f)}}
\caption{Results for the toughness-toughness distribution with unbalanced layering ($h=0.01$, Case 1, see Table.~\ref{Table:toughness}). Relative differences for the {\bf (a)}, {\bf (b)} fracture length $L(t)$, {\bf (c)}, {\bf (d)} fracture opening at the injection point $w(t,0)$, {\bf (e)}, {\bf (f)} fluid pressure at the mid-point $p(t, L(t)/2 )$.}
 \label{Uneven_Del100_10}
\end{figure}

\begin{figure}[t!]
 \centering
 \includegraphics[width=0.45\textwidth]{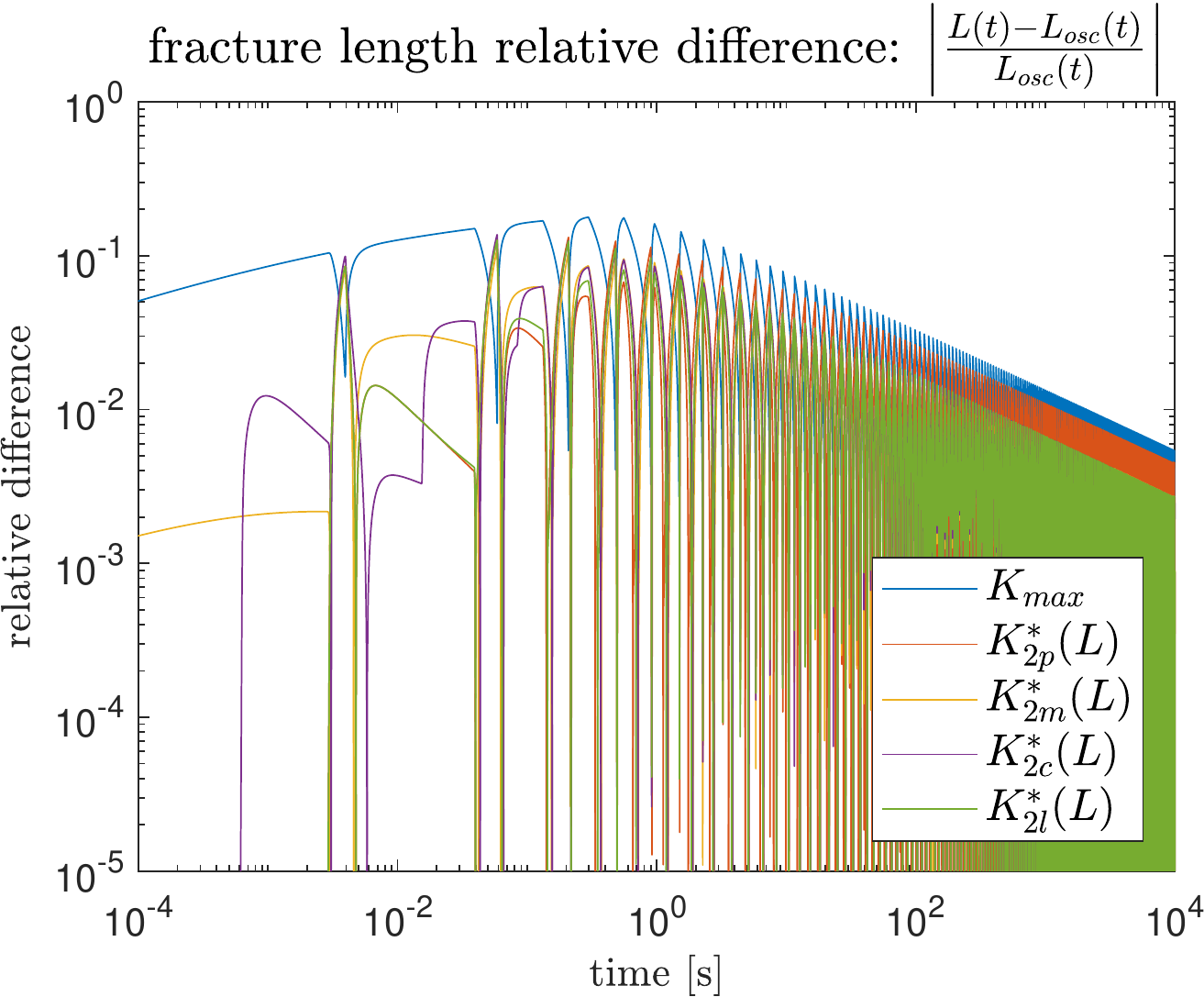}
\put(-225,155) {{\bf (a)}}
\hspace{12mm}
\includegraphics[width=0.45\textwidth]{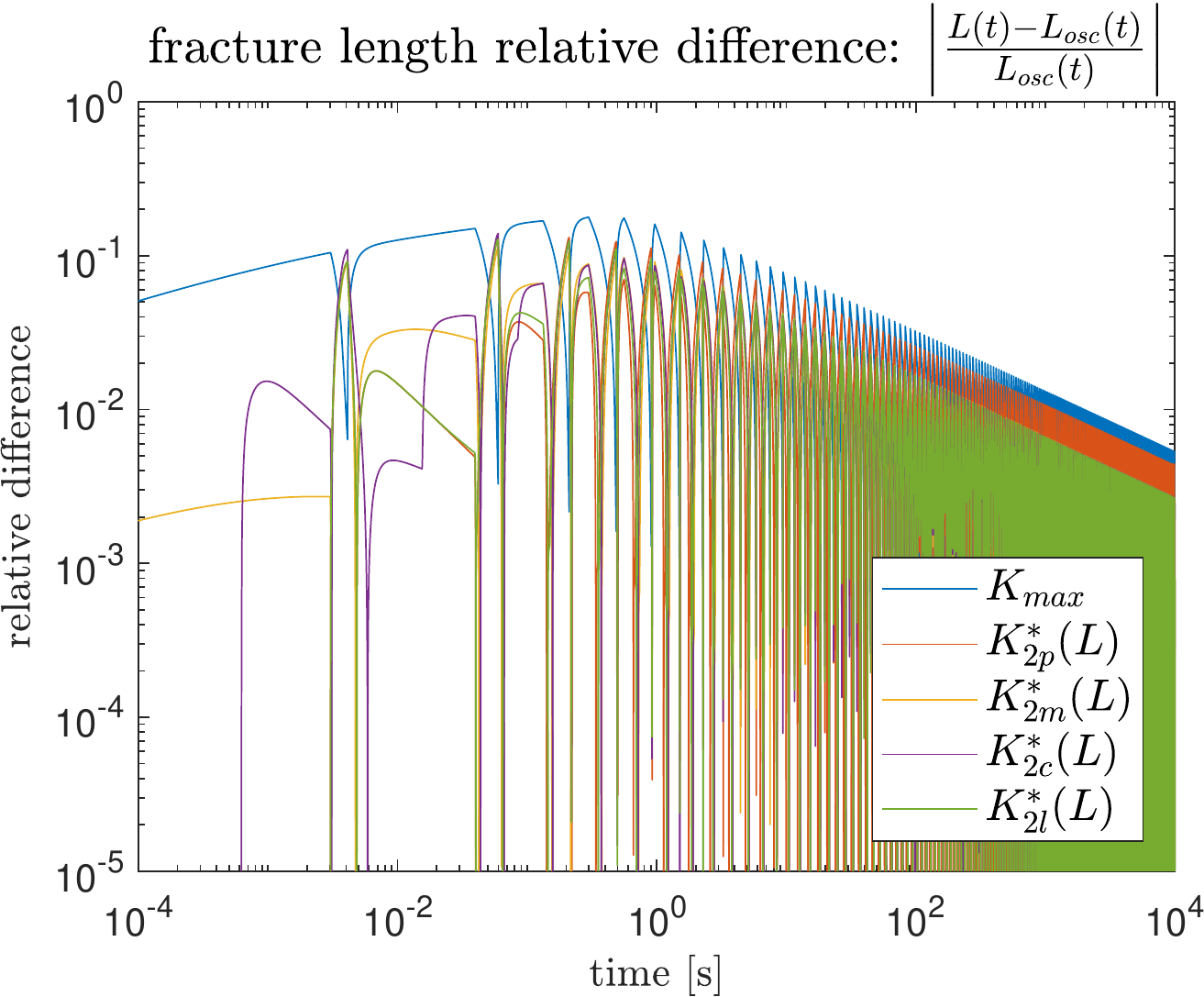}
\put(-225,155) {{\bf (b)}}
\\
\includegraphics[width=0.45\textwidth]{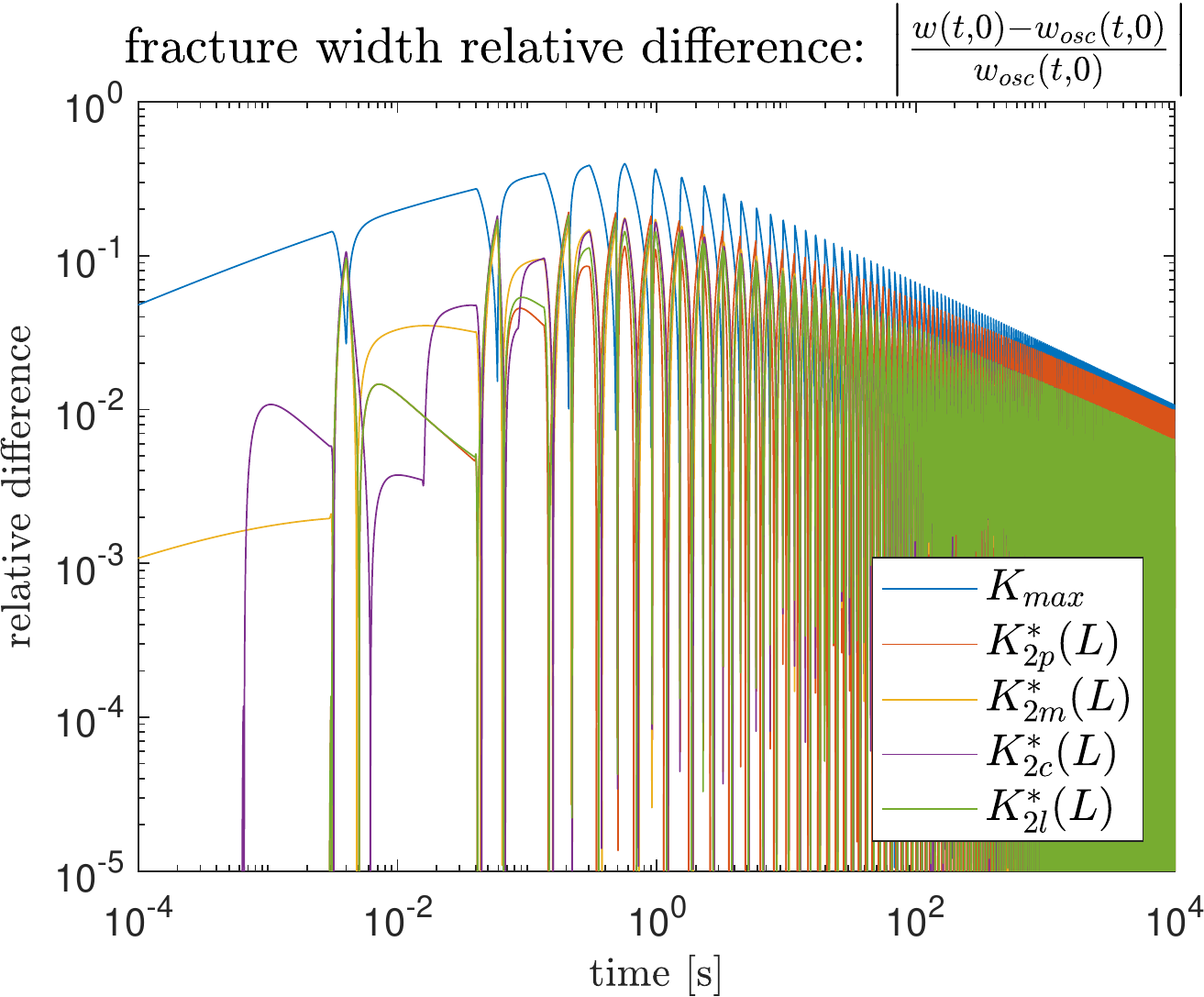}
\put(-225,155) {{\bf (c)}}
\hspace{12mm}
\includegraphics[width=0.45\textwidth]{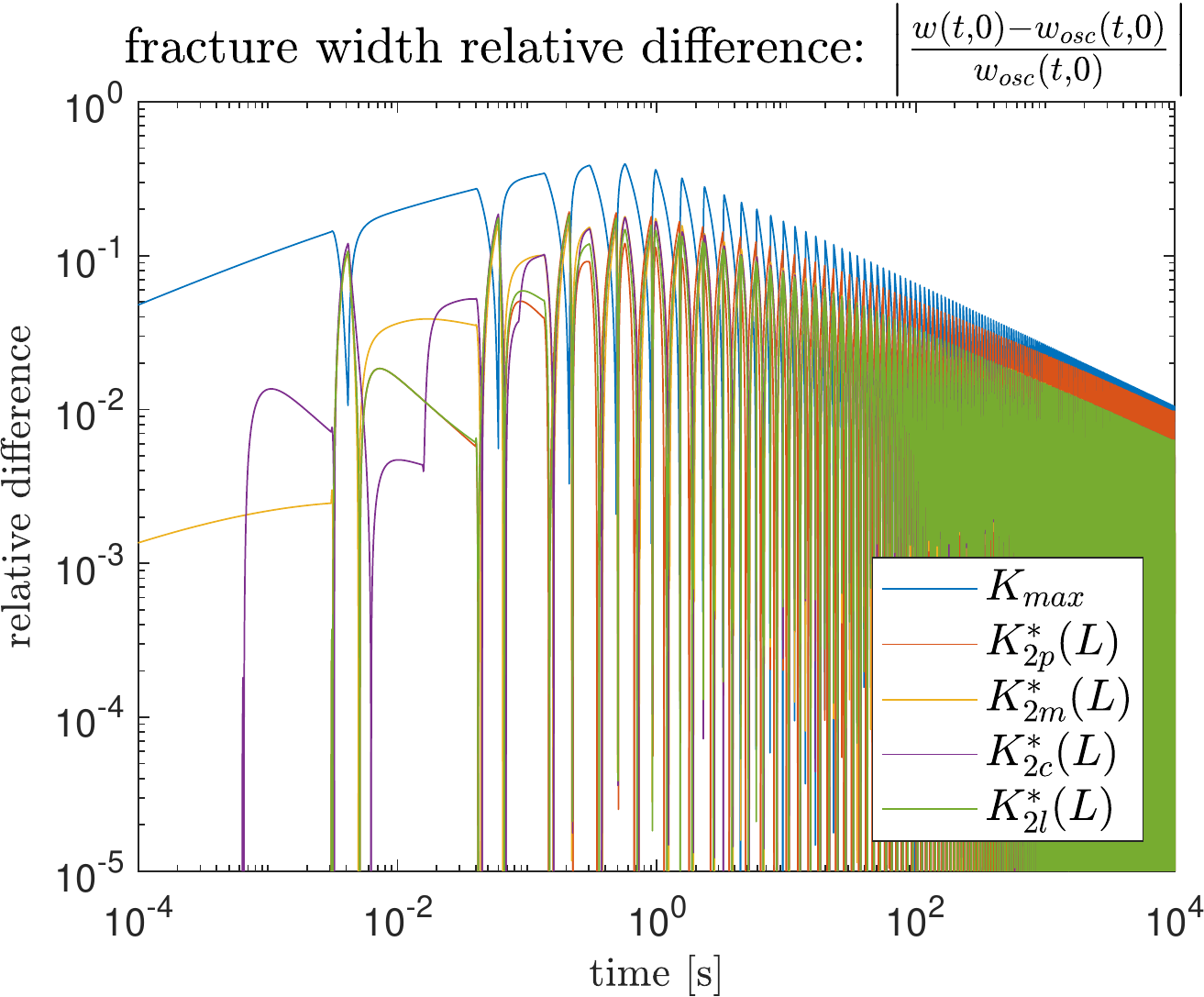}
\put(-225,155) {{\bf (d)}}
\\
\includegraphics[width=0.45\textwidth]{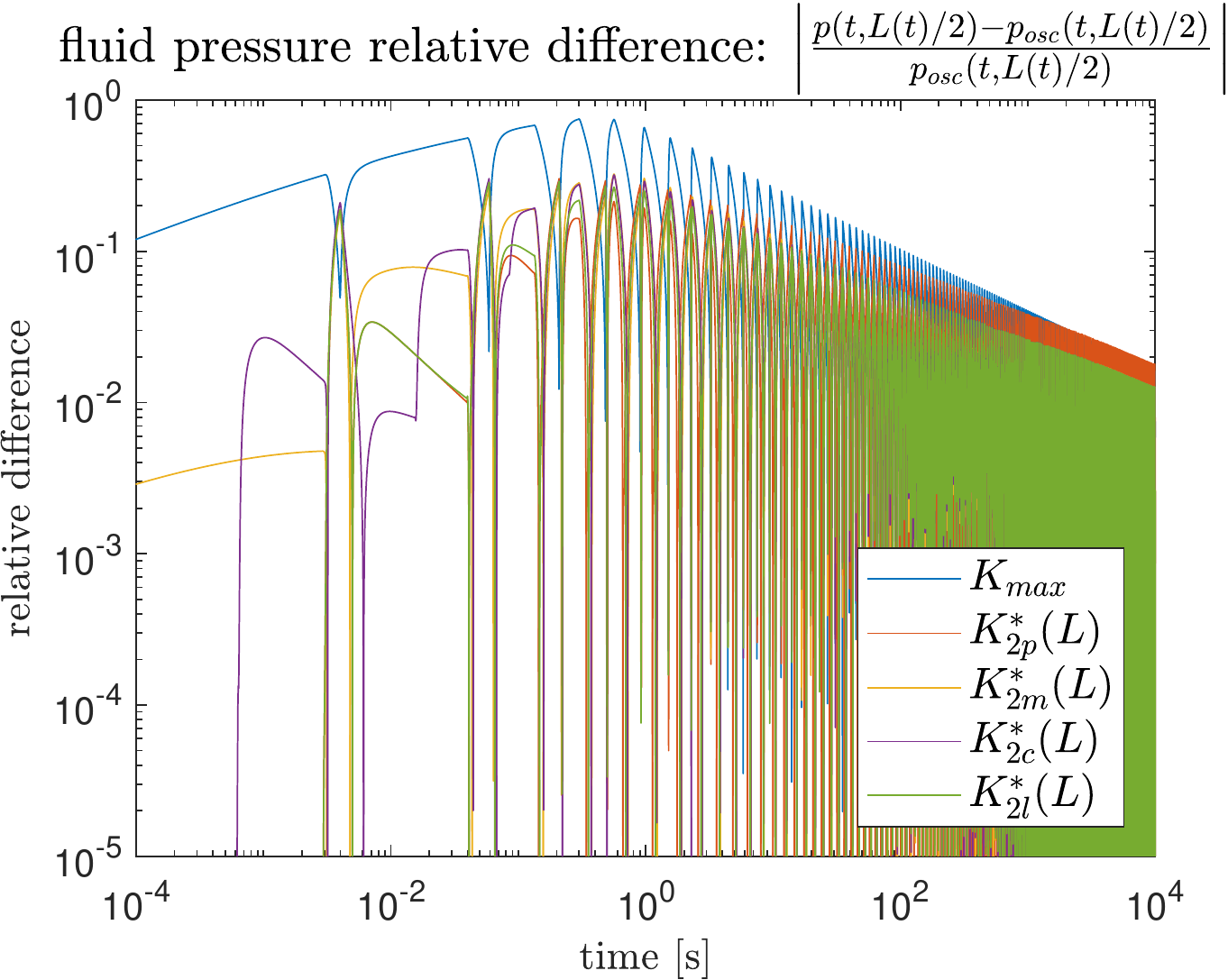}
\put(-225,155) {{\bf (e)}}
\hspace{12mm}
\includegraphics[width=0.45\textwidth]{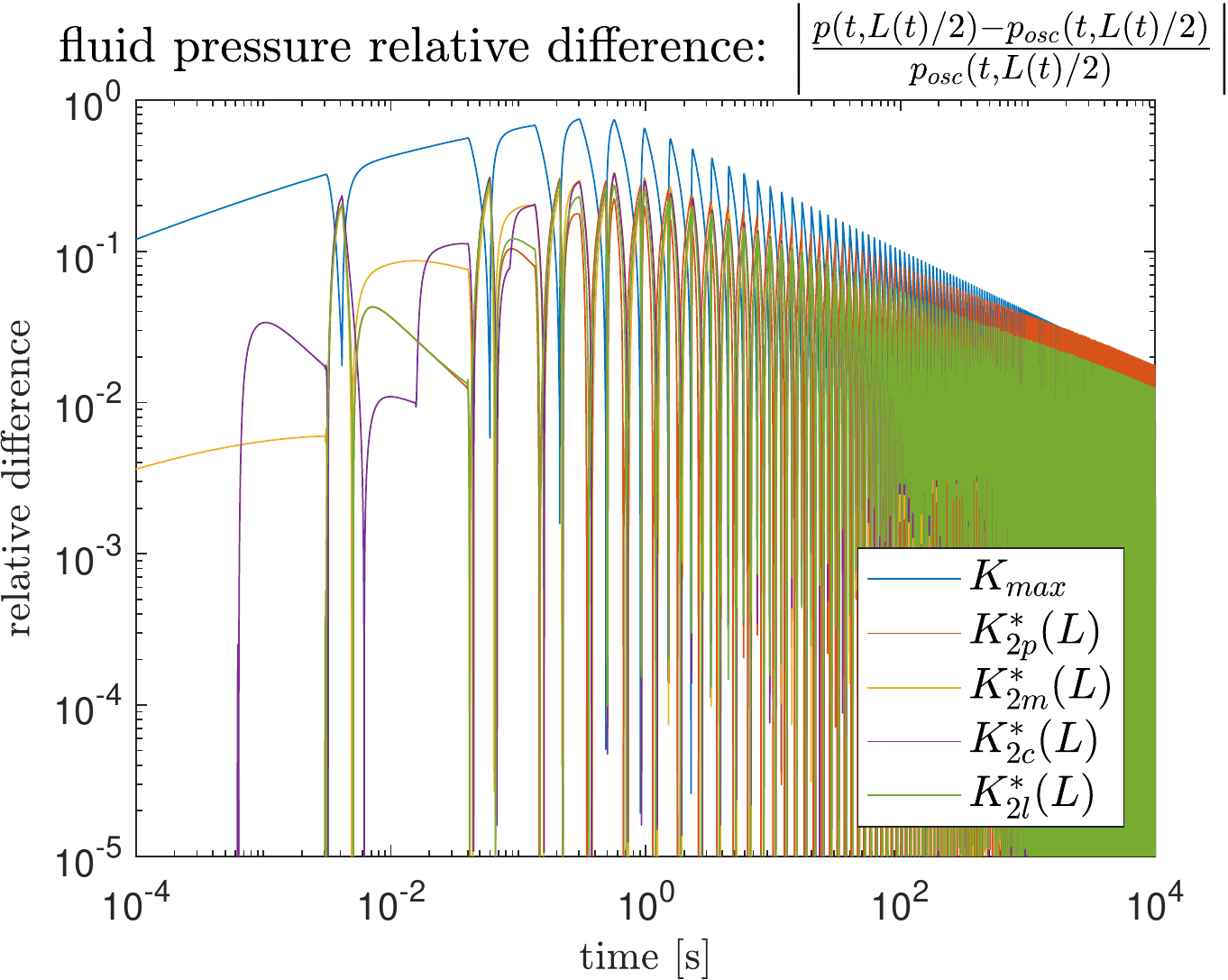}
\put(-225,155) {{\bf (f)}}
\caption{Results for the toughness-transient distribution with unbalanced layering ($h=0.01$, Case 2, see Table.~\ref{Table:toughness}). Relative differences for the {\bf (a)}, {\bf (b)} fracture length $L(t)$, {\bf (c)}, {\bf (d)} fracture opening at the injection point $w(t,0)$, {\bf (e)}, {\bf (f)} fluid pressure at the mid-point $p(t, L(t)/2 )$.}
 \label{Uneven_Del10_1}
\end{figure}

\begin{figure}[t!]
 \centering
 \includegraphics[width=0.45\textwidth]{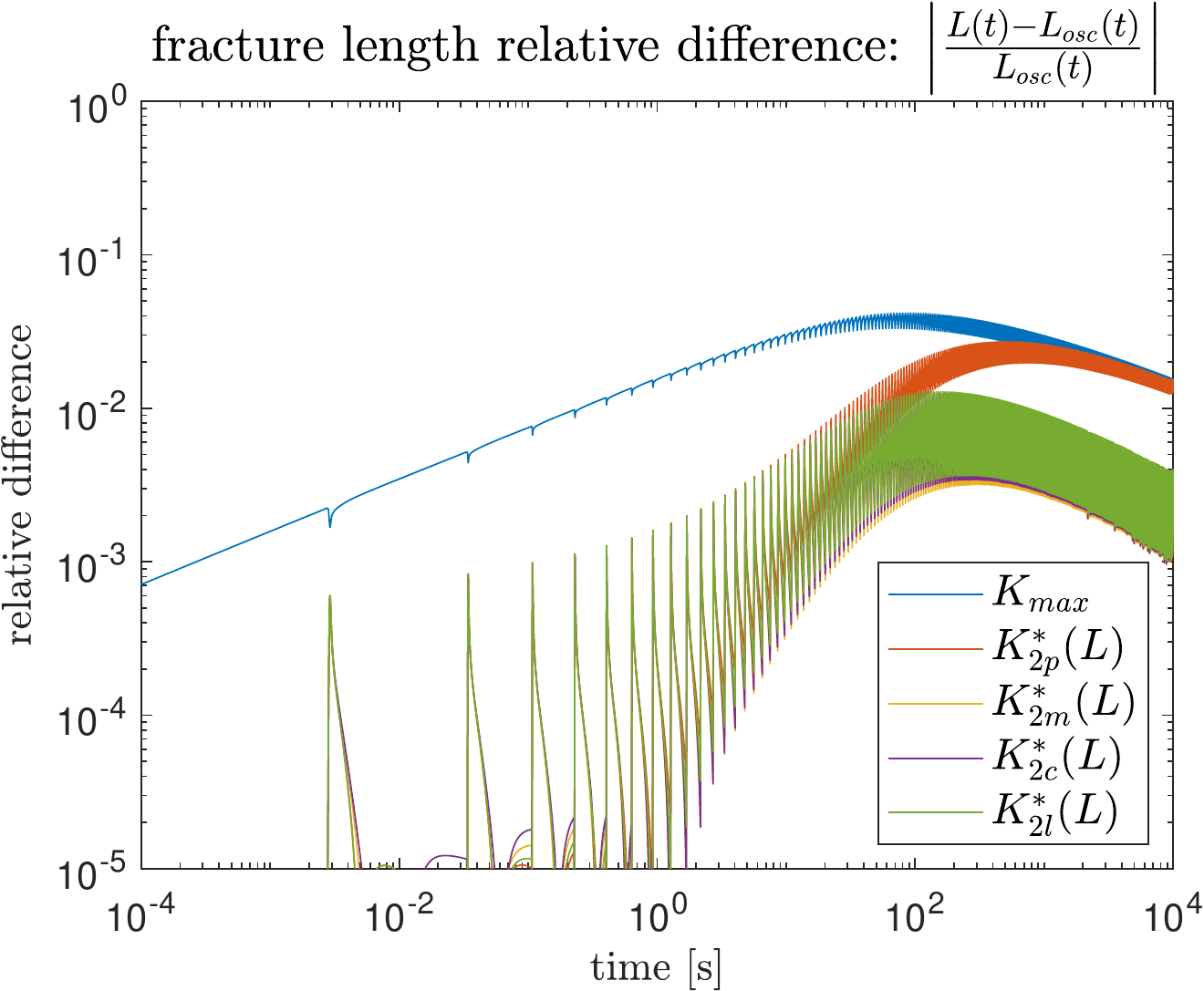}
\put(-225,155) {{\bf (a)}}
\hspace{12mm}
\includegraphics[width=0.45\textwidth]{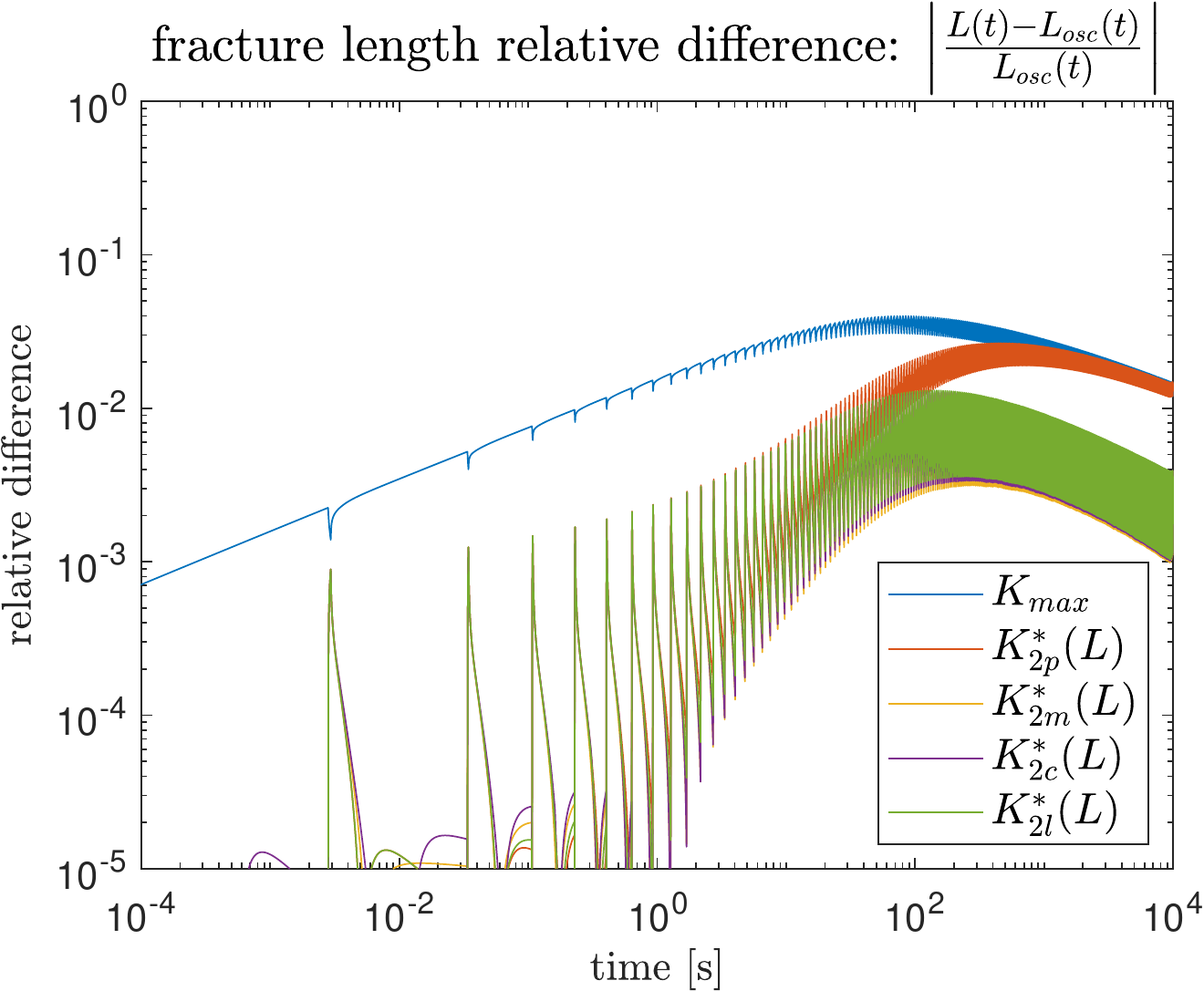}
\put(-225,155) {{\bf (b)}}
\\
\includegraphics[width=0.45\textwidth]{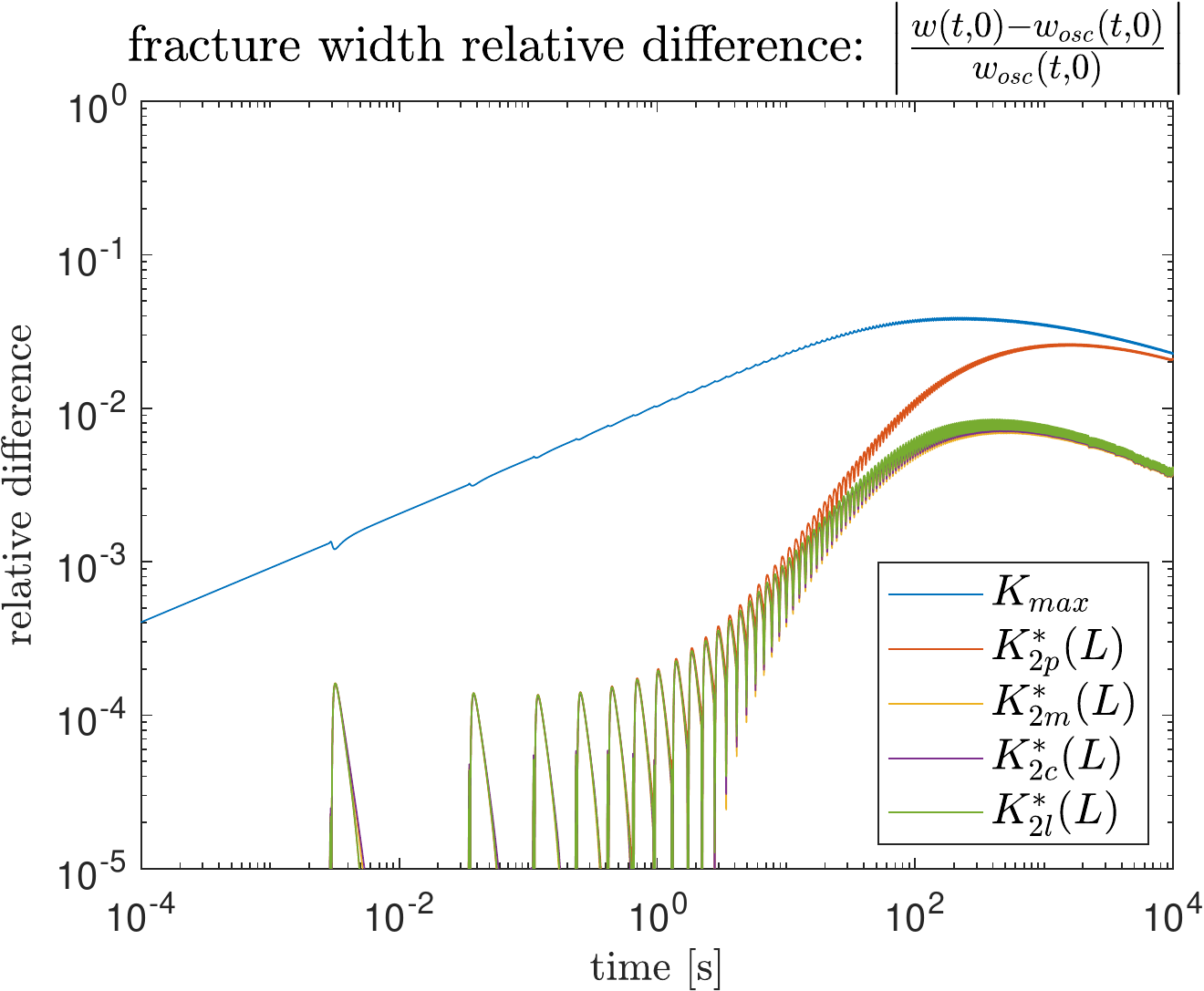}
\put(-225,155) {{\bf (c)}}
\hspace{12mm}
\includegraphics[width=0.45\textwidth]{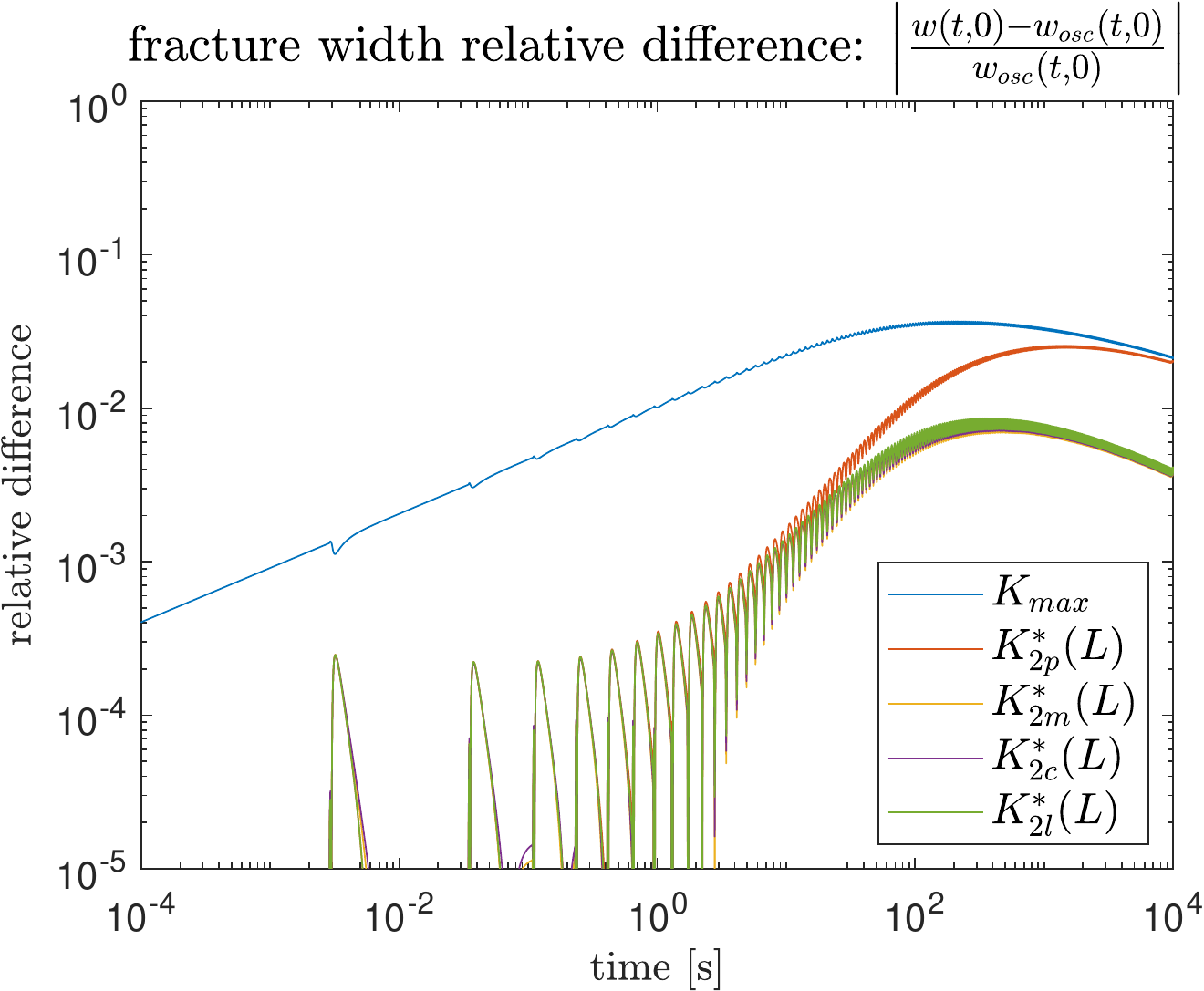}
\put(-225,155) {{\bf (d)}}
\\
\includegraphics[width=0.45\textwidth]{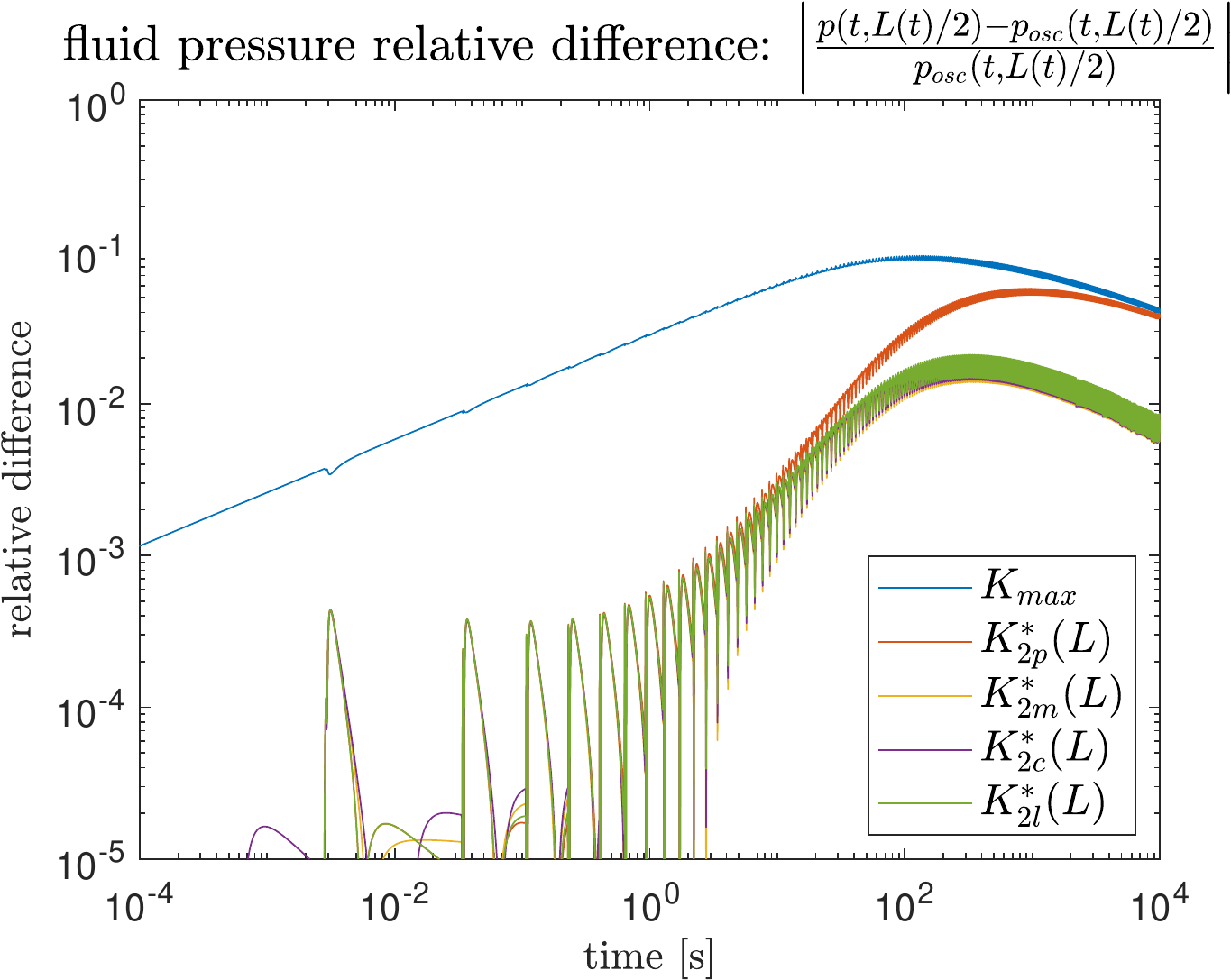}
\put(-225,155) {{\bf (e)}}
\hspace{12mm}
\includegraphics[width=0.45\textwidth]{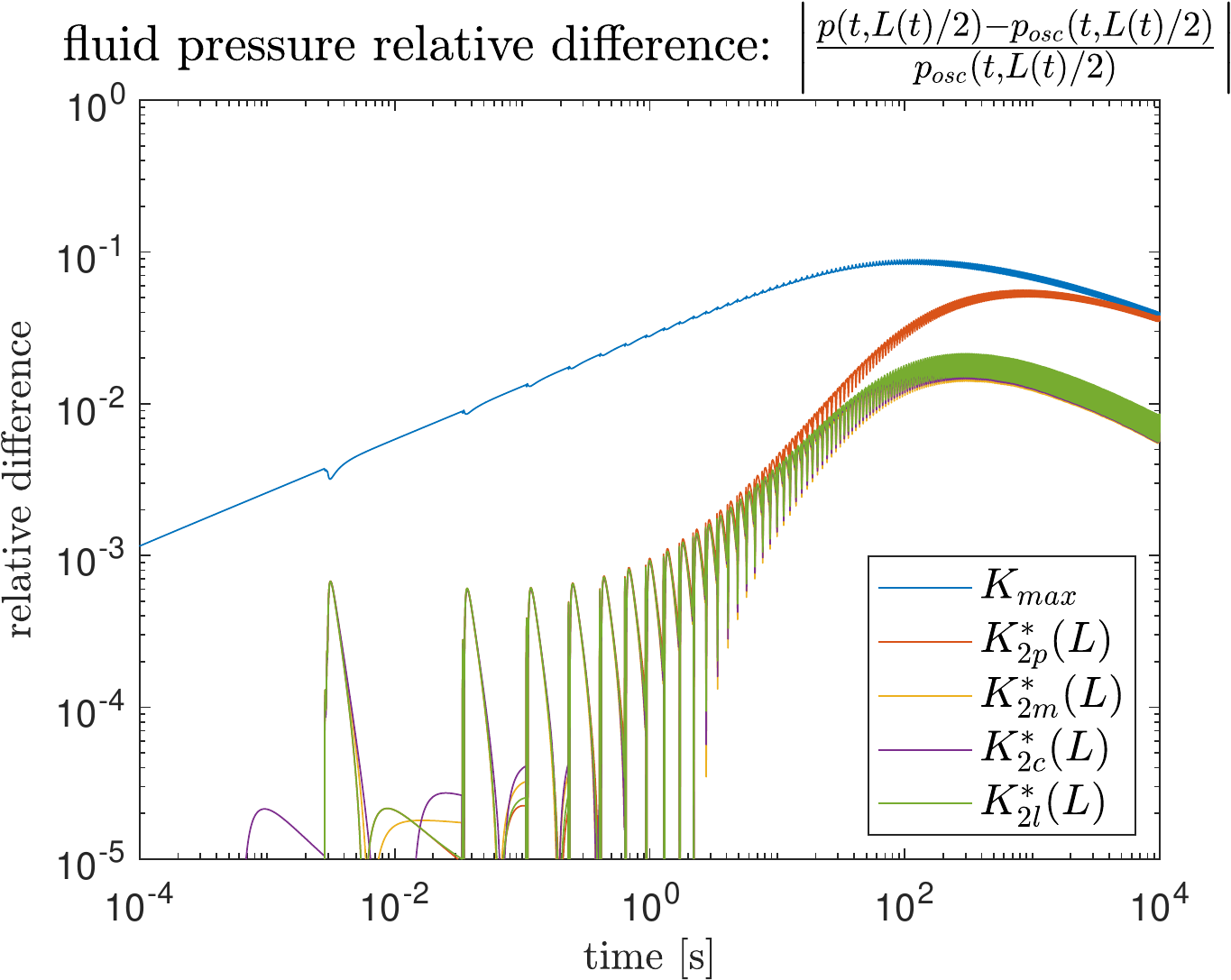}
\put(-225,155) {{\bf (f)}}
\caption{Results for the transient-viscosity distribution with unbalanced layering ($h=0.01$, Case 3, see Table.~\ref{Table:toughness}). Relative differences for the {\bf (a)}, {\bf (b)} fracture length $L(t)$, {\bf (c)}, {\bf (d)} fracture opening at the injection point $w(t,0)$, {\bf (e)}, {\bf (f)} fluid pressure at the mid-point $p(t, L(t)/2 )$.}
 \label{Uneven_Del1_01}
\end{figure}

Next, we examine the case where the average toughness of the heterogeneous material is almost identical to that of the minimum toughness, i.e.\ when the maximum toughness layer is so thin as to seem to be negligible. We do this by taking the fixed value of $h=0.01$ (see \eqref{defh}), and again investigating the relative difference between the process parameters for the periodic toughness simulation, and that estimated by the various homogenisation strategies. The resulting relative differences for the toughness-toughness, toughness-transient and transient-viscosity regimes are provided in Figs.~\ref{Uneven_Del100_10} - \ref{Uneven_Del1_01}.

It is clear from a comparison of Fig.~\ref{Even_Del100_10} - \ref{Even_Del10_1} with Fig.~\ref{Uneven_Del100_10} - \ref{Uneven_Del10_1} that the effect of unbalanced layering, even in the extreme example considered here, is negligible in both the toughness-toughness and (initially) toughness-transient regimes. This is inline with the results of the previous subsection, and the observations about the process behaviour given in Sect.~\ref{Sect:Behaviour}.

In the transient-viscosity case however, the difference is far more significant, as can be seen when comparing Fig.~\ref{Uneven_Del1_01} with Fig.~\ref{Even_Del1_01}. Alongside the expected `smoothing' of the relative error (due to reduced oscillation of the system parameters), the error of the maximum toughness strategy does increase slightly, for instance the relative error for the pressure exceeds $10$\% for the unbalanced layering, but not for the balanced layering (compare Fig.~\ref{Even_Del1_01}e,f with Fig.~\ref{Uneven_Del1_01}e,f). 

A similar impact can be seen for the temporal-averaging measures, however the relative error for all parameters never exceeds $3$\% (where previously it was $2$\%). However, there is also a significant improvement in the effectiveness of measures \eqref{MeasureK2r}-\eqref{MeasureK2p} during the initial stages of the fracture, with the relative error for the aperture and pressure after the first period remaining below $0.1$\% for the first second, where previously it was below $1$\%, while the error of the length only slightly exceeds $0.1$\% during this time.

Combining the results of the previous two subsections, we can conclude that the temporal-averaging approach remains effective irrespective of the unbalanced layering. The maximum toughness strategy is typically only negligibly effected, except in the transient-viscosity dominated regime where there is a slight decrease in effectiveness.


$\quad$
\newpage	


$\quad$
\newpage	


$\quad$
\newpage

\section{Conclusions}

A variety of homogenisation strategies for the material toughness have been considered, in the context of hydraulic fracture (HF) of rock with a periodic toughness distribution. The strategies considered took two different forms. The first was the maximum toughness strategy proposed in \cite{DONTSOV2021108144}, while the second utilized the concept of temporal-averaging previously proposed by the authors \cite{Gaspare2022}, and provided here in \eqref{MeasureK2r}-\eqref{MeasureK2p}. A wide range of periodic toughness distributions were considered, including differing fracture regimes, unbalanced material distributions, and taking both step-wise and sinusoidal forms.

The investigation was able to demonstrate the following:
\begin{itemize}
 \item The temporal-averaging approach to homogenising the fracture toughness was consistently more effective than the maximum toughness strategy for the radial model of hydraulic fracture (as has previously been demonstrated for plane strain HF \cite{Gaspare2022}). The largest difference between the two was seen for a fracture starting in the transient-viscosity dominated regime, while that for the toughness-toughness distribution was far smaller (see Figs.~\ref{Even_Del100_10} - \ref{Even_Del1_01}).
 \item The effectiveness of both measures improved with time (after the first 2-3 toughness periods) for the toughness-toughness and toughness-transient distributions, but not for the transient-viscosity distribution. Here, the transition from the viscosity to transient distribution caused an increase in the relative error of key process parameters for the maximum toughness strategy, while the temporal-averaging approach remained more consistent.
 \item The results above also hold for the case with unbalanced layering, where the maximum toughness layer only makes up a small portion of the material. The likely physical explanation for this was outlined in Sect.~\ref{Sect:Behaviour}. It was demonstrated that even in the case of an extreme imbalance ($h=0.01$, see \eqref{defh}), the only effect was a small increase in the relative error of the approximation, with the maximum toughness strategy more adversely affected than the temporal-averaging approach. It is however unlikely that these result would hold if the maximum toughness material instead made up the majority of the heteogeneous body ($h>0.5$).
 \item Interestingly, in the case with an extremely unbalanced layering of the rock strata, it was still the transition between viscosity and toughness dominated regimes which played the crucial role in determining the effectiveness of the homogenisation method. As such, the only significant effect of unbalanced layering on the homogenisation methods was through it's influence on the fracture evolution (for cracks with one layer starting in the viscosity dominated regime).
\end{itemize}
It should be restated that the temporal averaging homogenisation strategy utilized in this work is not dependent upon the periodic distributions examined here, and should continue to be effective for disordered toughness distributions as well.

One factor of the presented analysis that should be highlighted is that the principles behind the toughness homogenisation strategy outlined here are not particular to the hydraulic fracture process. This means that they should continue to provide an effective homogenisation strategy for any steady state propagation process within a heterogeneous media.

While these results indicate the effectiveness of the temporal averaging homogenisation strategy for a system only experiencing toughness heterogeneity, there are a number of crucial effects which should still be considered for HF. Most notably, materials with toughness heterogeneity are also likely to experience heterogeneity of the elastic parameters (a significant challenge, see e.g. \cite{HOSSAIN201415}), the fluid leak-off, and potential differences in the in-situ stress (stress barrier, see e.g. \cite{DONTSOV2022110841}). Additionally, the step-wise fracture advancement present even in homogeneous media (see e.g. \cite{CAO201724}) could have an even larger impact when coupled with the effect of toughness heterogeneity. The fact that the utilized measures \eqref{MeasureK2r}-\eqref{MeasureK2p} incorporate the instantaneous fracture velocity, $v(L)$, is however promising in this regard.

The dependence on the instantaneous velocity does however lead to the most significant open question regarding the proposed strategy. For this work, the results of existing numerical simulations were used to provide the instantaneous velocity used to compute the average toughness. In HF however, such data is not typically available in real-time. Developing an effective approach to computing (or approximating) measures \eqref{MeasureK2r}-\eqref{MeasureK2p} in the general case remains an open problem.


\section*{Funding}
The authors have been funded by the European Union's Horizon 2020 research and innovation programme under the Marie Sklodowska-Curie grant agreement EffectFact No 101008140 and the Welsh Government via S$\hat{\mbox {e}}$r Cymru Future Generations Industrial Fellowship grant AU224.

\section*{Acknowledgements}
MD acknowledges the Royal Academy of Engineering for the Industrial Fellowship. The authors would like to thank Prof. Gennady Mishuris for his continuous interest and helpful discussions during the course of this work.


\bibliography{Penny_Bib}
\bibliographystyle{apalike}

\end{document}